\documentclass[11pt]{article}
\usepackage{fancyvrb,authblk}
\usepackage[utf8]{inputenc}
\usepackage[english]{babel}
\usepackage{multicol}
\usepackage{multirow}
\usepackage{setspace}
\usepackage{natbib}
\usepackage[compact]{titlesec}
\usepackage{makecell}
\usepackage[table]{xcolor}
\usepackage{booktabs} 

\titlespacing*{\subsection}{0pt}{10pt}{0pt}
\titlespacing*{\subsubsection}{0pt}{10pt}{0pt}
\usepackage{epsfig,amssymb,amsmath, euscript,color, mathrsfs, graphicx}
\usepackage{hyperref}
\renewcommand{\baselinestretch}{1.3}
\makeatletter

 \@addtoreset{equation}{section}
 
\newtheorem{theorem}{Theorem}
\newtheorem{lemma}{Lemma} 
\newtheorem{proposition}{Proposition}
\newtheorem{cd}{Condition}

\renewcommand{\hat}{\widehat}
\def\singlespace{\def\baselinestretch{1}\@normalsize}

\newcommand{\cov}{{\rm Cov}}

\newcommand{\bA}{{\mathbf A}}
\newcommand{\bB}{{\mathbf B}}
\newcommand{\bF}{{\mathbf F}}
\newcommand{\bE}{{\mathbf E}}
\newcommand{\bG}{{\mathbf G}}
\newcommand{\bH}{{\mathbf H}}
\newcommand{\bI}{{\mathbf I}}
\newcommand{\bK}{{\mathbf K}}
\newcommand{\bJ}{{\mathbf J}}

\newcommand{\bP}{{\mathbf P}}
\newcommand{\bR}{{\mathbf R}}
\newcommand{\bS}{{\mathbf S}}
\newcommand{\bT}{{\mathbf T}}
\newcommand{\bU}{{\mathbf U}}
\newcommand{\bV}{{\mathbf V}}
\newcommand{\bW}{{\mathbf W}}
\newcommand{\bX}{{\mathbf X}}
\newcommand{\bY}{{\mathbf Y}}
\newcommand{\bZ}{{\mathbf Z}}
\newcommand{\ba}{{\mathbf a}}
\newcommand{\bb}{{\mathbf b}}

\newcommand{\bc}{{\mathbf c}}
\newcommand{\be}{{\mathbf e}}

\newcommand{\bg}{{\mathbf g}}

\newcommand{\bk}{{\mathbf k}}

\newcommand{\bu}{{\mathbf u}}
\newcommand{\bv}{{\mathbf v}}
\newcommand{\bw}{{\mathbf w}}

\newcommand{\by}{{\mathbf y}}
\newcommand{\bz}{{\mathbf z}}
\newcommand{\balpha} {\boldsymbol{\alpha}}
\newcommand{\bbeta}  {\boldsymbol{\beta}}

\newcommand{\bOmega}{\boldsymbol{\Omega}}

\newcommand{\bSigma}{\boldsymbol{\Sigma}}

\newcommand{\bgamma}{\boldsymbol{\gamma}}

\newcommand{\bTheta} {\boldsymbol{\Theta}}

\newcommand{\bPsi} {\boldsymbol{\Psi}}

\newcommand{\btheta} {\boldsymbol{\theta}}

\newcommand{\bzeta} {\boldsymbol{\zeta}}

\newcommand{\bC}{{\mathbf C}}
\newcommand{\bD}{{\mathbf D}}

\def\6bullets{\bullet\bullet\bullet\bullet\bullet\bullet}

\begin{document}

\title{Spatio-Temporal Autoregressions for High Dimensional Matrix-Valued Time Series}

\author[1]{\small Baojun Dou}
\author[2]{\small Jing He}
\author[3]{\small Sudhir Tiwari}
\author[4]{\small Qiwei Yao$^*$}
\affil[1]{\it \small Department of Decision Analytics and Operations, City University of Hong Kong, Hong Kong SAR}
\affil[2]{\it \small Joint Laboratory of Data Science and Business Intelligence, Southwestern University of Finance and Economics, Chengdu, China}
\affil[3]{\it \small Commodities and Global Markets Division, Macquarie, Hong Kong SAR}
\affil[4]{\it \small Department of Statistics, London School of Economics and Political Science, London, WC2A 2AE, U.K.}

\footnotetext[1]{Correspondence to: Department of Statistics, London School of Economics, Houghton Street, London, WC2A 2AE, United Kingdom. E-mail address: q.yao@lse.ac.uk (Q. Yao).}

%\date{}
\maketitle

\begin{abstract}
Motivated by predicting intraday trading volume curves, we consider two 
spatio-temporal autoregressive models for matrix time series, in which 
each column may represent daily trading volume curve of one asset, and each row captures synchronized 5-minute volume intervals across multiple assets. While traditional matrix time series focus mainly on temporal evolution, our approach incorporates both spatial and temporal dynamics, enabling simultaneous analysis of interactions across multiple dimensions. The inherent endogeneity in spatio-temporal autoregressive models renders ordinary least squares estimation inconsistent. To overcome this difficulty while simultaneously estimating two distinct weight matrices with banded structure, we develop an iterated generalized Yule-Walker estimator by adapting a generalized method of moments framework based on Yule-Walker equations. Moreover, unlike conventional models that employ a single bandwidth parameter,  the dual-bandwidth specification in our framework requires a new two-step, ratio-based sequential estimation procedure.
\end{abstract}

%\noindent {\bf JEL classification}:
%C13,
%C23,
%C32.


\newpage
\section{Introduction}\label{sec_intro}
In the big data era, more complex data types and diverse information sources have become increasingly accessible. While the wealth of data offers rich insights that researchers and practitioners in various fields can leverage more efficiently, it also poses growing challenges in capturing the intricate dependency structures embedded within different data types.
This paper examines matrix-valued time series data and their modeling, a type of data commonly encountered in practice. We initiate our study with a practical application - predicting trading volume curves - to demonstrate the importance of modeling matrix-valued data. Volume prediction is crucial for trading and execution as it influences market dynamics, affects market makers, and enhances strategy effectiveness. 

Motivated by the task of predicting the intraday volume curve across multiple assets simultaneously in financial market, we consider a matrix representation of raw trading volumes. Let $\mathbf{V}_t$ denote the $p \times q$ matrix of raw trading volumes on day $t$, where $p$ is the number of intraday time buckets (e.g., 5-minute intervals) and $q$ is the number of stocks. The entry $V_{t,i,j}$ represents the volume of stock $j$ during bucket $i$ on day $t$. For a fixed stock $j$, its intraday volume curve on day $t$ corresponds to the $j$-th column of $\mathbf{V}_t$, denoted $(V_{t,1,j}, \ldots, V_{t,p,j})'$. 

\begin{figure}[ht]
\centering
  \includegraphics[width=1.00\linewidth]{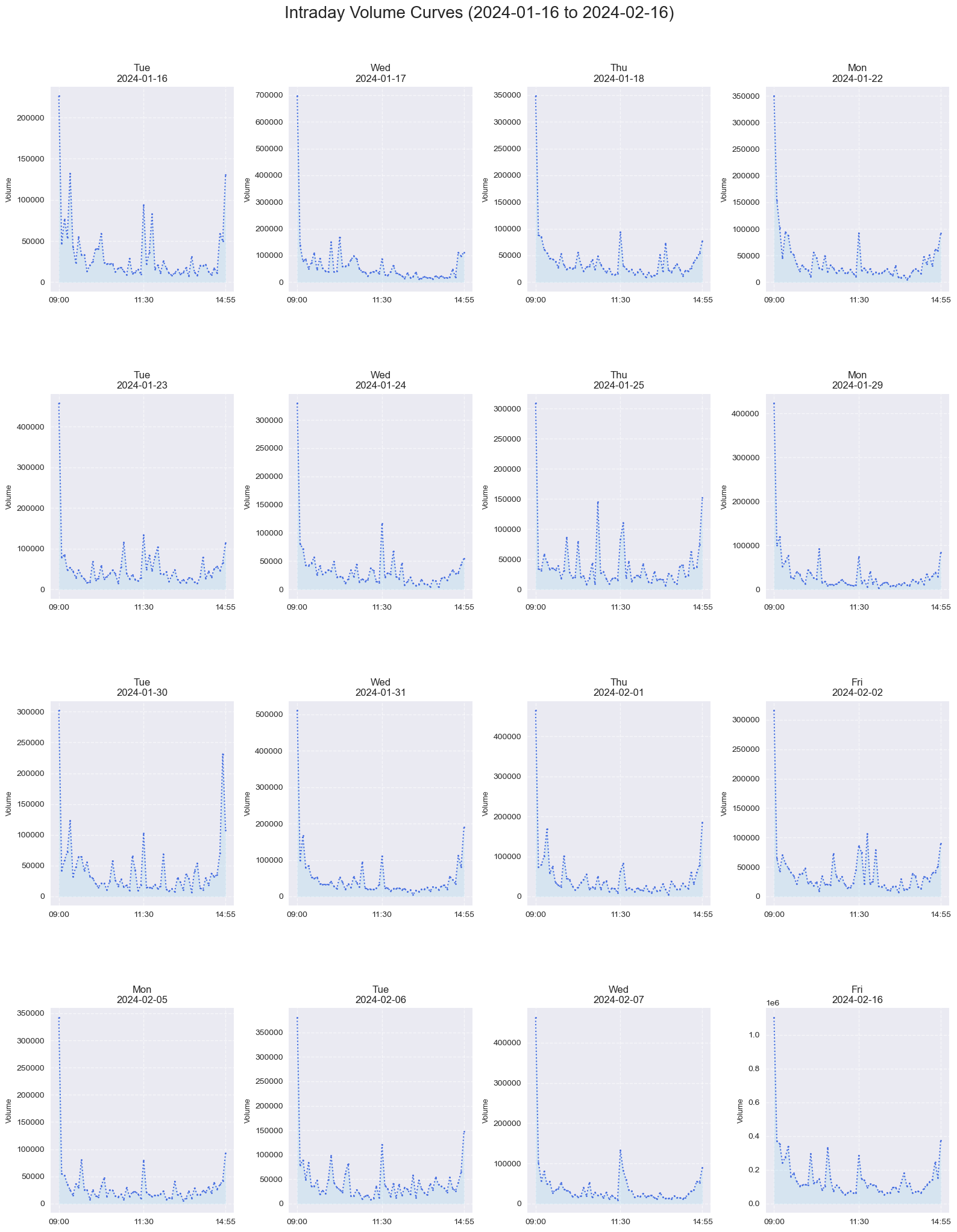}
 \caption{Time series plots of 5 minutes daily trading volume curves for Sony Group Corporation, dates ranges from 16/01/2024 to 16/02/2024.}
\label{sonyTS}
\end{figure}

We analyze 5-minute trading volumes for several liquid Japanese stocks. The 5-minute sampling frequency is a common industry standard for intraday volume analysis, offering a balance between capturing the U-shaped diurnal pattern and providing sufficient liquidity to yield meaningful volume measures while retaining intraday resolution. The selected stocks are Sony (6758), SoftBank (9984), Nintendo (7974), Tokyo Electron (8035), and Renesas (6723). To illustrate the key features of the data, Figure~\ref{sonyTS} plots the intraday volume curves for a single representative stock, Sony Group Corporation, in the Japanese market from 16 January 2024 to 16 February 2024. A consistent bowl-shaped pattern emerges across days, suggesting strong intraday seasonality and temporal persistence within each bucket. This single-stock case corresponds to one column of $\mathbf{V}_t$, and the matrix framework naturally accommodates the cross-section of $q$ stocks.

Intraday trading volume curves are known to exhibit bowl-shaped and potentially nonstationary patterns. As illustrated in Figure~\ref{sonyTS}, intraday volumes follow a deterministic, bowl-shaped seasonal pattern. Moreover, the shape of the bowl itself may evolve over time -- for example, the morning peak might become more pronounced or the closing rush might intensify. However, beyond this predictable within-day pattern, additional stochastic fluctuations remain that are not captured by the bowl shape alone. To capture both the time-varying seasonal pattern and the remaining stochastic dynamics, we decompose $\mathbf{V}_t$ into a time-varying seasonal component and a stationary anomaly component:
\[
\mathbf{V}_t = \mathbf{S}_t + \mathbf{X}_t,
\]
where:
\begin{itemize}
    \item $\mathbf{S}_t$ is a $p \times q$ matrix of deterministic intraday bowl-shaped seasonal patterns that may evolve over time. Each column $\mathbf{s}_{t,j} = (s_{t,1,j}, \ldots, s_{t,p,j})'$ contains the expected volume for stock $j$ at each intraday bucket on day $t$.
    \item $\mathbf{X}_t$ is a $p \times q$ matrix representing the \textit{volume anomaly}. This stochastic component captures day-specific deviations from the expected pattern: a positive value indicates higher-than-usual volume for that stock and time of day, while a negative value indicates lower-than-usual volume.
\end{itemize}

The deterministic bowl-shaped pattern $\mathbf{S}_t$ has been widely documented in the literature, particularly in the context of modeling intraday volatility curves \citep{Andersen2019, Andersen2024, Andersen2025a, Andersen2025b}. A well-established link between volume and volatility stems from the mixture-of-distributions hypothesis (MDH), which posits that both price changes and trading volumes are jointly governed by the same latent information arrival process \citep{Andersen1996}. This implies that volume and volatility should exhibit similar intraday patterns. One key distinction between volatility and volume modeling, however, is that volatility is latent and must be estimated from high-frequency returns, whereas trading volumes are directly observable. For example, \cite{Todorov2012} proposed a jump-robust nonparametric estimator for the Laplace transform of latent volatility, which, under long-span asymptotics, also estimates the integrated joint Laplace transform of volatility across different time points. We further recognize that the time-varying advanced methodologies for intraday volatility curves introduced by \cite{Andersen2019, Andersen2024, Andersen2025a, Andersen2025b} could, in principle, be adapted to our intraday volume curves.

The shape of the intraday deterministic volume pattern can evolve over time. To capture this potential time variation, we adopt a simple industry-standard baseline: a rolling moving average. We employ this approach in our volume prediction empirical analysis for two primary reasons. First, although it has potential limitations (e.g., slow adaptation to sudden structural changes), it continues to serve as the industry standard used by execution desks, offering a straightforward and transparent benchmark. Second, our main modelling focus is not on $\mathbf{S}_t$ but on $\mathbf{X}_t$. The time-varying behavior of $\mathbf{S}_t$ has already been thoroughly explored in the intraday volatility curve literature \citep{Andersen2019, Andersen2024, Andersen2025a, Andersen2025b}; extending this line of inquiry to volume curves lies beyond the scope of our study. In contrast, the anomaly component $\mathbf{X}_t$ has received far less attention, yet it may contain valuable information about news arrivals, sentiment shifts, liquidity shocks, and other transient trading activities.

Specifically, for each stock $j$ and intraday bucket $i$, the deterministic component is estimated as
\[
\hat{S}_{t,i,j} = \frac{1}{L} \sum_{l=1}^{L} V_{t-l,i,j},
\]
where $L$ denotes the lookback window (e.g., $L = 20$ trading days). Subtracting this estimated time-varying bowl-shaped pattern from the raw volumes yields the volume anomaly process $\mathbf{X}_t$. Thus, the nonstationarity in trading volumes is captured by $\mathbf{S}_t$, which we estimate adaptively, leaving $\mathbf{X}_t$ stationary and ready for subsequent modeling. After removing the deterministic pattern, $\mathbf{X}_t$ becomes stationary, allowing us to concentrate our efforts on its dependence structures through a matrix-valued spatio-temporal time series model. 

\begin{figure}[ht]
\centering
  \includegraphics[width=1.00\linewidth]{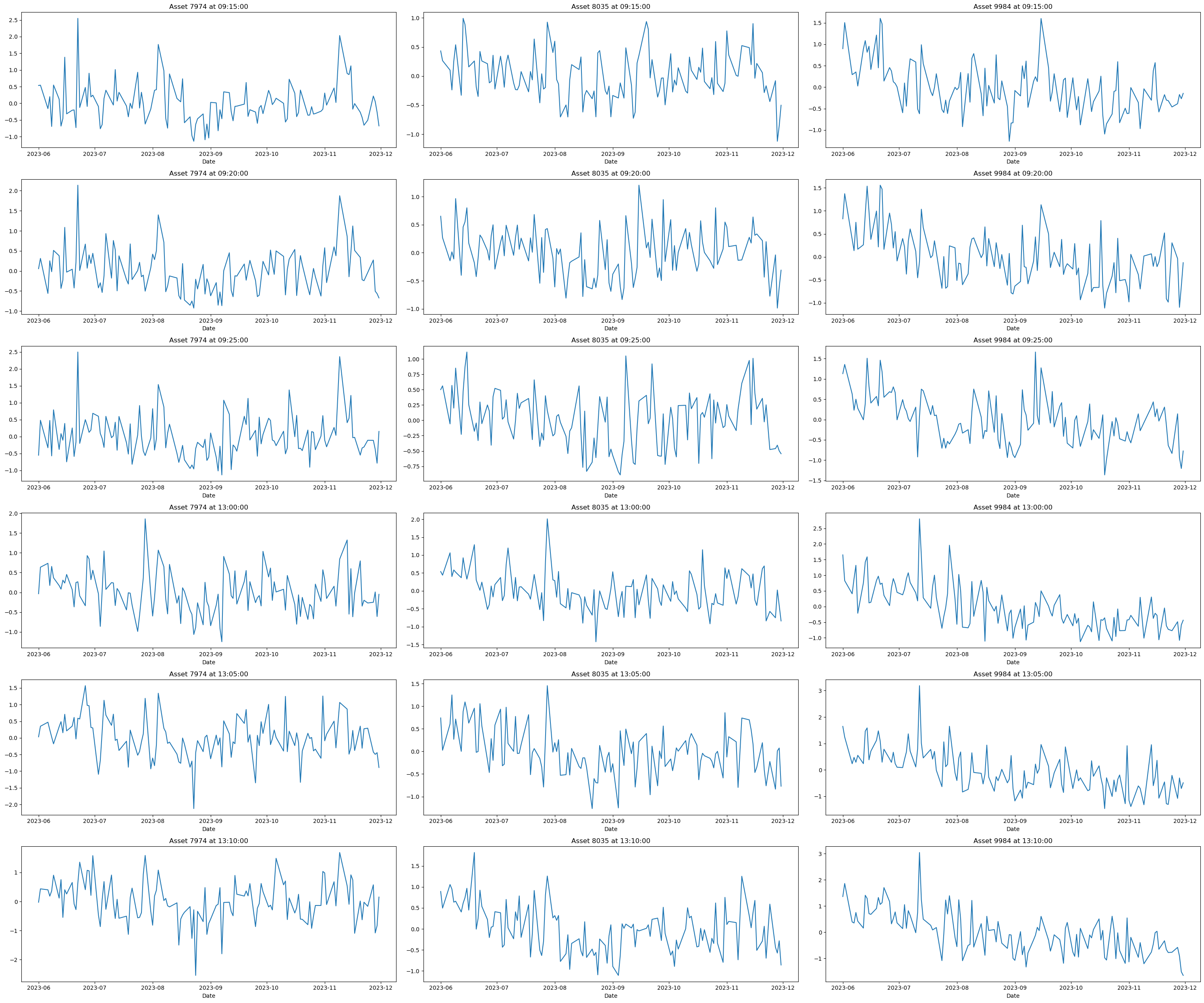}
  
 \caption{Time series plots of six particular 5 minutes time bucket trading volume for the period of 01/06/2023 to 01/12/2023 for Nintendo Co. Ltd, Tokyo Electron Limited, SoftBank Group Corp. 1st column is for  Nintendo Co. Ltd, 2nd column is for Tokyo Electron Limited and 3rd column is for SoftBank Group Corp. 1st row is for 9:15 - 9:20 bucket, 2nd row is for 9:20 - 9:25 bucket, 3rd row is for 9:25 - 9:30 bucket, 4th row is for 13:00 - 13:05 bucket, 5th row is for 13:05 - 13:10 bucket, 6th row is for 13:10 - 13:15 bucket.}
\label{figTS}
\end{figure}

\begin{figure}[ht]
\centering
  \includegraphics[width=1.00\linewidth]{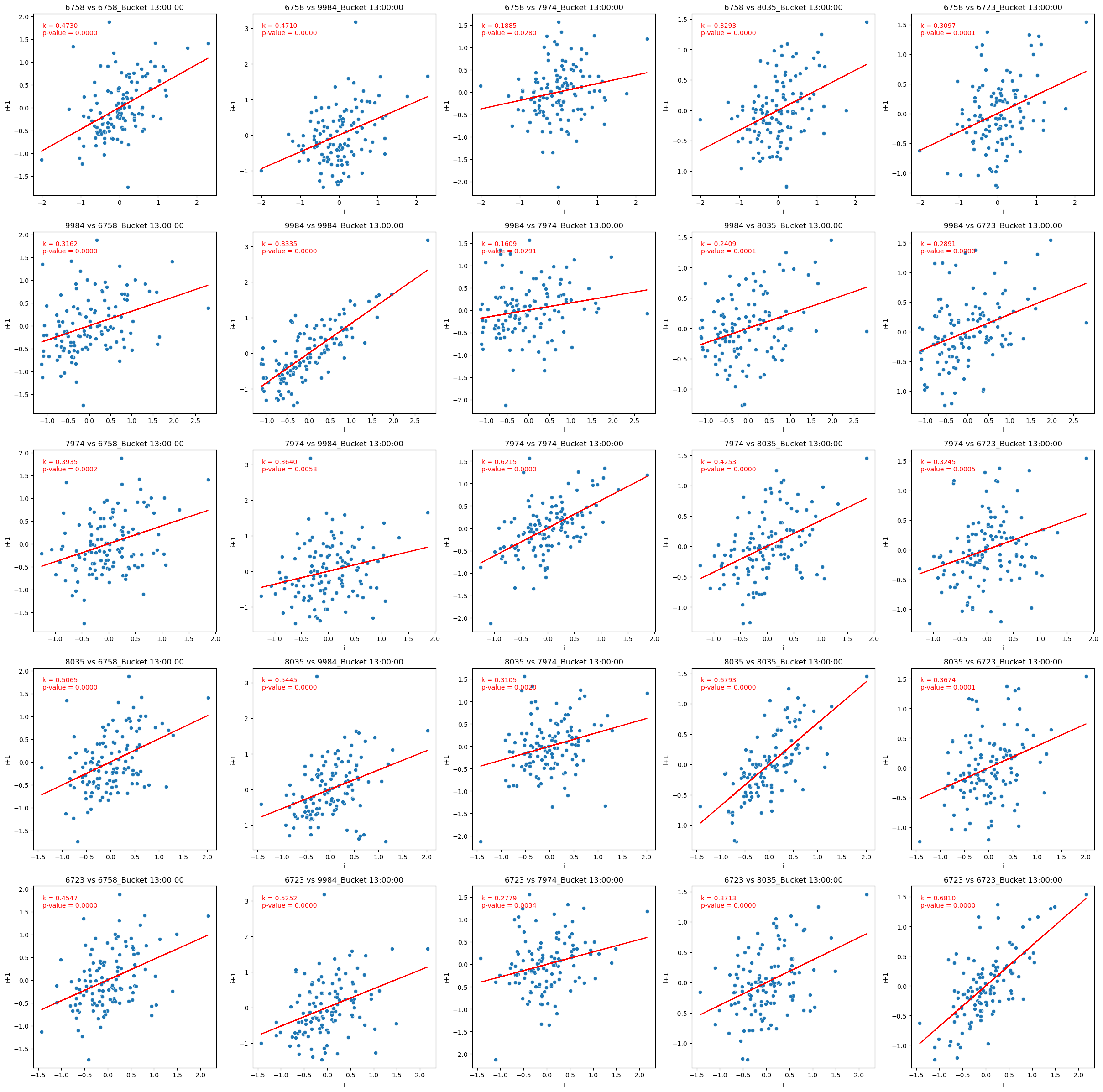}
  \caption{Scatter plots of trading volumes of 5 minutes bucket 13:00 to 13:05 versus its neighbouring 5 minutes bucket for 13:05 to 13:10 for among five assets in Japan market: Sony Group Corporation, SoftBank Group Corp, Nintendo Co., Ltd, Tokyo Electron Limited, Renesas Electronics Corporation, respectively. Data ranges from 01/06/2023 to 01/12/2023.}
\label{figSC}
\end{figure}

In the real data analysis, we first apply a log transformation to the raw data and then subtract the rolling moving average estimate of $\mathbf{S}_t$, producing the volume anomaly process $\mathbf{X}_t$. The subsequent analysis aims to provide a preliminary understanding of the types of dependence structures present in $\mathbf{X}_t$. Figure~\ref{figTS} displays the 5-minute bucket volume anomaly data across days for three assets (columns: Nintendo, Tokyo Electron, SoftBank), with rows representing different time intervals (9:15--9:20, 9:20--9:25, 9:25--9:30, 13:00--13:05, 13:05--13:10, 13:10--13:15). The plots illustrate stronger co-movement among neighboring time buckets (e.g., within morning or afternoon sessions) and weaker interactions between more distant intervals. In addition, cross-asset dependencies are evident, characterized by synchronized spikes (e.g., late June and mid-September for morning sessions; late July for afternoon sessions). Further evidence is provided in Figure~\ref{figSC}, which plots the anomalies for neighboring 5-minute buckets (13:00--13:05 vs. 13:05--13:10) across five assets, with fitted regression lines. The diagonal plots exhibit the strongest correlations, confirming high intraday dependence for individual assets. Off-diagonal plots (e.g., Renesas vs. Tokyo Electron in subplot (5,4), and Tokyo Electron vs. Sony in subplot (4,1)) also reveal significant cross-asset relationships.

The term ``anomaly'' is deliberate. Just as meteorologists study temperature anomalies to understand weather dynamics apart from seasonal norms, we study volume anomalies to isolate stochastic trading activity that deviates from the regular intraday rhythm. The empirical evidence in Figures~\ref{figTS} and~\ref{figSC} reveals that $\mathbf{X}_t$ exhibits rich dependence structures such as cross-sectional (across assets) and intraday (across time buckets) correlations. These findings confirm that $\mathbf{X}_t$ contains systematically predictable variation overlooked by existing approaches, which either focus solely on $\mathbf{S}_t$ or treat the anomalies as unstructured noise. Our modeling effort thus focuses on understanding and forecasting $\mathbf{X}_t$, and the techniques we develop in this paper are specifically proposed to capture its cross-sectional, intraday, and temporal dependencies.

Recent advances in matrix time series analysis have introduced several key methodologies for modeling $\bX_t$. These include bilinear autoregressive models \citep{Chen2021,Xiao2022,Jiang2025}
%(Chen et al., 2021; Xiao et al., 2022; Jiang et al., 2025) 
and factor models based on Tucker decomposition \citep{Chen2020a,Chen2020b, Wang2019, Han2024a},
%(Chen et al., 2020a,b; Wang et al., 2019; Han et al., 2024a), 
which extend the vector time series factor models \citep{Lam2011,Lam2012,Chang2015,Qiao2025}.
%(Lam et al., 2011; Lam and Yao, 2012; Chang et al., 2015; Qiao et al., 2025). 
Additionally, factor models utilizing tensor CP decomposition have been developed \citep{Chang2023, Chang2025, Han2024b}.
%(Chang et al., 2023; 2025; Han et al., 2024b). 
Other contributions within the factor model framework include the works of \cite{Yu2022} and \cite{Gao2023}.
%Yu et al. (2022) and Gao and Tsay (2023). 
Meanwhile, spatio-temporal data analysis has seen rapid progress, with developments including spatial dynamic panel data models (SDPD) \citep{Yu2008, Yu2012, Dou2016}
%(Yu et al., 2008; 2012; Dou et al., 2016) 
and banded spatio-temporal autoregressive models (BSAR) for vector processes \citep{Gao2019}.
%(Gao et al., 2019). 
We introduce two classes of linear spatio-temporal autoregressive models designed for matrix-valued time series. The first integrates known interaction structures through predefined weight matrices, which is a common approach in spatial econometrics \citep{Yu2008, Yu2012, Dou2016}.
%(Yu et al., 2008; 2012; Dou et al., 2016). 
For a detailed overview of the SDPD model with a known weight matrix, refer to \cite{Lee2010a} and \cite{Lee2010b}.
%Lee and Yu (2010). 
The second treats weight matrices as unknown but assumes a banded structure \citep{Guo2016,Gao2019},
%(Guo et al., 2016; Gao et al., 2019), 
offering a balance between flexibility and parsimony. To our knowledge, this is among the first efforts to model high-dimensional matrix-valued time series in an spatio-temporal autoregressive framework that accounts for both spatial and temporal dependencies. Estimating matrix-valued spatio-temporal autoregressive models presents several key challenges: inherent endogeneity effects and the simultaneous estimation of two distinct weight matrices capturing row-wise and column-wise dependencies. To address these issues, we develop an iterative generalized method of moments estimator derived from Yule-Walker equations. The estimation complexity is further compounded by requiring the determination of two bandwidth parameters, unlike existing approaches \citep{Guo2016,Gao2019}
%(Guo et al., 2016; Gao et al., 2019) 
that involve only a single bandwidth parameter. To address this more complex scenario, we develop a two-step ratio-based estimation procedure that sequentially estimates each bandwidth parameter. Empirically, we provide an ordering method that reorders both spatial and panel variables to induce a banded structure in the weight matrices, using conditional dependence information.

The rest of this paper is organized as follows. In Section 2, we present the matrix spatio-temporal autoregressive model with diagonal coefficients. Section 3 is on the matrix spatio-temporal autoregression with banded weight matrices. We evaluate the model’s performance through simulations in Section 4 and demonstrate its practical utility in Section 5 with an application to trade execution services in financial markets. Technical proofs and additional supporting figures are provided in the supplementary material.

%We present the notation conventions adopted in this work. 
{\it Notation.} For any positive integer $d \ge 1$, let $\bI_d$ denote the $d \times d$ identity matrix. For any $a,b \in \mathbb{R}$, let $a \vee b$ denote the larger number between $a$ and $b$. For a $d$-dimensional vector $\bu = (u_1,\ldots,u_d)'$, we use $\|\bu\|_2 = (\sum_{i=1}^d u_i^2)^{1/2}$ to denote its Euclidean norm. For any $m_1 \times m_2$ matrix $\bH = (h_{i,j})$, let ${\rm vec}(\bH)$ denote the $(m_1m_2)$-dimensional vector obtained by stacking the columns of $\bH$, and $\|\bH\|_F = (\sum_{i=1}^{m_1}\sum_{j=1}^{m_2}h_{i,j}^2)^{1/2}$ %{\color{blue}$\|\bH\|_2 = \lambda_{\max}^{1/2}(\bH'\bH)$} and 
%$\|\bH\|_F = \{{\rm tr}(\bH'\bH)\}^{1/2}$ 
denote %the operator norm and 
the Frobenius norm of $\bH$.
%, respectively, where $\lambda_{\max}(\cdot)$ is the largest eigenvalue of a matrix and ${\rm tr}(\cdot)$ is the trace of a square matrix. 
Let $\otimes$ denote the Kronecker product. A $d$-dimensional strictly stationary process $\{\by_t\}$ is $\alpha$-mixing if 
\begin{align}\label{eq:mixingcoeff}
    \alpha(k) \equiv \sup_{A \in \mathcal{F}_{-\infty}^{0},B \in \mathcal{F}_{k}^{+\infty}}|\mathbb{P}(A)\mathbb{P}(B)-\mathbb{P}(AB)| \to 0\,, ~\mbox{as}~k \to \infty\,, 
\end{align}
where $\mathcal{F}_{i}^{j}$ denotes the $\sigma$-algebra generated by $\{\by_t, i \le t \le j\}$. See, e.g., Section 2.6 of \cite{Fan2003} for a compact review of $\alpha$-mixing processes.

\section{Spatio-temporal regression with diagonal coefficients}\label{sec:model}
\subsection{Models} \label{subsec:model_known_matrix}
Let $\bX_t = (X_{t,i,j}) \equiv (\bX_{t, \cdot 1}, \ldots, \bX_{t, \cdot q})
\equiv (\bX_{t,1 \cdot}', \ldots, \bX_{t,p\cdot}')' 
$ be a $p \times q$ matrix representing the observations collected at time $t$,
where $\bX_{t,i\cdot}, \bX_{t,\cdot j}$ and $X_{t,i,j}$ denote, respectively, the $i$-th row vector, the $j$-th column vector and the $(i,j)$-th element of $\bX_t$. We always assume that vectors are in the form of columns. For any $p$-dimensional vector $\balpha = (\alpha_1, \ldots, \alpha_p)'$, denote by $D(\balpha)$ the $p\times p$ diagonal matrix with $\alpha_i$ as its $i$-th main diagonal element. 

We consider the following model:
\begin{equation}\label{model_class1_iter}
\bX_t = D(\balpha_0) \bW_0 \bX_t \bV_0' D(\bbeta_0) + D(\balpha_1) \bW_1 \bX_{t-1} \bV_1' D(\bbeta_1)+ \bE_t\,,
\end{equation}
where $\bW_0$ is the known $p \times p$ weight matrix with zero diagonal elements and $\bV_0$ is the known $q \times q$ weight matrix, $\bW_1$ and $\bV_1$ are known $p \times p$ and $q \times q$ weight matrices, and $\balpha_0, \balpha_1$ are unknown $p$-dimensional vectors, and $\bbeta_0, \bbeta_1$ are unknown $q$-dimensional vectors. In above expression, $\bE_t$ is the $p \times q$ innovation matrix at time $t$ which satisfies condition $ {\mathbb{E}}(\bE_t) = {\bf{0}} $ and $\cov(\bX_{t-m, \cdot i},\bE_{t, \cdot j}) = {\bf{0}}, \, 1\le i, j \le q, m > 0$, where $\bE_{t, \cdot j}$ denotes the $j$-th column vector of $\bE_t$. %{\color{blue} Note that all the innovations are uncorrelated with past observations, a common assumption in time series models.}

Model (\ref{model_class1_iter}) is a matrix extension
of vector spatial-temporal autoregression of \cite{Dou2016}.
%Dou et al. (2016).
When all the elements of each  $\balpha_0, \bbeta_0, \balpha_1$ and $\bbeta_1$ are the same, model (\ref{model_class1_iter}) can be seen as a matrix extension of the popular SDPD
model for vector processes, see, for example, \cite{Yu2008}
%Yu et al. (2008) 
and the references within. Like SDPD models,
spatial weight matrices $\bW_i$ and $\bV_i$ are assumed to be known. In practice, the weight matrices can be constructed based on geographic proximity or economic distances -- such as trade flows, transportation costs, or input-output linkages, see \cite{Conley1999}.
%Conley (1999). 
%Ge et al. (2025) 
\cite{Ge2023} employ news-based connections among firms as a proxy for inter-firm linkages to build these matrices. %Zhang and Yu (2018) 
\cite{Zhang2018} address the common scenario where multiple theoretically motivated weight matrices are available and propose a Mallows-type criterion for selecting among them or combining them through averaging. In our empirical analysis, we construct the weight matrices $\bW_i$ and $\bV_i$ for the spatial and panel variables, respectively, using a conditional correlation-based approach. The weighting structure is also guided by bandwidth values that are empirically derived from the data. Details of constructing $\bW_i$ and $\bV_i$ can be found in Section \ref{G::known_weight_mat_spec} in the supplementary material. In most SDPD specifications, the unknown parameters occur as a scalar coefficient in front of each spatial weight matrix. We replace each scalar by a diagonal matrix to allow the heterogeneity in mean values in regressing different rows and columns.  

Model (\ref{model_class1_iter}) can be represented in the form of below vector spatio-temporal autoregressive model: 
\begin{equation}\label{class1_iter_vec}
\text{vec}(\bX_t) = \{(D(\bbeta_0) \bV_0) \otimes (D(\balpha_0) \bW_0 )\}  \text{vec}(\bX_t) +  \{(D(\bbeta_1) \bV_1) \otimes (D(\balpha_1) \bW_1 )\}\text{vec}(\bX_{t-1}) + \text{vec}(\bE_t)\,.
\end{equation}
%where $\otimes$ denotes the matrix Kronecker product. 
The model is not identified since the model remains the same if $D(\balpha_k)$ and $D(\bbeta_k)$, $k = 0, 1$, are replaced by $cD(\balpha_k)$ and $c^{-1}D(\bbeta_k)$ with a nonzero constant $c$. However, note that the Kronecker product $D(\bbeta_k) \otimes D(\balpha_k)$ is unique and therefore identifiable. 

Let $\bI_{pq} - (D(\bbeta_0) \bV_0) \otimes (D(\balpha_0) \bW_0 )$ be invertible and all the eigenvalues of $\{\bI_{pq} - (D(\bbeta_0) \bV_0) \otimes (D(\balpha_0) \bW_0 )\}^{-1}\{(D(\bbeta_1) \bV_1) \otimes (D(\balpha_1) \bW_1 )\}$ be less than 1 in modulus. Model (\ref{class1_iter_vec}) becomes 
\begin{equation}\label{class1_iter_vec_reduced}
\begin{split}
\text{vec}(\bX_t) = &  \{\bI_{pq} - (D(\bbeta_0) \bV_0) \otimes (D(\balpha_0) \bW_0 )\}^{-1}\{(D(\bbeta_1) \bV_1) \otimes (D(\balpha_1) \bW_1 )\}\text{vec}(\bX_{t-1}) \\
&+ \{\bI_{pq} - (D(\bbeta_0) \bV_0) \otimes (D(\balpha_0) \bW_0 )\}^{-1}\text{vec}(\bE_t)\,,
\end{split}
\end{equation}
which admits a stationary solution.
%Throughout the paper, when discussing known weight matrices models, $\bX_t$ is referred to as the (weakly) stationary process defined by (\ref{class1_iter_vec_reduced}). 
For this stationary process, $\mathbb{E}(\bX_t) = {\bf 0}$, and the $j$-th column of $\bX_t$, $1\le j \le q$, admits the form  
\begin{equation}\label{class1_col}
\bX_{t, \cdot j} = \sum_{i=1}^q D(\balpha_0) \bW_0 \bX_{t, \cdot, i} v_{j, i}^{(0)} \beta_{j}^{(0)} + \sum_{i=1}^q D(\balpha_1) \bW_1 \bX_{t-1, \cdot, i} v_{j, i}^{(1)} \beta_{j}^{(1)} + \bE_{t, \cdot j}\,,
\end{equation}
where $v_{j, i}^{(0)}$ and $v_{j, i}^{(1)}$ are the ($j, i$)-th element of the known matrices $\bV_0$ and $\bV_1$, and $\beta_{j}^{(0)}$ and $\beta_{j}^{(1)}$ are the $j$-th elements of $\bbeta_0$ and $\bbeta_1$, respectively. 

\subsection{Estimation}\label{sec:est_diag}
Due to the innate endogeneity since $\bX_t$ appears on the right hand side of model (\ref{model_class1_iter}) and $\bX_t$ is correlated with $\bE_t$, the least squares estimation leads to inconsistent estimators. Instead we introduce a generalized Yule-Walker estimators for the parameters $\balpha_0, \balpha_1, \bbeta_0$ and $\bbeta_1$. 

Multiplying both sides of (\ref{class1_col}) by $\bX_{t-1, \cdot k}^{'}$ from the back and taking expectations on the both sides,   we obtain the Yule-Walker equations
\begin{equation}\label{class1_YW}
\bSigma_{jk}(1) = \sum_{i=1}^q D(\balpha_0) \bW_0 \bSigma_{ik}(1) v_{j, i}^{(0)} \beta_{j}^{(0)} + \sum_{i=1}^q D(\balpha_1) \bW_1 \bSigma_{ik}(0) v_{j, i}^{(1)} \beta_{j}^{(1)}, \quad 1\le j, k \le q\,,\end{equation}
where $\bSigma_{jk}(1) = \cov(\bX_{t, \cdot j}, \bX_{t-1, \cdot k})$ and $\bSigma_{jk}(0) = \cov(\bX_{t, \cdot j}, \bX_{t, \cdot k})$. 
Denote sample estimators of (auto)covariance $\bSigma_{jk}(1)$ and $\bSigma_{jk}(0)$, $1 \le j, k \le q$,  as
\begin{equation*}
\hat\bSigma_{jk}(1) = \frac{1}{n}\sum_{t=2}^n \bX_{t, \cdot j} \bX_{t-1, \cdot k}' \quad \text{and} \quad
\hat\bSigma_{jk}(0) = \frac{1}{n}\sum_{t=1}^n \bX_{t, \cdot j} \bX_{t, \cdot k}'\,.
\end{equation*}
Following Yule-Walker equations (\ref{class1_YW}), the generalized Yule-Walker estimator is obtained by solving the following optimization problem:
\begin{equation}\label{eq:Opt1}
\min_{\balpha_0, \bbeta_0, \balpha_1, \bbeta_1}\frac{1}{p^2q^2}\sum_{j, k=1}^q \Big\|\hat\bSigma_{jk}(1) - \sum_{i=1}^q D(\balpha_0) \bW_0 \hat \bSigma_{ik}(1) v_{j, i}^{(0)} \beta_{j}^{(0)} - \sum_{i=1}^q D(\balpha_1) \bW_1 \hat \bSigma_{ik}(0) v_{j, i}^{(1)} \beta_{j}^{(1)} \Big\|_F^2\,.
\end{equation}
We propose an iterated method to estimate $\balpha_k$ and $\bbeta_k$, $k = 0, 1$ in a recursive manner.
% , that is, 
% \begin{equation} \label{class1_opt}
% \min_{\balpha_0, \balpha_1} \left\{ \min_{\bbeta_0, \bbeta_1} {\color{red}\frac{1}{p^2q^2}}\sum_{j, k=1}^q \Big\|\hat\bSigma_{jk}(1) - \sum_{i=1}^q D(\balpha_0) \bW_0 \hat \bSigma_{ik}(1) v_{j, i}^{(0)} \beta_{j}^{(0)} - \sum_{i=1}^q D(\balpha_1) \bW_1 \hat \bSigma_{ik}(0) v_{j, i}^{(1)} \beta_{j}^{(1)} \Big\|_F^2 \right\}\,.
% \end{equation}
\paragraph{Estimating $\bbeta_0$ and $\bbeta_1$ given $\balpha_0$ and $\balpha_1$:} Given $\balpha_0$ and $\balpha_1$, we solve the following optimization problem with respect to $\bbeta_0$ and $\bbeta_1$:
%(\ref{eq:Opt1}) becomes 
\[
\min_{\bbeta_0, \bbeta_1}\frac{1}{p^2q^2} \sum_{j, k=1}^q \Big\|\hat\bSigma_{jk}(1) - \sum_{i=1}^q D(\balpha_0) \bW_0 \hat \bSigma_{ik}(1) v_{j, i}^{(0)} \beta_{j}^{(0)} - \sum_{i=1}^q D(\balpha_1) \bW_1 \hat \bSigma_{ik}(0) v_{j, i}^{(1)} \beta_{j}^{(1)} \Big\|_F^2\,,
\]
%which is equivalent to solve the following optimization problem for each $j=1,\ldots,q$:
which can be solved by decomposing it into $q$ least-squares problems. Specifically, for each $j=1,\ldots,q$, we solve
\begin{equation} \label{class1_opt_vec}
\min_{\beta_{j}^{(0)}, ~\beta_{j}^{(1)}} \frac{1}{p^2q^2}\sum_{k=1}^q \Big\|  \text{vec} \{\hat\bSigma_{jk}(1)\} - \sum_{i=1}^q \text{vec}\{D(\balpha_0) \bW_0 \hat \bSigma_{ik}(1) v_{j, i}^{(0)}\} \beta_{j}^{(0)} - \sum_{i=1}^q \text{vec}\{D(\balpha_1) \bW_1 \hat \bSigma_{ik}(0) v_{j, i}^{(1)}\} \beta_{j}^{(1)} \Big\|_2^2\,,
\end{equation}
% \begin{equation} \label{class1_opt_vec}
% {\color{red}\frac{1}{p^2q^2}}\sum_{j=1}^q \min_{\beta_{j}^{(0)}, \beta_{j}^{(1)}} \sum_{k=1}^q \Big\|  \text{vec} (\hat\bSigma_{jk}(1)) - \sum_{i=1}^q \text{vec}(D(\balpha_0) \bW_0 \hat \bSigma_{ik}(1) v_{j, i}^{(0)}) \beta_{j}^{(0)} - \sum_{i=1}^q \text{vec}(D(\balpha_1) \bW_1 \hat \bSigma_{ik}(0) v_{j, i}^{(1)}) \beta_{j}^{(1)} \Big\|_2^2\,,
% \end{equation}
Stacking across $k$, (\ref{class1_opt_vec}) becomes 
\begin{equation} \label{LS_beta}
\min_{\bb_j} \frac{1}{p^2q^2}\| \bY_{\bb_j} - \bX_{\bb_j}\bb_j\|_2^2\,,
\end{equation}
% \begin{equation} \label{LS_beta}
% {\color{red}\frac{1}{p^2q^2}}\sum_{j=1}^q \min_{\bb_j} \| \bY_{\bb_j} - \bX_{\bb_j}\bb_j\|_2^2\,,
% \end{equation}
where $\bb_j = (\beta_{j}^{(0)}, \beta_{j}^{(1)})'$, $\bY_{\bb_j} = (\text{vec}\{\hat \bSigma_{j1}(1)\}',\ldots,\text{vec}\{\hat \bSigma_{jq}(1)\})'$ 
% \begin{equation*}
% \bb_j = 
% \begin{pmatrix}
% \beta_{j}^{(0)} \\
% \beta_{j}^{(1)}
% \end{pmatrix}
% \end{equation*}
%is the $2 \times 1$ vector of unknown parameters, that is the $j$-th elements of $\bbeta_0$ and $\bbeta_1$, 
% \begin{equation*}
% \bY_{\bb_j} = 
% \begin{pmatrix}
% \text{vec}(\hat \bSigma_{j1}(1)) \\
% \text{vec}(\hat \bSigma_{j2}(1)) \\
% \vdots\\
% \text{vec}(\hat \bSigma_{jq}(1))
% \end{pmatrix}
% \end{equation*}
is a $p^2 q \times 1$ vector and 
\begin{equation*}
\bX_{\bb_j} = 
\begin{pmatrix}
\sum_{i=1}^q \text{vec}\{D(\balpha_0) \bW_0 \hat \bSigma_{i1}(1) v_{j, i}^{(0)}\}, ~ \sum_{i=1}^q \text{vec}\{D(\balpha_1) \bW_1 \hat \bSigma_{i1}(0) v_{j, i}^{(1)}\}\\
\sum_{i=1}^q \text{vec}\{D(\balpha_0) \bW_0 \hat \bSigma_{i2}(1) v_{j, i}^{(0)}\}, ~ \sum_{i=1}^q \text{vec}\{D(\balpha_1) \bW_1 \hat \bSigma_{i2}(0) v_{j, i}^{(1)}\}\\
\vdots \\
\sum_{i=1}^q \text{vec}\{D(\balpha_0) \bW_0 \hat \bSigma_{iq}(1) v_{j, i}^{(0)}\}, ~ \sum_{i=1}^q \text{vec}\{D(\balpha_1) \bW_1 \hat \bSigma_{iq}(0) v_{j, i}^{(1)}\}\\
\end{pmatrix}
\end{equation*}
is a $p^2 q \times 2$ matrix.  Hence (\ref{LS_beta}) leads to the least squares estimator 
\begin{equation*}
\hat \bb_j = 
\begin{pmatrix}
\hat \beta_{j}^{(0)} \\
\hat \beta_{j}^{(1)}
\end{pmatrix} = ({\bX_{\bb_j}}'\bX_{\bb_j})^{-1}{\bX_{\bb_j}}'\bY_{\bb_j}\,, \, ~~ 1 \le j \le q\,.
\end{equation*}

\paragraph{Estimating $\balpha_0$ and $\balpha_1$ given $\bbeta_0$ and $\bbeta_1$:} Given $\bbeta_0$ and $\bbeta_1$, we solve the following optimization problem with respect to $\balpha_0$ and $\balpha_1$:
%(\ref{class1_opt}) becomes 
\[
\min_{\balpha_0, \balpha_1} \frac{1}{p^2q^2}\sum_{j, k=1}^q \Big\|\hat\bSigma_{jk}(1)^{'} - \sum_{i=1}^q  v_{j, i}^{(0)} \beta_{j}^{(0)} \hat \bSigma_{ik}(1)^{'} \bW_0^{'} D(\balpha_0) - \sum_{i=1}^q  v_{j, i}^{(1)} \beta_{j}^{(1)} \hat \bSigma_{ik}(0)^{'} \bW_1^{'} D(\balpha_1) \Big\|_F^2\,,
\]
%which is equivalent to solve the following optimization problem for each $m=1,\ldots,p$:
 which is equivalent to 
\[
\min_{\balpha_0, \balpha_1} \frac{1}{p^2q^2}\sum_{j, k=1}^q \sum_{m=1}^p \Big\|\hat\bSigma_{jk}(1)^{'} \be_m - \sum_{i=1}^q  v_{j, i}^{(0)} \beta_{j}^{(0)} \hat \bSigma_{ik}(1)^{'} \bW_0^{'}{\be_m} \alpha_{m}^{(0)} - \sum_{i=1}^q  v_{j, i}^{(1)} \beta_{j}^{(1)} \hat \bSigma_{ik}(0)^{'} \bW_1^{'}{\be_m} \alpha_{m}^{(1)} \Big\|_2^2\,,
\]
where $\be_m$ is the unit vector with the $m$-th element to be 1, $\alpha_{m}^{(0)}$ and $\alpha_{m}^{(1)}$ are the $m$-th elements of $\balpha_0$ and $\balpha_1$, respectively.
% where $\be_m$ is the unit vector with the $m$-th element to be 1, $\alpha_{m}^{(0)}$ and $\alpha_{m}^{(1)}$ are the $m$-th elements of $\balpha_0$ and $\balpha_1$, respectively. Hence we have
% \begin{equation} \label{class1_opt_alpha}
% {\color{red}\frac{1}{p^2q^2}}\sum_{m=1}^p \min_{\alpha_{m}^{(0)}, \alpha_{m}^{(1)}} \sum_{j, k=1}^q \Big\|\hat\bSigma_{jk}(1)^{'} \be_m - \sum_{i=1}^q  v_{j, i}^{(0)} \beta_{j}^{(0)} \hat \bSigma_{ik}(1)^{'} \bW_0^{'} {\be_m}\alpha_{m}^{(0)} - \sum_{i=1}^q  v_{j, i}^{(1)} \beta_{j}^{(1)} \hat \bSigma_{ik}(0)^{'} \bW_1^{'} {\be_m}\alpha_{m}^{(1)} \Big\|_2^2\,, 
% \end{equation}
Analogously, the above problem can be decomposed into $p$ least-squares problems as follows:
\begin{equation} \label{class1_opt_alpha}
\min_{\alpha_{m}^{(0)}, \alpha_{m}^{(1)}} \frac{1}{p^2q^2}  \sum_{j, k=1}^q \Big\|\hat\bSigma_{jk}(1)^{'} \be_m - \sum_{i=1}^q  v_{j, i}^{(0)} \beta_{j}^{(0)} \hat \bSigma_{ik}(1)^{'} \bW_0^{'} {\be_m}\alpha_{m}^{(0)} - \sum_{i=1}^q  v_{j, i}^{(1)} \beta_{j}^{(1)} \hat \bSigma_{ik}(0)^{'} \bW_1^{'} {\be_m}\alpha_{m}^{(1)} \Big\|_2^2
\end{equation}
for each $m=1,\ldots,p$. 
Staking across $j$ and $k$, (\ref{class1_opt_alpha}) becomes 
 \begin{equation} \label{LS_alpha}
 \min_{\ba_m} \frac{1}{p^2q^2}\| \bY_{\ba_m} - \bX_{\ba_m}\ba_m\|_2^2\,,
\end{equation}
%  \begin{equation} \label{LS_alpha}
% {\color{red}\frac{1}{p^2q^2}}\sum_{m=1}^p \min_{\ba_m} \| \bY_{\ba_m} - \bX_{\ba_m}\ba_m\|_2^2\,,
% \end{equation}
where $\ba_m =(\alpha_{m}^{(0)}, \alpha_{m}^{(1)})'$, 
% \begin{equation*}
% \ba_m = 
% \begin{pmatrix}
% \alpha_{m}^{(0)} \\
% \alpha_{m}^{(1)}
% \end{pmatrix}
% \end{equation*}
% is the $2 \times 1$ vector of unknown parameters, that is the $m$-th elements of $\balpha_0$ and $\balpha_1$,
\begin{equation*}
\bY_{\ba_m} = 
\begin{pmatrix}
\hat \bSigma_{11}(1)' \be_m \\
\hat \bSigma_{12}(1)' \be_m \\
\vdots \\
\hat \bSigma_{qq}(1)' \be_m,
\end{pmatrix}
\end{equation*}
is a $pq^2 \times 1$ vector, and
\begin{equation*}
\bX_{\ba_m} = 
\begin{pmatrix}
\sum_{i=1}^q  v_{1, i}^{(0)} \beta_{1}^{(0)} \hat \bSigma_{i1}(1)^{'} \bW_0^{'}{\be_m}\,, ~ \sum_{i=1}^q  v_{1, i}^{(1)} \beta_{1}^{(1)} \hat \bSigma_{i1}(0)^{'} \bW_1^{'}{\be_m} \\
\sum_{i=1}^q  v_{1, i}^{(0)} \beta_{1}^{(0)} \hat \bSigma_{i2}(1)^{'} \bW_0^{'}{\be_m}\,, ~ \sum_{i=1}^q  v_{1, i}^{(1)} \beta_{1}^{(1)} \hat \bSigma_{i2}(0)^{'} \bW_1^{'}{\be_m} \\
\vdots \\
\sum_{i=1}^q  v_{q, i}^{(0)} \beta_{q}^{(0)} \hat \bSigma_{iq}(1)^{'} \bW_0^{'}{\be_m}\,, ~ \sum_{i=1}^q  v_{q, i}^{(1)} \beta_{q}^{(1)} \hat \bSigma_{iq}(0)^{'} \bW_1^{'} {\be_m}
\end{pmatrix}
\end{equation*}
is a $pq^2 \times 2$ matrix.  This leads to the least squares estimator
\begin{equation*}
\hat \ba_m = 
\begin{pmatrix}
\hat \alpha_{m}^{(0)} \\
\hat \alpha_{m}^{(1)}
\end{pmatrix} = ({\bX_{\ba_m}}'\bX_{\ba_m})^{-1}{\bX_{\ba_m}}'\bY_{\ba_m}\,, \, ~~ 1 \le m \le p\,.
\end{equation*}

The iterative estimation process will stop if 
\[
\sum_{k=0}^1 \|D(\hat \bbeta_k^{(l)}) \otimes D(\hat \balpha_k^{(l)}) -D(\hat \bbeta_k^{(l-1)}) \otimes D(\hat \balpha_k^{(l-1)}) \|_F
\]
is close to zero, where $D(\hat \bbeta_k^{(l-1)})$ and $D(\hat \balpha_k^{(l)})$ are the estimated $D(\bbeta_k)$ and $D(\balpha_k)$ from the $(l-1)$-th and $l$-th iterations. 

\subsection{Asymptotic properties}\label{sec:theory_diag}

%We impose the following regularity conditions on the matrix time series $\{\bX_t\}$ satisfying %the spatio-temporal autoregressive models \eqref{model_class1_iter} and \eqref{model_all}.

Let $\bgamma = (\ba_1',\ldots,\ba_p',\bb_1',\ldots,\bb_q')'$ with $\ba_m = (\alpha_{m}^{(0)},\alpha_{m}^{(1)})'$, $\bb_j = (\beta_{j}^{(0)},\beta_{j}^{(1)})'$, where $\alpha_{m}^{(k)}$ and $\beta_{j}^{(k)}$ denote the $m$-th element of $\balpha_{k}$ and $j$-th element of $\bbeta_{k}$ for $k = 0, 1$, respectively.
%$\balpha_{k} = (\alpha_{1}^{(k)},\ldots, \alpha_{p}^{(k)})'$ and $\bbeta_{k} = (\beta_{1}^{(k)},\ldots, \beta_{q}^{(k)})'$ for $k = 0, 1$. 
Due to the identifiability issue of $\balpha_k$ and $\bbeta_k$,
analogous to the assumption in \cite{Chen2021},
%Chen et al. (2021), 
we make the convention that, for $k = 0,1$, $\|\balpha_k\|_2 = 1$ and the first nonzero entry of $\balpha_k$ is positive. Let $\hat{\bgamma} = (\hat{\ba}_1',\ldots,\hat{\ba}_p',\hat{\bb}_1',\ldots,\hat{\bb}_q')'$ denote the generalized Yule-Walker estimator obtained by minimizing \eqref{eq:Opt1}, where $\hat{\alpha}_{m}^{(k)}$ and $\hat{\beta}_{j}^{(k)}$ are defined analogously. For $k=0,1$, denote $\hat{\balpha}_k = (\hat{\alpha}_{1}^{(k)},\ldots,\hat{\alpha}_{p}^{(k)})'$ and $\hat{\bbeta}_k = (\hat{\beta}_{1}^{(k)},\ldots,\hat{\beta}_{q}^{(k)})'$, and $\hat{\balpha}_k$ is rescaled so that the same normalization holds.
%which is rescaled so that for $k = 0,1$, $\|\hat{\balpha}_k\|_2 = 1$ and the first nonzero entry of $\hat{\balpha}_k$ is positive.
%$\|\hat{\balpha}_0\|_2 = 1$ and $\|\hat{\balpha}_1\|_2 = 1$. 

To establish the theoretical guarantee of the proposed method, we impose the following regularity conditions on the the process $\{{\rm vec}(\bX_t)\}$ and model \eqref{model_class1_iter}.

\begin{cd}\label{cond:mixing}
     %{\rm (i)} The innovation $\bE_t$ satisfies that $\cov(\bX_{t-m, \cdot i},\bE_{t, \cdot j}) = {\bf{0}}$ for any $1\le i, j \le q$ and $m > 0$. 
     {\rm (i)} The process $\{{\rm vec}(\bX_t)\}$ 
     %in model \eqref{class1_iter_vec} and \eqref{model_vec} 
     is strictly stationary and $\alpha$-mixing with its mixing coefficient $\alpha(k)$, defined in \eqref{eq:mixingcoeff}, satisfying that $\sum_{k=1}^{\infty}\alpha(k)^{\kappa/(4+\kappa)} < \infty$ for some constant $\kappa \in (0,2)$. {\rm (ii)} For $\kappa$ specified in {\rm (i)},  $\max_{i \in [p]}\max_{j \in [q]}\max_{t \in [n]}\mathbb{E}|E_{t,i,j}|^{4+\kappa} < \infty$, and there exists some universal constant $C>0$ such that $\max_{i \in [p]}\max_{j \in [q]}\max_{t \in [n]} \mathbb{P}(|X_{t,i,j}|> x) \le Cx^{-(8+2\kappa)/\kappa}$ for any $x>0$. 
    %{\rm (iii)} For $\kappa > 0$ specified in {\rm (ii)} above, 
    % \begin{align*}
    %     \sup_{1 \le l \le pq}\mathbb{E}(|\be_{l}{\rm vec}(\bX_{t})|^{4+\kappa}) < \infty\,,
    %     \sup_{1 \le l \le 2p+2q}\mathbb{E}\big|\be_{l}'\tilde{\bK} \{\bI_q \otimes {\rm vec}(\bX_{t-1})\otimes \bI_q\}\big|_2^{4+\kappa} < \infty\,,
    % \end{align*}
    % where $\be_l$ is the unit vector with the $l$-th element being 1.
\end{cd}

\begin{cd}\label{cond:ident}
    {\rm (i)} Let $\bI_{pq} - (D(\bbeta_0) \bV_0) \otimes (D(\balpha_0) \bW_0 )$ be invertible and all the eigenvalues of $\bPsi_0 := \{\bI_{pq} - (D(\bbeta_0) \bV_0) \otimes (D(\balpha_0) \bW_0 )\}^{-1}\{(D(\bbeta_1) \bV_1) \otimes (D(\balpha_1) \bW_1 )\}$ be smaller than 1 in modulus. {\rm (ii)} $\bW_{0}$, $\bW_{1}$, $\bV_{0}$ and $\bV_{1}$ are uniformly bounded in both row and column sums in absolute value.  
    %{\rm (iii)} For all $m$ and $j$, $\tilde{\bX}_{\ba_m}'\tilde{\bX}_{\ba_m}$ and $\tilde{\bX}_{\bb_j}'\tilde{\bX}_{\bb_j}$ are invertible. 
    {\rm (iii)} Let $\boldsymbol{\Theta}$ be the compact and convex parameter space. Assume $\boldsymbol{\gamma}$ is the unique solution of the Yule-Walker equation \eqref{class1_YW} in the interior of the parameter space $\boldsymbol{\Theta}$. 
\end{cd}

% {\color{red}
%  \begin{cd}\label{cond:moments}
%     For $\kappa$ specified in Condition \ref{cond:mixing}{\rm (i)} above, $\mathbb{E}(|{\rm vec}(\bX_{t})|_2^{4+\kappa}) < \infty$.
%     %$\sup_{1 \le l \le pq}\mathbb{E}(|\be_{l}{\rm vec}(\bX_{t})|^{4+\kappa}) < \infty$, where $\be_l$ is the $(pq)$-dimensional unit vector with the $l$-th element being 1.
%     % \begin{align*}
%     %     \sup_{1 \le l \le pq}\mathbb{E}(|\be_{l}{\rm vec}(\bX_{t})|^{4+\kappa}) < \infty\,,
%  \end{cd}
% }

Condition \ref{cond:mixing} specifies regularity conditions on the temporal dependence structure of the process $\{{\rm vec}(\bX_t)\}$ and as well as the tail behavior of its components. Such assumptions are standard in the literature on high-dimensional time series data analysis. Specifically,  Condition \ref{cond:mixing} {\rm (i)} specifies the rate of decay for the $\alpha$-mixing coefficients, which is also assumed in \cite{Dou2016}. See also Lemma 1 of \cite{Gao2019} that Condition \ref{cond:mixing} {\rm (i)} holds for the spatio-temporal models of vector time series data under certain regularity conditions. Condition \ref{cond:mixing} {\rm (ii)} is required to establish suitable tail probability bounds for the relevant statistics when $p$ and $q$ diverge with the sample size $n$. Condition \ref{cond:ident} imposes regularity conditions for the proposed model \eqref{model_class1_iter}. 
%provides the identification condition of model \eqref{model_class1_iter}. 
Specifically, Condition \ref{cond:ident}{\rm (i)} is assumed for obtaining the causality of the proposed model \eqref{model_class1_iter}. As we have mentioned in Section \ref{subsec:model_known_matrix}, after vectorization, model \eqref{model_class1_iter} can be represented in the form of the VAR(1) model in \eqref{class1_iter_vec}. Condition \ref{cond:ident}{\rm (i)} ensures that model \eqref{class1_iter_vec} is stationary and causal, which substantially facilitates the theoretical analysis and is necessary for establishing the main asymptotic results. Similar assumptions are commonly adopted in the related literature; see, for example, \cite{Dou2016}, \cite{Gao2019}, \cite{Chen2021} and \cite{Jiang2025}.
%Dou et al. (2016), Gao et al. (2019), Chen et al. (2021) and Jiang et al. (2025). 
In addition, since the generalized Yule–Walker estimator may become unstable in such near-unit-root settings, this condition also helps guarantee numerical stability in implementation and avoid ill-conditioned estimation problems.
Condition \ref{cond:ident}{\rm (ii)} imposes a restriction on the spatial correlation to ensure it remains at a manageable level, which is a standard assumption for spatial econometric models. See, for example, \cite{Kelejian2001} and \cite{Lee2010b}. Condition \ref{cond:ident}{\rm (iii)} %and {\rm (iv)} 
ensures that $\balpha_k$ and $\bbeta_k$, $k=0,1$, are uniquely identifiable based on the Yule-Walker equation \eqref{class1_YW}.
Proposition \ref{pn:consistency} shows that $\hat{\bgamma}$ is consistent.
\begin{proposition}\label{pn:consistency}
    Assume that Conditions \ref{cond:mixing}-\ref{cond:ident} hold. Let $d = p \vee q$. Then, %when $p,q \ge 1$ are fixed, 
    we have $\|\hat{\bgamma}-\bgamma\|_2 \stackrel{\mathrm{p}}{\to} 0$
    % \begin{align*}
    %     \|\hat{\bgamma}-\bgamma\|_2 \stackrel{\mathrm{p}}{\to} 0
    % \end{align*}
    as $n \to \infty$, provided that $d= o\{n^{1/(1+3\kappa)}\}$.
    %$d= o\{n^{2/(16+13\kappa)}\}$.
    %$d= o\{n^{2/(12+11\kappa)}\}$.  %$d = o\{n^{(4+\kappa)/(24+17\kappa)}\}$.
    %$|\hat{\bgamma}-\bgamma|_2 \stackrel{p}{\to} 0$ as $n \to \infty$.
\end{proposition}

To further investigate the asymptotic normality of $\hat{\bgamma}$, we introduce the following notation and regularity conditions.
Let
% \begin{align*}
%     \tilde{\bSigma}_{\bX,\bE} = \bSigma_{\bX,\bE}(0) + \sum_{k=1}^{\infty}\big\{ \bSigma_{\bX,\bE}(k) + \bSigma_{\bX,\bE}(k)'\big\}\,,
% \end{align*}
% where $\bSigma_{\bX,\bE}(k) = {\rm Cov}(\tilde{\boldsymbol{\mathcal{E}}}_{\bX,t},\tilde{\boldsymbol{\mathcal{E}}}_{\bX,t-k})$ for $k=0,1,\ldots$, with
\begin{align*}
   \tilde{\boldsymbol{\mathcal{E}}}_{\bX,t} = \big({\rm vec}'\{\bE_{t,\cdot 1}\bX_{t-1,\cdot 1}'\},\ldots,{\rm vec}'\{\bE_{t,\cdot 1}\bX_{t-1,\cdot q}'\},\ldots,{\rm vec}'\{\bE_{t,\cdot q}\bX_{t-1,\cdot 1}'\},\ldots,{\rm vec}'\{\bE_{t,\cdot q}\bX_{t-1,\cdot q}'\} \big)'\,.
\end{align*}
For any $m \in [p]$ and $j \in [q]$, let $\tilde{\bX}_{\ba_m}$ and $\tilde{\bX}_{\bb_j}$
% \begin{equation}\label{eq:Xam}
% \tilde{\bX}_{\ba_m} = 
% \begin{pmatrix}
% \sum_{i=1}^q  v_{1, i}^{(0)} \beta_{1}^{(0)} \bSigma_{i1}(1)^{'} \bW_0^{'}\be_m\,, & \sum_{i=1}^q  v_{1, i}^{(1)} \beta_{1}^{(1)} \bSigma_{i1}(0)^{'} \bW_1^{'}\be_m \\
% \sum_{i=1}^q  v_{1, i}^{(0)} \beta_{1}^{(0)} \bSigma_{i2}(1)^{'} \bW_0^{'}\be_m\,, & \sum_{i=1}^q  v_{1, i}^{(1)} \beta_{1}^{(1)} \bSigma_{i2}(0)^{'} \bW_1^{'}\be_m \\
% \vdots & \vdots \\
% \sum_{i=1}^q  v_{q, i}^{(0)} \beta_{q}^{(0)} \bSigma_{iq}(1)^{'} \bW_0^{'}\be_m\,, & \sum_{i=1}^q  v_{q, i}^{(1)} \beta_{q}^{(1)} \bSigma_{iq}(0)^{'} \bW_1^{'}\be_m 
% \end{pmatrix}
% \end{equation}
% and 
% \begin{equation}\label{eq:Xbj}
% \tilde{\bX}_{\bb_j} = 
% \begin{pmatrix}
% \sum_{i=1}^q \text{vec}(D(\balpha_0) \bW_0 \bSigma_{i1}(1) v_{j, i}^{(0)})\,,&  \sum_{i=1}^q \text{vec}(D(\balpha_1) \bW_1 \bSigma_{i1}(0) v_{j, i}^{(1)})\\
% \sum_{i=1}^q \text{vec}(D(\balpha_0) \bW_0 \bSigma_{i2}(1) v_{j, i}^{(0)})\,,&  \sum_{i=1}^q \text{vec}(D(\balpha_1) \bW_1 \bSigma_{i2}(0) v_{j, i}^{(1)})\\
% \vdots & \vdots \\
% \sum_{i=1}^q \text{vec}(D(\balpha_0) \bW_0 \bSigma_{iq}(1) v_{j, i}^{(0)})\,,&  \sum_{i=1}^q \text{vec}(D(\balpha_1) \bW_1 \bSigma_{iq}(0) v_{j, i}^{(1)})\\
% \end{pmatrix}
% \end{equation}
be $(pq^2) \times 2$ and $(p^2q) \times 2$ matrices obtained from $\bX_{\ba_m}$ and $\bX_{\bb_j}$, respectively, by replacing each $\hat{\bSigma}_{ik}(1)$ and $\hat{\bSigma}_{ik}(0)$ with their population counterparts $\bSigma_{ik}(1)$ and $\bSigma_{ik}(0)$.
%where $\be_m$ is the $p$-dimensional unit vector with the $m$-th element being 1. 
%Let $\tilde{\bP} = \tilde{\bK}\tilde{\bSigma}_{\bX,\bE}\tilde{\bK}'$, where $\tilde{\bK} = (\tilde{\bK}_{\ba}',\tilde{\bK}_{\bb}')'$ with
Let $\tilde{\bK} = (\tilde{\bK}_{\ba}', \tilde{\bK}_{\bb}')'$ and $\bS_n = n^{-1}\mathrm{Cov}(\sum_{t=2}^n \tilde{\bK}\tilde{\boldsymbol{\mathcal{E}}}_{\bX,t})$, where
\begin{align*}
    \tilde{\bK}_{\ba} = \begin{pmatrix}
        \tilde{\bX}_{\ba_1}'(\bI_{pq^2}\otimes\be_1')  \\
        \vdots \\
        \tilde{\bX}_{\ba_p}'(\bI_{pq^2}\otimes\be_p')
    \end{pmatrix} \in \mathbb{R}^{(2p)\times (pq)^2} ~\mbox{and}~     \tilde{\bK}_{\bb} = \begin{pmatrix}
        \tilde{\bX}_{\bb_1}' &   &  \\
        &   \ddots  & \\
        &   &  \tilde{\bX}_{\bb_q}'
    \end{pmatrix} \in \mathbb{R}^{(2q)\times (pq)^2}\,
\end{align*} 
with $\be_m$ denoting the $p$-dimensional unit vector whose $m$-th element is 1. Define 
\begin{align*}
    \tilde{\bZ}_{\ba_m,j} = \begin{pmatrix}
        {\bf 0}_{(j-1)pq \times 1}\,, &  {\bf 0}_{(j-1)pq \times 1} \\
        \alpha_{m}^{(0)}\sum_{i=1}^q v_{j, i}^{(0)}\bSigma_{i1}(1)' \bW_0'\be_m\,, & \alpha_{m}^{(1)}\sum_{i=1}^q v_{j, i}^{(1)}\bSigma_{i1}(0)' \bW_1'\be_m \\
        \vdots & \vdots  \\
        \alpha_{m}^{(0)}\sum_{i=1}^q v_{j, i}^{(0)}\bSigma_{iq}(1)' \bW_0'\be_m\,, & \alpha_{m}^{(1)}\sum_{i=1}^q v_{j, i}^{(1)}\bSigma_{iq}(0)' \bW_1'\be_m \\
        {\bf 0}_{(q-j)pq \times 1}\,, &  {\bf 0}_{(q-j)pq \times 1}
    \end{pmatrix}\,,
\end{align*}
where ${\bf 0}_{d_1 \times d_2}$ denote a $d_1 \times d_2$ matrix with all elements being zero, and
\begin{align*}
    \tilde{\bZ}_{\bb_j,m} = \begin{pmatrix}
        \beta_{j}^{(0)} \big\{ \sum_{i=1}^q (v_{j, i}^{(0)}\bSigma_{i1}(1)'\bW_0'\be_m)\otimes \be_m \big\}\,, & \beta_{j}^{(1)}\sum_{i=1}^q (v_{j, i}^{(1)}\bSigma_{i1}(0)'\bW_1'\be_m)\otimes \be_m \\
        \vdots & \vdots\\
        \beta_{j}^{(0)} \big\{ \sum_{i=1}^q (v_{j, i}^{(0)}\bSigma_{iq}(1)'\bW_0'\be_m)\otimes \be_m \big\}\,, & \beta_{j}^{(1)}\sum_{i=1}^q (v_{j, i}^{(1)}\bSigma_{iq}(0)'\bW_1'\be_m)\otimes \be_m
    \end{pmatrix}\,.
\end{align*}

Then, let
\begin{align}\label{eq:Umatrix}
    \bU_{\rm diag} = \begin{pmatrix}
        \tilde{\bX}_{\ba_1}'\tilde{\bX}_{\ba_1}\,, & \cdots & {\bf 0}_{2 \times 2}\,, & \tilde{\bX}_{\ba_1}'\tilde{\bZ}_{\ba_1,1}\,, & \cdots & \tilde{\bX}_{\ba_1}'\tilde{\bZ}_{\ba_1,q} \\
           &   \ddots  &  & \vdots  &  & \vdots  \\
        {\bf 0}_{2 \times 2}\,, & \cdots & \tilde{\bX}_{\ba_p}'\tilde{\bX}_{\ba_p}\,, & \tilde{\bX}_{\ba_p}'\tilde{\bZ}_{\ba_p,1}\,, & \cdots & \tilde{\bX}_{\ba_p}'\tilde{\bZ}_{\ba_p,q} \\
        \tilde{\bX}_{\bb_1}'\tilde{\bZ}_{\bb_1,1}\,,  & \cdots & \tilde{\bX}_{\bb_1}'\tilde{\bZ}_{\bb_1,p}\,, & \tilde{\bX}_{\bb_1}'\tilde{\bX}_{\bb_1}& \cdots & {\bf 0}_{2 \times 2}\\
         \vdots  &  & \vdots  &  &  \ddots  &    \\
        \tilde{\bX}_{\bb_q}'\tilde{\bZ}_{\bb_q,1}\,, & \cdots & \tilde{\bX}_{\bb_q}'\tilde{\bZ}_{\bb_q,p}\,, & {\bf 0}_{2 \times 2}& \cdots & \tilde{\bX}_{\bb_q}'\tilde{\bX}_{\bb_q}
    \end{pmatrix}\,.
\end{align}

%Now we impose the following regularity conditions on model \eqref{model_class1_iter}.
%\eqref{class1_iter_vec}.
%Define $\bPsi_0 = \{\bI_{pq} - (D(\bbeta_0) \bV_0) \otimes (D(\balpha_0) \bW_0 )\}^{-1}\{(D(\bbeta_1) \bV_1) \otimes (D(\balpha_1) \bW_1 )\}$.

{
\begin{cd}\label{cond:mmt_diagonalcase}
    %For $\kappa$ specified in Condition \ref{cond:mixing}{\rm (i)}, the following condition holds uniformly in $p$, $q$:
    {\rm (i)} For some universal constant $\rho_1 \in (2,3]$ satisfying $2\rho_1(\rho_1-1)/(\rho_1-2)^2 < (4+\kappa)/\kappa$ for $\kappa$ specified in Condition \ref{cond:mixing}{\rm (i)}, the following condition holds:
    \begin{align*}
        &~~~~~~~~ \max_{1\le t \le n}\sup_{\bu \in \mathbb{R}^{2p+2q}}\mathbb{E}|\bu'\tilde{\bK}\tilde{\boldsymbol{\mathcal{E}}}_{\bX,t}|^{\rho_1} <  \xi_n\,, 
    \end{align*}
    where $\xi_n$ is a sequence possibly depending on $n$. 
    {\rm (ii)} There exists a universal constant $c_1>0$ such that, for any $n \ge 1$,  $\lambda_{\min}(\bS_n) \ge c_1$, where $\lambda_{\min}(\bS_n)$ denotes the smallest eigenvalue of $\bS_n$.  {\rm (iii)} $\bU_{\rm diag}$ specified in \eqref{eq:Umatrix} is invertible.
    %For any $\bu \in \mathbb{R}^{2p+2q}$ and $n \ge 1$, there is a constant $c_1>0$ such that $\bu'\bS_n\bu \ge c_1 n$.
\end{cd}
}

% \begin{cd}\label{cond:mmt_diagonalcase}
%     For $\kappa$ specified in Condition \ref{cond:mixing}{\rm (i)}, the following condition holds uniformly in $p$, $q$:
%     \begin{align*}
%         &~~~~~~~~ \sup_{\bu \in \mathbb{R}^{2p}}\mathbb{E}|\bu'\tilde{\bK}_{\ba}\tilde{\boldsymbol{\mathcal{E}}}_{\bX,t}|^{\frac{4+\kappa}{2}} < {\color{red} \xi_n}\,, ~ \sup_{\bu \in \mathbb{R}^{2q}}\mathbb{E}|\bu'\tilde{\bK}_{\bb}\tilde{\boldsymbol{\mathcal{E}}}_{\bX,t}|^{\frac{4+\kappa}{2}} < {\color{red} \xi_n}\,, 
%     \end{align*}
%     {\color{red} where $\xi_n$ is a sequence possibly depending on $n$.}
% \end{cd}

%in the sense that $D(\bbeta_k) \otimes D(\balpha_k)$ is unique. 
%Analogous to the assumption in Chen et al. (2021), we use the convention that $\balpha_k$ is normalized with  $\|\balpha_k\|_2=1$. 
Condition \ref{cond:mmt_diagonalcase}(i) specifies the moment bound on $\tilde{\boldsymbol{\mathcal{E}}}_{\bX,t}$, which controls the spatial dependence between different elements of $\bX_t$. When the dimensions $p$ and $q$ diverge, the bound $\xi_n$ is allowed to grow with $n$. As a result, the condition on $\xi_n$ in Theorem \eqref{tm:mixing} implicitly requires a restriction on the growth rate of $p$ and $q$. Condition \ref{cond:mmt_diagonalcase}(ii) and (iii) ensure a non-degenerate limiting distribution. 
%Condition \ref{cond:mmt_diagonalcase}(ii) ensures that $\bS_n$ is uniformly non-degenerate, which is a standard assumption in high-dimensional analysis, and Condition \ref{cond:mmt_diagonalcase}(iii) ensures that $\bU_{\rm diag}$ is invertible. 

\begin{theorem}\label{tm:mixing}
    Assume that Conditions \ref{cond:mixing}-\ref{cond:mmt_diagonalcase} hold. Let $d = p \vee q$ and $\beta_1 = \{(4+\kappa)(\rho_1-2)^2\}/\{2\kappa \rho_1(\rho_1-1)\}$. Then, if $d = o(n^{\nu})$ with $\nu = \min\{1/11, 2/(10+9\kappa)\}$, and $\xi_n = o\{n^{(\beta_1-1)(\rho_1-2)/(2\beta_1+2)}\}$, as $n \to \infty$,
    %if $d = o(n^{\nu})$ with $\nu = \max\{1/9, 2/(16+13\kappa)\}$,
    %when $p,q \ge 1$ are fixed, 
    we have 
    \begin{align*}
    \sqrt{n}\bS_n^{-1/2}\bU_{\rm diag} (\hat{\bgamma}-\bgamma) \stackrel{\mathrm{d}}{\to} \mathcal{N}(0,\bI_{2(p+q)})\,.
\end{align*}
\end{theorem}

\noindent {\bf Remark 1} 
    If $p$ and $q$ are fixed, the limiting form of $\bS_n$ can be characterized explicitly. See Proposition \ref{CLT:diag_fixdim} in Section \ref{sec:diag_fixdim} of the supplementary material for details.

%In Theorem \ref{tm:mixing}, we do not impose any structural assumptions on $\bW_k$ and $\bV_k$ for $k=0,1$. If there is prior information of the correlations among different rows and columns in $\bX_t$, for example, $\bW_k$ and $\bV_k$ are banded, the condition $d = o\{n^{\nu}\}$ with $\nu = \max\{1/9, 2/(16+13\kappa)\}$ can be further improved.

\section{Spatio-temporal regression with banded weight matrices}\label{sec:model_banded}
\subsection{Models} \label{subsec:model_unknown_matrix}
Consider the matrix spatio-temporal autoregression 
\begin{equation}\label{model_all}
\bX_t = \bA_0 \bX_t \bB_0' + \bA_1 \bX_{t-1} \bB_1' + \bE_t\,,
\end{equation}
where weight matrices $\bA_k = (a_{i, j}^{(k)})$ and $\bB_k = (b_{i, j}^{(k)})$, $k = 0, 1$, are unknown with the banded structures in the sense that
\begin{equation} \label{band}
a_{i, j}^{(k)} = 0 \, \, \text{for all } |i-j| > k_0^{(\bA)}, \,\, b_{i, j}^{(k)} = 0 \, \, \text{for all } |i-j| > k_0^{(\bB)}\,,
\end{equation}
and $ k_0^{(\bA)} (< p)$ and $ k_0^{(\bB)} (< q)$ are
two unknown bandwidth parameters. Similar to model (\ref{model_class1_iter}), the main diagonal elements of $\bA_0$ are assumed to be 0.

Model (\ref{model_all}) can be represented in the form of a vector autoregressive model
\begin{equation}\label{model_vec}
\text{vec}(\bX_t) = (\bB_0 \otimes \bA_0)\text{vec}(\bX_t) +  (\bB_1 \otimes \bA_1)\text{vec}(\bX_{t-1}) + \text{vec}(\bE_t)\,,
\end{equation}
where $\otimes$ is the matrix Kronecker product. Note that the parameters in the model are not uniquely identifiable as, for example,  $\bA_k$ and $\bB_k$ can be replaced by $c\bA_k$ and $c^{-1}\bB_k$ for any non-zero constant $c$. But Kronecker product $\bB_k \otimes \bA_k$ is unique and identifiable. Following Chen et al. (2021), we use the convention that, for $k=0,1$, $\bA_k$ is normalized with Frobenius norm equals to 1 and its first nonzero entry is positive. Let $\bI_{pq} - \bB_0 \otimes \bA_0$ be invertible and all eigenvalues of $(\bI_{pq} - \bB_0 \otimes \bA_0)^{-1}(\bB_1 \otimes \bA_1)$ be smaller than 1 in modulus. Model (\ref{model_vec}) can then be written as
\begin{equation}\label{model_reduced}
\text{vec}(\bX_t) = (\bI_{pq} - \bB_0 \otimes \bA_0)^{-1}(\bB_1 \otimes \bA_1)\text{vec}(\bX_{t-1}) + (\bI_{pq} - \bB_0 \otimes \bA_0)^{-1}\text{vec}(\bE_t)\,,
\end{equation}
and it defines a (weakly) stationary process. For this stationary process, $\mathbb{E}(\bX_t) = {\bf 0}$, and the $j$-th column of $\bX_t$, $1\le j \le q$, admits the form 
\begin{equation}\label{model_col}
\bX_{t, \cdot j} = \sum_{i=1}^q \bA_0 \bX_{t, \cdot i} b_{j,i}^{(0)} + \sum_{i=1}^q \bA_1 \bX_{t-1, \cdot i} b_{j,i}^{(1)} + \bE_{t, \cdot j}\,.
\end{equation}
Multiplying both sides of (\ref{model_col}) by $\bX_{t-1, .k}'$ from the back, and taking expectation on both sides, we obtain the following Yule-Walker equations
\begin{equation}\label{YW}
\bSigma_{jk}(1) = \sum_{i=1}^q \bA_0\bSigma_{ik}(1)b_{j,i}^{(0)} + \sum_{i=1}^q \bA_1 \bSigma_{ik}(0) b_{j,i}^{(1)}\,, \quad 1\le j,k \le q\,,
\end{equation}
where $\bSigma_{jk}(1) = \cov(\bX_{t, \cdot j}, \bX_{t-1, \cdot k})$ and $\bSigma_{jk}(0) = \cov(\bX_{t, \cdot j}, \bX_{t, \cdot k})$. 

Although \eqref{model_reduced} (and similarly \eqref{class1_iter_vec_reduced}) resembles a matrix autoregressive model upon vectorization of the matrix data, our model is fundamentally different. The reduced form is merely an algebraic equivalence and does not reflect the underlying data-generating process. The key distinction lies in how spatial and temporal dependencies are treated. In the matrix autoregressive model proposed by \cite{Chen2021}, the dynamic behavior is governed entirely by a homogeneous autoregressive process. In contrast, our model explicitly separates spatial interactions from temporal evolution — just as traditional SDPD models can be rewritten as reduced-form VAR processes without becoming purely temporal models. Moreover, the reduced form would involve different and more complicated coefficient matrices and innovation structures than those of a standard matrix AR model after being written in a reduced AR form.

\subsection{Estimation}\label{sec:est_band}
We assume in this subsection that bandwidth parameters $k_0^{(\bA)}$ and $k_0^{(\bB)}$ are known. Following Yule-Walker equations \eqref{YW}, the generalized Yule-Walker estimator is obtained by solving the following optimization problem:
\begin{align}\label{eq:Banded_opt1}
%\min_{\bA_0\in \tilde{\mathcal{B}}(k_0^{(\bA)}),\bA_1\in \mathcal{B}(k_0^{(\bA)}), \bB_0, \bB_1\in \mathcal{B}(k_0^{(\bB)})} 
\frac{1}{p^2q^2}\sum_{j,k=1}^q\Big\|\hat{\bSigma}_{jk}(1) - \sum_{i=1}^q \bA_0\hat{\bSigma}_{ik}(1)b_{j,i}^{(0)} - \sum_{i=1}^q \bA_1 \hat{\bSigma}_{ik}(0) b_{j,i}^{(1)} \Big\|_F^2 \,,
\end{align}
with respect to the nonzero entries of $\bA_0$, $\bA_1$, $\bB_0$ and $\bB_1$, which are banded as defined in \eqref{band}.  We propose an iterated method to estimate $\bA_k$ and $\bB_k$ in a recursive manner. 
\paragraph{Estimating $\bB_0$ and $\bB_1$ given $\bA_0$ and $\bA_1$:}
Denote the $j$-th row of $\bB_0$ by $\bb_j^{(0)} = (b_{j,1}^{(0)}, \ldots,  b_{j,q}^{(0)})'$ and the $j$-th row of $\bB_1$ by $\bb_j^{(1)} = (b_{j,1}^{(1)}, \ldots,  b_{j,q}^{(1)})'$, that is, $\bB_0 = ({\bb_1^{(0)}}', \ldots, {\bb_q^{(0)}}')'$ and $\bB_1 = ({\bb_1^{(1)}}', \ldots, {\bb_q^{(1)}}')'$. As $\bB_k$, $k=0, 1$, is banded as defined in (\ref{band}), denote $\mathcal{B}_{j}^{(0)} = \{j_{1}^{(0)}, \ldots, j_{{\ell_{j,0}}}^{(0)}\}$
%$S_{\bb_j^{(0)}} = \{j_{1}^{(0)}, \ldots, j_{{|S_{\bb_j^{(0)}}|}}^{(0)}\}$ 
as the set of indices of nonzero elements in $\bb_j^{(0)}$ and $\ell_{j,0}$
%$|S_{\bb_j^{(0)}}|$ 
as its cardinality and denote $\mathcal{B}_{j}^{(1)} = \{j_{1}^{(1)}, \ldots, j_{{\ell_{j,1}}}^{(1)}\}$
%$S_{\bb_j^{(1)}} = \{j_{1}^{(1)}, \ldots, j_{{|S_{\bb_j^{(1)}}|}}^{(1)}\}$ 
as the set of indices of nonzero elements in $\bb_j^{(1)}$ and $\ell_{j,1}$
%$|S_{\bb_j^{(1)}}|$ 
as its cardinality. For simplicity, write $\ell_{j} = \ell_{j,0}+\ell_{j,1}$. Given $\bA_0, \bA_1$, by \eqref{eq:Banded_opt1}, we %aim to minimize
solve the following optimization problem with respect to $\bB_0$ and $\bB_1$:
\begin{equation*}
\min_{\bB_0, \bB_1}\frac{1}{p^2q^2}\sum_{j, k=1}^q \Big\|\hat\bSigma_{jk}(1) - \sum_{i \in   \mathcal{B}_{j}^{(0)}} \bA_0\hat\bSigma_{ik}(1)b_{j,i}^{(0)} - \sum_{i \in \mathcal{B}_{j}^{(1)}} \bA_1 \hat\bSigma_{ik}(0) b_{j,i}^{(1)} \Big\|_F^2\,.
\end{equation*}
% Yule-Walker equation (\ref{YW}) implies  
% \begin{equation*}
% \bSigma_{jk}(1) = \sum_{i \in  {\color{red}\mathcal{B}_{j}^{(0)}}} \bA_0\bSigma_{ik}(1)b_{j,i}^{(0)} + \sum_{i \in  {\color{red}\mathcal{B}_{j}^{(1)}}} \bA_1 \bSigma_{ik}(0) b_{j,i}^{(1)}\,. 
% \end{equation*}
Denote $\tilde \bb_j^{(0)}$ as the vector of nonzero elements in $\bb_j^{(0)}$ and $\tilde \bb_j^{(1)}$ as the vector of nonzero elements in $\bb_j^{(1)}$. The above problem lead to solve
%which can be rewritten as 
% \begin{equation*}
% {\color{red}\frac{1}{p^2q^2}}\sum_{j=1}^q \min_{\tilde \bb_j^{(0)}, \tilde \bb_j^{(1)}} \sum_{k=1}^q \Big\|\hat\bSigma_{jk}(1) - \sum_{i \in   {\color{red}\mathcal{B}_{j}^{(0)}}} \bA_0\hat\bSigma_{ik}(1)b_{j,i}^{(0)} - \sum_{i \in  {\color{red}\mathcal{B}_{j}^{(1)}}} \bA_1 \hat\bSigma_{ik}(0) b_{j,i}^{(1)} \Big\|_F^2\,,
% \end{equation*}
% and is equivalent to
\begin{equation}\label{opt_B}
 \min_{\tilde \bb_j^{(0)}, \tilde \bb_j^{(1)}} \frac{1}{p^2q^2}\sum_{k=1}^q \Big\|\text{vec}\{\hat\bSigma_{jk}(1)\} - \sum_{i \in  \mathcal{B}_{j}^{(0)}} \text{vec}\{\bA_0\hat\bSigma_{ik}(1)\}b_{j,i}^{(0)} - \sum_{i \in  \mathcal{B}_{j}^{(1)}} \text{vec}\{\bA_1 \hat\bSigma_{ik}(0)\} b_{j,i}^{(1)} \Big\|_2^2
\end{equation}
for each $j=1,\ldots,q$.
Stacking across $k$, (\ref{opt_B}) becomes 
\begin{equation} \label{LS_B}
 \min_{\tilde \bb_j} \frac{1}{p^2q^2} \| \bY_{\tilde \bb_j} - \bX_{\tilde \bb_j}\tilde \bb_j\|_2^2\,,
\end{equation}
where
\begin{equation*}
\bY_{\tilde \bb_j} = 
\begin{pmatrix}
\text{vec}(\hat \bSigma_{j1}(1)) \\
\text{vec}(\hat \bSigma_{j2}(1)) \\
\vdots\\
\text{vec}(\hat \bSigma_{jq}(1))
\end{pmatrix}
\end{equation*}
is a $p^2 q \times 1$ vector,
\begin{equation*}
\bX_{\tilde{\bb}_j} = 
\begin{pmatrix}
\text{vec}\{\bA_0\hat\bSigma_{j_{1}^{(0)}1}(1)\}, \cdots, \text{vec}\{\bA_0\hat\bSigma_{j_{\ell_{j,0}}^{{(0)}}1}(1)\}, \text{vec}\{\bA_1\hat\bSigma_{j_{1}^{(1)}1}(0)\}, \cdots, \text{vec}\{\bA_1\hat\bSigma_{j_{\ell_{j,1}}^{{(1)}}1}(0)\} \\
\text{vec}\{\bA_0\hat\bSigma_{j_{1}^{(0)}2}(1)\}, \cdots, \text{vec}\{\bA_0\hat\bSigma_{j_{\ell_{j,0}}^{{(0)}}2}(1)\}, \text{vec}\{\bA_1\hat\bSigma_{j_{1}^{(1)}2}(0)\}, \cdots, \text{vec}\{\bA_1\hat\bSigma_{j_{\ell_{j,1}}^{{(1)}}2}(0)\}\\
\vdots \\
\text{vec}\{\bA_0\hat\bSigma_{j_{1}^{(0)}q}(1)\}, \cdots, \text{vec}\{\bA_0\hat\bSigma_{j_{\ell_{j,0}}^{{(0)}}q}(1)\}, \text{vec}\{\bA_1\hat\bSigma_{j_{1}^{(1)}q}(0)\}, \cdots, \text{vec}\{\bA_1\hat\bSigma_{j_{\ell_{j,1}}^{{(1)}}q}(0)\}
\end{pmatrix}
\end{equation*}
is a $p^2 q \times \ell_j$ matrix, and 
\begin{equation*}
\tilde \bb_j = 
\begin{pmatrix}
\tilde \bb_j^{(0)} \\
\tilde \bb_j^{(1)}
\end{pmatrix}
\end{equation*}
is a $\ell_j \times 1$ vector of nonzero parameters. Hence (\ref{LS_B}) leads to the least squares estimator 
\begin{equation*}
\hat {\tilde \bb}_j = 
\begin{pmatrix}
\hat {\tilde \bb}_j^{(0)} \\
\hat {\tilde \bb}_j^{(1)}
\end{pmatrix} = ({\bX_{\tilde \bb_j}}'\bX_{\tilde \bb_j})^{-1}{\bX_{\tilde \bb_j}}'\bY_{\tilde \bb_j}, \, ~~ 1 \le j \le q\,. 
\end{equation*}

\paragraph{Estimating $\bA_0$ and $\bA_1$ given $\bB_0$ and $\bB_1$:}
Denote the $m$-th row of $\bA_0$ by $\ba_m^{(0)} = (a_{m,1}^{(0)}, \cdots,  a_{m,p}^{(0)})'$ and the $m$-th row of $\bA_1$ by $\ba_m^{(1)} = (a_{m,1}^{(1)}, \cdots,  a_{m,p}^{(1)})'$, that is, $\bA_0 = ({\ba_1^{(0)}}, \cdots, {\ba_p^{(0)}})'$ and $\bA_1 = ({\ba_1^{(1)}}, \cdots, {\ba_p^{(1)}})'$. Due to $\bA_k$ is banded as defined in (\ref{band}), for $m = 1, \cdots, p$, denote $\mathcal{A}_m^{(0)} = \{m_1^{(0)},\ldots,m_{r_{m,0}}^{(0)}\}$
%$S_{\ba_m^{(0)}} = \{ m_1^{(0)}, m_2^{(0)}, \cdots, m_{|S_{\ba_m^{(0)}}|}^{(0)} \}$ 
as the set of indices of nonzero elements in $\ba_m^{(0)}$ and $r_{m,0}$ is its cardinality. Analogously, denote $\mathcal{A}_m^{(1)} = \{m_1^{(1)},\ldots,m_{r_{m,1}}^{(1)}\}$
%$S_{\ba_m^{(1)}} = \{ m_1^{(1)}, m_2^{(1)}, \cdots, m_{|S_{\ba_m^{(1)}}|}^{(1)} \}$ 
as the set of indices of nonzero elements in $\ba_m^{(1)}$ and $r_{m,1}$ is its cardinality. For simplicity, write $r_{m} = r_{m,0}+r_{m,1}$. Denote $\tilde \ba_m^{(0)}$ as the vector of nonzero elements in $\ba_m^{(0)}$ and $\tilde \ba_m^{(1)}$ as the vector of nonzero elements in $\ba_m^{(1)}$.  Denote $\be_h$ as the $p \times 1$ unit vector with the $h$-th element being zero. Given $\bB_0$ and $\bB_1$, by \eqref{eq:Banded_opt1}, we %aim to minimize
solve the following optimization problem with respect to $\bA_0$ and $\bA_1$:
\begin{equation*}
\min_{\bA_0, \bA_1}\frac{1}{p^2q^2}\sum_{j, k=1}^q\sum_{m=1}^p \bigg\|\hat\bSigma_{jk}(1)'\be_m - \left(\sum_{i=1}^q b_{j,i}^{(0)}\hat \bSigma_{ik}(1)' \bE_{{ m}}^{(0)} \right) \tilde \ba_m^{(0)} - \left(\sum_{i=1}^q  b_{j,i}^{(1)}\hat \bSigma_{ik}(0)' \bE_{{ m}}^{(1)} \right) \tilde \ba_m^{(1)}  \bigg\|_2^2\,,
\end{equation*}
% Yule-Walker equation (\ref{YW}) implies
% \[
% \bSigma_{jk}(1)' \be_m = \left(\sum_{i=1}^q b_{j,i}^{(0)}\bSigma_{ik}(1)' \bE_{{ m}}^{(0)} \right) \tilde \ba_m^{(0)} + \left(\sum_{i=1}^q  b_{j,i}^{(1)}\bSigma_{ik}(0)' \bE_{{ m}}^{(1)} \right) \tilde \ba_m^{(1)}\,,\quad  m = 1, \ldots, p\,. 
% \]
where $\bE_{{ m}}^{(0)} = (\be_{m_1^{(0)}}, \ldots, \be_{m^{(0)}_{r_{m,0}}})$ is a $p \times r_{m,0}$ matrix and $\bE_{{ m}}^{(1)} = (\be_{m_1^{(1)}}, \ldots, \be_{m^{(1)}_{r_{m,1}}} )$ is  a $p \times {r_{m,1}}$ matrix. The above problem lead to solve
%which is equivalent to 
\begin{equation}\label{opt_A}
\min_{\tilde \ba_m^{(0)}, \tilde \ba_m^{(1)}}\frac{1}{p^2q^2} \sum_{j, k=1}^q \bigg\|\hat\bSigma_{jk}(1)'\be_m - \left(\sum_{i=1}^q b_{j,i}^{(0)}\hat \bSigma_{ik}(1)' \bE_{{m}}^{(0)} \right) \tilde \ba_m^{(0)} - \left(\sum_{i=1}^q  b_{j,i}^{(1)}\hat \bSigma_{ik}(0)' \bE_{{ m}}^{(1)} \right) \tilde \ba_m^{(1)}  \bigg\|_2^2
\end{equation}
for each $m=1,\ldots,p$. 
Stacking across $j$ and $k$, (\ref{opt_A}) becomes
\begin{equation} \label{LS_A}
\min_{\tilde \ba_m} \frac{1}{p^2q^2}  \| \bY_{\tilde \ba_m} - \bX_{\tilde \ba_m}\tilde \ba_m\|_2^2\,,
\end{equation}
where
\begin{equation*}
\bY_{\tilde \ba_m} = 
\begin{pmatrix}
\hat \bSigma_{11}(1)' \be_m \\
\hat \bSigma_{12}(1)' \be_m \\
\vdots \\
\hat \bSigma_{qq}(1)' \be_m,
\end{pmatrix}
\end{equation*}
is a $pq^2 \times 1$ vector,
\begin{equation*}
\bX_{\tilde \ba_m} = 
\begin{pmatrix}
\sum_{i=1}^q b_{1,i}^{(0)}\hat \bSigma_{i1}(1)' \bE_{{ m}}^{(0)}\,,~ \sum_{i=1}^q  b_{1,i}^{(1)}\hat \bSigma_{i1}(0)' \bE_{{ m}}^{(1)} \\
\sum_{i=1}^q b_{1,i}^{(0)}\hat \bSigma_{i2}(1)' \bE_{{ m}}^{(0)}\,,~ \sum_{i=1}^q  b_{1,i}^{(1)}\hat \bSigma_{i2}(0)' \bE_{{ m}}^{(1)} \\
\vdots \\
\sum_{i=1}^q b_{q,i}^{(0)}\hat \bSigma_{iq}(1)' \bE_{{ m}}^{(0)}\,,~ \sum_{i=1}^q  b_{q,i}^{(1)}\hat \bSigma_{iq}(0)' \bE_{{ m}}^{(1)} 
\end{pmatrix}
\end{equation*}
is a $pq^2 \times r_{m}$ matrix, and 
\begin{equation*}
\tilde \ba_m = 
\begin{pmatrix}
\tilde \ba_m^{(0)} \\
\tilde \ba_m^{(1)}
\end{pmatrix}
\end{equation*}
is a $r_{m} \times 1$ vector of parameters. This leads to the least squares estimator
%The least squares estimator is 
\begin{equation*}
\hat {\tilde \ba}_m = 
\begin{pmatrix}
\hat { \tilde \ba}_m^{(0)} \\
\hat { \tilde \ba}_m^{(1)}
\end{pmatrix} = ({\bX_{\tilde \ba_m}}'\bX_{\tilde \ba_m})^{-1}{\bX_{\tilde \ba_m}}'\bY_{\tilde \ba_m}, \,~~ 1 \le m \le p\,.
\end{equation*}

The iterative estimation process will stop if 
\[
\sum_{k=0}^1 \|\hat \bB_k^{(l)} \otimes \hat \bA_k^{(l)} -\hat\bB_k^{(l-1)} \otimes \hat\bA_k^{(l-1)} \|_F
\]
is close to zero, where $\hat \bB_k^{(l)}$ and $\hat \bA_k^{(l)}$ are the estimators of $\bB_k$ and $\bA_k$ from the $(l-1)$-th and $l$-th iterations. 

\subsection{Initialization based on NKP}
We initialize the iterated least squares estimation through a two-stage procedure utilizing the nearest Kronecker product (NKP) algorithm. 

The first stage involves recasting the matrix spatio-temporal autoregression model (\ref{model_all}) in vectorized form
\begin{equation}\label{model_vec_phi}
\text{vec}(\bX_t) = \bC_0 \text{vec}(\bX_t) +  \bC_1 \text{vec}(\bX_{t-1}) + \text{vec}(\bE_t)\,,
\end{equation}
where $\bC_0 = \bB_0 \otimes \bA_0$ and $\bC_1 = \bB_1 \otimes \bA_1$. Given $k_0^{(\bA)}$ and $k_0^{(\bB)}$, $\bC_0$ and $\bC_1$ can be estimated by applying the generalized Yule-Walker estimation method from \cite{Dou2016} and \cite{Gao2019},
%Dou et al. (2016) and Gao et al. (2019), 
and we denote their estimators by $\hat \bC_0, \hat \bC_1$. 

In the second stage, we estimate $\bA_0, \bB_0$ and $\bA_1, \bB_1$ by solving the optimization problem
\begin{equation}\label{NKP0}
(\tilde \bA_0, \tilde \bB_0) = \arg \min_{\bA_0, \bB_0} \|\hat \bC_0 -  \bB_0 \otimes \bA_0\|_F^2\,,
\end{equation}
and 
\begin{equation}\label{NKP1}
(\tilde \bA_1, \tilde \bB_1) = \arg \min_{\bA_1, \bB_1} \|\hat \bC_1 -  \bB_1 \otimes \bA_1\|_F^2\,,
\end{equation}
Minimization problems (\ref{NKP0}) and (\ref{NKP1}) can be addressed by the algorithm for NKP problem in matrix computation, see \cite{VanLoan2000}.
%Van Loan (2000). 
We use (\ref{NKP0}) to illustrate its idea. Note that all the elements in $\bB_0 \otimes \bA_0$ and $\text{vec}(\bA_0)\text{vec}(\bB_0)'$ are the same but the location of each element is different. And there exists a one to one mapping between the location of the same element in $\bB_0 \otimes \bA_0$ and $\text{vec}(\bA_0)\text{vec}(\bB_0)'$. By applying such mapping from $\bB_0 \otimes \bA_0$ to $\text{vec}(\bA_0)\text{vec}(\bB_0)'$ on $\hat \bC_0$, we obtain $\tilde \bC_0$, which shares the same elements as $\hat \bC_0$ but the locations of the elements are different than $\hat \bC_0$. And (\ref{NKP0}) is equivalent to
\begin{equation}\label{NKP0_vec}
(\tilde \bA_0, \tilde \bB_0) = \arg \min_{\bA_0, \bB_0} \|\tilde \bC_0 -  \text{vec}(\bA_0)\text{vec}(\bB_0)'\|_F^2\,,
\end{equation}
and the solution of (\ref{NKP0_vec}) can be obtained from the singular value decomposition of $\tilde \bC_0$ and satisfies 
\[
\text{vec}(\tilde \bA_0)\text{vec}(\tilde \bB_0)' = d_1 \bu_1\bv_1'\,,
\]
where $d_1$ is the largest singular value of $\tilde \bC_0$, $\bu_1$ and $\bv_1$ are the corresponding left and right singular vectors of $\tilde \bC_0$. $\tilde \bA_0$ and $\tilde \bB_0$ are obtained by transforming $\text{vec}(\tilde \bA_0)$ and $\text{vec}(\tilde \bB_0)$ into corresponding matrices with constraint 
%either
$\|\tilde \bA_0\|_F =1$. 
%{\color{blue}or $\|\tilde \bB_0\|_F=1$}. 
$\tilde \bA_1$ and $\tilde \bB_1$ can be obtained in the same way. 

While the proposed method can theoretically function as a standalone estimator, its practical implementation faces challenges when dealing with high-dimensional problems. The initial estimation step requires computing large $pq \times pq$ matrices, which may lead to suboptimal accuracy in finite samples (particularly when $p$ and $q$ are large relative to $n$), despite the estimator's asymptotic consistency. Nevertheless, our numerical experiments demonstrate that using this estimate as an initialization for the iterated least squares procedure significantly enhances the final estimation accuracy.  

\subsection{Determination of bandwidth parameters}\label{sc:bandwidth}
Starting from the vectorized spatio-temporal specification in (\ref{model_vec_phi}), we adapt the generalized Yule-Walker estimation framework established by \cite{Dou2016} and \cite{Gao2019}
%Dou et al. (2016) and Gao et al. (2019) 
to estimate $\bC_0$ and $\bC_1$ as below
\begin{equation}
\arg\min_{\bC_0, \bC_1} \sum_{i=1}^{pq} \| \hat \bSigma_{\bX}(1)' \be_i - \hat \bSigma_{\bX}(1)' \bc^{(0)}_i -  \hat \bSigma_{\bX}(0)' \bc^{(1)}_i\|_2^2 \,,  
\end{equation}
where 
\[
\hat \bSigma_{\bX}(1) = \frac{1}{n} \sum_{t=2}^n \text{vec}(\bX_t) \text{vec}(\bX_{t-1})', \; \; \hat \bSigma_{\bX}(0) = \frac{1}{n} \sum_{t=1}^n \text{vec}(\bX_t) \text{vec}(\bX_{t})'\,,
\]
$\bc^{(0)}_i$ and $\bc^{(1)}_i$ are the $i$-th row of $\bC_0$ and $\bC_1$, respectively, and $\be_i$ denotes the unit vector with 1 as the $i$-th element.   

For any $k^{(\bA)}$ and $k^{(\bB)}$, define
\begin{equation} \label{RSS_vec}
\text{RSS}_i(k^{(\bA)}, k^{(\bB)}) = \| \hat \bSigma_{\bX}(1)' \be_i - \hat \bSigma_{\bX}(1)' \bc^{(0)}_i -  \hat \bSigma_{\bX}(0)' \bc^{(1)}_i\|_2^2\,.
\end{equation}
Let $K > 0$ be a known upper bound for both $k_0^{(\bA)}$ and $k_0^{(\bB)}$. 
%Observe that if replacing $\hat \bSigma_{\bX}(1)$ and $\hat %\bSigma_{\bX}(0)$ by their true values $\bSigma_{\bX}(1)$ and %$\bSigma_{\bX}(0)$ in (\ref{RSS_vec}), we have
%\begin{equation*}
%\text{RSS}_i(k^{(\bA)}, k^{(\bB)}) = 0 \iff k_0^{(\bA)} \le %k^{(\bA)} \le K \text{ and } k_0^{(\bB)} \le k^{(\bB)} \le K
%\end{equation*}
%while
%\begin{equation}
%\text{RSS}_i(k^{(\bA)}, k^{(\bB)}) \in (0, \infty) \text{ if } %k^{(\bA)} < k_0^{(\bA)} \text{ or } k^{(\bB)} < k_0^{(\bB)}
%\end{equation}
We define the RSS difference metric as:
\begin{align} \label{eq:Delta_RSS}
\Delta\text{RSS}_i(k^{(\bA)}, k^{(\bB)}) &:= \frac{1}{2}\Big[
    \underbrace{\text{RSS}_i(k^{(\bA)}, k^{(\bB)}) - \text{RSS}_i(k^{(\bA)}+1, k^{(\bB)})}_{\text{A-direction change}} \notag\\
    &~~~~~~~~~~~~~~+ \underbrace{\text{RSS}_i(k^{(\bA)}, k^{(\bB)}) - \text{RSS}_i(k^{(\bA)}, k^{(\bB)}+1)}_{\text{B-direction change}}
\Big]\,.
\end{align}
Using the true covariance matrices $\bSigma_{\bX}(0)$ and $\bSigma_{\bX}(1)$ rather than their empirical estimates, the RSS difference metric exhibits the following properties:
\begin{equation}
\Delta\text{RSS}_i(k^{(\bA)}, k^{(\bB)}) = 
\begin{cases} 
0, & \text{if } k^{(\bA)} \in [k_0^{(\bA)}, K] \text{ and } k^{(\bB)} \in [k_0^{(\bB)}, K]\,, \\ \text{finite}, & \text{if } k^{(\bA)} < k_0^{(\bA)} \text{ or } k^{(\bB)} < k_0^{(\bB)}\,,
\end{cases}
\end{equation} 
For a fixed $k^{(\bA)}$, the ratio of successive $\Delta\text{RSS}_i$ terms exhibits distinct behavior:
\begin{itemize}
    \item When $k^{(\bA)} < k_0^{(\bA)}$:
    \begin{equation}
        \frac{\Delta\text{RSS}_i(k^{(\bA)}, k^{(\bB)} - 1)}{\Delta\text{RSS}_i(k^{(\bA)}, k^{(\bB)})} \text{ remains finite for all } k^{(\bB)}
    \end{equation}

    \item When $k^{(\bA)} \geq k_0^{(\bA)}$:
    \begin{equation}
        \frac{\Delta\text{RSS}_i(k^{(\bA)}, k^{(\bB)} - 1)}{\Delta\text{RSS}_i(k^{(\bA)}, k^{(\bB)})} =
        \begin{cases}
            \text{finite}, & k^{(\bB)} < k_0^{(\bB)} \\
            \text{excessively large}, & k^{(\bB)} = k_0^{(\bB)} \\
            0/0 , & k^{(\bB)} > k_0^{(\bB)}
        \end{cases}
    \end{equation}
\end{itemize}
Hence we can estimate $k_0^{(\bA)}$ and $k_0^{(\bB)}$ in the following way. Define
\begin{align*}
    (\hat{k}_{i}^{(\bA)}, \hat{k}_{i}^{(\bB)}) = \arg\max_{1 \le k^{(\bA)},k^{(\bB)} \le K} \frac{\Delta\text{RSS}_i(k^{(\bA)}, k^{(\bB)} - 1) + \omega_n}{\Delta\text{RSS}_i(k^{(\bA)}, k^{(\bB)}) + \omega_n}\,,
\end{align*}
where $\omega_n$ is a small factor to avoid the $0 / 0$ singularities. A ratio based estimator for $ k_0^{(\bB)}$ is 
\begin{equation} \label{kB}
\hat  k_0^{(\bB)} = \max_{1 \le i \le pq} \hat{k}_{i}^{(\bB)} \,.
\end{equation}
Similarly, $ k_0^{(\bA)}$ can be estimated by
\begin{equation} \label{kA}
\hat  k_0^{(\bA)} = \max_{1 \le i \le pq} \arg\max_{1 \le k^{(\bA)} \le K} \frac{\Delta\text{RSS}_i(k^{(\bA)} - 1, \hat{k}_{0}^{(\bB)}) + \omega_n}{\Delta\text{RSS}_i(k^{(\bA)}, \hat{k}_{0}^{(\bB)}) + \omega_n}\,.
\end{equation}
In practice, let $K$ be a prescribed positive integer. Simulation results in Section \ref{subsec:sim_banded} indicate the estimator is not sensitive to the choice of $K$ given $K \ge \max(k_0^{(\bA)}, k_0^{(\bB)})$.

\subsection{Asymptotic properties}\label{sec:theory_band}

To establish the theoretical guarantee of the proposed method, we begin by introducing some notation.

%Due to the identifiability issue of $\bA_k$ and $\bB_k$, $k=0,1$, we make the convention that $\|\bA_0\|_F = 1$ and $\|\bA_1\|_F = 1$. 
We first consider the case when $k_0^{(A)}$ and $k_0^{(B)}$ are know. 
%Recall that $\tilde \ba_m^{(0)}$ and $\tilde \ba_m^{(1)}$ represent the vectors consisting of the nonzero entries of the $m$-th rows of $\bA_0$ and $\bA_1$, respectively, with $S_{\ba_m^{(0)}}$ and $S_{\ba_m^{(1)}}$ denoting the corresponding sets of indices. Similarly, $\tilde \bb_j^{(0)}$, $\tilde \bb_j^{(1)}$, $S_{\bb_j^{(0)}}$, and $S_{\bb_j^{(1)}}$ are defined analogously for the $j$-th rows of $\bB_0$ and $\bB_1$. 
Let $\btheta = (\tilde{\ba}_1',\ldots,\tilde{\ba}_p',\tilde{\bb}_1',\ldots,\tilde{\bb}_q')'$, where the dimension of $\btheta$ is denoted by $s=\sum_{j=1}^q\ell_{j}+\sum_{m=1}^p r_{m}$.
%, where $\tilde{\ba}_m$ and $\tilde{\bb}_j$ are defined \eqref{LS_A} and \eqref{LS_B}, respectively.  
%Let $\btheta = (\tilde{\ba}_1',\ldots,\tilde{\ba}_p',\tilde{\bb}_1',\ldots,\tilde{\bb}_q')'$, where $\tilde{\ba}_m$ and $\tilde{\bb}_j$ are defined \eqref{LS_A} and \eqref{LS_B}, respectively. 
% We assume $k_0^{(\bA)}$ and $k_0^{(\bB)}$ are known. 
% Hence, the true sets of indices of nonzero elements in $\bA_k$ and $\bB_k$, i.e. $S_{\ba_1^{(k)}}, \ldots, S_{\ba_p^{(k)}}$ and $S_{\bb_1^{(k)}}, \ldots, S_{\bb_q^{(k)}}$, for $k=0,1$, are known. 
%Define $\bPhi_0 = (\bI_{pq} - \bB_0 \otimes \bA_0)^{-1}(\bB_1 \otimes \bA_1)$. 
Write $\bA_k = (a_{i,j}^{(k)})_{p \times p}$ and $\bB_k = (b_{i,j}^{(k)})_{q \times q}$ for $k=0,1$. Let $\hat{\btheta} = (\widehat{\tilde{\ba}}_1',\ldots,\widehat{\tilde{\ba}}_p',\widehat{\tilde{\bb}}_1',\ldots,\widehat{\tilde{\bb}}_q')'$ denote the generalized Yule-Walker estimator obtained by solving the optimization problem in \eqref{eq:Banded_opt1}.
% \begin{align}\label{eq:Banded_opt1}
% %\min_{\bA_0\in \tilde{\mathcal{B}}(k_0^{(\bA)}),\bA_1\in \mathcal{B}(k_0^{(\bA)}), \bB_0, \bB_1\in \mathcal{B}(k_0^{(\bB)})} 
% {\color{red}\frac{1}{p^2q^2}}\sum_{j,k=1}^q\Big\|\hat{\bSigma}_{jk}(1) - \sum_{i=1}^q \bA_0\hat{\bSigma}_{ik}(1)b_{j,i}^{(0)} - \sum_{i=1}^q \bA_1 \hat{\bSigma}_{ik}(0) b_{j,i}^{(1)} \Big\|_F^2 \,,
% \end{align}
% with respect to the nonzero entries of $\bA_0$, $\bA_1$, $\bB_0$ and $\bB_1$, which are banded as defined in \eqref{band}.
% where 
% \begin{align*}
%     \tilde{\mathcal{B}}(k) = \{ \bW = (w_{i,j}): w_{i, j} = 0 \,\mbox{for all } |i-j| > k \, \mbox{and} \, w_{i, i} = 0 \}
% \end{align*}
% and
% \begin{align*}
%     \mathcal{B}(k) = \{ \bW = (w_{i,j}): w_{i, j} = 0 \,\mbox{for all } |i-j| > k \}
% \end{align*}
% for any given positive integer $k$.
Write $\hat{\bA}_0 = (\hat{a}_{i,j}^{(0)})_{p \times p}$, $\hat{\bA}_1 = (\hat{a}_{i,j}^{(1)})_{p \times p}$ and $\hat{\bB}_k = (\hat{b}_{i,j}^{(k)})_{q \times q}$ for $k=0,1$, where 
%$\hat{a}_{i,j}^{(k)}=0$ for all $|i-j|>k_0^{(\bA)}$ and $\hat{b}_{i,j}^{(k)}=0$ for all $|i-j|>k_0^{(\bB)}$, and 
the nonzero elements of $\hat{\bA}_k$ and $\hat{\bB}_k$ take their corresponding values from $\hat{\tilde{\ba}}_1,\ldots, \hat{\tilde{\ba}}_p$ and $\hat{\tilde{\bb}}_1,\ldots, \hat{\tilde{\bb}}_q$, respectively. The estimator is also rescaled so that for $k=0,1$, $\|\hat{\bA}_k\|_F = 1$ and the first nonzero entry of $\hat{\bA}_k$ is positive. We need the following regularity conditions on the process $\{\mathrm{vec}(\bX_t)\}$ and model \eqref{model_all}.

\begin{cd}\label{cond:ident_2}
    {\rm (i)} Let $\bI_{pq} - \bB_0 \otimes \bA_0$ be invertible and all the eigenvalues of $(\bI_{pq} - \bB_0 \otimes \bA_0)^{-1}(\bB_1 \otimes \bA_1)$ be smaller than 1 in modulus. 
    %{\rm (ii)} For all $m$ and $j$, $\tilde{\bX}_{\tilde{\ba}_m}'\tilde{\bX}_{\tilde{\ba}_m}$ and $\tilde{\bX}_{\tilde{\bb}_j}'\tilde{\bX}_{\tilde{\bb}_j}$ are invertible. 
    {\rm (ii)} Let $\boldsymbol{\Omega}$ be the compact and convex parameter space. Assume $\boldsymbol{\theta}$ is the unique solution of the Yule-Walker equation \eqref{YW} in the interior of $\boldsymbol{\Omega}$.
\end{cd}

Condition \ref{cond:ident_2} specifies regularity conditions for the proposed model \eqref{model_all}. By vectorization, model \eqref{model_all} can be represented in the form of the VAR(1) model in \eqref{model_vec}. Condition \ref{cond:ident_2}{\rm (i)} ensures that model \eqref{model_vec} is stationary and causal, which substantially facilitates the theoretical analysis and is necessary for establishing the asymptotic results. This condition also helps avoid ill-conditioned Yule-Walker estimation problems. See detailed discussion in Section \ref{sec:theory_diag}. Condition \ref{cond:ident_2}{\rm (ii)} ensures that $\bA_k$ and $\bB_k$, $k=0,1$, are uniquely identifiable based on the Yule-Walker equation \eqref{class1_YW}. %in the sense that $\bB_k \otimes \bA_k$ is unique. 
%Analogous to the assumption in Chen et al. (2021), we use the convention that $\bA_k$ is normalized with  $\|\bA_k\|_F=1$.}

In practice, however, $k_0^{(A)}$ and $k_0^{(B)}$ are typically unknown and need to be estimated. We therefore first establish the consistency of their estimators $\hat{k}_0^{(\bA)}$ and $\hat{k}_0^{(\bB)}$ specified in \eqref{kA} and \eqref{kB}. To do this, we introduce some additional notations and Conditions \ref{cond:proj_matx} and \ref{cond:min}.  Let $\bV_{\bX} = (\bSigma_{\bX}(1)', \bSigma_{\bX}(0))$ with $\bSigma_{\bX}(1) = \mathbb{E}\{{\rm vec}(\bX_{t}){\rm vec}(\bX_{t-1})'\}$ and $\bSigma_{\bX}(0) = \mathbb{E}\{{\rm vec}(\bX_{t}){\rm vec}(\bX_{t})'\}$. For any $i = 1, \ldots, p$, denote $\mathcal{C}_i^{(0)}=\{i_1^{(0)},\ldots,i^{(0)}_{\tau_{i,0}}\}$
%$S_{\bc_i^{(0)}} = \{i_1^{(0)},\ldots,i_{|S_{\bc_i^{(0)}}|}^{(0)}\}$ 
as the set of the indices of nonzero elements in $\bc_i^{(0)}$ and $\tau_{i,0}$ is its cardinality. Analogously, denote $\mathcal{C}_i^{(1)}=\{i_1^{(1)},\ldots,i^{(1)}_{\tau_{i,1}}\}$
%$S_{\bc_i^{(1)}} = \{i_1^{(1)},\ldots,i_{|S_{\bc_i^{(1)}}|}^{(1)}\}$ 
as the set of the indices of nonzero elements in $\bc_i^{(1)}$ and $\tau_{i,1}$ is its cardinality.
%, respectively, which are determined by the bandwidth parameters $k_0^{(\bA)}$ and $k_0^{(\bB)}$. 
Let $\bI_{\mathcal{C}_i^{(0)}} = (\be_{i_1^{(0)}},\ldots, \be_{i^{(0)}_{\tau_{i,0}}})$ be a $(pq)\times \tau_{i,0}$ matrix and $\bI_{\mathcal{C}_i^{(1)}} = (\be_{i_1^{(1)}},\ldots, \be_{i^{(1)}_{\tau_{i,1}}})$ be a $(pq)\times \tau_{i,1}$ matrix, respectively. Define $\bV_{i} = (\bSigma_{\bX}(1)'{\bI_{\mathcal{C}_i^{(0)}}}, \bSigma_{\bX}(0){\bI_{\mathcal{C}_i^{(1)}}})$.

\begin{cd}\label{cond:proj_matx}
     For any $i = 1, \ldots, pq$, let $\bF_i$ be a matrix consisting of any subset of columns from $\bV_{\bX}$ including some columns of $\bV_{i}$ such that $\bF_i'\bF_i$ is full rank. Define $\bH_i = \bF_i(\bF_i'\bF_i)^{-1}\bF_i'$. Let $\tilde{\bR}_i=(\bI_{pq}-\bH_i)\bR_i$, where $\bR_i$ is a matrix consisting of a finite subset of columns from $\bV_{\bX}$ such that $\bR_i \neq \bF_i$.  For any matrix $\bG_i$, consisting of a finite number of columns of $\bV_i$ distinct from $\bF_i$ and $\bR_i$, there exist positive constants $\lambda_1 \le \lambda_2$ such that
    \begin{align*}
        \lambda_1 \le \lambda_{\min}\{\bG_i'(\tilde{\bR}_i(\tilde{\bR}_i'\tilde{\bR}_i)^{-1}\tilde{\bR}_i')\bG_i\} \le \lambda_{\max}\{\bG_i'(\tilde{\bR}_i(\tilde{\bR}_i'\tilde{\bR}_i)^{-1}\tilde{\bR}_i')\bG_i\} \le \lambda_2\,,
    \end{align*}
    where $\lambda_{\min}(\cdot)$ and $\lambda_{\max}(\cdot)$ denote the smallest and largest eigenvalues of a matrix, respectively.
\end{cd}

\begin{cd}\label{cond:min}
    The elements of $\bA_0$, $\bA_1$, $\bB_0$ and $\bB_1$ are bounded uniformly. For each $i=1,\ldots,p$, $|a_{i,i-k_0^{(\bA)}}^{(0)}|$, $|a_{i,i+k_0^{(\bA)}}^{(0)}|$, $|a_{i,i-k_0^{(\bA)}}^{(1)}|$ and $|a_{i,i+k_0^{(\bA)}}^{(1)}|$ are greater than $\{k_0^{(\bA)}\}^{1/2}[C_n n^{-1}\log\{(pq) \vee n\}]^{1/4}$, and $|b_{i,i-k_0^{(\bB)}}^{(0)}|$, $|b_{i,i+k_0^{(\bB)}}^{(0)}|$, $|b_{i,i-k_0^{(\bB)}}^{(1)}|$ and $|b_{i,i+k_0^{(\bB)}}^{(1)}|$ are greater than $\{k_0^{(\bB)}\}^{1/2}[C_n n^{-1}\log\{(pq) \vee n\}]^{1/4}$, where $C_n/n \to 0$ and $C_n^2/(npq) \to \infty$ as $n \to \infty$.
\end{cd}

% \begin{cd}\label{cond:ABbound}
%     For each $i,j=1,\ldots,p$, $a_{i,j}^{(0)}|$, $|a_{i,j}^{(1)}|$, $|b_{i,j}^{(0)}|$ and $b_{i,j}^{(1)}$ are bounded uniformly.
% \end{cd}

Conditions \eqref{cond:proj_matx} and \ref{cond:min} are imposed to prove the consistency of our ratio estimators $\hat{k}_0^{(\bA)}$ and $\hat{k}_0^{(\bB)}$. Condition \eqref{cond:proj_matx} is similar to Condition A4 in \cite{Gao2019}. 
Condition \ref{cond:min} specifies the minimum orders of the non-zero elements on the corresponding super-diagonals and sub-diagonals of $\bA_0$, $\bA_1$, $\bB_0$ and $\bB_1$ to ensure the identifiability of the bandwidths $k_0^{(\bA)}$ and $k_0^{(\bB)}$. Further discussion can be found in \cite{Gao2019} and the references therein.
Theorem \ref{tm:bandwidth} shows that $\hat{k}_0^{(\bA)}$ and $\hat{k}_0^{(\bB)}$ are consistent.

\begin{theorem}\label{tm:bandwidth}
    Let Conditions \ref{cond:mixing} and \ref{cond:ident_2}-\ref{cond:min} hold. Let $d = p \vee q$ and $k_{0,max} = k_0^{(\bA)} \vee k_0^{(\bB)}$. Choose $\omega_n = Cpq/n$ for some universal constant $C > 0$. If $C_n\log(d\vee n)k_{0,max}^{2} = o(n)$ and $d^{2+3\kappa}=o(n)$, then it holds that $\mathbb{P}(\hat{k}^{(\bA)}_0 = k_0^{(\bA)}, \hat{k}^{(\bB)}_0 = k_0^{(\bB)}) \to 1$ as $n \to \infty$.
    %$\mathbb{P}(\hat{k}^{(\bA)}_0 = k_0^{(\bA)}) \to 1$ and $\mathbb{P}(\hat{k}^{(\bB)}_0 = k_0^{(\bB)}) \to 1$ as $n \to \infty$.
\end{theorem}

Since the true bandwidths $k_0^{(A)}$ and $k_0^{(B)}$ are unknown in practice, in the sequel, we replace them in the estimation procedure by $\hat{k}_0^{(\bA)}$ and $\hat{k}_0^{(\bB)}$, respectively, and still denote the resulting generalized Yule-Walker estimator by $\hat{\btheta}$.
Then, Proposition \ref{pn:consistency_banded} shows that $\hat{\btheta}$ is consistent.

\begin{proposition}\label{pn:consistency_banded}
    Assume that Conditions \ref{cond:mixing} and \ref{cond:ident_2}-\ref{cond:min} hold. Let $d = p \vee q$ and $k_{0,max} = k_0^{(\bA)} \vee k_0^{(\bB)}$. Then, we have
    \begin{align*}
        \|\hat{\btheta}-\btheta\|_2 \stackrel{\mathrm{p}}{\to} 0\,, ~\mbox{as}~ n \to \infty\,,
    \end{align*}
    provided that $C_n\log(d\vee n)k_{0,max}^{2} = o(n)$, $p^{-1}(k_0^{(\bA)})^3(k_0^{(\bB)})^4 + q^{-1}(k_0^{(\bA)})^4(k_0^{(\bB)})^3 = O(1)$, $d^{2+3\kappa}=o(n)$ and $d^{(2+11\kappa)/4}(k_{0,max})^{7(2+\kappa)/4} = o(n)$.
    %$k_{0,max}^{(10+5\kappa)/4}d^{(22+21\kappa)/4} = o(n)$.
    %$k_{0,max}^{(6+3\kappa)/4}d^{(18+19\kappa)/4} = o(n)$.
    %$d^{18+13\kappa}\{k_0^{(\bA)}\}^{4+3\kappa}\{k_0^{(\bB)}\}^{2+\kappa} = o(n^{4+\kappa})$.
    %$|\hat{\btheta}-\btheta|_2 \stackrel{p}{\to} 0$ as $n \to \infty$.
\end{proposition}

\noindent\textbf{Remark 2} It is worth noting that the banded structure is mainly used to characterize local dependence patterns while reducing model complexity. In many applications, interactions between distant components are expected to be weak or negligible. Therefore, the bandwidths $k_0^{(\bA)}$ and $k_0^{(\bB)}$ are usually assumed to be small or moderately large in practical applications. 
Hence, it is of interest to consider the case where $k_0^{(\bA)}$ and $k_0^{(\bB)}$ are fixed. In this case, the restrictions on the dimensions $p$ and $q$ in Proposition \ref{pn:consistency_banded} reduce to $d = o\{n^{1/(2+3\kappa)}\}$.

Then we establish the asymptotic normality of $\hat{\btheta}$. 
%For simplicity, write $s_{\ba_m} = |S_{\ba_m^{(0)}}|+|S_{\ba_m^{(1)}}|$ and $s_{\bb_j} = |S_{\bb_j^{(0)}}|+|S_{\bb_j^{(1)}}|$. 
For any $m \in [p]$ and $j \in [q]$, let $\tilde{\bX}_{\tilde{\ba}_m}$ and $\tilde{\bX}_{\tilde{\bb}_j}$ be $(pq^2) \times r_{m}$ and $(p^2q) \times \ell_{j}$ matrices obtained from $\bX_{\tilde{\ba}_m}$ and $\bX_{\tilde{\bb}_j}$, respectively, by replacing each $\hat{\bSigma}_{ik}(1)$ and $\hat{\bSigma}_{ik}(0)$ with their population counterparts $\bSigma_{ik}(1)$ and $\bSigma_{ik}(0)$.
Let
\begin{align*}
    \tilde{\bZ}_{\tilde{\ba}_m,j} = \begin{pmatrix}
        \tilde{\boldsymbol{\psi}}_{\tilde{\ba}_m,j,j_1^{(0)}}^{(0)}\,, & \cdots\,, & \tilde{\boldsymbol{\psi}}_{\tilde{\ba}_m,j,j^{(0)}_{\ell_{j,0}}}^{(0)}\,, &  \tilde{\boldsymbol{\psi}}_{\tilde{\ba}_m,j,j_1^{(1)}}^{(1)}\,, & \cdots\,, & \tilde{\boldsymbol{\psi}}_{\tilde{\ba}_m,j,j^{(1)}_{\ell_{j,1}}}^{(1)}
    \end{pmatrix}
\end{align*}
be a $pq^2\times \ell_{j}$ matrix, where
\begin{align*}
    \tilde{\boldsymbol{\psi}}_{\tilde{\ba}_m,j,l}^{(0)} = \big({\bf 0}_{1 \times (j-1)pq}\,, ~ \tilde{\ba}_m^{(0)'}\bE_{m}^{(0)'}\bSigma_{l1}(1)\,,\cdots\,,\tilde{\ba}_m^{(0)'}\bE_{m}^{(0)'}\bSigma_{lq}(1)\,,~ {\bf 0}_{1 \times (q-j)pq} \big)'\,,\\
    \tilde{\boldsymbol{\psi}}_{\tilde{\ba}_m,j,l}^{(1)} = \big({\bf 0}_{1 \times (j-1)pq}\,, ~ \tilde{\ba}_m^{(1)'}\bE_{m}^{(1)'}\bSigma_{l1}(0)\,,\cdots\,,\tilde{\ba}_m^{(1)'}\bE_{m}^{(1)'}\bSigma_{lq}(0)\,,~ {\bf 0}_{1 \times (q-j)pq} \big)'\,.
\end{align*}
Analogously, let
\begin{align*}
    \tilde{\bZ}_{\tilde{\bb}_j,m} = \begin{pmatrix}
        \tilde{\boldsymbol{\psi}}_{\tilde{\bb}_j,m,m_1^{(0)}}^{(0)}\,, & \cdots\,, & \tilde{\boldsymbol{\psi}}_{\tilde{\bb}_j,m,m^{(0)}_{r_{m,0}}}^{(0)}\,, &  \tilde{\boldsymbol{\psi}}_{\tilde{\bb}_j,m,m_1^{(1)}}^{(1)}\,, & \cdots\,, & \tilde{\boldsymbol{\psi}}_{\tilde{\bb}_j,m,m^{(1)}_{r_{m,1}}}^{(1)} 
    \end{pmatrix}
\end{align*}
be a $p^2 q \times r_{m}$ matrix, where 
%which are $pq^2\times s_{\bb_j}$ and $p^2 q \times s_{\ba_m}$ matrices, respectively, 
\begin{align*}
    % \tilde{\boldsymbol{\psi}}_{\tilde{\ba}_m,j,l}^{(0)} = \big({\bf 0}_{1 \times (j-1)pq}\,, ~ \tilde{\ba}_m^{(0)'}\bE_{m}^{(0)'}\bSigma_{l1}(1)\,,\cdots\,,\tilde{\ba}_m^{(0)'}\bE_{m}^{(0)'}\bSigma_{lq}(1)\,,~ {\bf 0}_{1 \times (q-j)pq} \big)'\,,\\
    % \tilde{\boldsymbol{\psi}}_{\tilde{\ba}_m,j,l}^{(1)} = \big({\bf 0}_{1 \times (j-1)pq}\,, ~ \tilde{\ba}_m^{(1)'}\bE_{m}^{(1)'}\bSigma_{l1}(0)\,,\cdots\,,\tilde{\ba}_m^{(1)'}\bE_{m}^{(1)'}\bSigma_{lq}(0)\,,~ {\bf 0}_{1 \times (q-j)pq} \big)'\,,\\
    \tilde{\boldsymbol{\psi}}_{\tilde{\bb}_j,m,l}^{(0)} = \begin{pmatrix}
        (\bSigma_{j_1^{(0)}1}(1)'\be_l)\otimes \be_m\,, & \cdots\,, & (\bSigma_{j^{(0)}_{\ell_{j,0}} 1}(1)'\be_l)\otimes \be_m \\
        \vdots  &  &  \vdots \\
        (\bSigma_{j_1^{(0)}q}(1)'\be_l)\otimes \be_m\,, & \cdots\,, & (\bSigma_{j^{(0)}_{\ell_{j,0}} q}(1)'\be_l)\otimes \be_m
    \end{pmatrix}\tilde{\bb}_j^{(0)}
\end{align*}
and
\begin{align*}
    \tilde{\boldsymbol{\psi}}_{\tilde{\bb}_j,m,l}^{(1)} = \begin{pmatrix}
        (\bSigma_{j_1^{(1)}1}(0)'\be_l)\otimes \be_m\,, & \cdots\,, & (\bSigma_{j^{(1)}_{\ell_{j,1}} 1}(0)'\be_l)\otimes \be_m \\
        \vdots  &  &  \vdots \\
        (\bSigma_{j_1^{(1)}q}(0)'\be_l)\otimes \be_m\,, & \cdots\,, & (\bSigma_{j^{(1)}_{\ell_{j,1}} q}(0)'\be_l)\otimes \be_m
    \end{pmatrix}\tilde{\bb}_j^{(1)}\,.
\end{align*}
Here, $\be_l$ is the $p$-dimensional unit vector with the $l$-th element being 1. 
Then, let
\begin{align}\label{eq:Umatrix_band}
    \bU_{\rm band} = \begin{pmatrix}
        \tilde{\bX}_{\tilde{\ba}_1}'\tilde{\bX}_{\tilde{\ba}_1}\,, & \cdots & {\bf 0}_{r_{p} \times r_{p}}\,, & \tilde{\bX}_{\tilde{\ba}_1}'\tilde{\bZ}_{\tilde{\ba}_1,1}\,, & \cdots & \tilde{\bX}_{\tilde{\ba}_1}'\tilde{\bZ}_{\tilde{\ba}_1,q} \\
           &   \ddots  &  & \vdots  &  & \vdots  \\
        {\bf 0}_{r_{1} \times r_{1}}\,, & \cdots & \tilde{\bX}_{\tilde{\ba}_p}'\tilde{\bX}_{\tilde{\ba}_p}\,, & \tilde{\bX}_{\tilde{\ba}_p}'\tilde{\bZ}_{\tilde{\ba}_p,1}\,, & \cdots & \tilde{\bX}_{\tilde{\ba}_p}'\tilde{\bZ}_{\tilde{\ba}_p,q} \\
        \tilde{\bX}_{\tilde{\bb}_1}'\tilde{\bZ}_{\tilde{\bb}_1,1}\,,  & \cdots & \tilde{\bX}_{\tilde{\bb}_1}'\tilde{\bZ}_{\tilde{\bb}_1,p}\,, & \tilde{\bX}_{\tilde{\bb}_1}'\tilde{\bX}_{\tilde{\bb}_1}& \cdots & {\bf 0}_{\ell_{q} \times \ell_{q}}\\
         \vdots  &  & \vdots  &  &  \ddots  &    \\
        \tilde{\bX}_{\tilde{\bb}_q}'\tilde{\bZ}_{\tilde{\bb}_q,1}\,, & \cdots & \tilde{\bX}_{\tilde{\bb}_q}'\tilde{\bZ}_{\tilde{\bb}_q,p}\,, & {\bf 0}_{\ell_{1} \times \ell_{1}}& \cdots & \tilde{\bX}_{\tilde{\bb}_q}'\tilde{\bX}_{\tilde{\bb}_q}
    \end{pmatrix}\,.
\end{align}
% \begin{align}\label{eq:Umatrix_band}
%     \bU_{\rm band} = \begin{pmatrix}
%         \tilde{\bX}_{\tilde{\ba}_1}'\tilde{\bX}_{\tilde{\ba}_1}\,, & \cdots & {\bf 0}_{s_{\ba_p} \times s_{\ba_p}}\,, & \tilde{\bX}_{\tilde{\ba}_1}'\tilde{\bZ}_{\tilde{\ba}_1,1}\,, & \cdots & \tilde{\bX}_{\tilde{\ba}_1}'\tilde{\bZ}_{\tilde{\ba}_1,q} \\
%            &   \ddots  &  & \vdots  &  & \vdots  \\
%         {\bf 0}_{s_{\ba_1} \times s_{\ba_1}}\,, & \cdots & \tilde{\bX}_{\tilde{\ba}_p}'\tilde{\bX}_{\tilde{\ba}_p}\,, & \tilde{\bX}_{\tilde{\ba}_p}'\tilde{\bZ}_{\tilde{\ba}_p,1}\,, & \cdots & \tilde{\bX}_{\tilde{\ba}_p}'\tilde{\bZ}_{\tilde{\ba}_p,q} \\
%         \tilde{\bX}_{\tilde{\bb}_1}'\tilde{\bZ}_{\tilde{\bb}_1,1}\,,  & \cdots & \tilde{\bX}_{\tilde{\bb}_1}'\tilde{\bZ}_{\tilde{\bb}_1,p}\,, & \tilde{\bX}_{\tilde{\bb}_1}'\tilde{\bX}_{\tilde{\bb}_1}& \cdots & {\bf 0}_{s_{\bb_q} \times s_{\bb_q}}\\
%          \vdots  &  & \vdots  &  &  \ddots  &    \\
%         \tilde{\bX}_{\tilde{\bb}_q}'\tilde{\bZ}_{\tilde{\bb}_q,1}\,, & \cdots & \tilde{\bX}_{\tilde{\bb}_q}'\tilde{\bZ}_{\tilde{\bb}_q,p}\,, & {\bf 0}_{s_{\bb_1} \times s_{\bb_1}}& \cdots & \tilde{\bX}_{\tilde{\bb}_q}'\tilde{\bX}_{\tilde{\bb}_q}
%     \end{pmatrix}\,.
% \end{align}

Write $s=s_1+s_2$ with $s_1 = \sum_{m=1}^p r_{m}$ and $s_2 = \sum_{j=1}^q \ell_{j}$. Define $\bT_n = n^{-1}\mathrm{Cov}(\sum_{t=2}^n \tilde{\bJ}\tilde{\boldsymbol{\mathcal{E}}}_{\bX,t})$, where $\tilde{\bJ} = (\tilde{\bJ}_{\ba}', \tilde{\bJ}_{\bb}')'$ with
\begin{align*}
    \tilde{\bJ}_{\ba} = \begin{pmatrix}
        \tilde{\bX}_{\tilde{\ba}_1}'(\bI_{pq^2}\otimes\be_1')  \\
        \vdots \\
        \tilde{\bX}_{\tilde{\ba}_p}'(\bI_{pq^2}\otimes\be_p')
    \end{pmatrix} \in \mathbb{R}^{s_1\times (pq)^2} ~\mbox{and}~ 
    \tilde{\bJ}_{\bb} = \begin{pmatrix}
        \tilde{\bX}_{\tilde{\bb}_1}' &   &  \\
        &   \ddots  & \\
        &   &  \tilde{\bX}_{\tilde{\bb}_q}'
    \end{pmatrix} \in \mathbb{R}^{s_2\times (pq)^2}\,.
\end{align*}

Now we impose the following regularity condition.

\begin{cd}\label{cond:mmt_bandedcase}
    {\rm (i)} For some universal constant $\rho_2 \in (2,3]$ satisfying that $2\rho_2(\rho_2-1)/(\rho_2-2)^2 < (4+\kappa)/\kappa$ for $\kappa$ specified in Condition \ref{cond:mixing}{\rm (i)}, the following condition holds uniformly in $p$, $q$:
    \begin{align*}
        &~~~~~~~~ \max_{1\le t\le n}\sup_{\bu \in \mathbb{R}^{s}}\mathbb{E}|\bu'\tilde{\bJ}\tilde{\boldsymbol{\mathcal{E}}}_{\bX,t}|^{\rho_2} <  \eta_n\,, 
    \end{align*}
    where $\eta_n$ is a sequence possibly depending on $n$. 
    {\rm (ii)} There exists a universal constant $c_2>0$ such that, for any $n \ge 1$,  $\lambda_{\min}(\bT_n) \ge c_2$, where $\lambda_{\min}(\bT_n)$ denotes the smallest eigenvalue of $\bT_n$. {\rm (iii)} $\bU_{\rm band}$ specified in \eqref{eq:Umatrix_band} is invertible.
\end{cd}

Condition \eqref{cond:mmt_bandedcase} specifies the moment condition for $\tilde{\boldsymbol{\mathcal{E}}}_{\bX,t}$, which controls the spatial dependence between different elements of $\bX_t$. Analogous to the discussion for Condition \ref{cond:mmt_diagonalcase}, when the dimensions $p$ and $q$ diverge, the condition on $\eta_n$ in Theorem \eqref{tm:mixing_banded} implicitly requires a restriction on the growth rate of $p$ and $q$. Condition \ref{cond:mmt_bandedcase} (ii) and (iii) ensure a non-degenerate limiting distribution. 
%Condition \ref{cond:mmt_bandedcase} (ii) ensures that $\bT_n$ is uniformly non-degenerate, which is a standard assumption in high-dimensional analysis, and Condition \ref{cond:mmt_bandedcase} (iii) ensures that $\bU_{\rm band}$ is invertible.

\begin{theorem}\label{tm:mixing_banded}
    Assume that Conditions \ref{cond:mixing} and \ref{cond:ident_2}-\ref{cond:min} hold. Let $d = p \vee q$, $k_{0,max} = k_0^{(\bA)} \vee k_0^{(\bB)}$ and $\beta_2 = \{4+\kappa)(\rho_2-2)^2\}/\{2\kappa \rho_2(\rho_2-1)\}$. Then, if $C_n\log(d\vee n)k_{0,max}^{2} = o(n)$, $p^{-1}(k_0^{(\bA)})^3(k_0^{(\bB)})^4 + q^{-1}(k_0^{(\bA)})^4(k_0^{(\bB)})^3 = O(1)$, $k_{0,max}^{\delta_1}d^{\delta_2} = o(n)$ with $\delta_1 = \max\{ 5, (6+3\kappa)/2 \}$, $\delta_2 = \max\{ 11, (10+9\kappa)/2 \}$ and  $\eta_n = o\{n^{(\beta_2-1)(\rho_2-2)/(2\beta_2+2)}\}$,
    %$k_{0,max}^{(6+3\kappa)/2}d^{(22+21\kappa)/4} = o(n)$,
    %$k_{0,max}^{2+\kappa}d^{\varpi/4} = o(n)$ with $\varpi = \max\{20, 18+19\kappa\}$, 
    %$k_{0,max}^{6+4\kappa}d^{18+13\kappa} = o(n^{4+\kappa})$,
    %when  $p,q \ge 1$, $k_0^{(\bA)}$ and $k_0^{(\bB)}$ are fixed, 
    we have 
    \begin{align}\label{eq:CLT_band}
    \sqrt{n} \bT_n^{-1/2} \bU_{\rm band}(\hat{\btheta}-\btheta) \stackrel{\mathrm{d}}{\to} \mathcal{N}(0,\bI_{s}) ~\, \mbox{as} \,~ n \to \infty\,.
\end{align}
\end{theorem}

\noindent {\bf Remark 3} 
    If $p$ and $q$ are fixed, the limiting form of $\bT_n$ can be explicitly characterized. See Proposition \ref{CLT:band_fixdim} in Section \ref{sec:band_fixdim} of the supplementary material for details.

\section{Simulation}
\subsection{Known weight matrix with unknown diagonal coefficients}
We set the known weight matrices $\bW_0, \bV_0$ ($\bW_0$ has zero diagonal values) such that the main and first two sub-diagonal elements are all 1 and all remaining elements are zeros. And we set the known weight matrices $\bW_1, \bV_1$ such that the main and first two sub-diagonal elements are all 0.5 and all remaining elements are zeros. $\balpha_0, \balpha_1, \bbeta_0, \bbeta_1$ are randomly generated from uniform distribution $U[-0.5, 0.5]$. We apply the iterated generalized Yule-Walker estimation with different sample sizes $n$ equal to 200, 500, 1000, 2000, 5000, and with simulation times equals to 200. Initial values of the iterated estimation procedure is randomly generated from standard normal distribution. We compute and report below error
\begin{equation} \label{err0} 
{\| D(\hat{\bbeta}_0) \otimes   D(\hat{\balpha}_0)  -  D(\bbeta_0) \otimes    D(\balpha_0)\|_F} \; \;
\text{and} \; \;
{\| D(\hat{\bbeta}_1) \otimes   D(\hat{\balpha}_1)  -  D(\bbeta_1) \otimes    D(\balpha_1)\|_F}.
\end{equation}
As can be seen from Table \ref{known_mat_table} for a fixed $(p, q)$ equals to $(10, 10), (20, 20), (30, 30)$, as the sample size $n$ increases, both the mean and standard deviations of the errors for $D(\bbeta_0) \otimes    D(\balpha_0)$ and $D(\bbeta_1) \otimes    D(\balpha_1)$ decreases. 

%For a fixed sample size, as the dimension $(p, q)$ increases, mean and standard deviations of the errors tend to increase. 

\subsection{Unknown weight matrix with banded structures}\label{subsec:sim_banded}
\paragraph{Scenario setup}
We set the bandwidth parameters $k_0^{(\bA)}=2$ and $k_0^{(\bB)}=2$. The nonzero elements of $\bA_0, \bB_0, \bA_1, \bB_1$ are randomly generated from uniform distribution $U(-0.5, 0.5)$. For different $p$ and $q$ equals to $(10, 10), (20, 20), (30, 30)$, we apply the iterated generalized Yule-Walker estimation and the nearest Kronecker product method with different sample sizes $n$ equal to 200, 500, 1000, 2000, 5000, and with simulation times equals to 200. Initial values of the iterated estimation procedure is set to be the estimator from nearest Kronecker product method. We compute and report below error
\begin{equation} \label{err1} 
{\| \hat \bB_0 \otimes   \hat \bA_0  -  \bB_0 \otimes  \bA_0\|_F} \; \;
\text{and} \; \;
{\| \hat \bB_1 \otimes   \hat \bA_1  -  \bB_1 \otimes \bA_1\|_F}.
\end{equation}
\paragraph{Performance of iterative least square method given bandwidth parameters} 
Table \ref{unknown_mat_known_band_table} in the appendix reports the mean and standard deviations of error (\ref{err1}) from 200 simulations with different $(p, q, n)$ for iterated generalized Yule-Walker estimation method and nearest Kronecker product method with known $k_0^{(\bA)}$ and $k_0^{(\bB)}$. As can be seen from the table, for each fixed set of $(p, q)$, as the sample size $n$ increases, both the mean and standard deviation of the errors for $\bB_0 \otimes \bA_0$ and $\bB_1 \otimes \bA_1$ decreases, no matter for iterated generalized Yule-Walker estimation method or nearest Kronecker product method. Furthermore, it can be seen that the performance of  iterated generalized Yule-Walker estimation method is always better than the  nearest Kronecker product method. 

%Also note as the dimension $(p, q)$ increases, performance of the proposed estimator become worse with larger mean and standard deviation. 

\paragraph{Performance of ratio based method for bandwidth parameters}
Table \ref{bandwidth_est_table} reports the relative frequency of the occurrence of the event $\{\hat k_0^{(\bA)}  = k_0^{(\bA)}\}$ and $\{\hat k_0^{(\bB)}  = k_0^{(\bB)}\}$, that is the percentage of correct estimates for $k_0^{(\bA)}$ and $k_0^{(\bB)}$, respectively, from 200 simulations, with different $(p, q, n)$ combinations. Furthermore, to check how sensitive the ratio based estimator is to the choice of $K$, we report the  relative frequencies for different $K = 4, 6, 8$. From the table, we can see that as the sample size $n$ increases, the accuracy of the ratio based estimator for both $k_0^{(\bA)}$ and $k_0^{(\bB)}$ improve for different $(p, q)$. At the same time, it can be seen that the relative frequencies with different $K$ are relatively stable with the same $(p, q, n)$, especially with large $n$, which implies the ratio based estimator is not sensitive to $K$. The simulation uses $\omega_n = 0.1pq/n$.  

\paragraph{Performance of iterative least square method with unknown bandwidth parameters} 
Table \ref{unknown_mat_unknown_band_table} in the appendix  reports the mean and standard deviations of error (\ref{err1}) from 200 simulations with different $(p, q, n)$ for iterated generalized Yule-Walker estimation method and nearest Kronecker product method with unknown $k_0^{(\bA)}$ and $k_0^{(\bB)}$, which are obtained from the ratio based estimation method. As can be seen from the table, for each fixed set of $(p, q)$, as the sample size $n$ increases, both the mean and standard deviation of the errors for $\bB_0 \otimes \bA_0$ and $\bB_1 \otimes \bA_1$ decreases, no matter for iterated generalized Yule-Walker estimation method or nearest Kronecker product method. Furthermore, it can be seen that the performance of  iterated generalized Yule-Walker estimation method is always better than the  nearest Kronecker product method. Comparing table \ref{unknown_mat_known_band_table} and \ref{unknown_mat_unknown_band_table}, it can be noted that both mean and standard deviations of errors when the bandwidth parameters are unknown tend to be larger, especially for smaller sample sizes, this is due to the fact that estimating bandwidth parameters further introduces estimating error. As sample size increases, such extra estimating error reduces, hence we can see that the performance for large samples as $n=2000, 5000$ are similar no matter the bandwidth parameters need to be estimated or not. 

%Also note as the dimension $(p, q)$ increases, performance of the proposed estimator become worse with larger mean and standard deviation.   

\section{Forecasting Equity Trading Volume in Japan with Applications to POV Strategies}
The motivation for this study stems from real-world trade execution services in financial markets. Execution services encompass the end-to-end process of executing buy or sell orders for financial instruments (e.g., equities, fixed income, derivatives) on behalf of clients such as institutional investors. Commonly used execution algorithms - such as Volume-Weighted Average Price (VWAP) and Percentage of Volume (POV), also referred to as Volume Inline \citep[see][]{UBS2015,CLSA2018}
%(see  UBS, 2015, and CLSA, 2018) 
- rely heavily on precise volume forecasts for each security. Thus, the predicted volume curve is a key driver of execution strategy performance. Moreover, intraday volume forecasting has gained increasing relevance in helping execution desks optimize trade execution, as shown in \cite{Bialkowski2008}.
%Białkowski et al. (2008).

Standard industry methods forecast volume curves based on historical daily data, whereas our approach - employing a spatio-temporal autoregressive model for matrix data that incorporates intraday dynamics and cross-asset information - demonstrates superior empirical performance across two critical dimensions:
\begin{itemize}
    \item Forecasting accuracy for trading volumes across multiple intraday time horizons
    \item Enhanced execution quality in volume-based execution algorithms (exemplified by POV strategy)
\end{itemize}

In this empirical data analysis, we examine the trading volumes of 5 liquid stocks in the Japanese market: Sony Group Corporation (6758), SoftBank Group Corp (9984), Nintendo Co., Ltd (7974), Tokyo Electron Limited (8035), and Renesas Electronics Corporation (6723). The analysis follows the industry practice of considering 5-minute trading volumes. The prediction horizon spans three months, specifically from 01/11/2023 to 31/01/2024. For the panel variables, we apply the ordering procedure described in Appendix~\ref{ordering_method} to the set of stocks, yielding the following sequence for analysis: 9984, 6758, 8035, 6723, and 7974. For the spatial variables, the natural chronological order of intraday buckets is retained. Appendix~\ref{justify_panel} and ~\ref{G::spatial} provides a comprehensive justification for the banded structures assumed for both the spatial and panel variables.

\subsection{Analysis of Raw Volume Predictions}\label{subsec:vol_pred}
For each stock $j$, we evaluate six different methods to predict the trading volume $V_{t+1, i, j}$ for any arbitrary 5-minute interval $i$ on trading day $t+1$:
\begin{itemize}
    \item \textbf{Spatio-temporal Model with banded weight matrices (YW)}: see section \ref{sec:model_banded}.

    \item \textbf{Spatio-temporal Model with diagonal coefficients (known$\_$YW)}: see section \ref{sec:model}. 

    \item \textbf{Matrix Autoregressive Model (MAR)}: see \cite{Chen2021}.

    \item \textbf{Adjusted MAR (adj$\_$MAR)}: An intraday-adjusted version incorporating dynamic scaling to MAR:
    \[
    \hat{V}_{t+1, i, j}^{\text{adj\_MAR}} = r_{i, j} \times \hat{V}_{t+1, i, j}^{\text{MAR}},
    \]
    with scaling factor $r_{i, j}$ computed as:
    $
    r_{i, j} = \frac{\sum_{k < i} V_{t+1, k, j}}{\sum_{k < i} \hat{V}_{t+1, k, j}^{\text{MAR}}}.
    $
    
    \item \textbf{Simple Moving Average (SMA)}: 
    \[
    \hat{V}_{t+1, i, j}^{\text{SMA}} = \frac{1}{L} \sum_{d = t-L+1}^{t} V_{d, i, j},
    \]
    where $L$ is the lookback window length (chosen as 22 days following industry standard).

    \item \textbf{Adjusted SMA (adj$\_$SMA)}: An intraday-adjusted version incorporating dynamic scaling to SMA:
    \[
    \hat{V}_{t+1, i, j}^{\text{adj\_SMA}} = r_{i, j} \times \hat{V}_{t+1, i, j}^{\text{SMA}},
    \]
    with scaling factor $r_{i, j}$ computed as:
    $
    r_{i, j} = \frac{\sum_{k < i} V_{t+1, k, j}}{\sum_{k < i} \hat{V}_{t+1, k, j}^{\text{SMA}}}.
    $
\end{itemize}

For approaches that require fitting, specifically spatio-temporal and MAR models, we operate on the anomaly process. For each bucket $i$ and stock $j$, we compute the deterministic pattern as ${s}_{i,j} = \frac{1}{T}\sum_{t=1}^T U_{t,i,j}$, where $U_{t,i,j}$ denotes the log trading volume (log of $V_{t, i, j}$), and $T$ is the training sample size used in each rolling window estimation and we choose it to be 60, representing 60 trading days in each estimation window. The ``demeaning'' procedure subtracts this estimated bucket-stock-specific mean from the log-transformed data ${\mathbf{X}}_t = \mathbf{U}_t - {\mathbf{S}}$.

We then generate predictions for ${\mathbf{X}}_{t+1}$. Using the spatio-temporal model as an illustration, the $s$-th row of $\mathbf{X}_{t+1}$ is predicted via
\begin{equation}
\hat{\mathbf{X}}_{t+1, s\cdot} = \mathbf{A}_{0, s\cdot} \mathbf{X}_{t+1} \mathbf{B}_0' + \mathbf{A}_{1, s\cdot} \mathbf{X}_{t} \mathbf{B}_1',
\end{equation}
where $\mathbf{A}_{0, s\cdot}$ and $\mathbf{A}_{1, s\cdot}$ are the $s$-th rows of $\mathbf{A}_0$ and $\mathbf{A}_1$, respectively. Note that by the $s$-th time bucket of day $t+1$, the rows $\mathbf{X}_{t+1, i\cdot}$ for $i < s$ have already been observed. This is incorporated into $\mathbf{A}_{0, s\cdot}$ by allowing only its first $s-1$ entries to be nonzero, ensuring that no future information from the same day is used in the prediction.

After obtaining predictions for the anomaly process $\hat{\mathbf{X}}_{t+1}$, we add back the deterministic pattern ${\mathbf{S}}$ and then apply the exponential transformation to revert the log transformation, thereby reconstructing the final predictions in raw trading volumes $\hat{\mathbf{V}}_{t+1} = \exp(\hat{\mathbf{X}}_{t+1} + {\mathbf{S}})$.

It is worth noting that for the spatio-temporal model with known weight matrices, the specification of the weight matrix matters. Although not presented in this paper, detailed analysis has been performed, indicating that more meaningful weight matrices generally yield better prediction results. In our analysis, we first order the panel variable following the procedure in Appendix~\ref{ordering_method}. The weight matrices for the spatial and panel variables are specified as in Appendix~\ref{G::known_weight_mat_spec}, originating from conditional correlation matrices. Supported by the banded structure justification analysis in Appendix~\ref{sec:ordering_var} for both the spatial and panel variables, we subsequently truncate these matrices in a banded manner such that the bandwidths are chosen to be 4 for the spatial variable and 2 for the panel variable. The truncated matrices serve as the final input weight matrices.

For the spatio-temporal model with unknown weight matrices, the choice of the bandwidths matters. Given that underestimating the optimal bandwidths leads to more severe consequences than using a value that is overestimated -- provided the overestimation is not excessive -- we offer the following practical recommendation. We suggest first selecting a conservative initial bandwidth, denoted $\bar{k}_A$ and $\bar{k}_B$, which is large enough to capture relevant dependencies but not so large as to compromise computational feasibility. For instance, one may set $\bar{k}_A$ and $\bar{k}_B$ based on the bandwidths implied by the conditional adjacency matrices (see Appendix~\ref{sec:ordering_var}) for spatial and panel variables. The final bandwidth can then be chosen as
\[
\tilde{k}_A = \max\{\hat{k}_A, \bar{k}_A\}, \qquad
\tilde{k}_B = \max\{\hat{k}_B, \bar{k}_B\},
\]
where $\hat{k}_A$ and $\hat{k}_B$ are the ratio-based estimates in section~\ref{sc:bandwidth}. This estimator strikes a balance: it retains as much information as the banded structure permits, while preventing over-parameterization and maintaining computational efficiency. In our analysis we choose $\bar{k}_A$ and $\bar{k}_B$ to be 4 and 2 respectively, as implied from the analysis in Appendix~\ref{sec:ordering_var}. A final observation is that adherence to the original ratio-based estimator alone entails only a minor degradation in prediction accuracy, particularly when set against the performance gains delivered by SMA and MAR-based methodologies.

For each 5-minute interval, we calculate the relative prediction error for a given stock $j$:

\begin{equation}
\frac{1}{n} \sum_{t=1}^n \frac{1}{m} \sum_{i=1}^m \frac{|\hat{V}_{t, i, j} - V_{t, i, j}|}{V_{t, i, j}},
\end{equation}
where $n$ is the number of trading days in our evaluation period, and $m$ is the number of 5-minute intervals within each trading session under consideration.

We analyze the following sessions: full day (09:00--15:00), morning (09:00--11:30), afternoon (12:30--15:00), and five hourly windows: P1 (09:00--10:00), P2 (10:00--11:00), P3 (11:00--13:00), P4 (13:00--14:00), and P5 (14:00--15:00). This granularity is crucial because clients may submit orders at different times of the day, requiring accurate volume predictions across all intraday intervals.

As shown in Table~\ref{vol_ret_table}, the spatio-temporal models employing banded weight matrices (YW) and diagonal coefficients (known$\_$YW) achieve statistically significant outperformance, with the YW model delivering the best overall results. This superior performance is theoretically supported by the greater flexibility inherent to the YW methodology. Furthermore, the improvements observed when comparing the adjusted SMA and MAR methods against their standard counterparts underscore the importance of incorporating intraday information for volume prediction. We also observe that while the SMA- and MAR-based methods perform similarly overall, the MAR model holds a slight edge. This marginal advantage is potentially attributable to the MAR model's ability to capture autocorrelation patterns across different days. Notably, these observations remain consistent across all intraday forecasting horizons. 

Accurate volume predictions are critical for transaction price efficiency, as evidenced by their impact on the Volume-Weighted Average Price (VWAP)---a key benchmark that computes the average price weighted by trading volume over a specified period. Table~\ref{vwap_ret_table} quantifies the relative discrepancy between the realized VWAP (using predicted volumes) and the true VWAP, revealing a direct correlation between raw volume prediction accuracy and VWAP performance. The results indicate that the $YW$ and $known\_YW$ models continue to outperform the alternatives, while the MAR- and SMA-based methods exhibit similar performance. Interestingly, the adjusted SMA and MAR methods do not consistently outperform their standard versions in terms of VWAP, despite yielding superior raw volume predictions. This discrepancy arises because VWAP places greater emphasis on the accurate allocation of volume weights across intraday buckets; a more accurate but volatile raw volume forecast can lead to distorted weight distributions, ultimately degrading VWAP accuracy. Nevertheless, the YW-based methods deliver strong performance in both raw volume and VWAP metrics, further reinforcing the advantages of our proposed approach. 

As indicated by the red highlighting in Tables~\ref{vol_ret_table} and~\ref{vwap_ret_table}, our models (YW and known$\_$YW) achieve superior forecasting accuracy. The improvement of the model primarily stems from the incorporation of intraday information from preceding neighboring time buckets. As illustrated through the scatter plot of neighboring buckets within the same day and the analysis of the banded structure of spatial variables, intraday information serves as a key driver of performance gains. This observation is further supported by comparison with the MAR model, which does not utilize intraday information in its predictions. Additionally, panel variables contribute further explanatory power through conditional cross-asset dependencies, as justified by the conditional adjacency analysis in the Appendix. Finally, the fitting procedure produces more smoothed estimated values that mitigate abrupt temporal changes within a given day, unlike approaches such as adjusted SMA where intraday adjustments rely solely on that day's information. In contrast, our fitting method ensures that adjustments are informed by all past intraday relationships, yielding a smoother intraday adjustment process. We further show that these gains in raw volume and VWAP prediction translate directly into improved execution quality for Percentage of Volume (POV) strategies, highlighting the real-world applicability of our methodology.

\subsection{Backtesting Analysis of POV Execution Strategy}
POV execution algorithms are among the most widely used strategies by execution desks. Accordingly, we assess the POV strategy on five Japanese equities, utilizing the volume forecasts generated by the methods in section ~\ref{subsec:vol_pred}. The POV algorithm implements the following procedure:

\begin{itemize}
    \item \textbf{Order Initiation}: The execution process begins when a client submits a request at time $t_0$ (e.g., 09:30) to trade $Z$ shares using a POV strategy on day $t+1$.
    
    \item \textbf{Participation Rate}: The trading desk sets the target participation rate $\alpha$ (for example $\alpha = 10\%$) for POV algorithm, which specifies the fraction of total market volume to target during execution.
    
    \item \textbf{Dynamic Order Sizing}: For each subsequent 5-minute interval $s+1$ on day $t+1$, the system calculates the order quantity as:
    \[
    Q_{t+1,s+1} = \alpha \cdot \hat{V}_{t+1,s+1}
    \]
    where $\hat{V}_{t+1,s+1}$ represents the forecasted total market volume for the next interval. The algorithm then executes an order of size $Q_{t+1,s+1}$ in the market.
    
    \item \textbf{Termination Condition}: The algorithm continues this process iteratively until the total executed volume matches the client's original order size $Z$.
\end{itemize}

The POV execution process described above highlights the critical role of each subsequent 5-minute volume prediction in achieving target participation rates. As the mechanism demonstrates, both overprediction and underprediction fundamentally impact execution outcomes:

\begin{itemize}
    \item \textbf{Overprediction Consequences}: When volumes are overpredicted, the algorithm submits oversized orders at the fixed participation rate. While this reduces time-based price risk through faster execution, the resultant increase in trade visibility and order size often elevates market impact and adverse price movements, ultimately raising total transaction costs.
    
    \item \textbf{Underprediction Consequences}: When volumes are underpredicted, the algorithm generates smaller-than-optimal orders, extending execution duration and amplifying exposure to adverse price movements. This may prevent timely trade completion within the specified window.
\end{itemize}

In light of these volume prediction impacts on execution outcomes, we systematically evaluate POV performance across different volume forecasting methods using two core metrics:
\begin{itemize}
    \item \textbf{Market Impact}: The relative deviation between target and realized participation rates, which quantifies relative participation rate errors:
    \[
     \frac{{\alpha_{\text{actual}} - \alpha_{\text{target}}}}{\alpha_{\text{target}}}\times 100\%
    \]
    where:
    \begin{itemize}
        \item $\alpha_{\text{actual}} = \frac{\sum_{k=1}^N Q_k}{\sum_{k=1}^N V_k}$ is the actual achieved participation rate
        \item $Q_k$: Shares executed in interval $k$
        \item $V_k$: Actual market volume in interval $k$
        \item $N$: Total number of execution intervals
        \item $\alpha_{\text{target}}$: Target participation rate
    \end{itemize}

    \item \textbf{Execution Time}: The relative execution delay which measures the time penalty of prediction errors:
    \[
    \frac{T_{\text{actual}}(\hat{V}) - T_{\text{ideal}}(V)}{T_{\text{ideal}}(V)}\times 100\%
    \]
    where:
    \begin{itemize}
        \item $T_{\text{ideal}}(V)$: Execution time using perfect volume knowledge $V$
        \item $T_{\text{actual}}(\hat{V})$: Execution time using predicted volumes $\hat{V}$
    \end{itemize}
\end{itemize}

We present backtesting results for the period November 1, 2023 to January 31, 2024, generated under a standardized experimental protocol. Daily executions commence at 09:15am with order sizes fixed at $5\%$ of actual daily trading volume. The study evaluates four participation rate scenarios ($10\%$, $20\%$, $40\%$, and $60\%$) with rigorous separation of under-prediction and over-prediction cases throughout the analysis. In Table~\ref{market_impact_ret_table} and Table~\ref{exe_time_ret_table}, the top performers are highlighted in red.

Table~\ref{market_impact_ret_table} presents the results of participation rate deviations across different scenarios. Over-prediction consistently leads to positive errors, indicating that order sizes exceed targets and thereby amplify market impact. Conversely, under-prediction yields negative errors, reflecting order sizes that fall short of intended levels. Overall, among all methods, YW and known$\_$YW demonstrate greater accuracy in maintaining target participation rates, with YW performing slightly better. Similar to the raw volume analysis, the adjusted MAR and SMA methods outperform their baseline counterparts, underscoring the value of incorporating intraday information. In cases where non-YW based methods perform best, the differences are relatively minor, whereas substantial improvements are evident when YW-based methods dominate.

Table~\ref{exe_time_ret_table} reports execution timing metrics across all experimental conditions. Under-prediction results in positive timing errors associated with extended execution windows, whereas over-prediction yields negative errors corresponding to accelerated completion. The YW-based method again proves to be the most reliable approach, maintaining optimal execution speeds across all scenarios, with adjusted MAR and SMA methods consistently outperforming their baseline counterparts.

%Remarkably, the framework demonstrates robust stability regardless of parameter variations. Methodological rankings remain insensitive with respect to changes in either absolute order size or target participation rate. We observe a strong positive correlation between execution quality and raw volume prediction accuracy, with the performance hierarchy persisting consistently across all tested configurations and assets.

The analysis highlights two critical contributions of our spatio-temporal autoregressive framework. First, the model's distinctive capacity to capture both intraday volume patterns and cross-asset dependencies - relationships that conventional approaches fail to properly account for - generates significantly more accurate volume predictions. By explicitly modeling these complex temporal and spatial correlations, our framework overcomes fundamental limitations of traditional methods. Second, these modeling improvements translate directly into enhanced execution performance. The framework's superior forecasts enable POV strategies to maintain tighter adherence to target participation rates and minimize market impact through more accurate order sizing, and optimize execution timing via precise volume trajectory prediction. Crucially, the model achieves these gains consistently across different market scenarios, demonstrating both its robustness and practical utility for algorithmic trading applications.

\bigskip

\noindent
{\bf \large Acknowledgements}. Dou's research was partially supported by RGC of Hong Kong (ECS Grant No. 21508524), He's research was partially supported by the National Natural Science Foundation of China (grant nos. 72473114, 72495121), and Yao's research was partially supported by U.K. EPSRC grants EP/V0075561/1 and EP/X002195/1.

\newpage
%\section{Appendix}

\begin{table}[ht] 
\begin{center}
\begin{tabular}{|c|c|c|c|c|c|c|}
\hline 
\multicolumn{7}{|c|}{Error for $D(\bbeta_0) \otimes D(\balpha_0)$ and $D(\bbeta_1) \otimes D(\balpha_1)$} \\
\hline
 (p, q)& \multicolumn{2}{c|}{ (10, 10)} &\multicolumn{2}{c|}{ (20, 20)}&\multicolumn{2}{c|}{ (30, 30)} \\
\hline
n & $\bbeta_0, \balpha_0$ & $\bbeta_1, \balpha_1$ & $\bbeta_0, \balpha_0$ & $\bbeta_1, \balpha_1$ & $\bbeta_0, \balpha_0$ & $\bbeta_1, \balpha_1$\\\cline{2-7}
200 & 0.31 (0.065) & 0.15 (0.027) & 0.32 (0.064) & 0.20 (0.028) & 0.40 (0.196) & 0.24 (0.035) \\
500 & 0.28 (0.062) & 0.09 (0.017) & 0.27 (0.066) & 0.13 (0.017) & 0.35 (0.147) & 0.16 (0.030) \\
1000 & 0.25 (0.056) & 0.07 (0.012) & 0.24 (0.064) & 0.09 (0.010) & 0.30 (0.101) & 0.11 (0.011) \\
2000 & 0.23 (0.053) & 0.05 (0.008) & 0.22 (0.058) & 0.06 (0.007) & 0.28 (0.099) & 0.08 (0.008) \\
5000 & 0.19 (0.049) & 0.03 (0.005) & 0.19 (0.051) & 0.04 (0.005) & 0.25 (0.087) & 0.05 (0.005) \\
\hline
\end{tabular}
\caption{Error for $D(\bbeta_0) \otimes D(\balpha_0)$ and $D(\bbeta_1) \otimes D(\balpha_1)$ with iterated generalized Yule-Walker estimation method Mean and standard deviations (in parentheses) of $\| D(\hat \bbeta_0) \otimes   D(\hat \balpha_0)  -  D(\bbeta_0) \otimes  D(\balpha_0)\|_F$ and $\| D(\hat \bbeta_1) \otimes   D(\hat \balpha_1) -  D(\bbeta_1) \otimes  D(\balpha_1)\|_F$ are reported for different $(p, q, n)$.}
\label{known_mat_table}
\end{center}
\end{table}

\begin{table}[ht] 
\begin{center}
\begin{tabular}{|c|c|c|c|c|c|c|}
\hline 
\multicolumn{7}{|c|}{Error for $\bB_0 \otimes \bA_0$ given $k_0^{(\bA)}$ and $k_0^{(\bB)}$} \\
\hline
 (p, q)& \multicolumn{2}{c|}{ (10, 10)} &\multicolumn{2}{c|}{ (20, 20)}&\multicolumn{2}{c|}{ (30, 30)} \\
\hline
n & YW & NKP & YW & NKP & YW & NKP \\\cline{2-7}
200 & 1.32 (0.266) & 1.90 (0.369) & 2.06 (0.635) & 2.84 (0.794) & 3.12 (1.057) & 3.90 (1.219) \\
500 & 1.01 (0.206) & 1.60 (0.333) & 1.65 (0.542) & 2.52 (0.755) & 2.38 (0.952) & 3.58 (1.180) \\
1000 & 0.81 (0.157) & 1.41 (0.292) & 1.34 (0.419) & 2.26 (0.693) & 2.02 (0.824) & 3.32 (1.091) \\
2000 & 0.63 (0.114) & 1.24 (0.246) & 1.04 (0.298) & 1.98 (0.599) & 1.64 (0.645) & 3.01 (0.981) \\
5000 & 0.44 (0.075) & 1.06 (0.203) & 0.71 (0.180) & 1.59 (0.470) & 1.15 (0.424) & 2.51 (0.783) \\
\hline
\hline
\multicolumn{7}{|c|}{Error for $\bB_1 \otimes \bA_1$ given $k_0^{(\bA)}$ and $k_0^{(\bB)}$} \\
\hline
 (p, q)& \multicolumn{2}{c|}{ (10, 10)} &\multicolumn{2}{c|}{ (20, 20)}&\multicolumn{2}{c|}{ (30, 30)} \\
\hline
n & YW & NKP & YW & NKP & YW & NKP \\\cline{2-7}
200 & 0.70 (0.059) & 1.04 (0.093) & 1.00 (0.069) & 1.51 (0.112) & 1.61 (0.089) & 1.99 (0.140) \\
500 & 0.43 (0.038) & 0.69 (0.070) & 0.66 (0.055) & 1.14 (0.099) & 0.83 (0.082) & 1.60 (0.131) \\
1000 & 0.30 (0.025) & 0.55 (0.063) & 0.47 (0.040) & 0.96 (0.092) & 0.61 (0.066) & 1.43 (0.133) \\
2000 & 0.21 (0.019) & 0.45 (0.062) & 0.33 (0.029) & 0.82 (0.091) & 0.44 (0.051) & 1.29 (0.124) \\
5000 & 0.13 (0.010) & 0.36 (0.052) & 0.20 (0.016) & 0.66 (0.078) & 0.28 (0.031) & 1.10 (0.114) \\
\hline
\end{tabular}
\caption{Error for $\bB_0 \otimes \bA_0$ and $\bB_1 \otimes \bA_1$ with iterated generalized Yule-Walker estimation method and nearest Kronecker product method with known $k_0^{(\bA)} = 2$ and $k_0^{(\bB)} = 2$. Mean and standard deviations (in parentheses) of $\| \hat \bB_0 \otimes   \hat \bA_0  -  \bB_0 \otimes  \bA_0\|_F$ and $\| \hat \bB_1 \otimes   \hat \bA_1 -  \bB_1 \otimes  \bA_1\|_F$ are reported for different $(p, q, n)$.}
\label{unknown_mat_known_band_table}
\end{center}
\end{table}

\begin{table}[ht]
\begin{center}
\begin{tabular}{|c|c|c|c|c|c|c|}
\hline 
\multicolumn{7}{|c|}{$k_0^{(\bA)} = k_0^{(\bB)} = 2, K=4$} \\
\hline
 (p, q)& \multicolumn{2}{c|}{ (10, 10)} &\multicolumn{2}{c|}{ (20, 20)}&\multicolumn{2}{c|}{ (30, 30)} \\
\hline
n & $\hat k_0^{(\bA)}  = k_0^{(\bA)}$ & $\hat k_0^{(\bB)}  = k_0^{(\bB)}$ & $\hat k_0^{(\bA)}  = k_0^{(\bA)}$ & $\hat k_0^{(\bB)}  = k_0^{(\bB)}$ & $\hat k_0^{(\bA)}  = k_0^{(\bA)}$ & $\hat k_0^{(\bB)}  = k_0^{(\bB)}$ \\\cline{2-7}
200 &  0.56 & 0.34 &  0.59 &  0.46 &  0.58 &  0.76\\
500 & 0.68 & 0.63 &  0.74 &  0.78 & 0.66 &  0.82 \\
1000 &  0.77 &  0.92 &  0.83&  0.96 & 0.74 & 0.97\\
2000 &  0.90 &  0.99 &  0.89 &  0.99 & 0.89 &  0.97 \\
5000 &  0.94 & 1.00 &  0.96 & 0.99 & 0.96 &  1.00\\
\hline 
\hline
\multicolumn{7}{|c|}{$k_0^{(\bA)} = k_0^{(\bB)} = 2, K=6$} \\
\hline
 (p, q)& \multicolumn{2}{c|}{ (10, 10)} &\multicolumn{2}{c|}{ (20, 20)}&\multicolumn{2}{c|}{ (30, 30)} \\
\hline
n & $\hat k_0^{(\bA)}  = k_0^{(\bA)}$ & $\hat k_0^{(\bB)}  = k_0^{(\bB)}$ & $\hat k_0^{(\bA)}  = k_0^{(\bA)}$ & $\hat k_0^{(\bB)}  = k_0^{(\bB)}$ & $\hat k_0^{(\bA)}  = k_0^{(\bA)}$ & $\hat k_0^{(\bB)}  = k_0^{(\bB)}$ \\\cline{2-7}
200 & 0.54 & 0.30 & 0.61 & 0.37 & 0.58 &  0.69\\
500 & 0.68 & 0.59 & 0.74 & 0.74 &  0.66 &  0.82\\
1000 & 0.77 & 0.88 & 0.83 & 0.96 &  0.74 &  0.97\\
2000 & 0.89 & 0.99 & 0.89 & 0.99 &  0.89 & 0.97 \\
5000 & 0.95 & 1.00 & 0.96 & 0.99 & 0.96 &  1.00\\
\hline
\hline
\multicolumn{7}{|c|}{$k_0^{(\bA)} = k_0^{(\bB)} = 2, K=8$} \\
\hline
 (p, q)& \multicolumn{2}{c|}{ (10, 10)} &\multicolumn{2}{c|}{ (20, 20)}&\multicolumn{2}{c|}{ (30, 30)} \\
\hline
n & $\hat k_0^{(\bA)}  = k_0^{(\bA)}$ & $\hat k_0^{(\bB)}  = k_0^{(\bB)}$ & $\hat k_0^{(\bA)}  = k_0^{(\bA)}$ & $\hat k_0^{(\bB)}  = k_0^{(\bB)}$ & $\hat k_0^{(\bA)}  = k_0^{(\bA)}$ & $\hat k_0^{(\bB)}  = k_0^{(\bB)}$ \\\cline{2-7}
200 & 0.57 & 0.25  &  0.57 & 0.26 & 0.56 & 0.51  \\
500 & 0.64 & 0.49 &  0.72 &  0.67 &  0.66 & 0.78  \\
1000 & 0.76 &  0.84 & 0.82 &  0.94 & 0.74 & 0.97 \\
2000 & 0.89 &  0.99 &  0.89 &  0.99 &  0.89 & 0.97  \\
5000 & 0.94 & 1.00  & 0.96 &  0.99 & 0.96 &  1.00 \\
\hline
\end{tabular}
\caption{Relative frequency of the occurrence of the event $\{\hat k_0^{(\bA)}  = k_0^{(\bA)}\}$ and $\{\hat k_0^{(\bB)}  = k_0^{(\bB)}\}$ are reported for different $(p, q, n)$.}
\label{bandwidth_est_table}
\end{center}
\end{table}

\begin{table}[ht] 
\begin{center}
\begin{tabular}{|c|c|c|c|c|c|c|}
\hline 
\multicolumn{7}{|c|}{Error for $\bB_0 \otimes \bA_0$ with unkown  $k_0^{(\bA)}$ and $k_0^{(\bB)}$} \\
\hline
 (p, q)& \multicolumn{2}{c|}{ (10, 10)} &\multicolumn{2}{c|}{ (20, 20)}&\multicolumn{2}{c|}{ (30, 30)} \\
\hline
n & YW & NKP & YW & NKP & YW & NKP \\\cline{2-7}
200 & 2.28 (0.713) & 3.52 (3.029) & 2.69 (1.250) & 4.14 (1.333) & 3.22 (1.163) & 6.06 (1.760)\\
500 & 1.62 (0.938) & 2.30 (1.642) & 1.98 (1.174) & 2.87 (1.225) & 2.54 (0.932) & 4.04 (1.370)\\
1000 & 0.99 (0.648) & 1.60 (0.888) & 1.38 (0.663) & 2.30 (0.838) & 2.10 (0.798) & 3.49 (1.159)\\
2000 & 0.64 (0.152) & 1.27 (0.359) & 1.06 (0.439) & 1.97 (0.669) & 1.70 (0.629) & 3.17 (1.035) \\
5000 & 0.44 (0.090) & 1.07 (0.242) & 0.71 (0.195)  & 1.60 (0.471) & 1.18 (0.403) & 2.63 (0.823) \\
\hline
\hline
\multicolumn{7}{|c|}{Error for $\bB_1 \otimes \bA_1$ with unkown  $k_0^{(\bA)}$ and $k_0^{(\bB)}$} \\
\hline
 (p, q)& \multicolumn{2}{c|}{ (10, 10)} &\multicolumn{2}{c|}{ (20, 20)}&\multicolumn{2}{c|}{ (30, 30)} \\
\hline
n & YW & NKP & YW & NKP & YW & NKP \\\cline{2-7}
200 & 1.82 (0.811) & 2.64 (2.384) & 1.77 (1.360) & 2.73 (1.097) & 1.57 (0.596) & 3.37 (0.789)\\
500 & 1.03 (0.889) & 1.32 (1.025) & 1.04 (1.176) & 1.49 (1.024) & 0.87 (0.115) & 1.71 (0.248)\\
1000 & 0.46 (0.507)& 0.72 (0.587) & 0.53 (0.522) & 1.02 (0.457) & 0.62 (0.073) & 1.45 (0.147)\\
2000 & 0.22 (0.156) & 0.48 (0.158) & 0.35 (0.293) & 0.84 (0.274) & 0.45 (0.058) & 1.30 (0.136)\\
5000 & 0.13 (0.011) & 0.37 (0.070) & 0.20 (0.028) & 0.66 (0.085) & 0.28 (0.034) & 1.11 (0.120)\\
\hline
\end{tabular}
\caption{Error for $\bB_0 \otimes \bA_0$ and $\bB_1 \otimes \bA_1$ with iterated generalized Yule-Walker estimation method and nearest Kronecker product method with unknown $k_0^{(\bA)}$ and $k_0^{(\bB)}$. Mean and standard deviations (in parentheses) of $\| \hat \bB_0 \otimes   \hat \bA_0  -  \bB_0 \otimes  \bA_0\|_F$ and $\| \hat \bB_1 \otimes   \hat \bA_1 -  \bB_1 \otimes  \bA_1\|_F$ are reported for different $(p, q, n)$.}
\label{unknown_mat_unknown_band_table}
\end{center}
\end{table}

%\begin{figure}[ht]
%\centering
%  \includegraphics[width=0.70\linewidth]{figures/vol_plots.png}
%  \caption{Volume plots for all assets on a sample date 29/12/2023. Blue: %SMA, Orange: adjusted SMA, Green: YW, Red: Realized.}
%\label{figVOL}
%\end{figure}

%\newpage

\begin{table}[ht] 
\begin{center}
\begin{tabular}{|c|c|c|c|c|c|c|c|c|c|}
\hline 
\multicolumn{2}{|c|}{VOLUME} & 
\makecell{full \\ trading day} & 
\makecell{morning \\ session} & 
\makecell{afternoon \\ session} & 
p1 & p2 & p3 & p4 & p5 \\
\hline
\hline
\multirow{5}{*}{9984} 
 & YW & \cellcolor{red!20}0.39  & \cellcolor{red!20}0.38 & \cellcolor{red!20}0.41  & \cellcolor{red!20}0.34 & \cellcolor{red!20}0.39 & \cellcolor{red!20}0.43 & \cellcolor{red!20}0.45 & \cellcolor{red!20}0.36 \\ 
 & known$\_$YW & 0.41 & 0.40  & 0.42  & 0.36 & 0.41 & 0.44 & 0.47 & 0.37 \\ 
 & MAR & 0.52  & 0.52  & 0.53  & 0.49 & 0.52 & 0.54 & 0.59 & 0.48 \\ 
 & adj$\_$MAR & 0.43  & 0.41  & 0.44  & 0.35 & 0.43 & 0.46 & 0.49 & 0.40 \\ 
 & SMA & 0.56  & 0.56  & 0.56  & 0.51 & 0.56 & 0.60 & 0.63 & 0.50 \\ 
 & adj$\_$SMA & 0.44  & 0.43  & 0.45  & 0.36 & 0.45 & 0.49 & 0.50 & 0.39 \\ 
\hline
\hline
\multirow{5}{*}{6758}  
 & YW & 0.43  & 0.42  & 0.45  & 0.38 & 0.44 & \cellcolor{red!20}0.44 & \cellcolor{red!20}0.48 & 0.42 \\ 
 & known$\_$YW & \cellcolor{red!20}0.42  & \cellcolor{red!20}0.40  & \cellcolor{red!20}0.44  & \cellcolor{red!20}0.36 & \cellcolor{red!20}0.43 & \cellcolor{red!20}0.44 & \cellcolor{red!20}0.48 & \cellcolor{red!20}0.39 \\ 
 & MAR & 0.51  & 0.50  & 0.53  & 0.45 & 0.54 & 0.53 & 0.56 & 0.49 \\ 
 & adj$\_$MAR & 0.48  & 0.45  & 0.51  & 0.38 & 0.49 & 0.49 & 0.54 & 0.49 \\ 
 & SMA & 0.63  & 0.61  & 0.65  & 0.54 & 0.67 & 0.67 & 0.70 & 0.56 \\ 
 & adj$\_$SMA & 0.49  & 0.47  & 0.52  & 0.41  & 0.52 & 0.53 & 0.57 & 0.45  \\ 
\hline
\hline
\multirow{5}{*}{8035} 
 & YW & \cellcolor{red!20}0.33  & \cellcolor{red!20}0.33  & \cellcolor{red!20}0.34  & 0.31 & 0.35 & \cellcolor{red!20}0.34 & \cellcolor{red!20}0.36 & \cellcolor{red!20}0.32 \\ 
 & known$\_$YW & 0.34  & \cellcolor{red!20}0.33  & 0.35  & \cellcolor{red!20}0.30 & \cellcolor{red!20}0.34 & \cellcolor{red!20}0.34 & 0.37 & 0.33 \\ 
 & MAR & 0.43  & 0.41  & 0.45  & 0.39 & 0.43 & 0.42 & 0.47 & 0.44 \\ 
 & adj$\_$MAR & 0.39  & 0.37  & 0.41  & 0.31 & 0.40 & 0.38 & 0.42 & 0.42 \\ 
 & SMA & 0.48  & 0.46  & 0.49  & 0.44 & 0.48 & 0.49 & 0.53 & 0.46 \\ 
 & adj$\_$SMA & 0.38  & 0.37  & 0.39  & 0.33 & 0.39 & 0.39 & 0.42 & 0.36 \\ 
\hline
\hline
\multirow{5}{*}{6723} 
 & YW & \cellcolor{red!20}0.31  & \cellcolor{red!20}0.32  & \cellcolor{red!20}0.30  & \cellcolor{red!20}0.32 & \cellcolor{red!20}0.32 & \cellcolor{red!20}0.34 & \cellcolor{red!20}0.32 & \cellcolor{red!20}0.27 \\ 
 & known$\_$YW & 0.34  & 0.34  & 0.33  & 0.34 & 0.34 & 0.36 & 0.35 & 0.28  \\ 
 & MAR & 0.49  & 0.49  & 0.48  & 0.49 & 0.49 & 0.51  & 0.51 & 0.42 \\ 
 & adj$\_$MAR & 0.38  & 0.39  & 0.38  & 0.35 & 0.41 & 0.42 & 0.40 & 0.33 \\ 
 & SMA & 0.62  & 0.62  & 0.62  & 0.62 & 0.63 & 0.65 & 0.65 & 0.54 \\ 
 & adj$\_$SMA &  0.38 & 0.39  & 0.37  & 0.36 & 0.41 & 0.42 & 0.41 & 0.31 \\  
\hline
\hline
\multirow{5}{*}{7974} 
 & YW & \cellcolor{red!20}0.37  & \cellcolor{red!20}0.39  & \cellcolor{red!20}0.36  & \cellcolor{red!20}0.36 & 0.39 & 0.43 & \cellcolor{red!20}0.38 & \cellcolor{red!20}0.31 \\ 
 & known$\_$YW & 0.38  & \cellcolor{red!20}0.39  & 0.37  & 0.37 & \cellcolor{red!20}0.38 & \cellcolor{red!20}0.42 & 0.39  & 0.32 \\ 
 & MAR & 0.51  & 0.48  & 0.53  & 0.46 & 0.48 & 0.58 & 0.56 & 0.47 \\ 
 & adj$\_$MAR & 0.45  & 0.41  & 0.49  & 0.39 & 0.38 & 0.51 & 0.50 & 0.45 \\ 
 & SMA &  0.60 & 0.60  & 0.60  & 0.60 & 0.60 & 0.67 & 0.64 & 0.50 \\ 
 & adj$\_$SMA & 0.44  & 0.42  & 0.45  & 0.44 & 0.39 & 0.51  & 0.46  & 0.38 \\ 
\hline
\end{tabular}
\caption{Mean absolute relative error of the raw volume prediction for each asset across different intraday periods: 9am-3pm (full trading day), 9am-11:30am (morning session), 12:30pm-3pm (afternoon session), 9am-10am (p1), 10am-11am (p2), 11am-1pm (p3), 1pm-2pm (p4), 2pm-3pm (p5). The smallest value in each column per stock is highlighted in red.}
\label{vol_ret_table}
\end{center}
\end{table}

\newpage

\begin{table}[ht] 
\begin{center}
\begin{tabular}{|c|c|c|c|c|c|c|c|c|c|}
\hline 
\multicolumn{2}{|c|}{VWAP} & whole day & morning & afternoon & p1 &  p2 &  p3 &  p4 &  p5\\
\hline
\hline
\multirow{5}{*}{9984} 
 & YW & \cellcolor{red!20}3.92  & \cellcolor{red!20}4.31 & 1.67  & \cellcolor{red!20}4.09 & \cellcolor{red!20}3.03 & \cellcolor{red!20}2.29 & \cellcolor{red!20}1.50 & 1.06 \\ 
 & known$\_$YW & 4.49 & 4.69  & \cellcolor{red!20}1.64  & 4.17 & 3.10 & 2.39 & 1.64 & \cellcolor{red!20}1.05 \\ 
 & MAR & 6.16  & 6.15  & 2.16  & 4.77 & 3.71 & 3.75 & 1.86 & 1.21 \\ 
 & adj$\_$MAR & 5.48  & 5.80  & 2.05  & 5.38 & 3.65 & 3.57 & 1.83 & 1.21 \\ 
 & SMA & 6.32 & 6.13  & 2.25  & 4.82 & 3.61 & 3.52 & 1.94 & 1.29 \\ 
 & adj$\_$SMA & 5.84  & 6.04  & 2.16  & 5.62 & 3.56 & 3.33 & 1.91 & 1.29\\ 
\hline
\hline
\multirow{5}{*}{6758}  
 & YW & 3.61  & \cellcolor{red!20}3.58 & 1.55  & 3.38 & \cellcolor{red!20}1.97 & \cellcolor{red!20}2.42 & \cellcolor{red!20}1.23 & 1.36 \\ 
 & known$\_$YW & \cellcolor{red!20}3.59 & 3.61  & \cellcolor{red!20}1.30  & \cellcolor{red!20}3.17 & 2.10 & 2.44 & 1.24 & \cellcolor{red!20}1.25 \\ 
 & MAR & 4.94  & 4.70  & 1.56  & 3.83 & 2.38 & 3.42 & 1.56 & 1.55 \\ 
 & adj$\_$MAR & 3.98  & 4.21  & 1.60  & 3.94 & 2.32 & 3.32 & 1.57 & 1.58 \\ 
 & SMA & 5.13  & 4.61  & 1.85  & 3.69 & 2.61 & 3.18 & 1.58 & 1.72 \\ 
 & adj$\_$SMA & 4.40  & 4.65  & 1.92  & 4.22 & 2.56 & 3.05 & 1.61 & 1.73\\  
\hline
\hline
\multirow{5}{*}{8035} 
 & YW & \cellcolor{red!20}4.26  & \cellcolor{red!20}4.66 & \cellcolor{red!20}1.33  & 4.69 & 2.55 & \cellcolor{red!20}2.39 & \cellcolor{red!20}1.41 & 0.97 \\ 
 & known$\_$YW & 5.16 & 5.01  & 1.52  & 4.70 & \cellcolor{red!20}2.41 & 2.82 & 1.51 & \cellcolor{red!20}0.80 \\ 
 & MAR & 6.00  & 5.37  & 2.06  & \cellcolor{red!20}4.08 & 3.08 & 3.96 & 1.94 & 0.94 \\ 
 & adj$\_$MAR & 6.42  & 6.53  & 1.90  & 5.92 & 2.96 & 3.84 & 1.92 & 0.93 \\ 
 & SMA & 6.24  &  5.63 & 2.32  & 4.15 & 3.32 & 3.90 & 2.08 & 1.01 \\ 
 & adj$\_$SMA & 6.49  & 6.65  & 2.12  & 5.96 & 3.18 & 3.79 & 2.05 & 1.00\\ 
\hline
\hline
\multirow{5}{*}{6723} 
 & YW & \cellcolor{red!20}4.75  & \cellcolor{red!20}5.21 & \cellcolor{red!20}2.29  & \cellcolor{red!20}5.79 & \cellcolor{red!20}2.80 & \cellcolor{red!20}2.81 & \cellcolor{red!20}1.97 & \cellcolor{red!20}2.32 \\ 
 & known$\_$YW & 6.80 & 6.08  & 2.52  & 6.05 & 2.88 & 3.22 & 2.06 & 2.33 \\ 
 & MAR & 9.63  & 7.28  & 3.25  & 5.91 & 3.44 & 5.45 & 2.36 & 2.57 \\ 
 & adj$\_$MAR & 8.38  & 8.25  & 3.21  & 6.96 & 3.33 & 5.24 & 2.37 & 2.61 \\ 
 & SMA & 9.71  & 7.36  & 3.47  & 6.14 & 3.58 & 5.41 & 2.48 & 2.65 \\ 
 & adj$\_$SMA & 8.96  & 8.38  & 3.39  & 7.03 & 3.45 & 5.19 & 2.48 & 2.67\\  
\hline
\hline
\multirow{5}{*}{7974} 
 & YW & 3.61  & \cellcolor{red!20}4.08 & 1.70  & 4.35 & \cellcolor{red!20}1.70 & 2.09 & 1.22 & \cellcolor{red!20}1.01 \\ 
 & known$\_$YW & \cellcolor{red!20}3.57 & 4.22  & \cellcolor{red!20}1.62  & 4.43 & 1.71 & \cellcolor{red!20}1.79 & 1.24 & 1.07 \\ 
 & MAR & 4.85  & 4.86  & 2.21  & 4.58 & 2.41 & 2.87 & \cellcolor{red!20}1.18 & 1.26 \\ 
 & adj$\_$MAR & 5.29  & 5.62  & 2.27  & 6.22 & 2.35 & 2.78 & 1.20 & 1.26 \\ 
 & SMA &  5.02 & 4.60  & 2.49  & \cellcolor{red!20}4.17 & 2.40 & 2.86 & 1.18 & 1.39 \\ 
 & adj$\_$SMA &  5.94 & 6.10  & 2.52  & 6.66 & 2.35 & 2.76 & 1.18 & 1.37\\  
\hline
\end{tabular}
\caption{Mean absolute relative error (in bps) of VWAP for each asset across different intraday periods: 9am-3pm (whole day), 9am-11:30am (morning), 12:30pm-3pm (afternoon), 9am-10am (p1), 10am-11am (p2), 11am-1pm (p3), 1pm-2pm (p4), 2pm-3pm (p5). The smallest value in each column per stock is highlighted in red.}
\label{vwap_ret_table}
\end{center}
\end{table}

%\newpage

%\begin{figure}[ht]
%\centering
%  \includegraphics[width=0.90\linewidth]{figures/6758_overest_pr.png}
%  \caption{cumulative execution rate curves for for 6758 on 14/11/2023 from %volume inline}
%\label{6758_overest_pr}
%\end{figure}

%\begin{figure}[ht]
%\centering
%  \includegraphics[width=0.90\linewidth]{figures/6758_underest_pr.png}
%  \caption{cumulative execution rate curves for for 6758 on 17/01/2024 from %volume inline}
%\label{6758_underest_pr}
%\end{figure}

%\newpage

%\begin{figure}[ht]
%\centering
%  \includegraphics[width=0.90\linewidth]{figures/6723_overest_pr.png}
%  \caption{cumulative execution rate curves for for 6723 on 29/12/2023 from %volume inline}
%\label{6723_overest_pr}
%\end{figure}

%\begin{figure}[ht]
%\centering
%  \includegraphics[width=0.90\linewidth]{figures/6723_underest_pr.png}
%  \caption{cumulative execution rate curves for for 6723 on 22/01/2024 from %volume inline}
%\label{6723_underest_pr}
%\end{figure}

%\newpage

\begin{table}[ht] 
\begin{center}
\begin{tabular}{|c|c|c|c|c|c|c|c|c|c|}
\hline
\multicolumn{2}{|c|}{MARKET IMPACT} &\multicolumn{4}{|c|}{over prediction of volume} & \multicolumn{4}{|c|}{under prediction of volume}\\ 
\hline
\hline
\multicolumn{2}{|c|}{Participation Rate} & 0.1 & 0.2 & 0.4 & 0.6 &  0.1 &  0.2 & 0.4 & 0.6\\
\hline
\multirow{5}{*}{9984} 
 & YW & \cellcolor{red!20}6.71  & \cellcolor{red!20}10.39  & \cellcolor{red!20}11.60 & \cellcolor{red!20}13.67 & \cellcolor{red!20}-11.59 & \cellcolor{red!20}-13.54 & -19.71 & -23.26 \\ 
 & known$\_$YW & 14.10  & 15.47  & 14.82 & 16.67 & -15.51 & -14.84 & \cellcolor{red!20}-18.67 & \cellcolor{red!20}-22.18 \\
 & MAR & 32.43  & 36.88  & 40.10 & 32.98 & -28.75 & -25.30 & -24.55 & -28.36 \\ 
 & adj$\_$MAR & 13.14  & 16.35  & 14.91 & 22.65 & -14.75 & -17.11 & -19.25 & -24.87 \\ 
 & SMA & 37.43  & 41.96  & 41.10 & 37.76 & -23.25 & -23.08 & -23.36 & -27.86 \\ 
 & adj$\_$SMA & 13.04  & 15.61  & 18.64 & 22.34 & -16.48 & -18.37 & -19.74 & -23.72 \\  
\hline
\hline
\multirow{5}{*}{6758} 
 & YW & \cellcolor{red!20}11.25  & 16.72  & 17.14 & 19.34 & -16.00 & \cellcolor{red!20}-17.45 & -22.34 & \cellcolor{red!20}-24.83 \\ 
 & known$\_$YW & 11.41  & \cellcolor{red!20}13.54  & \cellcolor{red!20}14.62 & \cellcolor{red!20}14.80 & -17.80 & -20.02 & -22.39 & -25.15 \\
 & MAR & 46.59  & 33.59  & 39.51 & 39.50 & -26.09 & -32.77 & -31.70 & -32.23 \\ 
 & adj$\_$MAR & 19.59  & 19.76  & 24.49 & 19.16 & \cellcolor{red!20}-14.97 & -23.55 & -23.11 & -27.54 \\ 
 & SMA & 40.40  & 41.66  & 37.89 & 36.38 & -25.98 & -31.18 & -33.83 & -33.60 \\ 
 & adj$\_$SMA & 18.90 & 21.90  & 25.97 & 22.26 & -16.65 & -20.92 & \cellcolor{red!20}-21.23 & -26.28 \\
\hline
\hline
\multirow{5}{*}{8035} 
 & YW & \cellcolor{red!20}8.56  & \cellcolor{red!20}9.07  & \cellcolor{red!20}14.64 & \cellcolor{red!20}13.25 & \cellcolor{red!20}-11.87 & \cellcolor{red!20}-14.39 & \cellcolor{red!20}-17.15 & \cellcolor{red!20}-21.24 \\ 
 & known$\_$YW & 9.56  & 10.70  & 15.70 & 18.14 & -15.05 & -14.98 & -18.01 & -22.36 \\
 & MAR & 31.23  & 39.70  & 26.70 & 32.39 & -24.22 & -25.83 & -28.71 & -29.18 \\ 
 & adj$\_$MAR & 20.11  & 21.27  & 16.33 & 34.11 & -12.63 & -14.91 & -20.74 & -22.09 \\ 
 & SMA & 36.47  & 43.57  & 34.87 & 31.21 & -19.16 & -17.97 & -24.08 & -25.75 \\ 
 & adj$\_$SMA &  18.32 & 22.00  & 17.56 & 38.59 & -14.35 & -14.59  & -19.80 & -23.36 \\
\hline
\hline
\multirow{5}{*}{6723} 
 & YW & \cellcolor{red!20}7.09  & \cellcolor{red!20}8.31  & \cellcolor{red!20}14.67 & 19.90 & \cellcolor{red!20}-11.84 & \cellcolor{red!20}-11.76 & \cellcolor{red!20}-16.83 & \cellcolor{red!20}-23.05  \\ 
 & known$\_$YW & 11.27  & 11.26  & 16.92 & 19.71 & -17.17 & -16.73 & -19.58 & -23.59  \\
 & MAR & 51.99  & 51.92  & 48.20 & 59.17 & -31.17 & -33.72  & -38.28 & -39.05 \\ 
 & adj$\_$MAR & 18.62  & 21.29  & 24.97 & \cellcolor{red!20}19.07 & -14.09  & -15.14  & -18.67 & -23.58 \\ 
 & SMA & 56.57  & 63.12  & 64.09 & 63.23 & -28.98 &  -29.07 & -32.01 & -34.34 \\ 
 & adj$\_$SMA & 18.65  & 21.23  & 24.49 & 21.64 & -14.90 &  -15.73 & -20.00 & -24.80 \\
\hline
\hline
\multirow{5}{*}{7974} 
 & YW & \cellcolor{red!20}8.13  & \cellcolor{red!20}7.35  & \cellcolor{red!20}18.40 & 24.04 & -14.41  & -18.00 & \cellcolor{red!20}-21.53 & -24.80 \\ 
 & known$\_$YW & 8.46  & 12.17  & 20.97 & \cellcolor{red!20}19.98 & -17.19 & -16.93  & -21.91  & \cellcolor{red!20}-23.53 \\
 & MAR & 31.23  & 35.84  & 32.75 & 43.04 & -33.52 & -37.10  & -41.15 & -39.54 \\ 
 & adj$\_$MAR & 24.59  & 26.35  & 31.44 & 30.13 & \cellcolor{red!20}-14.33  & -17.45  & -22.15 & -27.01 \\ 
 & SMA & 51.77  & 50.63  & 56.93 & 53.51 & -26.66 & -32.24  & -33.82 & -34.66 \\ 
 & adj$\_$SMA & 24.06  & 28.64  & 35.26 & 35.76 & -14.68  & \cellcolor{red!20}-16.83  & -22.81 & -26.36 \\
\hline
\end{tabular}
\caption{Difference between actual participation rate and target participation rate: mean relative error (in $\%$) across different prediction dates. The smallest absolute value in each column per stock is highlighted in red.}
\label{market_impact_ret_table}
\end{center}
\end{table}

\begin{table}[ht] 
\begin{center}
\begin{tabular}{|c|c|c|c|c|c|c|c|c|c|}
\hline
\multicolumn{2}{|c|}{EXECUTION TIME} &\multicolumn{4}{|c|}{over prediction of volume} & \multicolumn{4}{|c|}{under prediction of volume}\\ 
\hline
\hline
\multicolumn{2}{|c|}{Participation Rate} & 0.1 & 0.2 & 0.4 & 0.6 &  0.1 &  0.2 & 0.4 & 0.6\\
\hline
\multirow{5}{*}{9984} 
 & YW & \cellcolor{red!20}-9.91 & \cellcolor{red!20}-14.71 & \cellcolor{red!20}-23.76 & \cellcolor{red!20}-27.78 & \cellcolor{red!20}16.78 & \cellcolor{red!20}16.08  & \cellcolor{red!20}17.26 & \cellcolor{red!20}11.62 \\ 
 & known$\_$YW & -16.30  & -18.61  & -26.91 & -31.25 & 26.16 & 22.84  & 19.11 & \cellcolor{red!20}11.62 \\
 & MAR & -28.31  & -32.34  & -39.44 & -35.93 & 66.41 & 63.23  & 42.82 & 28.57 \\ 
 & adj$\_$MAR &  -14.92 & -20.21  & -26.03 & -32.29 & 19.73 & 21.84  & 20.96 & 15.99 \\ 
 & SMA & -32.68  & -35.87  & -39.91 & -38.08 & 52.53  & 56.00  & 39.20 & 27.16 \\ 
 & adj$\_$SMA & -15.00  & -19.10  & -28.67 & -32.35 & 22.00 & 24.23  & 21.11 & 13.66 \\  
\hline
\hline
\multirow{5}{*}{6758} 
 & YW & -9.93  & -20.59  & -24.31 & -31.22 & \cellcolor{red!20}19.81 & \cellcolor{red!20}25.23  & \cellcolor{red!20}27.29 & 28.51 \\ 
 & known$\_$YW & \cellcolor{red!20}-9.31  & \cellcolor{red!20}-16.62  & \cellcolor{red!20}-22.11 & \cellcolor{red!20}-29.05 & 23.60 & 30.45  & 29.61 & \cellcolor{red!20}26.71 \\
 & MAR & -28.86  & -28.82  & -34.69 & -42.22 & 52.42 &  82.87 & 74.90 & 61.79 \\ 
 & adj$\_$MAR & -17.37  & -20.24  & -25.28 & -30.18 & 17.04  & 33.29  & 36.43 & 37.75 \\ 
 & SMA & -27.66  & -33.14  & -33.02 & -37.73 & 55.41 & 73.26  & 75.96 & 55.65 \\ 
 & adj$\_$SMA & -16.57  & -22.79  & -27.04 & -30.61 & 18.41 &  28.98 & 35.13 & 34.41 \\
\hline
\hline
\multirow{5}{*}{8035} 
 & YW & -12.74  & \cellcolor{red!20}-15.80  & \cellcolor{red!20}-25.02 & \cellcolor{red!20}-30.77 & 18.50 & \cellcolor{red!20}19.54  & \cellcolor{red!20}13.33 & \cellcolor{red!20}10.83 \\ 
 & known$\_$YW & \cellcolor{red!20}-11.30  & -16.56  & -26.35 & -32.29 & 25.94 & 22.92  & 14.21 & 12.61 \\
 & MAR & -30.64  & -35.38  & -31.60 & -34.09 & 52.39 & 56.31  & 49.19 & 30.16 \\ 
 & adj$\_$MAR & -21.63  & -23.11  & -25.61 & -39.58 & \cellcolor{red!20}17.90 & 20.96  & 20.74 & 14.07 \\ 
 & SMA & -32.68  & -36.02  & -32.75 & -35.56 & 37.62 & 35.28  & 38.28 & 23.25 \\ 
 & adj$\_$SMA & -19.77  & -23.56  & -25.83 & -38.54 & 21.02 & 20.87  & 17.70 & 16.67 \\
\hline
\hline
\multirow{5}{*}{6723} 
 & YW & \cellcolor{red!20}-11.83  & \cellcolor{red!20}-15.43  & \cellcolor{red!20}-23.20 & -32.78 & \cellcolor{red!20}14.06 & \cellcolor{red!20}12.81 & \cellcolor{red!20}14.34 & \cellcolor{red!20}25.81 \\ 
 & known$\_$YW & -15.23  & -16.93  & -26.31 & -32.02 & 29.24 & 25.81 & 21.95 & 26.92 \\
 & MAR & -38.45  & -34.84  & -40.11 & -37.89 & 69.83 & 104.53 & 109.52 & 100.00 \\ 
 & adj$\_$MAR & -22.78  & -24.94  & -28.40 & \cellcolor{red!20}-31.76 & 15.65 & 19.60 & 20.67 & 27.31 \\ 
 & SMA & -38.74  & -40.25  & -45.83 & -42.62 & 69.22 & 92.91 & 80.32 & 83.07 \\ 
 & adj$\_$SMA &  -22.25 & -24.37  & -28.02 & -31.85 & 16.93 & 21.42 & 23.76 & 30.24 \\
\hline
\hline
\multirow{5}{*}{7974} 
 & YW & \cellcolor{red!20}-9.43  & \cellcolor{red!20}-13.18  & \cellcolor{red!20}-22.84 & -29.76 & 16.25 & \cellcolor{red!20}25.58 & \cellcolor{red!20}19.14 & 19.66 \\ 
 & known$\_$YW & -9.94  & -18.38  & -25.01 & \cellcolor{red!20}-28.57 & 28.46 & 30.93 & 19.96 & \cellcolor{red!20}16.24 \\
 & MAR &  -23.11 & -30.12  & -26.53 & -34.56 & 66.50 & 137.14 & 128.10 & 103.51 \\ 
 & adj$\_$MAR & -16.84  & -24.80  & -26.00 & -29.90 & \cellcolor{red!20}13.67 & 30.07 & 23.57 & 27.25 \\ 
 & SMA & -36.16  & -39.21  & -36.31 & -39.77 & 47.79 & 111.40 & 97.64 & 80.11 \\ 
 & adj$\_$SMA & -15.23  & -26.06  & -28.40 & -31.27 & 14.03 & 27.82 & 23.57 & 22.66 \\   
\hline
\end{tabular}
\caption{Difference between actual execution time and ideal execution time: mean relative error (in $\%$) across different prediction dates. The smallest absolute value in each column per stock is highlighted in red.}
\label{exe_time_ret_table}
\end{center}
\end{table}

\clearpage
%\newpage
\onehalfspacing
\normalsize
\numberwithin{equation}{section}

\setcounter{equation}{0}
\setcounter{section}{0}
\setcounter{table}{0}
\setcounter{figure}{0}

\pagenumbering{arabic}
\setcounter{page}{1}
\renewcommand{\thepage}{S\arabic{page}}
\renewcommand{\thesection}{\Alph{section}}

\begin{center}
	{\bf  \Large
		Supplementary material for ``Spatio-Temporal Autoregressions for High Dimensional Matrix-Valued Time Series" by Baojun Dou, Jing He, Sudhir Tiwari, and Qiwei Yao}  \\
\end{center}

\bigskip

\bigskip

\allowdisplaybreaks

The supplementary material includes all technical proofs of the main results in this article. We begin by introducing some notation that will be used throughout the proofs. 
Let $C$, $C_1$, $\ldots$ denote generic positive constants that are independent of $n$, $p$, and $q$, and their values may vary in different contexts. For any positive integer $d$, write $[d] =\{1, \ldots, d\}$, and denote by $\bI_d$ the $d \times d$ identity matrix. Let $\be_l$ denote the unit vector with the $l$-th element being 1, where its dimension may be different in different places. For any positive integers $d_1$ and $d_2$, let ${\bf 0}_{d_1 \times d_2}$ denote a $d_1 \times d_2$ matrix consisting entirely of zeros. For a vector $\ba = (a_1,\ldots, a_k)' \in \mathbb{R}^k$, let $\|\ba\|_2 = (\sum_{i=1}^k a_i^2)^{1/2}$ denote its $L_2$-norm. For a matrix $\bA = (A_{i,j})_{k_1 \times k_2}$, we write $\|\bA\|_{\infty}=\max_{i \in [k_1], j \in [k_2]}|A_{i,j}|$, $\|\bA\|_2 = \lambda_{\max}^{1/2}(\bA'\bA)$ and $\|\bA\|_{F}=(\sum_{i=1}^{k_1}\sum_{j=1}^{k_2}A_{i,j}^2)^{1/2}$, where $\lambda_{\max}(\cdot)$ is the largest eigenvalue of a matrix. For any $d_1$-dimensional vector $\bZ=(Z_1,\ldots,Z_{d_1})'$ and $d_2$-dimensional vector $\boldsymbol{\eta} = (\eta_1, \ldots, \eta_{d_2})'$, write the derivative of $\bZ$ with respect to $\boldsymbol{\eta}$ as
\begin{align*}
    \frac{\partial \bZ}{\partial \boldsymbol{\eta}} = \begin{pmatrix}
        \frac{\partial Z_1}{\partial \eta_1}\,, & \cdots \,, & \frac{\partial Z_1}{\partial \eta_{d_2}} \\
        \vdots  &   &  \vdots \\
        \frac{\partial Z_{d_1}}{\partial \eta_1}\,, & \cdots \,, & \frac{\partial Z_{d_1}}{\partial \eta_{d_2}}
    \end{pmatrix}\,.
\end{align*}

Before presenting the main results, we first introduce a lemma that will be used in the following proofs. The proof of Lemma \ref{lem:Sigmajk} will be given in Section \ref{lem:1}.

\begin{lemma}\label{lem:Sigmajk}
Under Condition \ref{cond:mixing}, 
%if $p,q \ge 1$ are fixed, 
it holds that 
\begin{align*}
    \max_{j,k \in [q]}\|\hat{\bSigma}_{jk}(1)-\bSigma_{jk}(1)\|_{\infty} = O_{\rm p}\{(pq)^{2\kappa/(2+\kappa)}n^{-2/(2+\kappa)}\} = \max_{j,k \in [q]}\|\hat{\bSigma}_{jk}(0)-\bSigma_{jk}(0)\|_{\infty}\,,
\end{align*}
and 
\begin{align*}
    \max_{j,k \in [q]} \|\hat{\bSigma}_{jk}(1)-\bSigma_{jk}(1)\|_F = O_{\rm p}\{ p^{(2+3\kappa)/(2+\kappa)}q^{2\kappa/(2+\kappa)}n^{-2/(2+\kappa)}\} = \max_{j,k \in [q]} \|\hat{\bSigma}_{jk}(0)-\bSigma_{jk}(0)\|_F\,.
\end{align*}
%$\|\hat{\bSigma}_{jk}(1) - \bSigma_{jk}(1) \|_F = O_{\rm p}\{p^{2(1+\kappa)/(\kappa+2)}n^{-(4+\kappa)/(2\kappa+4)}\}$ and $\|\hat{\bSigma}_{jk}(0) - \bSigma_{jk}(0) \|_F = O_{\rm p}\{p^{2(1+\kappa)/(\kappa+2)}n^{-(4+\kappa)/(2\kappa+4)}\}$ for any $j,k \in [q]$.
%$\|\hat{\bSigma}_{jk}(1) - \bSigma_{jk}(1) \|_F = O_{\rm p}(n^{-1/2})$ and $\|\hat{\bSigma}_{jk}(0) - \bSigma_{jk}(0) \|_F = O_{\rm p}(n^{-1/2})$ for any $j,k \in [q]$ for $fixed $p,q$.
\end{lemma}

\section{Proof of Proposition \ref{pn:consistency}}

Write $\bar{\balpha}_{k} = (\bar{\alpha}_{1}^{(k)},\ldots, \bar{\alpha}_{p}^{(k)})'$ and $\bar{\bbeta}_{k} = (\bar{\beta}_{1}^{(k)},\ldots, \bar{\beta}_{q}^{(k)})'$ for $k = 0, 1$. Let $\bar{\bgamma} = (\bar{\ba}_1',\ldots,\bar{\ba}_p',\bar{\bb}_1',\ldots,\bar{\bb}_q')'$ be a $(2p+2q)$-dimensional parameter with $\bar{\ba}_m = (\bar{\alpha}_{m}^{(0)},\bar{\alpha}_{m}^{(1)})'$ and $\bar{\bb}_j = (\bar{\beta}_{j}^{(0)},\bar{\beta}_{j}^{(1)})'$.
Let 
\begin{align}\label{eq:hatSjk}
   \hat{\bS}_{j,k}(\bar{\bgamma}) & = \hat\bSigma_{jk}(1) - \sum_{i=1}^q D(\bar{\balpha}_0) \bW_0 \hat \bSigma_{ik}(1) v_{j, i}^{(0)} \bar{\beta}_{j}^{(0)} - \sum_{i=1}^q D(\bar{\balpha}_1) \bW_1 \hat \bSigma_{ik}(0) v_{j, i}^{(1)} \bar{\beta}_{j}^{(1)} \\
   & = \frac{1}{n}\sum_{t=2}^n \Big\{ \bX_{t,\cdot j}\bX_{t-1,\cdot k}' - \sum_{i=1}^q D(\bar{\balpha}_0) \bW_0 \bX_{t,\cdot i}\bX_{t-1,\cdot k}' v_{j, i}^{(0)} \bar{\beta}_{j}^{(0)} - \sum_{i=1}^q D(\bar{\balpha}_1) \bW_1 \bX_{t-1,\cdot i}\bX_{t-1,\cdot k}' v_{j, i}^{(1)} \bar{\beta}_{j}^{(1)} \Big\}\,.\notag
\end{align}
Define 
\begin{align*}
    \hat{Q}_{n}(\bar{\bgamma}) = \frac{1}{p^2 q^2}\sum_{j,k=1}^q\|\hat{\bS}_{j,k}(\bar{\bgamma})\|_F^2\,.
\end{align*}
%$\hat{Q}_{n}(\bar{\bgamma}) = \sum_{j,k=1}^q\|\hat{\bS}_{j,k}(\bar{\bgamma})\|_F^2$. 
By \eqref{eq:Opt1}, the generalized Yule-Walker estimator is obtained by solving the following optimization problem:
\begin{align}\label{eq:Opt2}
    \hat{\bgamma} = \arg\min_{\bar{\bgamma}\in \boldsymbol{\Theta}} \hat{Q}_{n}(\bar{\bgamma})\,.
\end{align}

Since $\{{\rm vec}(\bX_t)\}$ is stationary, for any $\bar{\bgamma}$, we define 
\begin{align*}
    \tilde{Q}(\bar{\bgamma}) = \frac{1}{p^2 q^2}\sum_{j,k=1}^q\|\bS_{j,k}(\bar{\bgamma})\|_F^2
\end{align*}
%$\tilde{Q}(\bar{\bgamma}) = \sum_{j,k=1}^q\|\bS_{j,k}(\bar{\bgamma})\|_F^2$ 
with
\begin{align}\label{eq:Sjk}
    \bS_{j,k}(\bar{\bgamma}) & = \mathbb{E}\{\hat{\bS}_{j,k}(\bar{\bgamma})\} \notag \\
    & = \bSigma_{jk}(1) - \sum_{i=1}^q D(\bar{\balpha}_0) \bW_0 \bSigma_{ik}(1) v_{j, i}^{(0)} \bar{\beta}_{j}^{(0)} - \sum_{i=1}^q D(\bar{\balpha}_1) \bW_1 \bSigma_{ik}(0) v_{j, i}^{(1)} \bar{\beta}_{j}^{(1)}\,.
\end{align}
% Define 
% \begin{align*}
%     \tilde{Q}(\bar{\bgamma}) = \sum_{j, k}\|\bSigma_{jk}(1) - \sum_{i=1}^q D(\bar{\balpha}_0) \bW_0 \bSigma_{ik}(1) v_{j, i}^{(0)} \bar{\beta}_{jj}^{(0)} - \sum_{i=1}^q D(\bar{\balpha}_1) \bW_1 \bSigma_{ik}(0) v_{j, i}^{(1)} \bar{\beta}_{jj}^{(1)} \|_F^2\,,
% \end{align*}
Recall that $\bgamma = (\ba_1',\ldots,\ba_p',\bb_1',$ $\ldots,\bb_q')'$ is the true parameter with $ \ba_m = (\alpha_{m}^{(0)}, \alpha_{m}^{(1)})'$, $\bb_j = (\beta_{j}^{(0)},\beta_{j}^{(1)})'$, , where $\alpha_{m}^{(k)}$ and $\beta_{j}^{(k)}$ denote the $m$-th element of $\balpha_{k}$ and $j$-th element of $\bbeta_{k}$ for $k = 0, 1$, respectively.
%$\balpha_{k} = (\alpha_{1}^{(k)},\ldots, \alpha_{p}^{(k)})'$ and $\bbeta_{k} = (\beta_{1}^{(k)},\ldots, \beta_{q}^{(k)})'$ for $k = 0, 1$. 
By \eqref{class1_YW}, it holds that $\bS_{j,k}(\bgamma) = 0$ for any $j,k \in [q]$. We then have $\tilde{Q}(\bgamma) = 0$.
%, and $\bgamma$ is the unique solution of the optimization problem $\min_{\bar{\bgamma}}\tilde{Q}(\bar{\bgamma})$.

As we will show in Section \ref{subsec:UniformConvergence}, the uniform convergence of $\hat{Q}_{n}(\bar{\bgamma})$ over the compact and convext set $\bTheta$ holds, i.e.
\begin{align}\label{eq:UniformConv}
    \sup_{\bar{\bgamma} \in \bTheta}|\hat{Q}_{n}(\bar{\bgamma}) - \tilde{Q}(\bar{\bgamma})| = o_{\mathrm{p}}(1)\,.
\end{align}
provided that $d= o\{n^{1/(1+3\kappa)}\}$. 
Under Condition \ref{cond:ident}(iii), for any $\varepsilon > 0$, there exists a constant $\delta > 0$ such that 
\begin{align*}
    \tilde{Q}(\bar{\bgamma}) \ge \delta
\end{align*}
for every $\bar{\boldsymbol{\gamma}} \in \boldsymbol{\Theta}$ satisfying $\|\bar{\boldsymbol{\gamma}} - \boldsymbol{\gamma}\|_{2} > \varepsilon$. Hence, if $\|\bar{\boldsymbol{\gamma}} - \boldsymbol{\gamma}\|_{2} > \varepsilon$, by \eqref{eq:UniformConv}, we have 
\begin{align*}
    \mathbb{P}(\hat{Q}_{n}(\bar{\bgamma}) \ge \delta/2 ) \ge \mathbb{P}(\hat{Q}_{n}(\bar{\bgamma}) \ge \tilde{Q}(\bar{\bgamma})-\delta/2 ) \ge \mathbb{P}(|\hat{Q}_{n}(\bar{\bgamma}) - \tilde{Q}(\bar{\bgamma})| \le \delta/2) \to 1\,,
\end{align*}
provided that $d= o\{n^{1/(1+3\kappa)}\}$.
%$d= o\{n^{2/(16+13\kappa)}\}$.
%$d= o\{n^{2/(12+11\kappa)}\}$.

On the other hand, by the definition of $\hat{\bgamma}$ in \eqref{eq:Opt2}, we have 
\begin{align*}
    \hat{Q}_{n}(\hat{\bgamma}) \le \hat{Q}_{n}(\bgamma)\,,
\end{align*}
where $\bgamma$ is the true value. Notice that $\tilde{Q}(\bgamma) = 0$. Hence, by \eqref{eq:UniformConv} again, for $\delta > 0$ specified above, it holds that 
\begin{align*}
    \mathbb{P}(\hat{Q}_{n}(\hat{\bgamma}) > \delta/2) \le \mathbb{P}(\hat{Q}_{n}(\bgamma) > \delta/2) = \mathbb{P}(|\hat{Q}_{n}(\bgamma) - \tilde{Q}(\bgamma)| > \delta/2) \to 0 \,,
\end{align*}
provided that $d= o\{n^{1/(1+3\kappa)}\}$,
which implies that 
\begin{align*}
    \mathbb{P}(\|\hat{\boldsymbol{\gamma}} - \boldsymbol{\gamma}\|_{2} > \varepsilon) \to 0\,.
\end{align*}
This completes the proof of Proposition \ref{pn:consistency}. $\hfill\Box$

\subsection{Proof of \eqref{eq:UniformConv}}\label{subsec:UniformConvergence}
We first show that, for any given $\bar{\bgamma} \in \boldsymbol{\Theta}$, it holds that
\begin{align}\label{eq:uniformConverge}
    \hat{Q}_{n}(\bar{\bgamma}) - \tilde{Q}(\bar{\bgamma}) = o_{\rm p}(1)\,,
\end{align}
provided that $d= o\{n^{1/(1+3\kappa)}\}$.
%$p^{8+6\kappa}q^{8+7\kappa} = o(n^2)$.
% we have
% \begin{align}\label{eq:uniformConverge}
%     \mathbb{P}\{|\hat{Q}_{n}(\bar{\bgamma}) - Q_{n}(\bar{\bgamma})| > x\} \le XXX
% \end{align}
% for any $x>0$. 
By the Bonferroni inequality, for any $x>0$, it holds that
\begin{align}\label{eq:Q_dev2}
    &\mathbb{P}\{|\hat{Q}_{n}(\bar{\bgamma}) - \tilde{Q}(\bar{\bgamma})| > x\} \notag\\
    & ~~~~~~ \le \mathbb{P}\Big\{\sum_{j,k=1}^q\big| \|\hat{\bS}_{j,k}(\bar{\bgamma})\|_F^2 - \|\bS_{j,k}(\bar{\bgamma})\|_F^2 \big| > p^2q^2 x \Big\} \notag\\
    & ~~~~~~\le \sum_{j,k=1}^q \mathbb{P} \Big\{\big| \|\hat{\bS}_{j,k}(\bar{\bgamma})\|_F^2 - \|\bS_{j,k}(\bar{\bgamma})\|_F^2 \big| > p^2x \Big\}\,. 
\end{align}
It then suffices to derive the upper bound of $\mathbb{P} \{| \|\hat{\bS}_{j,k}(\bar{\bgamma})\|_F^2 - \|\bS_{j,k}(\bar{\bgamma})\|_F^2 | > p^2 x\}$. Since 
\begin{align*}
   \big| \|\hat{\bS}_{j,k}(\bar{\bgamma})\|_F^2 - \|\bS_{j,k}(\bar{\bgamma})\|_F^2 \big| \le \|\hat{\bS}_{j,k}(\bar{\bgamma}) - \bS_{j,k}(\bar{\bgamma})\|_F  \|\hat{\bS}_{j,k}(\bar{\bgamma})+\bS_{j,k}(\bar{\bgamma})\|_F\,,
\end{align*}  
for any $x>0$ and sufficiently large $M > 0$, it holds that
\begin{align}\label{eq:Q_dev1}
    & \mathbb{P}\Big\{\big| \|\hat{\bS}_{j,k}(\bar{\bgamma})\|_F^2 - \|\bS_{j,k}(\bar{\bgamma})\|_F^2 \big| > p^2  x \Big\} \notag \\
    & ~~~~~~\le \mathbb{P}\Big\{\big| \|\hat{\bS}_{j,k}(\bar{\bgamma})\|_F^2 - \|\bS_{j,k}(\bar{\bgamma})\|_F^2 \big| > p^2  x\,, \|\hat{\bS}_{j,k}(\bar{\bgamma})+\bS_{j,k}(\bar{\bgamma})\|_F \le M \Big\} \notag \\
    & ~~~~~~~~~~~+ \mathbb{P}\Big\{\big| \|\hat{\bS}_{j,k}(\bar{\bgamma})\|_F^2 - \|\bS_{j,k}(\bar{\bgamma})\|_F^2 \big| > p^2  x \,, \|\hat{\bS}_{j,k}(\bar{\bgamma})+\bS_{j,k}(\bar{\bgamma})\|_F > M \Big\} \notag \\
    & ~~~~~~\le \mathbb{P}\Big(\|\hat{\bS}_{j,k}(\bar{\bgamma}) - \bS_{j,k}(\bar{\bgamma})\|_F > \frac{p^2 x}{M} \Big) + \mathbb{P}( \|\hat{\bS}_{j,k}(\bar{\bgamma})+\bS_{j,k}(\bar{\bgamma})\|_F > M )  \notag \\
    & ~~~~~~ := \mathrm{K}_1 +  \mathrm{K}_2\,.
\end{align}
To derive the upper bounds of $\mathrm{K}_1$ and $\mathrm{K}_2$, it suffices to bound $\mathbb{P}(\|\hat{\bS}_{j,k}(\bar{\bgamma}) - \bS_{j,k}(\bar{\bgamma})\|_F > z)$ for any $z>0$. By \eqref{eq:hatSjk} and \eqref{eq:Sjk}, it holds that 
\begin{align*}
    \hat{\bS}_{j,k}(\bar{\bgamma}) - \bS_{j,k}(\bar{\bgamma}) &= \{\hat\bSigma_{jk}(1) - \bSigma_{jk}(1)\} - \sum_{i=1}^q D(\bar{\balpha}_0) \bW_0 \{\hat{\bSigma}_{ik}(1) - \bSigma_{ik}(1)\} v_{j, i}^{(0)} \bar{\beta}_{j}^{(0)} \\
    &~~~~~~~~- \sum_{i=1}^q D(\bar{\balpha}_1) \bW_1 \{\hat{\bSigma}_{ik}(0) - \bSigma_{ik}(0)\} v_{j, i}^{(1)} \bar{\beta}_{j}^{(1)}\,. 
\end{align*}
Then we have
\begin{align}\label{eq:F-norm}
    \|\hat{\bS}_{j,k}(\bar{\bgamma}) - \bS_{j,k}(\bar{\bgamma})\|_F^2 & \le 3 \|\hat\bSigma_{jk}(1) - \bSigma_{jk}(1)\|_F^2 + 3\Big\|\sum_{i=1}^q D(\bar{\balpha}_0) \bW_0 \{\hat{\bSigma}_{ik}(1) - \bSigma_{ik}(1)\} v_{j, i}^{(0)} \bar{\beta}_{j}^{(0)}\Big\|_F^2 \notag\\
    &~~~~~~ + 3\Big\| \sum_{i=1}^q D(\bar{\balpha}_1) \bW_1 \{\hat{\bSigma}_{ik}(0) - \bSigma_{ik}(0)\} v_{j, i}^{(1)} \bar{\beta}_{j}^{(1)} \Big\|_F^2\,.
    % & \le 3 \|\hat\bSigma_{jk}(1) - \bSigma_{jk}(1)\|_F^2 + 3q \sum_{i=1}^q \{v_{j, i}^{(0)}\}^2 \{\bar{\beta}_{jj}^{(0)}\}^2\|D(\bar{\balpha}_0) \bW_0\|_F^2 \|\hat{\bSigma}_{ik}(1) - \bSigma_{ik}(1)\|_F^2 \\
    % &~~~~~~~~ + 3q \sum_{i=1}^q \{v_{j, i}^{(1)}\}^2 \{\bar{\beta}_{jj}^{(1)}\}^2 \|D(\bar{\balpha}_1) \bW_1\|_F^2 \|\hat{\bSigma}_{ik}(0) - \bSigma_{ik}(0)\|_F^2\,. 
\end{align} 
For any $z>0$, it then holds that 
\begin{align}\label{eq:hatSjk_dev1}
    \mathbb{P}(\|\hat{\bS}_{j,k}(\bar{\bgamma}) - \bS_{j,k}(\bar{\bgamma})\|_F^2 > z) & \le \mathbb{P}(3 \|\hat\bSigma_{jk}(1) - \bSigma_{jk}(1)\|_F^2 > z/3) \notag\\
    & ~~~~~ + \mathbb{P}\Big( 3\Big\|\sum_{i=1}^q D(\bar{\balpha}_0) \bW_0 \{\hat{\bSigma}_{ik}(1) - \bSigma_{ik}(1)\} v_{j, i}^{(0)} \bar{\beta}_{j}^{(0)}\Big\|_F^2 > z/3 \Big) \notag\\
    &~~~~~ + \mathbb{P}\Big( 3\Big\| \sum_{i=1}^q D(\bar{\balpha}_1) \bW_1 \{\hat{\bSigma}_{ik}(0) - \bSigma_{ik}(0)\} v_{j, i}^{(1)} \bar{\beta}_{j}^{(1)} \Big\|_F^2 > z/3 \Big) \notag\\
    & := \mathrm{I}_1 + \mathrm{I}_2 + \mathrm{I}_3\,.
\end{align}
It suffices to derive the upper bound of $\mathrm{I}_1$, $\mathrm{I}_2$, and $\mathrm{I}_3$, respectively. By using the similar arguments as in \eqref{eq:hatSjk_dev} in the proof of Lemma \ref{lem:Sigmajk} in Section \ref{lem:1}, we have
\begin{align*}
    \mathrm{I}_1 \lesssim p^{(2+3\kappa)/\kappa}n^{-2/\kappa}z^{-(2+\kappa)/(2\kappa)}\,.
\end{align*}
Notice that 
\begin{align}\label{eq:F-norm2}
    &\Big\|\sum_{i=1}^q D(\bar{\balpha}_0) \bW_0 \{\hat{\bSigma}_{ik}(1) - \bSigma_{ik}(1)\} v_{j, i}^{(0)} \bar{\beta}_{j}^{(0)}\Big\|_F^2 \notag\\ 
    &~~~~~~~~~~~~~ \le q \sum_{i=1}^q \{v_{j, i}^{(0)}\}^2 \{\bar{\beta}_{j}^{(0)}\}^2\|D(\bar{\balpha}_0) \bW_0\|_F^2 \|\hat{\bSigma}_{ik}(1) - \bSigma_{ik}(1)\|_F^2 \notag\\
    &~~~~~~~~~~~~~  \le C_1q \|D(\bar{\balpha}_0) \bW_0\|_F^2 \max_{i \in [q]} \|\hat{\bSigma}_{ik}(1) - \bSigma_{ik}(1)\|_F^2  \,,
\end{align}
and $\|D(\bar{\balpha}_0) \bW_0\|_F^2 = O(p)$
under the assumption that $\bW_{0}$ and $\bV_{0}$ are uniformly bounded in both row and column sums in absolute value.
By using the similar arguments as in \eqref{eq:hatSjk_dev} in the proof of Lemma \ref{lem:Sigmajk} in Section \ref{lem:1}, it then holds that
\begin{align*}
    \mathrm{I}_2 & \le \mathbb{P}\Big\{ \max_{i \in [q]}\|\hat{\bSigma}_{ik}(1) - \bSigma_{ik}(1)\|_F^2 > \frac{C_2 z}{q\|D(\bar{\balpha}_0) \bW_0\|_F^2} \Big\}  \\
    & \le \mathbb{P}\Big\{ \max_{i \in [q]}\|\hat{\bSigma}_{ik}(1) - \bSigma_{ik}(1)\|_F^2 > \frac{C_3 z}{pq} \Big\} \\
    & \lesssim p^{(6+7\kappa)/(2\kappa)}q^{(2+3\kappa)/(2\kappa)}n^{-2/\kappa}z^{-(2+\kappa)/(2\kappa)}\,.
\end{align*}
Using the similar arguments, we can show that the same upper bound holds for $\mathrm{I}_3$. By \eqref{eq:hatSjk_dev1}, for any $z >0$, we have
\begin{align}\label{eq:hatSjk_dev2}
    & \mathbb{P}(\|\hat{\bS}_{j,k}(\bar{\bgamma}) - \bS_{j,k}(\bar{\bgamma})\|_F^2 > z) \notag\\
    & ~~~~~~~~~~~\lesssim p^{(2+3\kappa)/\kappa}n^{-2/\kappa}z^{-(2+\kappa)/(2\kappa)} + p^{(6+7\kappa)/(2\kappa)}q^{(2+3\kappa)/(2\kappa)}n^{-2/\kappa}z^{-(2+\kappa)/(2\kappa)} \,.
    %p^{4(1+\kappa)/\kappa}n^{-(4+\kappa)/\kappa}z^{-(2+\kappa)/\kappa} + p^{(8+6\kappa)/\kappa}q^{(4+3\kappa)/\kappa}n^{-(4+\kappa)/\kappa}z^{-(2+\kappa)/\kappa}\,.
\end{align}

Hence, for $\mathrm{K}_2$ in \eqref{eq:Q_dev1}, we have
\begin{align*}
    \mathrm{K}_2 & = \mathbb{P}( \|\hat{\bS}_{j,k}(\bar{\bgamma})+\bS_{j,k}(\bar{\bgamma})\|_F^2 > M^2 ) = \mathbb{P}( \|\hat{\bS}_{j,k}(\bar{\bgamma}) - \bS_{j,k}(\bar{\bgamma}) + 2\bS_{j,k}(\bar{\bgamma})\|_F^2 > M^2 ) \\
    & \le \mathbb{P}( 2\|\hat{\bS}_{j,k}(\bar{\bgamma}) - \bS_{j,k}(\bar{\bgamma})\|_F^2 + 8\|\bS_{j,k}(\bar{\bgamma})\|_F^2 > M^2 ) \\
    & \le \mathbb{P}\Big( \|\hat{\bS}_{j,k}(\bar{\bgamma}) - \bS_{j,k}(\bar{\bgamma})\|_F^2 > \frac{M^2}{2} -  8\|\bS_{j,k}(\bar{\bgamma})\|_F^2 \Big)\,.
\end{align*}
%Analogous to \eqref{eq:F-norm} and \eqref{eq:F-norm2}, 
Since $\bW_{0}$, $\bW_{1}$, $\bV_{0}$ and $\bV_{1}$ are uniformly bounded in both row and column sums in absolute value, 
it holds that 
\begin{align*}
    \|\bS_{j,k}(\bar{\bgamma})\|_F^2 \lesssim p^2\,.
\end{align*}
Choose $M = Cp$ for some sufficiently large constant $C>0$. Hence, by \eqref{eq:hatSjk_dev2}, we have 
\begin{align*}
    \mathrm{K}_2 & \le \mathbb{P}( \|\hat{\bS}_{j,k}(\bar{\bgamma}) - \bS_{j,k}(\bar{\bgamma})\|_F^2 > C_4 p^2 ) \lesssim p^{2}n^{-2/\kappa} + p^{(2+5\kappa)/(2\kappa)}q^{(2+3\kappa)/(2\kappa)}n^{-2/\kappa}\,.
    %\lesssim p^{(\kappa-2)/\kappa}q^{-(2+\kappa)/\kappa}n^{-2/\kappa} + p^{(3\kappa-2)/(2\kappa)}q^{(\kappa-2)/(2\kappa)}n^{-2/\kappa}\,.
\end{align*}
For $\mathrm{K}_1$ in \eqref{eq:Q_dev1}, by \eqref{eq:hatSjk_dev2} and choosing $M = Cp$ specified in $\mathrm{K}_2$, it holds that
\begin{align*}
    \mathrm{K}_1 & = \mathbb{P}\Big( \|\hat{\bS}_{j,k}(\bar{\bgamma}) - \bS_{j,k}(\bar{\bgamma})\|_F^2 >  \frac{p^4 x^2}{M^2} \Big) \\
    &  \lesssim p^{2}n^{-2/\kappa}x^{-(2+\kappa)/\kappa} + p^{(2+5\kappa)/(2\kappa)}q^{(2+3\kappa)/(2\kappa)}n^{-2/\kappa}x^{-(2+\kappa)/\kappa}\,.
    % &  \lesssim p^{(2+3\kappa)/\kappa}q^{-(2+\kappa)/\kappa}n^{-2/\kappa}x^{-(2+\kappa)/\kappa} + p^{(6+7\kappa)/(2\kappa)}q^{(\kappa-2)/(2\kappa)}n^{-2/\kappa}x^{-(2+\kappa)/\kappa}\,.
    % \mathrm{K}_1 & = \mathbb{P}\Big( \|\hat{\bS}_{j,k}(\bar{\bgamma}) - \bS_{j,k}(\bar{\bgamma})\|_F^2 >  \frac{C_5p^4 q^4 x^2}{p^4q^2} \Big) \\
    % &  \lesssim p^{(2+3\kappa)/\kappa}q^{-(2+\kappa)/\kappa}n^{-2/\kappa}x^{-(2+\kappa)/\kappa} + p^{(6+7\kappa)/(2\kappa)}q^{(\kappa-2)/(2\kappa)}n^{-2/\kappa}x^{-(2+\kappa)/\kappa}\,.
\end{align*}
By \eqref{eq:Q_dev2} and \eqref{eq:Q_dev1}, we then have 
\begin{align}\label{eq:Q_dev3}
    & \mathbb{P}\{|\hat{Q}_{n}(\bar{\bgamma}) - \tilde{Q}(\bar{\bgamma})| > x\} \notag \\
    & ~~~~~~\lesssim p^{2}q^2n^{-2/\kappa} + p^{(2+5\kappa)/(2\kappa)}q^{(2+7\kappa)/(2\kappa)}n^{-2/\kappa} \\
    & ~~~~~~~~~~~ +  p^{2}q^2n^{-2/\kappa}x^{-(2+\kappa)/\kappa} + p^{(2+5\kappa)/(2\kappa)}q^{(2+7\kappa)/(2\kappa)}n^{-2/\kappa}x^{-(2+\kappa)/\kappa}\,. \notag
    % & ~~~~~~\lesssim p^{(\kappa-2)/\kappa}q^{(\kappa-2)/\kappa}n^{-2/\kappa} + p^{(3\kappa-2)/(2\kappa)}q^{(5\kappa-2)/(2\kappa)}n^{-2/\kappa} \\
    % & ~~~~~~~~~~~ +  p^{(2+3\kappa)/\kappa}q^{(\kappa-2)/\kappa}n^{-2/\kappa}x^{-(2+\kappa)/\kappa} + p^{(6+7\kappa)/(2\kappa)}q^{(5\kappa-2)/(2\kappa)}n^{-2/\kappa}x^{-(2+\kappa)/\kappa} \,. \notag
\end{align}
Hence, for any given $\bar{\bgamma} \in \boldsymbol{\Theta}$, \eqref{eq:uniformConverge} holds provided that $p^2q^2 = o(n^{2/\kappa})$ and $p^{(2+5\kappa)/(2\kappa)}q^{(2+7\kappa)/(2\kappa)}=o(n^{2/\kappa})$.
%$p^{3\kappa-2}q^{5\kappa-2} = o(n^4)$, $p^{2+3\kappa}q^{\kappa-2}=o(n^2)$ and $p^{6+7\kappa}q^{5\kappa-2}=o(n^4)$.
%$p^{6+5\kappa}q^{6+6\kappa} = o(n^2)$. 
Recall $d=p \vee q$ and $\kappa \in (0,2)$. It then holds that $|\hat{Q}_{n}(\bar{\bgamma}) - \tilde{Q}(\bar{\bgamma})| = o_{\rm p}(1)$, provided that $d= o\{n^{1/(1+3\kappa)}\}$.
%$d= o\{n^{2/(12+11\kappa)}\}$. 

Next, we show the uniform convergence of $\hat{Q}_{n}(\bar{\bgamma})$ over the compact and convext set $\bTheta$ in \eqref{eq:UniformConv}.
% \begin{align}\label{eq:UniformConv}
%     \sup_{\bar{\bgamma} \in \bTheta}|\hat{Q}_{n}(\bar{\bgamma}) - \tilde{Q}(\bar{\bgamma})| = o_{\mathrm{p}}(1)\,.
% \end{align}
Recall $\tilde{Q}(\bar{\bgamma}) = p^{-2}q^{-2}\sum_{j,k=1}^q\|\bS_{j,k}(\bar{\bgamma})\|_F^2$, where $\bS_{j,k}(\bar{\bgamma})$ is specified in \eqref{eq:Sjk}. 
By direct calculation, we have $\|\nabla \tilde{Q}(\bar{\bgamma})/\nabla \bar{\bgamma}\|_2 \le O(1)$,
% \begin{align*}
%     \left\| \frac{\nabla \tilde{Q}(\bar{\bgamma})}{\nabla \bar{\bgamma}} \right\|_2 \le O(1)\,,
% \end{align*}
% \begin{align*}
%     \frac{\partial \tilde{Q}(\bar{\bgamma})}{\partial \bar{\bgamma}} = \left( \frac{\partial \tilde{Q}(\bar{\bgamma})}{\partial \bar{\alpha}_{1}^{(1)}},\frac{\partial \tilde{Q}(\bar{\bgamma})}{\partial \bar{\alpha}_{1}^{(1)}},\ldots,\frac{\partial \tilde{Q}(\bar{\bgamma})}{\partial \bar{\beta}_{q}^{(0)}},\frac{\partial \tilde{Q}(\bar{\bgamma})}{\partial \bar{\beta}_{q}^{(1)}}\right)' \le O(1)\,,
% \end{align*}
which indicates that $\tilde{Q}(\bar{\bgamma})$ is equicontinuous on $\bTheta$. 
Recall $\hat{Q}_{n}(\bar{\bgamma}) = p^{-2}q^{-2}\sum_{j,k=1}^q\|\hat{\bS}_{j,k}(\bar{\bgamma})\|_F^2$, where $\hat{\bS}_{j,k}(\bar{\bgamma})$ is specified in \eqref{eq:hatSjk}. Write $\bW_k = (w_{i,j}^{(k)})_{p \times p}$ and $\bV_k = (v_{i,j}^{(k)})_{q \times q}$ for $k=0,1$. Hence, we have
\begin{align*}
    \frac{\partial \hat{Q}_{n}(\bar{\bgamma})}{\partial \bar{\alpha}_{m}^{(0)}} & = \frac{-2}{p^2 q^2}\sum_{j,k=1}^q \sum_{m_2=1}^p \hat{s}_{m,m_2}^{j,k}(\bar{\bgamma}) \bigg\{ \sum_{i=1}^q \sum_{l=1}^p w_{m,l}^{(0)}\hat{\sigma}_{l,m_2}^{(i,k,1)}v_{j,i}^{(0)}\bar{\beta}_j ^{(0)} \bigg\}\,,
\end{align*}
where 
\begin{align}
    \hat{s}_{m_1,m_2}^{j,k}(\bar{\bgamma}) & = \frac{1}{n}\sum_{t=2}^n \Big\{ X_{t,m_1,j}X_{t-1,m_2, k} - \bar{\alpha}_{m_1}^{(0)}\sum_{i=1}^q\sum_{l=1}^p w_{m_1,l}^{(0)} X_{t,l, i}X_{t-1,m_2, k} v_{j, i}^{(0)} \bar{\beta}_{j}^{(0)} \notag\\
    &~~~~~~~~~~~~~~~~~~~~~~~- \bar{\alpha}_{m_1}^{(1)}\sum_{i=1}^q\sum_{l=1}^p w_{m_1,l}^{(1)} X_{t-1,l, i}X_{t-1,m_2, k} v_{j, i}^{(1)} \bar{\beta}_{j}^{(1)} \Big\}\,, \label{eq:hatS}\\
    &~~~~~~~~~~~~~~~~~~~\hat{\sigma}_{l,m_2}^{(i,k,1)} = \frac{1}{n}\sum_{t=2}^n X_{t,l, i}X_{t-1,m_2, k}\notag
\end{align}
for any $i,j,k \in [q]$ and $m_1,m_2,l \in [p]$. Direct calculation shows that 
\begin{align}\label{eq:T_decompose}
    & \left| \frac{\partial \hat{Q}_{n}(\bar{\bgamma})}{\partial \bar{\alpha}_{m}^{(0)}} - \frac{\partial \tilde{Q}(\bar{\bgamma})}{\partial \bar{\alpha}_{m}^{(0)}} \right| \notag\\
    &~~~ \le \frac{2}{p^2q^2} \sum_{j,k=1}^q \sum_{m_2=1}^p \big| s_{m,m_2}^{j,k}(\bar{\bgamma}) \big| \bigg\{ \sum_{i=1}^q \sum_{l=1}^p \big|w_{m,l}^{(0)}\big|\big|\hat{\sigma}_{l,m_2}^{(i,k,1)}-\sigma_{l,m_2}^{(i,k,1)}\big|\big|v_{j,i}^{(0)}\bar{\beta}_j ^{(0)}\big| \bigg\}  \notag\\
    &~~~~~~~~~ + \frac{2}{p^2q^2} \sum_{j,k=1}^q \sum_{m_2=1}^p \big|\hat{s}_{m,m_2}^{j,k}(\bar{\bgamma}) -  s_{m,m_2}^{j,k}(\bar{\bgamma}) \big| \bigg\{ \sum_{i=1}^q \sum_{l=1}^p \big|w_{m,l}^{(0)}\sigma_{l,m_2}^{(i,k,1)}v_{j,i}^{(0)}\bar{\beta}_j ^{(0)}\big| \bigg\} \notag\\
    &~~~~~~~~~ + \frac{2}{p^2q^2} \sum_{j,k=1}^q \sum_{m_2=1}^p \big|\hat{s}_{m,m_2}^{j,k}(\bar{\bgamma}) -  s_{m,m_2}^{j,k}(\bar{\bgamma}) \big| \bigg\{ \sum_{i=1}^q \sum_{l=1}^p \big|w_{m,l}^{(0)}\big|\big|\hat{\sigma}_{l,m_2}^{(i,k,1)}-\sigma_{l,m_2}^{(i,k,1)}\big|\big|v_{j,i}^{(0)}\bar{\beta}_j ^{(0)}\big| \bigg\} \notag\\
    &~~~ = T_{1,m,j,k} + T_{2,m,j,k} + T_{3,m,j,k}\,.
\end{align}
For $T_{1,m,j,k}$, we have 
\begin{align*}
    \max_{m \in [p],j,k\in [q]}|T_{1,m,j,k}| \lesssim &~ \frac{1}{p}\max_{j,k\in [q]}\|\hat{\bSigma}_{jk}(1)-\bSigma_{jk}(1)\|_{\infty}
    %\frac{pq^3}{p^2q^4} \sum_{i=1}^q \left[\sum_{l=1}^p \big\{w_{m,l}^{(0)}\big\}^2\right]\left[\sum_{l=1}^p \big\{\hat{\sigma}_{l,m_2}^{(i,k,1)}-\sigma_{l,m_2}^{(i,k,1)}\big\}^2\right] \\
    %\lesssim &~ \frac{1}{pq} \sum_{i=1}^q \sum_{l=1}^p \big\{\hat{\sigma}_{l,m_2}^{(i,k,1)}-\sigma_{l,m_2}^{(i,k,1)}\big\}^2 \,,
\end{align*}
%where the second inequality holds 
under the assumption that $\bW_{0}$ and $\bV_{0}$ are uniformly bounded in both row and column sums in absolute value. By Lemma \ref{lem:Sigmajk}, it holds that $\max_{m \in [p],j,k\in [q]}|T_{1,m,j,k}| = O_{\rm p}\{d^{(4\kappa)/(2+\kappa)}n^{-2/(2+\kappa)}\}$.
% By \eqref{eq:elementwise}, it holds that 
% \begin{align}\label{eq:T1}
%     \mathbb{P}(T_1 > x) \le \sum_{l=1}^p \mathbb{P}\big(|\hat{\sigma}_{l,m_2}^{(i,k,1)}-\sigma_{l,m_2}^{(i,k,1)}| > x^{1/2} \big) \le px^{-(2+\kappa)/(2\kappa)}n^{-2\kappa}\,,
% \end{align}
% which implies that $T_1 = O_{\mathrm{p}}\{p^{2\kappa/(2+\kappa)}n^{-4/(2+\kappa)}\}$. 
Hence, we have $\max_{m \in [p],j,k\in [q]}|T_{1,m,j,k}| = o_{\mathrm{p}}(d^{-1/2})$ if $d=o(n^{4/(2+9\kappa)})$. %$p^{\kappa} = O(n^2)$.
For $T_{2,m,j,k}$, notice that
\begin{align*}
    \big|\hat{s}_{m,m_2}^{j,k}(\bar{\bgamma}) -  s_{m,m_2}^{j,k}(\bar{\bgamma}) \big| \le &~ \big| \hat{\sigma}_{m,m_2}^{(j, k,1)} - \sigma_{m,m_2}^{(j, k,1)} \big| + \big|\bar{\alpha}_{m}^{(0)}\big|\sum_{i=1}^q\sum_{l=1}^p \big|w_{m,l}^{(0)}\big| \big|\hat{\sigma}_{l, m_2}^{(i, k,1)} - \sigma_{l, m_2}^{(i, k,1)} \big|v_{j, i}^{(0)} \bar{\beta}_{j}^{(0)} \big| \notag\\
    &~~~~~~~+ \big|\bar{\alpha}_{m}^{(1)}\big|\sum_{i=1}^q\sum_{l=1}^p \big|w_{m,l}^{(1)}\big| \big|\hat{\sigma}_{l, m_2}^{(i, k,0)} - \sigma_{l, m_2}^{(i, k,0)} \big| \big|v_{j, i}^{(1)} \bar{\beta}_{j}^{(1)}\big|\,, \notag\\
    \lesssim &~ \max_{m_1,m_2}\big| \hat{\sigma}_{m_1,m_2}^{(j, k,1)} - \sigma_{m_1,m_2}^{(j, k,1)} \big| + \max_{m_1,m_2}\big| \hat{\sigma}_{m_1,m_2}^{(j, k,0)} - \sigma_{m_1,m_2}^{(j, k,0)} \big|\,,
\end{align*}
where 
\begin{align*}
    \hat{\sigma}_{l,m_2}^{(i,k,0)} = \frac{1}{n}\sum_{t=2}^n X_{t-1,l, i}X_{t-1,m_2, k}\,.
\end{align*}
Hence, by Lemma \ref{lem:Sigmajk}, we have that $\max_{m \in [p],j,k\in [q]}|T_{2,m,j,k}| = O_{\rm p}\{d^{(4\kappa)/(2+\kappa)}n^{-2/(2+\kappa)}\}$, which implies $\max_{m \in [p],j,k\in [q]}|T_{2,m,j,k}| = o_{\mathrm{p}}(d^{-1/2})$ if $d=o(n^{4/(2+9\kappa)})$.
%Analogous to the derivation of \eqref{eq:T1}, by \eqref{eq:elementwise}, it holds that $T_2 = O_{\mathrm{p}}\{p^{2\kappa/(2+\kappa)}n^{-4/(2+\kappa)}\}$. Hence, we have $T_2 = O_{\mathrm{p}}(1)$ if $p^{\kappa} = O(n^2)$. 
Similarly, it can be shown that $\max_{m \in [p],j,k\in [q]}|T_{3,m,j,k}| = o_{\mathrm{p}}(d^{-1/2})$ if $d=o(n^{4/(2+9\kappa)})$.
%$p^{\kappa} = O(n^2)$. 
By \eqref{eq:T_decompose}, it yields
\begin{align*}
    \left| \frac{\partial \hat{Q}_{n}(\bar{\bgamma})}{\partial \bar{\alpha}_{m}^{(0)}} \right| = o_{\mathrm{p}}(d^{-1/2})\,,
\end{align*}
provided that $d=o\{n^{4/(2+9\kappa)}\}$. %$p^{\kappa} = O(n^2)$.
The partial derivatives of $\hat{Q}_{n}(\bar{\bgamma})$ with respect to $\bar{\alpha}_{m}^{(1)}$, $\bar{\beta}_{j}^{(0)}$ and $\bar{\beta}_{j}^{(1)}$ can be similarly derived.
% , which are specified as follows:
% \begin{align*}
%     \frac{\partial \hat{Q}_{n}(\bar{\bgamma})}{\partial \bar{\alpha}_{m}^{(1)}} & = -\frac{2}{p^2q^2}\sum_{j,k=1}^q \sum_{m_2=1}^p \hat{s}_{m,m_2}^{j,k}(\bar{\bgamma}) \bigg\{ \sum_{i=1}^q \sum_{l=1}^p w_{m,l}^{(1)}\hat{\sigma}_{l,m_2}^{(i,k,0)}v_{j,i}^{(1)}\bar{\beta}_j ^{(1)} \bigg\}\,, \\
%     \frac{\partial \hat{Q}_{n}(\bar{\bgamma})}{\partial \bar{\beta}_{j}^{(0)}} & = -\frac{2}{p^2q^2}\sum_{k=1}^q \sum_{m_1,m_2=1}^p \hat{s}_{m_1,m_2}^{j,k}(\bar{\bgamma}) \bigg\{ \bar{\alpha}_{m_1}^{(0)}\sum_{i=1}^q \sum_{l=1}^p w_{m_1,l}^{(0)}\hat{\sigma}_{l,m_2}^{(i,k,1)}v_{j,i}^{(0)} \bigg\}\,, \\
%     \frac{\partial \hat{Q}_{n}(\bar{\bgamma})}{\partial \bar{\beta}_{j}^{(1)}} & = -\frac{2}{p^2q^2}\sum_{k=1}^q \sum_{m_1,m_2=1}^p \hat{s}_{m_1,m_2}^{j,k}(\bar{\bgamma}) \bigg\{ \bar{\alpha}_{m_1}^{(1)}\sum_{i=1}^q \sum_{l=1}^p w_{m_1,l}^{(1)}\hat{\sigma}_{l,m_2}^{(i,k,0)}v_{j,i}^{(1)} \bigg\}\,,
% \end{align*}
% where $\hat{s}_{m,m_2}^{j,k}(\bar{\bgamma})$ is specified in \eqref{eq:hatS}.
%for any $i,j,k \in [q]$ and $m_2,l \in [p]$. 
Hence, we have
\begin{align*}
    \left\| \frac{\nabla \hat{Q}_{n}(\bar{\bgamma})}{\nabla \bar{\bgamma}} \right\|_2 \le O_{\mathrm{p}}(1)
\end{align*}
% \begin{align*}
%     \frac{\partial \hat{Q}_{n}(\bar{\bgamma})}{\partial \bar{\bgamma}} = \left( \frac{\partial \hat{Q}_{n}(\bar{\bgamma})}{\partial \bar{\alpha}_{1}^{(1)}},\frac{\partial \hat{Q}_{n}(\bar{\bgamma})}{\partial \bar{\alpha}_{1}^{(1)}},\ldots,\frac{\partial \hat{Q}_{n}(\bar{\bgamma})}{\partial \bar{\beta}_{q}^{(0)}},\frac{\partial \hat{Q}_{n}(\bar{\bgamma})}{\partial \bar{\beta}_{q}^{(1)}}\right)' = O_{\mathrm{p}}(1)
% \end{align*}
if $d=o\{n^{4/(2+9\kappa)}\}$.
%if $p^{\kappa} \le O(n^2)$. 
By Corollary 2.2 of \cite{Newey1991}, we have \eqref{eq:UniformConv} hold, 
% \begin{align*}
% \sup_{\bar{\bgamma} \in \bTheta}|\hat{Q}_{n}(\bar{\bgamma}) - \tilde{Q}(\bar{\bgamma})| = o_{\rm p}(1)\,,
% \end{align*}
provided that $d= o\{n^{1/(1+3\kappa)}\}$. $\hfill\Box$

\section{Proof of Theorem \ref{tm:mixing}}\label{sec:proof_thm1}

%To prove Theorem \ref{tm:mixing}, we need Lemma \ref{lem:Sigmajk} with its proof given in Section 

Recall that the generalized Yule-Walker estimator aims to solve  
\begin{equation}\label{eq:YW_est_min}
\min_{\bar{\bgamma} \in \boldsymbol{\Theta}} \frac{1}{p^2q^2}\sum_{j, k=1}^q \Big\|\hat\bSigma_{jk}(1) - \sum_{i=1}^q D(\bar{\balpha}_0) \bW_0 \hat \bSigma_{ik}(1) v_{j, i}^{(0)} \bar{\beta}_{j}^{(0)} - \sum_{i=1}^q D(\bar{\balpha}_1) \bW_1 \hat \bSigma_{ik}(0) v_{j, i}^{(1)} \bar{\beta}_{j}^{(1)} \Big\|_F^2\,,
\end{equation}
where $\bar{\bgamma} = (\bar{\ba}_1',\ldots,\bar{\ba}_p',\bar{\bb}_1',\ldots,\bar{\bb}_q')'$ is a $(2p+2q)$-dimensional parameter with $\bar{\ba}_m = (\bar{\alpha}_{m}^{(0)},\bar{\alpha}_{m}^{(1)})'$ and $\bar{\bb}_j = (\bar{\beta}_{j}^{(0)},\bar{\beta}_{j}^{(1)})'$. 
%Write $\bar{\balpha}_{k} = (\bar{\alpha}_{11}^{(k)},\ldots, \bar{\alpha}_{pp}^{(k)})'$ and $\bar{\bbeta}_{k} = (\bar{\beta}_{11}^{(k)},\ldots, \bar{\beta}_{qq}^{(k)})'$ for $k = 0, 1$. 
For $m \in [p]$ and $j \in [q]$, recall 
\begin{align}\label{eq:def}
\bY_{\ba_m} = 
\begin{pmatrix}
\hat \bSigma_{11}(1)' \be_m \\
\hat \bSigma_{12}(1)' \be_m \\
\vdots \\
\hat \bSigma_{qq}(1)' \be_m
\end{pmatrix} ~ \mbox{and} ~
\bY_{\bb_j} = 
\begin{pmatrix}
\text{vec}\{\hat \bSigma_{j1}(1)\} \\
\text{vec}\{\hat \bSigma_{j2}(1)\} \\
\vdots\\
\text{vec}\{\hat \bSigma_{jq}(1)\}
\end{pmatrix}\,,
\end{align}
% \begin{equation*}
% \bb_j = 
% \begin{pmatrix}
% \beta_{jj}^{(0)} \\
% \beta_{jj}^{(1)}
% \end{pmatrix}\,,   ~~~  
% \ba_m = 
% \begin{pmatrix}
% \alpha_{mm}^{(0)} \\
% \alpha_{mm}^{(1)}
% \end{pmatrix}
% \end{equation*}
% and
% \begin{equation*}
% \bY_{\bb_j} = 
% \begin{pmatrix}
% \text{vec}(\hat \bSigma_{j1}(1)) \\
% \text{vec}(\hat \bSigma_{j2}(1)) \\
% \vdots\\
% \text{vec}(\hat \bSigma_{jq}(1))
% \end{pmatrix}\,, ~~~ 
% \bY_{\ba_m} = 
% \begin{pmatrix}
% \hat \bSigma_{11}(1)' \be_m \\
% \hat \bSigma_{12}(1)' \be_m \\
% \vdots \\
% \hat \bSigma_{qq}(1)' \be_m
% \end{pmatrix}\,,
% \end{equation*}
which are specified in \eqref{LS_alpha} and \eqref{LS_beta}, respectively. Let $\hat{\boldsymbol{\alpha}}_k = (\hat{\alpha}_{1}^{(k)},\ldots,\hat{\alpha}_{p}^{(k)})'$ and $\hat{\boldsymbol{\beta}}_k = (\hat{\beta}_{1}^{(k)},\ldots,\hat{\beta}_{q}^{(k)})'$, $k=0,1$, denote the generalized Yule-Walker estimators obtained by solving \eqref{eq:YW_est_min}. For $m \in [p]$ and $j\in [q]$, 
%write $\hat{\ba}_m =(\hat{\alpha}_{mm}^{(0)},\hat{\alpha}_{mm}^{(1)})'$ and $\hat{\bb}_j = (\hat{\beta}_{jj}^{(0)}, \hat{\beta}_{jj}^{(1)})'$. 
define
\begin{equation*}
\hat{\bX}_{\ba_m} = 
\begin{pmatrix}
\sum_{i=1}^q  v_{1, i}^{(0)} \hat{\beta}_{1}^{(0)} \hat \bSigma_{i1}(1)^{'} \bW_0^{'}\be_m\,, ~ \sum_{i=1}^q  v_{1, i}^{(1)} \hat{\beta}_{1}^{(1)} \hat \bSigma_{i1}(0)^{'} \bW_1^{'}\be_m \\
\sum_{i=1}^q  v_{1, i}^{(0)} \hat{\beta}_{1}^{(0)} \hat \bSigma_{i2}(1)^{'} \bW_0^{'}\be_m\,, ~ \sum_{i=1}^q  v_{1, i}^{(1)} \hat{\beta}_{1}^{(1)} \hat \bSigma_{i2}(0)^{'} \bW_1^{'}\be_m \\
\vdots \\
\sum_{i=1}^q  v_{q, i}^{(0)} \hat{\beta}_{q}^{(0)} \hat \bSigma_{iq}(1)^{'} \bW_0^{'}\be_m\,, ~ \sum_{i=1}^q  v_{q, i}^{(1)} \hat{\beta}_{q}^{(1)} \hat \bSigma_{iq}(0)^{'} \bW_1^{'}\be_m 
\end{pmatrix} \in \mathbb{R}^{p q^2 \times 2}
\end{equation*}
and 
\begin{equation*}
\hat{\bX}_{\bb_j} = 
\begin{pmatrix}
\sum_{i=1}^q \text{vec}(D(\hat{\balpha}_0) \bW_0 \hat \bSigma_{i1}(1) v_{j, i}^{(0)})\,,~  \sum_{i=1}^q \text{vec}(D(\hat{\balpha}_1) \bW_1 \hat \bSigma_{i1}(0) v_{j, i}^{(1)})\\
\sum_{i=1}^q \text{vec}(D(\hat{\balpha}_0) \bW_0 \hat \bSigma_{i2}(1) v_{j, i}^{(0)})\,,~  \sum_{i=1}^q \text{vec}(D(\hat{\balpha}_1) \bW_1 \hat \bSigma_{i2}(0) v_{j, i}^{(1)})\\
\vdots \\
\sum_{i=1}^q \text{vec}(D(\hat{\balpha}_0) \bW_0 \hat \bSigma_{iq}(1) v_{j, i}^{(0)})\,,~  \sum_{i=1}^q \text{vec}(D(\hat{\balpha}_1) \bW_1 \hat \bSigma_{iq}(0) v_{j, i}^{(1)})\\
\end{pmatrix} \in \mathbb{R}^{p^2 q \times 2}\,.
\end{equation*}
%which are $pq^2 \times 2$ and $p^2 q \times 2$ matrices, respectively. 
Write $\hat{\ba}_m =(\hat{\alpha}_{m}^{(0)},\hat{\alpha}_{m}^{(1)})'$ and $\hat{\bb}_j = (\hat{\beta}_{j}^{(0)}, \hat{\beta}_{j}^{(1)})'$. Taking partial derivatives of the objective function \eqref{eq:YW_est_min} with respect to the elements of $\bar{\bgamma}$, we obtain the first-order gradient condition:
\begin{equation}\label{eq:gradient_condition}
\begin{pmatrix}
{\hat{\bX}_{\ba_1}}'\hat{\bX}_{\ba_1}\hat{\ba}_1 - {\hat{\bX}_{\ba_1}}'\bY_{\ba_1} \\
\vdots \\
{\hat{\bX}_{\ba_p}}'\hat{\bX}_{\ba_p}\hat{\ba}_p - {\hat{\bX}_{\ba_p}}'\bY_{\ba_p} \\
{\hat{\bX}_{\bb_1}}'\hat{\bX}_{\bb_1}\hat{\bb}_1 - {\hat{\bX}_{\bb_1}}'\bY_{\bb_1} \\
\vdots \\
{\hat{\bX}_{\bb_q}}'\hat{\bX}_{\bb_q}\hat{\bb}_q - {\hat{\bX}_{\bb_q}}'\bY_{\bb_q}
\end{pmatrix}   = {\bf 0}_{2(p+q)\times 1}\,. 
% \begin{pmatrix}
% {\hat{\bX}_{\ba_1}}'\bY_{\ba_1} \\
% \vdots \\
% {\hat{\bX}_{\ba_p}}'\bY_{\ba_p} \\
% {\hat{\bX}_{\bb_1}}'\bY_{\bb_1} \\
% \vdots \\
% {\hat{\bX}_{\bb_q}}'\bY_{\bb_q}
% \end{pmatrix}\,,
\end{equation}
Recall that $d = p\vee q$, $\bgamma = (\ba_1',\ldots,\ba_p',\bb_1',\ldots,\bb_q')'$ is the true parameter and the generalized Yule-Walker estimator $\hat{\bgamma} = (\hat{\ba}_1',\ldots,\hat{\ba}_p',\hat{\bb}_1',\ldots,\hat{\bb}_q')'$ is obtained by minimizing \eqref{eq:YW_est_min}. By Proposition \ref{pn:consistency}, it holds that $\|\hat{\bgamma} - \bgamma\|_2 = o_{\rm p}(1)$, provided that $d= o\{n^{1/(1+3\kappa)}\}$.
%$d= o\{n^{2/(16+13\kappa)}\}$.
%$d = o\{n^{2/(12+11\kappa)}\}$. 
For any $m \in [p]$ and $j \in [q]$, define $\bg_m(\bar{\bgamma}) = \bar{\bX}_{\ba_m}'\bar{\bX}_{\ba_m}\bar{\ba}_m - \bar{\bX}_{\ba_m}'\bY_{\ba_m}$ and $\bg_{p+j}(\bar{\bgamma}) = \bar{\bX}_{\bb_j}'\bar{\bX}_{\bb_j}\bar{\bb}_j - \bar{\bX}_{\bb_j}'\bY_{\bb_j}$, where $\bar{\bX}_{\ba_m}$ and $\bar{\bX}_{\bb_j}$ are defined in the same way as $\bX_{\ba_m}$ and $\bX_{\bb_j}$ in \eqref{LS_alpha} and \eqref{LS_beta}, respectively, except that the true parameter $\bgamma$ is replaced with $\bar{\bgamma}$. By Taylor's Theorem, if $d= o\{n^{1/(1+3\kappa)}\}$,
%$d= o\{n^{2/(16+13\kappa)}\}$,
%if $d = o\{n^{2/(12+11\kappa)}\}$, 
we have 
\begin{align}\label{eq:Taylor}
    \bG_{\rm diag} =  \bH_{\rm diag} (\hat{\bgamma} - \bgamma) + o_{\rm p}(\|\hat{\bgamma} - \bgamma\|_2)\,,
\end{align}
where 
\begin{align*}
    \bG_{\rm diag} = \begin{pmatrix}
        \bX_{\ba_1}'\bY_{\ba_1} - \bX_{\ba_1}'\bX_{\ba_1}\ba_1  \\
        \vdots \\
        \bX_{\ba_p}'\bY_{\ba_p} - \bX_{\ba_p}'\bX_{\ba_p}\ba_p  \\
        \bX_{\bb_1}'\bY_{\bb_1} - \bX_{\bb_1}'\bX_{\bb_1}\bb_1  \\
        \vdots \\
        \bX_{\bb_q}'\bY_{\bb_q} - \bX_{\bb_q}'\bX_{\bb_q}\bb_q 
        \end{pmatrix}\,,~
    \bH_{\rm diag} = \begin{pmatrix}
        \frac{\partial \bg_1(\bgamma)}{\partial \bar{\bgamma}} \\
        \vdots \\
        \frac{\partial \bg_p(\bgamma)}{\partial \bar{\bgamma}} \\
        \frac{\partial \bg_{p+1}(\bgamma)}{\partial \bar{\bgamma}} \\
        \vdots \\
        \frac{\partial \bg_{p+q}(\bgamma)}{\partial \bar{\bgamma}}
    \end{pmatrix}\,,
\end{align*}
with $\bX_{\ba_m}$ and $\bX_{\bb_j}$ specified in \eqref{LS_alpha} and \eqref{LS_beta}, respectively.
For any $j,k \in [q]$, notice that
\begin{equation}\label{eq:cov_jk1}
\hat\bSigma_{jk}(1) = \frac{1}{n}\sum_{t=2}^n \bX_{t, \cdot j} \bX_{t-1, \cdot k}'\,.
\end{equation}
Replacing $\bX_{t, \cdot j}$ by the right hand side of \eqref{class1_col},
we have
\begin{equation}\label{eq:Sigmajk_1}
\hat\bSigma_{jk}(1) =  \sum_{i=1}^q D(\balpha_0) \bW_0 \hat{\bSigma}_{ik}(1) v_{j, i}^{(0)} \beta_{j}^{(0)} + \sum_{i=1}^q D(\balpha_1) \bW_1 \hat{\bSigma}_{ik}(0) v_{j, i}^{(1)} \beta_{j}^{(1)} + \frac{1}{n}\sum_{t=2}^n \bE_{t, \cdot j}\bX_{t-1, \cdot k}' \,.
\end{equation}
Write $\bX_{t, \cdot j} = (X_{t,1,j},\ldots,X_{t,p,j})'$ for any $t \in [n]$ and $j \in [q]$.
Hence, by \eqref{eq:def}, it holds that 
\begin{align}\label{eq:YamYbj}
\bY_{\ba_m}  
= \bX_{\ba_m}\ba_m + \boldsymbol{\mathcal{E}}_{\ba_m} ~\mbox{and}~ \bY_{\bb_j}  = \bX_{\bb_j}\bb_j + \boldsymbol{\mathcal{E}}_{\bb_j}\,,
\end{align}
where 
\begin{align*}
    \boldsymbol{\mathcal{E}}_{\ba_m} = \begin{pmatrix}
        \frac{1}{n}\sum_{t=2}^n \bX_{t-1, \cdot 1}\bE_{t, \cdot 1}'\be_m \\
        \frac{1}{n}\sum_{t=2}^n \bX_{t-1, \cdot 2}\bE_{t, \cdot 1}'\be_m \\
        \vdots \\
        \frac{1}{n}\sum_{t=2}^n \bX_{t-1, \cdot q}\bE_{t, \cdot q}'\be_m
    \end{pmatrix} \,,~~\boldsymbol{\mathcal{E}}_{\bb_j} = \begin{pmatrix}
        \frac{1}{n}\sum_{t=2}^n \text{vec}(\bE_{t, \cdot j}\bX_{t-1, \cdot 1}') \\
        \frac{1}{n}\sum_{t=2}^n \text{vec}(\bE_{t, \cdot j}\bX_{t-1, \cdot 2}') \\
        \vdots \\
        \frac{1}{n}\sum_{t=2}^n \text{vec}(\bE_{t, \cdot j}\bX_{t-1, \cdot q}')
    \end{pmatrix}\,.
\end{align*}
We then have 
\begin{align}\label{eq:CLT_term}
    \bG_{\rm diag} = \begin{pmatrix}
        \bX_{\ba_1}'\boldsymbol{\mathcal{E}}_{\ba_1} \\
        \vdots \\
        \bX_{\ba_p}'\boldsymbol{\mathcal{E}}_{\ba_p} \\
        \bX_{\bb_1}'\boldsymbol{\mathcal{E}}_{\bb_1} \\
        \vdots \\
        \bX_{\bb_q}'\boldsymbol{\mathcal{E}}_{\bb_q}
    \end{pmatrix} = \begin{pmatrix}
        \tilde{\bX}_{\ba_1}'\boldsymbol{\mathcal{E}}_{\ba_1} \\
        \vdots \\
        \tilde{\bX}_{\ba_p}'\boldsymbol{\mathcal{E}}_{\ba_p} \\
        \tilde{\bX}_{\bb_1}'\boldsymbol{\mathcal{E}}_{\bb_1} \\
        \vdots \\
        \tilde{\bX}_{\bb_q}'\boldsymbol{\mathcal{E}}_{\bb_q}
    \end{pmatrix} + \begin{pmatrix}
        (\bX_{\ba_1}-\tilde{\bX}_{\ba_1})'\boldsymbol{\mathcal{E}}_{\ba_1} \\
        \vdots \\
        (\bX_{\ba_p}-\tilde{\bX}_{\ba_p})'\boldsymbol{\mathcal{E}}_{\ba_p} \\
        (\bX_{\bb_1}-\tilde{\bX}_{\bb_1})'\boldsymbol{\mathcal{E}}_{\bb_1} \\
        \vdots \\
        (\bX_{\bb_q}-\tilde{\bX}_{\bb_q})'\boldsymbol{\mathcal{E}}_{\bb_q}
    \end{pmatrix} := \tilde{\bG}_{\rm diag} + \tilde{\boldsymbol{\mathcal{U}}}_{\rm diag}\,,
    %:= \tilde{\bG}_{\rm diag} + \begin{pmatrix}
    %     (\bX_{\ba_1}-\tilde{\bX}_{\ba_1})'\boldsymbol{\mathcal{E}}_{\ba_1} \\
    %     \vdots \\
    %     (\bX_{\ba_p}-\tilde{\bX}_{\ba_p})'\boldsymbol{\mathcal{E}}_{\ba_p} \\
    %     (\bX_{\bb_1}-\tilde{\bX}_{\bb_1})'\boldsymbol{\mathcal{E}}_{\bb_1} \\
    %     \vdots \\
    %     (\bX_{\bb_q}-\tilde{\bX}_{\bb_q})'\boldsymbol{\mathcal{E}}_{\bb_q}
    % \end{pmatrix}\,,
\end{align}
where $\tilde{\bX}_{\ba_m}$ and $\tilde{\bX}_{\bb_j}$ be $pq^2 \times 2$ and $p^2q \times 2$ matrices obtained from $\bX_{\ba_m}$ and $\bX_{\bb_j}$, respectively, by replacing each $\hat{\bSigma}_{ik}(1)$ and $\hat{\bSigma}_{ik}(0)$ with their population counterparts $\bSigma_{ik}(1)$ and $\bSigma_{ik}(0)$.
%are specified in \eqref{eq:Xam} and \eqref{eq:Xbj}, respectively. 
For any $m \in [p]$, notice that
\begin{align}\label{eq:Fnorm}
    \|(\bX_{\ba_m}-\tilde{\bX}_{\ba_m})'\boldsymbol{\mathcal{E}}_{\ba_m}\|_2^2 
    %\le \| \bX_{\ba_m}-\tilde{\bX}_{\ba_m} \|_2^2 \cdot \|\boldsymbol{\mathcal{E}}_{\ba_m}\|_2^2
    \le \| \bX_{\ba_m}-\tilde{\bX}_{\ba_m} \|_F^2 \cdot \|\boldsymbol{\mathcal{E}}_{\ba_m}\|_2^2 \,.
\end{align}
Under Condition \ref{cond:mixing}(ii), for any $i \in [p]$ and $j \in [q]$, we have 
\begin{align}\label{eq:moment_bounded}
    \mathbb{E}(|X_{t,i,j}|^{4+\kappa}) & = (4+\kappa)\int_{0}^{+\infty} x^{3+\kappa}\mathbb{P}(|X_{t,i,j}|>x) \,{\mathrm d}x \lesssim \int_{0}^{+\infty} x^{3+\kappa-\frac{8+2\kappa}{\kappa}} \,{\mathrm d}x < \infty
\end{align} 
with $\kappa \in (0,2)$ specified in Condition \ref{cond:mixing}(i). Let $E_{t,i,j}$ denote the $(i,j)$-th element of $\bE_t$, which is specified in \eqref{model_class1_iter} in Section \ref{sec:model}. Then, for any $i \in [p]$ and $j,k \in [q]$,  we have
\begin{align*}
    \mathbb{E}|X_{t-1,i,j}E_{t,m,k}|^{\frac{4+\kappa}{2}} \le \{\mathbb{E}|X_{t-1,i,j}|^{4+\kappa}\}^{1/2}\{\mathbb{E}|E_{t,m,k}|^{4+\kappa}\}^{1/2} < \infty\,.
\end{align*}
Hence, by Davydov inequality (Proposition 2.5 of \cite{FanYao2003}), it holds that
\begin{align*}
    & \mathbb{E}\Big(\frac{1}{n}\sum_{t=2}^{n}X_{t-1,i,j}E_{t,m,k} \Big)^2  \\
    &~~~~ = \frac{1}{n^2}\sum_{t=2}^{n}{\rm Var}(X_{t-1,i,j}E_{t,m,k})  + \frac{1}{n^2}\sum_{t\neq s}{\rm Cov}(X_{t-1,i,j}E_{t,m,k}, X_{s-1,i,j}E_{s,m,k}) \\
    &~~~~\le \frac{C_1}{n} + \frac{8}{n^2}\sum_{t\neq s}  \alpha(|t-s|)^{\frac{\kappa}{4+\kappa}} \big[\mathbb{E}|X_{t-1,i,j}E_{t,m,k}|^{\frac{4+\kappa}{2}}\big]^{\frac{2}{4+\kappa}}\big[\mathbb{E}|X_{s-1,i,j}E_{s,m,k}|^{\frac{4+\kappa}{2}}\big]^{\frac{2}{4+\kappa}} \\
    &~~~~ \le \frac{C_1}{n} + \frac{C_2}{n}\sum_{j=1}^{n}  \alpha(j)^{\frac{\kappa}{4+\kappa}} = O\Big(\frac{1}{n}\Big)\,.
\end{align*} 
Then, we have 
\begin{align}\label{eq:Epsilon_am}
    \|\boldsymbol{\mathcal{E}}_{\ba_m}\|_2^2 = O_{\rm p}\Big(\frac{pq^2}{n}\Big)\,,
\end{align}
which implies $\max_{m \in [p]}\|\boldsymbol{\mathcal{E}}_{\ba_m}\|_2^2 = O_{\rm p}(p^2q^2n^{-1})$.

Let $\xi_{l_1,l_2}^{(m)}$ denote the $(l_1,l_2)$-th element of $\bX_{\ba_m}-\tilde{\bX}_{\ba_m}$. Notice that 
\begin{align*}
    \bX_{\ba_m}-\tilde{\bX}_{\ba_m} =
\begin{pmatrix}
\sum_{i=1}^q  v_{1, i}^{(0)} \beta_{1}^{(0)} \{\hat\bSigma_{i1}(1) - \bSigma_{i1}(1)\}^{'} \bW_0^{'}\be_m\,, ~ \sum_{i=1}^q  v_{1, i}^{(1)} \beta_{1}^{(1)} \{\hat\bSigma_{i1}(0)-\bSigma_{i1}(0)\}^{'} \bW_1^{'}\be_m \\
\sum_{i=1}^q  v_{1, i}^{(0)} \beta_{1}^{(0)} \{\hat\bSigma_{i2}(1)-\bSigma_{i2}(1)\}^{'} \bW_0^{'}\be_m\,, ~ \sum_{i=1}^q  v_{1, i}^{(1)} \beta_{1}^{(1)} \{\hat\bSigma_{i2}(0)-\bSigma_{i2}(0)\}^{'} \bW_1^{'}\be_m \\
\vdots \\
\sum_{i=1}^q  v_{q, i}^{(0)} \beta_{q}^{(0)} \{\hat\bSigma_{iq}(1)-\bSigma_{iq}(1)\}^{'} \bW_0^{'}\be_m\,, ~ \sum_{i=1}^q  v_{q, i}^{(1)} \beta_{q}^{(1)} \{\hat \bSigma_{iq}(0)-\bSigma_{iq}(0)\}^{'} \bW_1^{'}\be_m 
\end{pmatrix}\,.
\end{align*}
Hence, by \eqref{eq:InfNorm} in the proof of Lemma \ref{lem:Sigmajk} in Section \ref{lem:1}, for any $x > 0$, we have
\begin{align}\label{eq:Xam_1}
    \mathbb{P}( \| \bX_{\ba_m}-\tilde{\bX}_{\ba_m}\|_F^2 > x) & = \mathbb{P}\Big(\sum_{l_1,l_2}\{\xi_{l_1,l_2}^{(m)}\}^2 > x \Big) \notag\\
    & \lesssim pq^2 \max_{l_1,l_2}\mathbb{P}\Big(|\xi_{l_1,l_2}^{(m)}| > \frac{C_3 x^{1/2}}{p^{1/2}q} \Big) \notag\\
    & \lesssim pq^2 \mathbb{P}\Big( \max_{i,k} \|\hat\bSigma_{ik}(1) - \bSigma_{ik}(1)\|_{\infty} > \frac{C_4x^{1/2}}{p^{1/2}q} \Big) \notag\\
    &~~~~~~ + pq^2 \mathbb{P}\Big( \max_{i,k} \|\hat\bSigma_{ik}(0) - \bSigma_{ik}(0)\|_{\infty} > \frac{C_5x^{1/2}}{p^{1/2}q} \Big) \notag\\
    & \lesssim x^{-(2+\kappa)/(2\kappa)}p^{(2+7\kappa)/(2\kappa)}q^{(2+5\kappa)/\kappa}n^{-2/\kappa}\,.
    %& \lesssim x^{-(2+\kappa)/(2\kappa)}p^{(6+9\kappa)/(2\kappa)}q^{(4+6\kappa)/\kappa}n^{-2/\kappa}\,.
\end{align}
It then holds that 
\begin{align*}
    \mathbb{P}(\max_{m}\| \bX_{\ba_m}-\tilde{\bX}_{\ba_m}\|_F^2 > x) & \lesssim x^{-(2+\kappa)/(2\kappa)}p^{(2+9\kappa)/(2\kappa)}q^{(2+5\kappa)/\kappa}n^{-2/\kappa}\,,
\end{align*}
which implies that $\max_{m \in [p]}\|\bX_{\ba_m}-\tilde{\bX}_{\ba_m}\|_{F}^2 = O_{\rm p}\{p^{(2+9\kappa)/(2+\kappa)}q^{(4+10\kappa)/(2+\kappa)}n^{-4/(2+\kappa)}\}$. 
%$\max_{m \in [p]}\|\bX_{\ba_m}-\tilde{\bX}_{\ba_m}\|_{F}^2 = O_{\rm p}\{p^{(6+11\kappa)/(2+\kappa)}q^{(8+12\kappa)/(2+\kappa)}n^{-4/(2+\kappa)}\}$. 
Recall $d = p \vee q$. 
Then \eqref{eq:Fnorm} and \eqref{eq:Epsilon_am} imply $\max_{m \in [p]}\|(\bX_{\ba_m}-\tilde{\bX}_{\ba_m})'\boldsymbol{\mathcal{E}}_{\ba_m}\|_2^2 = O_{\rm p}\{d^{(14+23\kappa)/(2+\kappa)}n^{-(6+\kappa)/(2+\kappa)}\}$. 
%$\max_{m \in [p]}\|(\bX_{\ba_m}-\tilde{\bX}_{\ba_m})'\boldsymbol{\mathcal{E}}_{\ba_m}\|_2 = O_{\rm p}\{d^{(10+13\kappa)/(2+\kappa)}n^{-(6+\kappa)/(4+2\kappa)}\}$. 
Using the similar arguments, we can also show that the same result holds for $\max_{j \in [q]}\|(\bX_{\bb_j}-\tilde{\bX}_{\bb_j})'\boldsymbol{\mathcal{E}}_{\bb_j}\|_2^2$. 
Then, we have $\|\tilde{\boldsymbol{\mathcal{U}}}_{\rm diag}\|_2^2 = O_{\rm p}\{d^{(16+24\kappa)/(2+\kappa)}n^{-(6+\kappa)/(2+\kappa)}\}$.
By \eqref{eq:CLT_term}, it holds that 
\begin{align}\label{eq:G_small}
    \bG_{\rm diag} = \tilde{\bG}_{\rm diag} + o_{\rm p}(n^{-1/2})\,,
\end{align}
provided that $d = o\{n^{(4+\kappa)/(8+12\kappa)}\}$.
%$d = o\{n^{(6+\kappa)/(18+25\kappa)}\}$.
%$d = o\{n^{(6+\kappa)/(20+26\kappa)}\}$. 

As we will show in Sections \ref{subsec:BE_bound_diag} and \ref{subsec:HtoU_diag}, as $n \to \infty$, the following two assertions holds:
\begin{align}
    &\sqrt{n}\bS_n^{-1/2}\tilde{\bG}_{\rm diag}  \stackrel{\mathrm{d}}{\to} \mathcal{N}(0,\bI_{2(p+q)})\,,\label{eq:BE_bound} \\
    &~~~~~\bU_{\rm diag}\bH_{\rm diag}^{-1} \stackrel{\mathrm{p}}{\to} \bI_{2(p+q)}\,,\label{eq:HtoU}
\end{align}
provided that $d = o(n^{\nu})$ with $\nu = \min\{1/11, 2/(10+9\kappa)\}$, and $\xi_n = o\{n^{(\beta_1-1)(\rho_1-2)/(2\beta_1+2)}\}$.
By \eqref{eq:Taylor}, the conclusion of Theorem \ref{tm:mixing} holds. We complete the proof of Theorem \ref{tm:mixing}. $\hfill\Box$

\subsection{Proof of \eqref{eq:BE_bound}}\label{subsec:BE_bound_diag}

Notice that
\begin{align*}
    \tilde{\bX}_{\ba_m}'\boldsymbol{\mathcal{E}}_{\ba_m} & =     \begin{pmatrix} 
        \sum_{j=1}^q\sum_{k=1}^q \big\{ \sum_{i=1}^q v_{j,i}^{(0)} \bar{\beta}_{j}^{(0)}\be_m'\bW_0\bSigma_{ik}(1) \cdot \frac{1}{n}\sum_{t=2}^n \bX_{t-1, \cdot k}\be_m'\bE_{t,\cdot j} \big\} \\
        \sum_{j=1}^q\sum_{k=1}^q \big\{ \sum_{i=1}^q v_{j,i}^{(1)} \bar{\beta}_{j}^{(1)}\be_m'\bW_1\bSigma_{ik}(0) \cdot \frac{1}{n}\sum_{t=2}^n \bX_{t-1,\cdot k}\be_m' \bE_{t,\cdot j} \big\}
    \end{pmatrix} \\
    & = \frac{1}{n}\sum_{t=2}^n \tilde{\bX}_{\ba_m}'\{\bI_{pq^2}\otimes\be_m'\} \cdot \tilde{\boldsymbol{\mathcal{E}}}_{\bX,t}\,,
\end{align*}
where $\tilde{\mathcal{E}}_{\bX,t} = ({\rm vec}'\{\bE_{t,\cdot 1}\bX_{t-1,\cdot 1}'\},{\rm vec}'\{\bE_{t,\cdot 1}\bX_{t-1,\cdot 2}'\},\ldots,{\rm vec}'\{\bE_{t,\cdot q}\bX_{t-1,\cdot q}'\} )'$ and $\be_m$ is the $p$-dimensional unit vector with the $m$-th element being 1.
Write 
\begin{align*}
    \tilde{\bK}_{\ba} = \begin{pmatrix}
        \tilde{\bX}_{\ba_1}'\{\bI_{pq^2}\otimes\be_1'\}  \\
        \vdots \\
        \tilde{\bX}_{\ba_p}'\{\bI_{pq^2}\otimes\be_p'\}
    \end{pmatrix} \in \mathbb{R}^{(2p)\times (pq)^2}\,.
\end{align*}
Hence, it holds that
\begin{align}\label{eq:XEam_reformulate}
    \tilde{\bG}_{{\rm diag},1} := \begin{pmatrix}
        \tilde{\bX}_{\ba_1}'\boldsymbol{\mathcal{E}}_{\ba_1} \\
        \vdots \\
        \tilde{\bX}_{\ba_p}'\boldsymbol{\mathcal{E}}_{\ba_p}
    \end{pmatrix} & = \frac{1}{n}\sum_{t=2}^n \tilde{\bK}_{\ba} \tilde{\boldsymbol{\mathcal{E}}}_{\bX,t}\,.
\end{align}
Analogously, we have
\begin{align*}
    \tilde{\bX}_{\bb_j}'\boldsymbol{\mathcal{E}}_{\bb_j} & = \begin{pmatrix} 
        \sum_{k=1}^q\big[ \sum_{i=1}^q {\rm vec}'\big\{D(\bar{\balpha}_0)\bW_0\bSigma_{ik}(1)v_{j,i}^{(0)}\big\}\cdot \frac{1}{n}\sum_{t=2}^n {\rm vec}(\bE_{t,\cdot j}\bX_{t-1,\cdot k}') \big] \\
        \sum_{k=1}^q\big[ \sum_{i=1}^q {\rm vec}'\big\{D(\bar{\balpha}_1)\bW_1\bSigma_{ik}(0)v_{j,i}^{(1)}\big\}\cdot \frac{1}{n}\sum_{t=2}^n {\rm vec}(\bE_{t,\cdot j}\bX_{t-1,\cdot k}') \big]
    \end{pmatrix} \\
    & = \frac{1}{n}\sum_{t=2}^n \tilde{\bX}_{\bb_j}' \begin{pmatrix}
        {\rm vec}(\bE_{t,\cdot j}\bX_{t-1,\cdot 1}') \\
        \vdots \\
        {\rm vec}(\bE_{t,\cdot j}\bX_{t-1,\cdot q}')
    \end{pmatrix}\,.
\end{align*}
Write 
\begin{align*}
    \tilde{\bK}_{\bb} = \begin{pmatrix}
        \tilde{\bX}_{\bb_1}' &   &  \\
        &   \ddots  & \\
        &   &  \tilde{\bX}_{\bb_q}'
    \end{pmatrix} \in \mathbb{R}^{(2q)\times (pq)^2}\,.
\end{align*}
Hence, we have
\begin{align}\label{eq:XEbj_reformulate}
    \tilde{\bG}_{{\rm diag},2} := \begin{pmatrix}
        \tilde{\bX}_{\bb_1}'\boldsymbol{\mathcal{E}}_{\bb_1} \\
        \vdots \\
        \tilde{\bX}_{\bb_q}'\boldsymbol{\mathcal{E}}_{\bb_q}
    \end{pmatrix} & = \frac{1}{n}\sum_{t=2}^n \tilde{\bK}_{\bb} \tilde{\boldsymbol{\mathcal{E}}}_{\bX,t}\,.
\end{align}
By the definition of $\tilde{\bG}_{\rm diag}$, it holds that
\begin{align*}
    \tilde{\bG}_{\rm diag} = \begin{pmatrix} \tilde{\bG}_{{\rm diag},1} \\ \tilde{\bG}_{{\rm diag},2} \end{pmatrix} =  \frac{1}{n}\sum_{t=2}^n \begin{pmatrix} \tilde{\bK}_{\ba} \\ \tilde{\bK}_{\bb}\end{pmatrix} \tilde{\boldsymbol{\mathcal{E}}}_{\bX,t} = \frac{1}{n}\sum_{t=2}^n \tilde{\bK}\tilde{\boldsymbol{\mathcal{E}}}_{\bX,t} \,.
\end{align*} 
Given a vector $\bu \in \mathbb{R}^{2p+2q}$,  by Assumption \ref{cond:mmt_diagonalcase}, we have $\mathbb{E}|\bu'\tilde{\bK}\tilde{\boldsymbol{\mathcal{E}}}_{\bX,t}|^{\rho_1} < \xi_n$ and $\bu'\bS_n\bu \ge c_1$. By the Berry-Esseen bound in Theorem 2 of \cite{Sunklodas1984}, we have
\begin{align*}
    \sup_{z \in \mathbb{R}}\left| \mathbb{P}\bigg( \frac{\sqrt{n}\bu'\tilde{\bG}_{\rm diag}}{\sqrt{\bu'\bS_n\bu}} \le z\bigg) - \Phi(z)  \right| \le \xi_n n^{-(\beta_1-1)(\rho_1-2)/(2\beta_1+2)}\,,
\end{align*}
where $\Phi(\cdot)$ is the cumulative distribution function of $\mathcal{N}(0,1)$.
It implies that \eqref{eq:BE_bound} holds, provided that %$d= o\{n^{(6+\kappa)/(18+25\kappa)}\}$ and 
$\xi_n = o\{n^{(\beta_1-1)(\rho_1-2)/(2\beta_1+2)}\}$. $\hfill\Box$

\subsection{Proof of \eqref{eq:HtoU}}\label{subsec:HtoU_diag}

To prove \eqref{eq:HtoU}, it suffices to show 
% \begin{align*}
%     \|\bU_{\rm diag} - \bH_{\rm diag}\|_{2} = o_{\rm p}(1)\,,
% \end{align*}
% where $\|\cdot\|_2$ denotes the matrix operator norm, which is implied by 
\begin{align}\label{eq:infty_norm}
    \|\bU_{\rm diag} - \bH_{\rm diag}\|_{\infty} = o_{\rm p}(d^{-1})\,.
\end{align}
%Hence, it suffices to show \eqref{eq:infty_norm}.
%that every element of $\bH_{\rm diag}$ converges in probability to the corresponding element of $\bU_{\rm diag}$. 

By the definition of $\bH_{\rm diag}$, we first calculate $\partial (\bar{\bX}_{\ba_m}'\bar{\bX}_{\ba_m}\bar{\ba}_m - \bar{\bX}_{\ba_m}'\bY_{\ba_m})/\partial \bar\bgamma$ and $\partial (\bar{\bX}_{\bb_j}'\bar{\bX}_{\bb_j}\bar{\bb}_j - \bar{\bX}_{\bb_j}'\bY_{\bb_j})/\partial \bar{\bgamma}$ for any $m \in [p]$ and $j \in [q]$. It follows from the direct computation that 
\begin{align*}
    \frac{\partial \bar{\bX}_{\ba_m}}{\partial \bar{\beta}_{j}^{(0)}} = \begin{pmatrix}
        {\bf 0}_{(j-1)pq \times 1}  & {\bf 0}_{(j-1)pq \times 1} \\
        \sum_{i=1}^q v_{j, i}^{(0)}\hat{\bSigma}_{i1}(1)' \bW_0'\be_m\,, & {\bf 0}_{p \times 1} \\
        \vdots & \vdots \\
        \sum_{i=1}^q v_{j, i}^{(0)}\hat{\bSigma}_{iq}(1)' \bW_0'\be_m\,, & {\bf 0}_{p \times 1} \\
        {\bf 0}_{(q-j)pq \times 1}  & {\bf 0}_{(q-j)pq \times 1}
    \end{pmatrix}\,.
\end{align*}
Then, it holds that
\begin{align}\label{eq:partial_am_1}
    \frac{\partial (\bar{\bX}_{\ba_m}'\bar{\bX}_{\ba_m}\bar{\ba}_m)}{\partial \bar{\beta}_{j}^{(0)} } & = \bigg( \frac{\partial\bar{\bX}_{\ba_m}}{\partial\bar{\beta}_{j}^{(0)}} \bigg)'\bar{\bX}_{\ba_m}\bar{\ba}_m + \bar{\bX}_{\ba_m}'\bigg( \frac{\partial\bar{\bX}_{\ba_m}}{\partial\bar{\beta}_{j}^{(0)}} \bigg)\bar{\ba}_m \notag\\
    & = \begin{pmatrix}
        2\xi_{\ba_m,j}^{(0)}\bar{\alpha}_{m}^{(0)}+\xi_{\ba_m,j}^{(1)}\bar{\alpha}_{m}^{(1)} \\
        \xi_{\ba_m,j}^{(1)}\bar{\alpha}_{m}^{(0)}
    \end{pmatrix}\,,
\end{align}
where 
\begin{align*}
    \xi_{\ba_m,j}^{(0)} & = \sum_{k=1}^q \bigg\{ \sum_{i=1}^q \be_m'\bW_0\hat{\bSigma}_{ik}(1) v_{j, i}^{(0)} \bigg\}\bigg\{ \sum_{i=1}^q v_{j, i}^{(0)}\bar{\beta}_{j}^{(0)}\hat{\bSigma}_{ik}(1)'\bW_0'\be_m \bigg\}\,, \\
   \xi_{\ba_m,j}^{(1)} & = \sum_{k=1}^q \bigg\{ \sum_{i=1}^q \be_m'\bW_0\hat{\bSigma}_{ik}(1) v_{j, i}^{(0)} \bigg\}\bigg\{ \sum_{i=1}^q v_{j, i}^{(1)}\bar{\beta}_{j}^{(1)}\hat{\bSigma}_{ik}(0)'\bW_1'\be_m \bigg\}\,. 
\end{align*}
In the same manner, we can show that 
\begin{align}\label{eq:partial_am_2}
    \frac{\partial (\bar{\bX}_{\ba_m}'\bY_{\ba_m})}{\partial \bar{\beta}_{j}^{(0)} } & = \bigg( \frac{\partial\bar{\bX}_{\ba_m}}{\partial\bar{\beta}_{j}^{(0)}} \bigg)'\bY_{\ba_m}  = \begin{pmatrix}
        \sum_{k=1}^q \big\{ \sum_{i=1}^q \be_m'\bW_0\hat{\bSigma}_{ik}(1) v_{j, i}^{(0)} \big\} \hat \bSigma_{jk}(1)'\be_m \\
        0
    \end{pmatrix}\,.
\end{align}
Combining \eqref{eq:partial_am_1} and \eqref{eq:partial_am_2}, by \eqref{eq:Sigmajk_1}, we have
\begin{align}\label{eq:derivative_am_j0_pre}
        \frac{\partial \bg_m(\bgamma)}{\partial \bar{\beta}_{j}^{(0)}} = 
        & \frac{\partial (\bar{\bX}_{\ba_m}'\bar{\bX}_{\ba_m}\bar{\ba}_m - \bar{\bX}_{\ba_m}'\bY_{\ba_m})}{\partial \bar{\beta}_{j}^{(0)}}\bigg|_{\bar{\bgamma} = \bgamma} \notag\\
    & = 
    \alpha_{m}^{(0)}\begin{pmatrix}
        \sum_{k=1}^q \big\{ \sum_{i=1}^q \be_m'\bW_0\hat{\bSigma}_{ik}(1) v_{j, i}^{(0)} \big\}\big\{ \sum_{i=1}^q  v_{j, i}^{(0)}\beta_{j}^{(0)}\hat{\bSigma}_{ik}(1)'\bW_0'\be_m \big\} \\
        \sum_{k=1}^q \big\{ \sum_{i=1}^q \be_m'\bW_0\hat{\bSigma}_{ik}(1) v_{j, i}^{(0)} \big\}\big\{ \sum_{i=1}^q v_{j, i}^{(1)}\beta_{j}^{(1)}\hat{\bSigma}_{ik}(0)'\bW_1'\be_m \big\}
    \end{pmatrix} \notag\\
    &~~~~~~~ - \begin{pmatrix}
        \sum_{k=1}^q \big\{ \sum_{i=1}^q \be_m'\bW_0\hat{\bSigma}_{ik}(1) v_{j, i}^{(0)} \big\}\big\{\frac{1}{n}\sum_{t=2}^n \bX_{t-1, \cdot k}E_{t, m, j} \big\} \\
        0
    \end{pmatrix} \,.
    %&~~~~~~~~~~ = \bX_{\ba_m}'\bpsi_{\ba_m,j}^{(0)} + (\mathcal{E}_{\ba_m}\,, ~ {\bf 0}_{pq^2 \times 1})'\bpsi_{\ba_m,j}^{(0)}\,,
\end{align}
Notice that 
\begin{align}\label{eq:dev1}
    &\sum_{k=1}^q \bigg\{ \sum_{i=1}^q \be_m'\bW_0\hat{\bSigma}_{ik}(1) v_{j, i}^{(0)} \bigg\}\bigg\{\frac{1}{n}\sum_{t=2}^n \bX_{t-1, \cdot k}E_{t, m, j} \bigg\} \notag\\
    &~~~~~~~ = \sum_{k=1}^q \bigg\{ \sum_{i=1}^q \be_m'\bW_0\bSigma_{ik}(1) v_{j, i}^{(0)} \bigg\}\bigg\{\frac{1}{n}\sum_{t=2}^n \bX_{t-1, \cdot k}E_{t, m, j}\bigg\} \notag\\
    &~~~~~~~~~~~~ +  \sum_{k=1}^q \bigg[ \bigg\{\sum_{i=1}^q \be_m'\bW_0\big\{\hat{\bSigma}_{ik}(1) - \bSigma_{ik}(1) \big\} v_{j, i}^{(0)} \bigg\}\bigg\{\frac{1}{n}\sum_{t=2}^n \bX_{t-1, \cdot k}E_{t, m, j}\bigg\}\bigg] \notag\\
    &~~~~~~~ := \mathrm{S}_{1,m,j}+\mathrm{S}_{2,m,j}\,.
\end{align}
We then derive the stochastic orders of $\mathrm{S}_{1,m,j}$ and $\mathrm{S}_{2,m,j}$, respectively. We notice that
\begin{align*}
    & \mathrm{S}_{1,m,j} = \be_m'\bW_0 \tilde{\bD}_j \boldsymbol{\mathcal{E}}_{\ba_m}\,,
\end{align*}
where $\boldsymbol{\mathcal{E}}_{\ba_m}$ is specified in \eqref{eq:YamYbj} and 
\begin{align*}
    \tilde{\bD}_j = \Big(\mathbf{0}_{ p \times (j-1)pq}\,, ~ \sum_{i=1}^{q}\bSigma_{i1}(1)v_{j,i}^{(0)}\,, \ldots, \sum_{i=1}^{q}\bSigma_{iq}(1)v_{j,i}^{(0)}\,, ~ \mathbf{0}_{ p \times (q-j)pq} \Big)\,.
\end{align*}
Under Conditions \ref{cond:mixing}, by \eqref{eq:Epsilon_am}, we have shown that $\|\boldsymbol{\mathcal{E}}_{\ba_m}\|_2^2 = O_{\rm p}(pq^2/n)$. Since $\bW_{0}$ are uniformly bounded in both row and column sums in absolute value, which implies $\|\be_m'\bW_0 \tilde{\bD}_j\|_2^2 \le O(pq)$, then it holds that 
\begin{align*}
    |\mathrm{S}_{1,m,j}| \le \|\be_m'\bW_0 \tilde{\bD}_j\|_2 \cdot \|\boldsymbol{\mathcal{E}}_{\ba_m}\|_2 \le O_{\rm p}\Big(\frac{pq^{3/2}}{n^{1/2}}\Big)\,.
\end{align*}
Hence, we have 
\begin{align*}
    \max_{m\in[p],j\in[q]}|\mathrm{S}_{1,m,j}| \le O_{\rm p}\Big(\frac{p^2q^{5/2}}{n^{1/2}}\Big)\,,
\end{align*}
which implies $\max_{m\in[p],j\in[q]}|\mathrm{S}_{1,m,j}| = o_{\rm p}(d^{-1})$ provided that $d = o(n^{1/11})$. 
% by \eqref{eq:longruncov_1}, we have shown that $\bu'\boldsymbol{\mathcal{E}}_{\ba_m} = O_{p}(n^{-1/2})$ for any $\bu \in \mathbb{R}^{pq^2}$. It then suffices to derive the order of the elements of $\be_m'\bW_0 \tilde{\bD}_j$. We have
% \begin{align*}
%     \max_{k \in [q]} \Big\|\sum_{i=1}^q \bW_0 \bSigma_{ik}(1) v_{j, i}^{(0)} \Big\|_{\infty} \le q \max_{i,k \in [q]} \|\bW_0\bSigma_{ik}(1)v_{j, i}^{(0)}\|_{\infty} \lesssim pq %\max_{i,k \in [q]} \|\bSigma_{ik}(1)\|_{\infty}
% \end{align*}
% for any $j \in [q]$. 
%Hence, we have $\mathrm{S}_1 = o_{\rm p}(1)$ provided that $d = o(n^{1/4})$. 
Using the similar arguments, by Lemma \ref{lem:Sigmajk}, it holds that 
\begin{align*}
    \max_{m\in[p],j\in[q]}|\mathrm{S}_{2,m,j}|^2  & \lesssim pq \Big\{\max_{i,k \in [q]} \|\hat{\bSigma}_{ik}(1) -\bSigma_{ik}(1)\|_{\infty}\Big\}^2 \max_{m\in[p]}\|\boldsymbol{\mathcal{E}}_{\ba_m}\|_2^2  \\
    & \le O_{\rm p}\{p^{(6+7\kappa)/(2+\kappa)}q^{(6+7\kappa)/(2+\kappa)}n^{-(6+\kappa)/(2+\kappa)}\}\,.
\end{align*}
% \begin{align*}
%     \max_{k \in [q]} \Big\|\sum_{i=1}^q \bW_0 \big( \hat{\bSigma}_{ik}(1) - \bSigma_{ik}(1) \big) v_{j, i}^{(0)} \Big\|_{\infty}  & \lesssim pq \max_{i,k \in [q]} \|\hat{\bSigma}_{ik}(1) -\bSigma_{ik}(1)\|_{\infty}  \\
%     & \le O_{\rm p}\{p^{(2\kappa)/(2+\kappa)}q^{(2+3\kappa)/(2+\kappa)}n^{-2/(2+\kappa)}\}\,.
% \end{align*}
Hence, we have $\mathrm{S}_2 = o_{\rm p}(d^{-1})$ provided that $d = o\{n^{(6+\kappa)/(16+16\kappa)}\}$.
%$d = o\{n^{(3+\kappa)/(4+2\kappa)}\}$. 
By \eqref{eq:dev1}, it then holds that 
% Similar to the derivation of \eqref{eq:varianceSnp}, under Conditions \ref{cond:mixing} and \ref{cond:mmt_diagonalcase}, we can show the variance of the term $\sum_{k=1}^q \{ \sum_{i=1}^q \be_m'\bW_0\bSigma_{ik}(1) v_{j, i}^{(0)} \}\{n^{-1}\sum_{t=2}^n \bX_{t-1, \cdot k}E_{t, m, j}\}$ is $O(n^{-1})$. By Lemma \ref{lem:Sigmajk}, it then implies that
\begin{align}\label{eq:derivative_am_j0}
    \frac{\partial \bg_m(\bgamma)}{\partial \bar{\beta}_{j}^{(0)}} = \alpha_{m}^{(0)}\begin{pmatrix}
        \sum_{k=1}^q \big\{ \sum_{i=1}^q \be_m'\bW_0\hat{\bSigma}_{ik}(1) v_{j, i}^{(0)} \big\}\big\{ \sum_{i=1}^q  v_{j, i}^{(0)}\beta_{j}^{(0)}\hat{\bSigma}_{ik}(1)'\bW_0'\be_m \big\} \\
        \sum_{k=1}^q \big\{ \sum_{i=1}^q \be_m'\bW_0\hat{\bSigma}_{ik}(1) v_{j, i}^{(0)} \big\}\big\{ \sum_{i=1}^q v_{j, i}^{(1)}\beta_{j}^{(1)}\hat{\bSigma}_{ik}(0)'\bW_1'\be_m \big\}
    \end{pmatrix} + o_{\rm p}(d^{-1})\,,
\end{align}
provided that $d=o(n^{1/11})$.
%provided that $d=o(n^{1/4})$. 

Analogously, we can show that the derivative of $\bar{\bX}_{\ba_m}$ with respect to $\bar{\beta}_{j}^{(1)}$ is
\begin{align*}
    \frac{\partial \bar{\bX}_{\ba_m}}{\partial \bar{\beta}_{j}^{(1)}} = \begin{pmatrix}
        {\bf 0}_{(j-1)pq \times 1}  & {\bf 0}_{(j-1)pq \times 1} \\
        {\bf 0}_{p \times 1}\,, & \sum_{i=1}^q v_{j, i}^{(1)}\hat{\bSigma}_{i1}(0)' \bW_1'\be_m \\
        \vdots & \vdots \\
        {\bf 0}_{p \times 1}\,, & \sum_{i=1}^q v_{j, i}^{(1)}\hat{\bSigma}_{iq}(0)' \bW_1'\be_m \\
        {\bf 0}_{(q-j)pq \times 1}  & {\bf 0}_{(q-j)pq \times 1}
    \end{pmatrix}\,.
\end{align*}
Using the similar arguments for the derivation of \eqref{eq:derivative_am_j0}, it holds that
\begin{align}\label{eq:derivative_am_j1}
        \frac{\partial \bg_m(\bgamma)}{\partial \bar{\beta}_{j}^{(1)}} & =  \frac{\partial (\bar{\bX}_{\ba_m}'\bar{\bX}_{\ba_m}\bar{\ba}_m-\bar{\bX}_{\ba_m}'\bY_{\ba_m})}{\partial \bar{\beta}_{j}^{(1)}}\bigg|_{\bar{\bgamma} = \bgamma} \\
    & = 
    \alpha_{m}^{(1)}\begin{pmatrix}
        \sum_{k=1}^q \big\{ \sum_{i=1}^q \be_m'\bW_1\hat{\bSigma}_{ik}(0) v_{j, i}^{(1)} \big\}\big\{ \sum_{i=1}^q  v_{j, i}^{(0)}\beta_{j}^{(0)}\hat{\bSigma}_{ik}(1)'\bW_0'\be_m \big\} \\
        \sum_{k=1}^q \big\{ \sum_{i=1}^q \be_m'\bW_1\hat{\bSigma}_{ik}(0) v_{j, i}^{(1)} \big\}\big\{ \sum_{i=1}^q v_{j, i}^{(1)}\beta_{j}^{(1)}\hat{\bSigma}_{ik}(0)'\bW_1'\be_m \big\}
    \end{pmatrix} + o_{\rm p}(d^{-1})\,,  \notag
    % &~~~~~~~~~~~~~~~~~~ - \begin{pmatrix}
    %     0 \\
    %     \sum_{k=1}^q \big\{ \sum_{i=1}^q \be_m'\bW_1\hat{\bSigma}_{ik}(0) v_{j, i}^{(1)} \big\}\big\{\frac{1}{n}\sum_{t=2}^n \bX_{t-1, \cdot k}E_{t, m, j} \big\} 
    % \end{pmatrix}
\end{align}
provided that $d=o(n^{1/11})$.
%provided that $d=o(n^{1/4})$. 
Combining \eqref{eq:derivative_am_j0} and \eqref{eq:derivative_am_j1}, we have
\begin{align*}%\label{eq:derv_am_j}
    \frac{\partial \bg_m(\bgamma)}{\partial \bar{\bb}_{j}} = \bX_{\ba_m}'\bZ_{\ba_m,j} + o_{\rm p}(d^{-1})\,,
\end{align*}
provided that $d=o(n^{1/11})$, where
%provided that $d=o(n^{1/4})$, where 
\begin{align*}
    \bZ_{\ba_m,j} = \begin{pmatrix}
        {\bf 0}_{(j-1)pq \times 1}\,, &  {\bf 0}_{(j-1)pq \times 1} \\
        \alpha_{m}^{(0)}\sum_{i=1}^q v_{j, i}^{(0)}\hat{\bSigma}_{i1}'(1) \bW_0'\be_m\,, & \alpha_{m}^{(1)}\sum_{i=1}^q v_{j, i}^{(1)}\hat{\bSigma}_{i1}'(0) \bW_1'\be_m \\
        \vdots & \vdots  \\
        \alpha_{m}^{(0)}\sum_{i=1}^q v_{j, i}^{(0)}\hat{\bSigma}_{iq}'(1) \bW_0'\be_m\,, & \alpha_{m}^{(1)}\sum_{i=1}^q v_{j, i}^{(1)}\hat{\bSigma}_{iq}'(0) \bW_1'\be_m \\
        {\bf 0}_{(q-j)pq \times 1}\,, &  {\bf 0}_{(q-j)pq \times 1}
    \end{pmatrix}
\end{align*}
is a $pq^2\times 2$ matrix. 
% By Lemma \ref{lem:Sigmajk}, it holds that $\mathcal{E}_{\ba_m}'\bZ_{\ba_m,j}= \mathcal{E}_{\ba_m}'\tilde{\bZ}_{\ba_m,j} + o_{\rm p}(1)$, where $\tilde{\bZ}_{\ba_m,j}$ is defined in the same manner as $\bZ_{\ba_m,j}$ except that $\hat{\bSigma}_{jk}(1)$ and $\hat{\bSigma}_{jk}(0)$ are replaced by $\bSigma_{jk}(1)$ and $\bSigma_{jk}(0)$, respectively. Since $\mathbb{E}(\mathcal{E}_{\ba_m}'\tilde{\bZ}_{\ba_m,j}) = 0$, by calculating its variance under Condition \ref{cond:mixing}, it can be shown that $\mathcal{E}_{\ba_m}'\tilde{\bZ}_{\ba_m,j}=o_{\rm p}(1)$.
Furthermore, we have
\begin{align*}
    \frac{\partial \bg_m(\bgamma)}{\partial \bar{\ba}_m} = \bX_{\ba_m}'\bX_{\ba_m}\,, ~\mbox{and}~ \frac{\partial \bg_m(\bgamma)}{\partial \bar{\ba}_k} = {\bf 0}_{2 \times 2} ~\mbox{for}~ k \neq m\,.
\end{align*}
Hence, if $d=o(n^{1/11})$, it holds that 
\begin{align}\label{eq:der_am}
    &\frac{\partial \bg_m(\bgamma)}{\partial \bar{\bgamma}} = \big({\bf 0}_{2 \times 2}\,, \cdots\,, \bX_{\ba_m}'\bX_{\ba_m}\,, \cdots\,, {\bf 0}_{2 \times 2}\,,  \bX_{\ba_m}'\bZ_{\ba_m,1}\,, \cdots\,, \bX_{\ba_m}'\bZ_{\ba_m,q}\big) + o_{\rm p}(d^{-1})\,.
\end{align} 
%provided that $d=o(n^{1/4})$.

On the other hand, for any $m \in [p]$ and $j \in [q]$, it follows from the direct computation that 
\begin{align*}
    \frac{\partial \bar{\bX}_{\bb_j}}{\partial \bar{\alpha}_{m}^{(0)}} = \begin{pmatrix}
        \sum_{i=1}^q \{v_{j, i}^{(0)}\hat{\bSigma}_{i1}(1)' \bW_0'\be_m\}\otimes \be_m\,, & {\bf 0}_{p^2 \times 1} \\
        \sum_{i=1}^q \{v_{j, i}^{(0)}\hat{\bSigma}_{i2}(1)' \bW_0'\be_m\}\otimes \be_m\,, & {\bf 0}_{p^2 \times 1} \\
        \vdots & \vdots \\
        \sum_{i=1}^q \{v_{j, i}^{(0)}\hat{\bSigma}_{iq}(1)' \bW_0'\be_m\}\otimes \be_m\,, & {\bf 0}_{p^2 \times 1} \\
    \end{pmatrix}\,.
\end{align*}
Using the similar arguments for the derivation of \eqref{eq:partial_am_1}, we have
\begin{align}\label{eq:partial_bj_1}
    \frac{\partial (\bar{\bX}_{\bb_j}'\bar{\bX}_{\bb_j}\bar{\bb}_j)}{\partial \bar{\alpha}_{m}^{(0)} } & = \bigg( \frac{\partial\bar{\bX}_{\bb_j}}{\partial\bar{\alpha}_{m}^{(0)}} \bigg)'\bar{\bX}_{\bb_j}\bar{\bb}_j + \bar{\bX}_{\bb_j}'\bigg( \frac{\partial\bar{\bX}_{\bb_j}}{\partial\bar{\alpha}_{m}^{(0)}} \bigg)\bar{\bb}_j \notag\\
    & = \begin{pmatrix}
        2\xi_{\bb_j,m}^{(0)}\bar{\beta}_{j}^{(0)}+\xi_{\bb_j,m}^{(1)}\bar{\beta}_{j}^{(1)} \\
        \xi_{\bb_j,m}^{(1)}\bar{\beta}_{j}^{(0)}
    \end{pmatrix}\,,
\end{align}
where 
\begin{align*}
    \xi_{\bb_j,m}^{(0)} & = \sum_{k=1}^q \bigg\{ \sum_{i=1}^q (\be_m'\bW_0\hat{\bSigma}_{ik}(1) v_{j, i}^{(0)})\otimes \be_m' \bigg\}\bigg\{ \sum_{i=1}^q \text{vec}(D(\bar{\balpha}_0) \bW_0 \hat{\bSigma}_{ik}(1) v_{j, i}^{(0)}) \bigg\}\,, \\
   \xi_{\bb_j,m}^{(1)} & = \sum_{k=1}^q \bigg\{ \sum_{i=1}^q (\be_m'\bW_0\hat{\bSigma}_{ik}(1) v_{j, i}^{(0)})\otimes \be_m' \bigg\}\bigg\{ \sum_{i=1}^q \text{vec}(D(\bar{\balpha}_1) \bW_1 \hat{\bSigma}_{ik}(0) v_{j, i}^{(1)}) \bigg\}\,. 
\end{align*}
In the same manner, we can show that 
\begin{align}\label{eq:partial_bj_2}
    \frac{\partial (\bar{\bX}_{\bb_j}'\bY_{\bb_j})}{\partial \bar{\alpha}_{m}^{(0)} } & = \bigg( \frac{\partial\bar{\bX}_{\bb_j}}{\partial\bar{\alpha}_{m}^{(0)}} \bigg)'\bY_{\bb_j} \notag\\
    & = \begin{pmatrix}
        \sum_{k=1}^q \big\{ \sum_{i=1}^q (\be_m'\bW_0\hat{\bSigma}_{ik}(1) v_{j, i}^{(0)})\otimes \be_m' \big\}\text{vec}\{\hat \bSigma_{jk}(1)\} \\
        0
    \end{pmatrix}\,.
\end{align}
Hence, combining \eqref{eq:partial_bj_1} and \eqref{eq:partial_bj_2}, by \eqref{eq:Sigmajk_1}, using similar arguments for deriving \eqref{eq:derivative_am_j0}, we have
\begin{align}\label{eq:derivative_bj_m0}
    \frac{\partial \bg_{p+j}(\bgamma)}{\partial \bar{\alpha}_{m}^{(0)}} & = \frac{\partial (\bar{\bX}_{\bb_j}'\bar{\bX}_{\bb_j}\bar{\bb}_j-\bar{\bX}_{\bb_j}'\bY_{\bb_j})}{\partial \bar{\alpha}_{m}^{(0)}}\bigg|_{\bar{\bgamma} = \bgamma} \\
    & = 
    \beta_{j}^{(0)}\begin{pmatrix}
        \sum_{k=1}^q \big\{ \sum_{i=1}^q (\be_m'\bW_0\hat{\bSigma}_{ik}(1) v_{j, i}^{(0)})\otimes \be_m' \big\}\big\{ \sum_{i=1}^q \text{vec}(D(\balpha_0) \bW_0 \hat{\bSigma}_{ik}(1) v_{j, i}^{(0)}) \big\} \\
        \sum_{k=1}^q \big\{ \sum_{i=1}^q (\be_m'\bW_0\hat{\bSigma}_{ik}(1) v_{j, i}^{(0)})\otimes \be_m' \big\}\big\{ \sum_{i=1}^q \text{vec}(D(\balpha_1) \bW_1 \hat{\bSigma}_{ik}(0) v_{j, i}^{(1)}) \big\}
    \end{pmatrix} + o_{\rm p}(d^{-1})\,, \notag
    % &~~~~~~ - \begin{pmatrix}
    %     \sum_{k=1}^q \big\{ \sum_{i=1}^q (\be_m'\bW_0\hat{\bSigma}_{ik}(1) v_{j, i}^{(0)})\otimes \be_m' \big\}\big\{\frac{1}{n}\sum_{t=2}^n {\rm vec}(\bE_{t, \cdot j}\bX_{t-1, \cdot k}')\big\} \\
    %     0
    % \end{pmatrix}\,.
\end{align}
provided that $d=o(n^{1/11})$.
%provided that $d=o(n^{1/4})$. 
Analogously, we can show that the derivative of $\bar{\bX}_{\bb_j}$ with respect to $\bar{\alpha}_{m}^{(1)}$ is
\begin{align*}
    \frac{\partial \bar{\bX}_{\bb_j}}{\partial \bar{\alpha}_{m}^{(1)}} = \begin{pmatrix}
        {\bf 0}_{p^2 \times 1}\,, & \sum_{i=1}^q \{v_{j, i}^{(1)}\hat{\bSigma}_{i1}(0)' \bW_1'\be_m\}\otimes \be_m \\
        {\bf 0}_{p^2 \times 1} \,, & \sum_{i=1}^q \{v_{j, i}^{(1)}\hat{\bSigma}_{i2}(0)' \bW_1'\be_m\}\otimes \be_m \\
        \vdots & \vdots \\
        {\bf 0}_{p^2 \times 1}\,, & \sum_{i=1}^q \{v_{j, i}^{(1)}\hat{\bSigma}_{iq}(0)' \bW_1'\be_m\}\otimes \be_m \\
    \end{pmatrix}\,.
\end{align*}
Using the similar arguments for the derivation of \eqref{eq:derivative_bj_m0}, it holds that
\begin{align}\label{eq:derivative_bj_m1}
    \frac{\partial \bg_{p+j}(\bgamma)}{\partial \bar{\alpha}_{m}^{(1)}} & =\frac{\partial (\bar{\bX}_{\bb_j}'\bar{\bX}_{\bb_j}\bar{\bb}_j-\bar{\bX}_{\bb_j}'\bY_{\bb_j})}{\partial \bar{\alpha}_{m}^{(1)}}\bigg|_{\bar{\bgamma} = \bgamma} \\
    & = 
    \beta_{j}^{(1)}\begin{pmatrix}
        \sum_{k=1}^q \big\{ \sum_{i=1}^q (\be_m'\bW_1\hat{\bSigma}_{ik}(0) v_{j, i}^{(1)})\otimes \be_m' \big\}\big\{ \sum_{i=1}^q \text{vec}(D(\balpha_0) \bW_0 \hat{\bSigma}_{ik}(1) v_{j, i}^{(0)}) \big\} \\
        \sum_{k=1}^q \big\{ \sum_{i=1}^q (\be_m'\bW_1\hat{\bSigma}_{ik}(0) v_{j, i}^{(1)})\otimes \be_m' \big\}\big\{ \sum_{i=1}^q \text{vec}(D(\balpha_1) \bW_1 \hat{\bSigma}_{ik}(0) v_{j, i}^{(1)}) \big\}
    \end{pmatrix} + o_{\rm p}(d^{-1})\,, \notag
    % &~~~~~~ - \begin{pmatrix}
    %     \sum_{k=1}^q \big\{ \sum_{i=1}^q (\be_m'\bW_1\hat{\bSigma}_{ik}(0) v_{j, i}^{(1)})\otimes \be_m' \big\}\big\{\frac{1}{n}\sum_{t=2}^n {\rm vec}(\bE_{t, \cdot j}\bX_{t-1, \cdot k}')\big\} \\
    %     0
    % \end{pmatrix}\,.
\end{align}
provided that $d=o(n^{1/11})$.
%provided that $d=o(n^{1/4})$. 
Combining \eqref{eq:derivative_bj_m0} and \eqref{eq:derivative_bj_m1}, we have
\begin{align*}%\label{eq:derv_bj_m}
    \frac{\partial \bg_{p+j}(\bgamma)}{\partial \bar{\ba}_{m}} = \bX_{\bb_j}'\bZ_{\bb_j,m} + o_{\rm p}(d^{-1})\,,
\end{align*}
provided that $d=o(n^{1/11})$, where
%provided that $d=o(n^{1/4})$, where
\begin{align*}
    \bZ_{\bb_j,m} = \begin{pmatrix}
        \beta_{j}^{(0)} \big\{ \sum_{i=1}^q (v_{j, i}^{(0)}\hat{\bSigma}_{i1}'(1)\bW_0'\be_m)\otimes \be_m \big\}\,, & \beta_{j}^{(1)}\sum_{i=1}^q (v_{j, i}^{(1)}\hat{\bSigma}_{i1}'(0)\bW_1'\be_m)\otimes \be_m \\
        \vdots & \vdots\\
        \beta_{j}^{(0)} \big\{ \sum_{i=1}^q (v_{j, i}^{(0)}\hat{\bSigma}_{iq}'(1)\bW_0'\be_m)\otimes \be_m \big\}\,, & \beta_{j}^{(1)}\sum_{i=1}^q (v_{j, i}^{(1)}\hat{\bSigma}_{iq}'(0)\bW_1'\be_m)\otimes \be_m
    \end{pmatrix}
\end{align*}
is a $p^2q\times 2$ matrix. 
%Analogously, by Condition \ref{cond:mixing} and Lemma \ref{lem:Sigmajk}, it holds that $\mathcal{E}_{\bb_j}'\bZ_{\bb_j,m}=o_{\rm p}(1)$.
Furthermore, we have
\begin{align*}
    \frac{\partial \bg_{p+j}(\bgamma)}{\partial \bar{\bb}_{j}} = \bX_{\bb_j}'\bX_{\bb_j}\,, ~\mbox{and}~ \frac{\partial \bg_{p+j}(\bgamma)}{\partial \bar{\bb}_k} = {\bf 0}_{2 \times 2} ~\mbox{for}~ k \neq j\,.
\end{align*}
Hence, if $d=o(n^{1/11})$, it holds that
%if $d=o(n^{1/4})$, it holds that 
\begin{align}\label{eq:der_bj}
    \frac{\partial \bg_{p+j}(\bgamma)}{\partial \bar{\bgamma}} = \big(\bX_{\bb_j}'\bZ_{\bb_j,1}\,,  \cdots\,,\bX_{\bb_j}'\bZ_{\bb_j,p}\,, {\bf 0}_{2 \times 2}\,, \cdots\,, \bX_{\bb_j}'\bX_{\bb_j}\,, \cdots\,, {\bf 0}_{2 \times 2}\big) + o_{\rm p}(d^{-1})\,.
\end{align} 
Recall that $\bH_{\rm diag}$ is specified in \eqref{eq:Taylor}.

For any $m \in [p]$, we have
\begin{align*}
    & \|\bX_{\ba_m}'\bX_{\ba_m} - \tilde{\bX}_{\ba_m}'\tilde{\bX}_{\ba_m}\|_{\infty} \\
    &~~~~~ \le \Big| \sum_{i_1,i_2,j,k=1}^q  v_{j, i_1}^{(0)}v_{j, i_2}^{(0)} \{\beta_{j}^{(0)}\}^2 \be_m'\bW_0 \{\hat\bSigma_{i_1k}(1)\hat\bSigma_{i_2k}(1)'- \bSigma_{i_1k}(1)\bSigma_{i_2k}(1)'\} \bW_0^{'}\be_m  \Big| \\
    & ~~~~~~~~~ + \Big| \sum_{i_1,i_2,j,k=1}^q  v_{j, i_1}^{(1)} v_{j, i_2}^{(1)} \{\beta_{j}^{(1)}\}^2 \be_m'\bW_1 \{\hat\bSigma_{i_1k}(0)\hat\bSigma_{i_2k}(0)'- \bSigma_{i_1k}(0)\bSigma_{i_2k}(0)'\} \bW_1^{'}\be_m  \Big| \\
    & ~~~~~~~~~ + 2 \Big| \sum_{i_1,i_2,j,k=1}^q  v_{j, i_1}^{(0)}v_{j, i_2}^{(1)} \beta_{j}^{(0)}\beta_{j}^{(1)} \be_m'\bW_0 \{\hat\bSigma_{i_1k}(1)\hat\bSigma_{i_2k}(0)'- \bSigma_{i_1k}(1)\bSigma_{i_2k}(0)'\} \bW_1^{'}\be_m  \Big| \\
    &~~~~~ := \mathrm{R}_{1,m} + \mathrm{R}_{2,m} + \mathrm{R}_{3,m}\,.
\end{align*}
For $\mathrm{R}_{1,m}$, notice that
\begin{align}\label{eq:R1}
    |\mathrm{R}_{1,m}| & \lesssim q^2 \max_{i_1,i_2,k \in [q]}|\be_m'\bW_0 \{\hat\bSigma_{i_1k}(1)\hat\bSigma_{i_2k}(1)'- \bSigma_{i_1k}(1)\bSigma_{i_2k}(1)'\} \bW_0^{'}\be_m| \notag \\
    & \lesssim q^2 \max_{i_1,i_2,k \in [q]}\|\hat\bSigma_{i_1k}(1)\hat\bSigma_{i_2k}(1)'- \bSigma_{i_1k}(1)\bSigma_{i_2k}(1)'\|_{\infty}
\end{align}
and 
\begin{align*}
    &  \max_{i_1,i_2,k \in [q]}\|\hat\bSigma_{i_1k}(1)\hat\bSigma_{i_2k}(1)'- \bSigma_{i_1k}(1)\bSigma_{i_2k}(1)'\|_{\infty} \\
    &~~~~~ = \max_{i_1,i_2,k \in [q]}\max_{l_1,l_2 \in [p]}\Big|\sum_{l=1}^p \big\{\hat{\sigma}_{l_1,l}^{(i_1,k,1)}\hat{\sigma}_{l_2,l}^{(i_2,k,1)} - \sigma_{l_1,l}^{(i_1,k,1)}\sigma_{l_2,l}^{(i_2,k,1)} \big\} \Big| \\
    &~~~~~ \le p\max_{i_1,i_2,k \in [q]}\max_{l_1,l_2,l \in [p]}|\hat{\sigma}_{l_1,l}^{(i_1,k,1)}\hat{\sigma}_{l_2,l}^{(i_2,k,1)} - \sigma_{l_1,l}^{(i_1,k,1)}\sigma_{l_2,l}^{(i_2,k,1)} | \,,
\end{align*}
where $\bSigma_{jk}(1) = (\sigma_{l_1,l_2}^{(j,k,1)})_{p \times p}$ and $\hat\bSigma_{jk}(1) = (\hat{\sigma}_{l_1,l_2}^{(j,k,1)})_{p \times p}$. It holds that 
\begin{align*}
    \hat{\sigma}_{l_1,l}^{(i_1,k,1)}\hat{\sigma}_{l_2,l}^{(i_2,k,1)} - \sigma_{l_1,l}^{(i_1,k,1)}\sigma_{l_2,l}^{(i_2,k,1)} & = (\hat{\sigma}_{l_2,l}^{(i_2,k,1)} - \sigma_{l_2,l}^{(i_2,k,1)})\sigma_{l_1,l}^{(i_1,k,1)} + (\hat{\sigma}_{l_1,l}^{(i_1,k,1)}-\sigma_{l_1,l}^{(i_1,k,1)})\sigma_{l_2,l}^{(i_2,k,1)} \\
    & ~~~~~ + (\hat{\sigma}_{l_1,l}^{(i_1,k,1)}-\sigma_{l_1,l}^{(i_1,k,1)})(\hat{\sigma}_{l_2,l}^{(i_2,k,1)} - \sigma_{l_2,l}^{(i_2,k,1)})\\
    & := \mathrm{R}_{l_1,l_2,l}^{(i_1,i_2,k,1)} + \mathrm{R}_{l_1,l_2,l}^{(i_1,i_2,k,2)} + \mathrm{R}_{l_1,l_2,l}^{(i_1,i_2,k,3)}\,.
\end{align*}
By \eqref{eq:InfNorm} in the proof of Lemma \ref{lem:Sigmajk} in Section \ref{lem:1}, we have 
\begin{align*}
    \max_{i_1,i_2,k \in [q]}\max_{l_1,l_2,l \in [q]}|\mathrm{R}_{l_1,l_2,l}^{(i_1,i_2,k,1)}| = O_{\rm p}\{(pq)^{2\kappa/(2+\kappa)}n^{-2/(2+\kappa)}\} = \max_{i_1,i_2,k \in [q]}\max_{l_1,l_2,l \in [q]}|\mathrm{R}_{l_1,l_2,l}^{(i_1,i_2,k,2)}| \,.
\end{align*}
%By \eqref{eq:InfNorm} in the proof of Lemma \ref{lem:Sigmajk} in Section \ref{lem:1},
Similarly, we have
\begin{align*}
    \mathbb{P}(\max_{i_1,i_2,k \in [q]}\max_{l_1,l_2,l \in [q]}|\mathrm{R}_{l_1,l_2,l}^{(i_1,i_2,k,3)} | > x ) & \le \mathbb{P}(\max_{i_1,k \in [q]}\max_{l_1,l \in [q]}|\hat{\sigma}_{l_1,l}^{(i_1,k,1)}-\sigma_{l_1,l}^{(i_1,k,1)} | > x^{1/2} ) \\
    & ~~~~~~+ \mathbb{P}(\max_{i_2,k \in [q]}\max_{l,l_2 \in [q]}| \hat{\sigma}_{l_2,l}^{(i_2,k,1)} - \sigma_{l_2,l}^{(i_2,k,1)}| > x^{1/2} ) \\
    & \lesssim x^{-(2+\kappa)/(2\kappa)} p^2 q^2 n^{-2/\kappa}\,,
\end{align*}
which implies that 
\begin{align*}
    \max_{i_1,i_2,k \in [q]}\max_{l_1,l_2,l \in [q]}|\mathrm{R}_{l_1,l_2,l}^{(i_1,i_2,k,3)} | = O_{\rm p}\{(pq)^{4\kappa/(2+\kappa)}n^{-4/(2+\kappa)}\}\,.
\end{align*}
Hence, we have
\begin{align}\label{eq:XX_consistency}
    \max_{i_1,i_2,k \in [q]}\|\hat\bSigma_{i_1k}(1)\hat\bSigma_{i_2k}(1)'- \bSigma_{i_1k}(1)\bSigma_{i_2k}(1)'\|_{\infty} & \le O_{\rm p}\{p^{(2+3\kappa)/(2+\kappa)}q^{2\kappa/(2+\kappa)}n^{-2/(2+\kappa)}\} \notag \\
    & ~~~ + O_{\rm p}\{p^{(2+5\kappa)/(2+\kappa)}q^{4\kappa/(2+\kappa)}n^{-4/(2+\kappa)}\}\,.
\end{align}
By \eqref{eq:R1}, it holds that $\max_{m \in [p]}|\mathrm{R}_{1,m}| = o_{\rm p}(d^{-1})$ provided that $d = o\{n^{2/(10+9\kappa)}\}$. Using the similar arguments, we show that the same results hold for $\max_{m \in [p]}|\mathrm{R}_{2,m}|$ and $\max_{m \in [p]}|\mathrm{R}_{3,m}|$, which yields $\max_{m \in [p]}\|\bX_{\ba_m}'\bX_{\ba_m} - \tilde{\bX}_{\ba_m}'\tilde{\bX}_{\ba_m}\|_{\infty} = o_{\rm p}(d^{-1})$ provided that $d = o\{n^{2/(10+9\kappa)}\}$. Using the similar arguments, we can show that $\max_{m \in [p], j \in [q]}\|\bX_{\ba_m}'\bZ_{\ba_m,j} - \tilde{\bX}_{\ba_m}'\tilde{\bZ}_{\ba_m,j}\|_{\infty} = o_{\rm p}(d^{-1})$, provided that $d = o\{n^{1/(4+4\kappa)}\}$. Similarly, it holds that $\max_{j \in [q]}\|\bX_{\bb_j}'\bX_{\bb_j} - \tilde{\bX}_{\bb_j}'\tilde{\bX}_{\bb_j}\|_{\infty} = o_{\rm p}(d^{-1}) = \max_{j \in [q], m\in [p]}\|\bX_{\bb_j}'\bZ_{\bb_j,m} - \tilde{\bX}_{\bb_j}'\tilde{\bZ}_{\bb_j,m}\|_{\infty}$, provided that $d = o\{n^{2/(10+9\kappa)}\}$. We complete the proof of \eqref{eq:HtoU}. $\hfill\Box$
%By \eqref{eq:R1}, it holds that $|\mathrm{R}_1| = o_{\rm p}(1)$ provided that $d = o\{n^{2/(14+11\kappa)}\}$. Using the similar arguments, we show that the same results hold for $|\mathrm{R}_2|$ and $|\mathrm{R}_3|$, which yields $\|\bX_{\ba_m}'\bX_{\ba_m} - \tilde{\bX}_{\ba_m}'\tilde{\bX}_{\ba_m}\|_{\infty} = o_{\rm p}(1)$ provided that $d = o\{n^{2/(14+11\kappa)}\}$. Using the similar arguments, we can show that $\|\bX_{\ba_m}'\bZ_{\ba_m,j} - \tilde{\bX}_{\ba_m}'\tilde{\bZ}_{\ba_m,j}\|_{\infty} = o_{\rm p}(1)$ for any $m \in [p]$ and $j \in [q]$, provided that $d = o\{n^{1/(6+5\kappa)}\}$. Similarly, it holds that $\|\bX_{\bb_j}'\bX_{\bb_j} - \tilde{\bX}_{\bb_j}'\tilde{\bX}_{\bb_j}\|_{\infty} = o_{\rm p}(1) = \|\bX_{\bb_j}'\bZ_{\bb_j,m} - \tilde{\bX}_{\bb_j}'\tilde{\bZ}_{\bb_j,m}\|_{\infty}$ for any $j \in [q]$ and $m \in [p]$, provided that $d = o\{n^{2/(24+11\kappa)}\}$.

\section{Proposition \ref{CLT:diag_fixdim}}\label{sec:diag_fixdim}

%Recall that $\tilde{\bX}_{\ba_m}$ and $\tilde{\bX}_{\bb_j}$ are defined in \ref{eq:Xam} and \ref{eq:Xbj}, respectively. 

\begin{proposition}\label{CLT:diag_fixdim}
Assume that Conditions \ref{cond:mixing}-\ref{cond:mmt_diagonalcase} hold. 
%Let $d = p \vee q$. Then, if $d = o(n^{\nu})$ with $\nu = \max\{1/9, 2/(16+13\kappa)\}$,
    When $p,q \ge 1$ are fixed, if 
        \begin{align*}
        &~~~~~~~~ \sup_{\bu \in \mathbb{R}^{2p}}\mathbb{E}|\bu'\tilde{\bK}_{\ba}\tilde{\boldsymbol{\mathcal{E}}}_{\bX,t}|^{\frac{4+\kappa}{2}} < \infty ~~\,\mbox{and}\,~ \sup_{\bu \in \mathbb{R}^{2q}}\mathbb{E}|\bu'\tilde{\bK}_{\bb}\tilde{\boldsymbol{\mathcal{E}}}_{\bX,t}|^{\frac{4+\kappa}{2}} < \infty\,, 
        \end{align*}
then we have 
    \begin{align*}
    \sqrt{n} \tilde{\bP}_{\rm diag}^{-1/2} \bU_{\rm diag}(\hat{\bgamma}-\bgamma) \stackrel{\mathrm{d}}{\to} \mathcal{N}(0,\bI_{2(p+q)})\,,
\end{align*}
where 
\begin{align}\label{eq:Pdiag}
    \tilde{\bP}_{\rm diag} = \begin{pmatrix}
       \tilde{\bK}_{\ba}\tilde{\bSigma}_{\bX,\bE}\tilde{\bK}_{\ba}'\,, & \tilde{\bK}_{\ba}\tilde{\bSigma}_{\bX,\bE}\tilde{\bK}_{\bb}'\\
       \tilde{\bK}_{\bb}\tilde{\bSigma}_{\bX,\bE}\tilde{\bK}_{\ba}'\,, & \tilde{\bK}_{\bb}\tilde{\bSigma}_{\bX,\bE}\tilde{\bK}_{\bb}'
    \end{pmatrix}\,.
\end{align}
\end{proposition}

\subsection{Proof of Proposition \ref{CLT:diag_fixdim}}

Analogous to the proof of Theorem \ref{tm:mixing} in Section \ref{sec:proof_thm1}, by \eqref{eq:Taylor},
Proposition \ref{CLT:diag_fixdim} can be shown by proving the following two assertions:
% Hence, by \eqref{eq:Taylor}, if {\color{red}$d = o\{n^{(6+\kappa)/(18+25\kappa)}\}$},
% %$d= o\{n^{2/(16+13\kappa)}\}$,
% %if $d = o\{n^{2/(12+11\kappa)}\}$, 
% it suffices to prove the assertions \eqref{eq:CLT_mixing} and \eqref{eq:HtoU} below:
\begin{align}
    & \sqrt{n} \tilde{\bP}_{\rm diag}^{-1/2} \tilde{\bG}_{\rm diag} \stackrel{\mathrm{d}}{\to} \mathcal{N}(0,\bI_{2(p+q)})\,,\label{eq:CLT_mixing} \\
    &~~~~~~  \bU_{\rm diag}\bH_{\rm diag}^{-1} \stackrel{\mathrm{p}}{\to} \bI_{2(p+q)}\,,\label{eq:HtoU_fixed}
\end{align}
where $\bU_{\rm diag}$ is specified in \eqref{eq:Umatrix} and $\tilde{\bP}_{\rm diag}$ is specified in \eqref{eq:Pdiag}.

Notice that \eqref{eq:HtoU_fixed} can be shown using the similar arguments for the proof of \eqref{eq:HtoU} in Section \ref{subsec:HtoU_diag} for fixed $p$ and $q$. It then suffices to show \eqref{eq:CLT_mixing}. 
To do so, we need to show that $\sqrt{n}\bu'\tilde{\bG}_{\rm diag}$ is asymptotically normal for any nonzero vector $\bu = (\bu_1',\bu_2')'$, where $\bu_1 \in \mathbb{R}^{2p}$ and $\bu_2\in \mathbb{R}^{2q}$. By \eqref{eq:XEam_reformulate} and \eqref{eq:XEbj_reformulate}, it holds that
\begin{align*}
    \sqrt{n}\bu'\tilde{\bG}_{\rm diag} = \sqrt{n}\bu_1'\tilde{\bG}_{{\rm diag},1} + \sqrt{n}\bu_2'\tilde{\bG}_{{\rm diag},2}\,.
\end{align*}

Now we calculate the variance of $\sqrt{n}\bu'\tilde{\bG}_{\rm diag}$. Notice that 
\begin{align*}
    {\rm Var}(\sqrt{n}\bu'\tilde{\bG}_{\rm diag}) & = {\rm Var}(\sqrt{n}\bu_1'\tilde{\bG}_{{\rm diag},1}) + {\rm Var}(\sqrt{n}\bu_2'\tilde{\bG}_{{\rm band},2}) + 2{\rm Cov}(\sqrt{n}\bu_1'\tilde{\bG}_{{\rm diag},1},\sqrt{n}\bu_2'\tilde{\bG}_{{\rm diag},2})\,.
\end{align*}
Next, we need to calculate each term separately and then sum them up. It holds that 
\begin{align}\label{eq:varianceSnp}
    {\rm Var}(\sqrt{n}\bu_1'\tilde{\bG}_{{\rm diag},1}) & = {\rm Var}(\frac{1}{\sqrt{n}}\sum_{t=2}^n \bu_1'\tilde{\bK}_{\ba}  \tilde{\boldsymbol{\mathcal{E}}}_{\bX,t}) \\
    & = \frac{n-1}{n}\bu_1'\tilde{\bK}_{\ba} \bSigma_{\bX,\bE}(0) \tilde{\bK}_{\ba}'\bu_1 + \sum_{k=1}^{n-2}\Big(1-\frac{k+1}{n}\Big)\bu_1'\tilde{\bK}_{\ba} \big\{ \bSigma_{\bX,\bE}(k) + \bSigma_{\bX,\bE}(k)'\big\}\tilde{\bK}_{\ba}'\bu_1\,, \notag
\end{align}
where $\bSigma_{\bX,\bE}(k) = {\rm Cov}(\tilde{\boldsymbol{\mathcal{E}}}_{\bX,t},\tilde{\boldsymbol{\mathcal{E}}}_{\bX,t-k})$. 
% Since 
% \begin{align}\label{eq:EXt}
%     \tilde{\boldsymbol{\mathcal{E}}}_{\bX,t} = \{\bI_q \otimes {\rm vec}(\bX_{t-1}) \otimes \bI_p\}{\rm vec}(\bE_{t})\,,
% \end{align}
% by Cauchy-Schwarz inequality and Conditions \ref{cond:moments} and \ref{cond:mmt_diagonalcase}, we have 
% \begin{align*}
%     \mathbb{E}|\be_l'\tilde{\bK}_{\ba}\tilde{\boldsymbol{\mathcal{E}}}_{\bX,t}|^{\frac{4+\kappa}{2}} \le \big[\mathbb{E}|\be_{l}'\tilde{\bK}_{\ba} \{\bI_q \otimes {\rm vec}(\bX_{t-1})\otimes \bI_q\}|_2^{4+\kappa}\big]^{1/2}\big\{\mathbb{E}|{\rm vec}(\bE_{t})|_2^{4+\kappa}\big\}^{1/2} < \infty\,.
% \end{align*}
% Here, $\be_l$ is the $2p$-dimensional unit vector with the $l$-th element being 1. 
Then, by Conditions \ref{cond:mixing}(i) and Davydov's inequality (Proposition 2.5 of \cite{FanYao2003}), it holds that
\begin{align}\label{eq:longruncov}
    & \sum_{k=1}^{\infty}|\bu_1'\tilde{\bK}_{\ba} \big\{ \bSigma_{\bX,\bE}(k) + \bSigma_{\bX,\bE}(k)'\big\}\tilde{\bK}_{\ba}'\bu_1| \notag\\
    &~~~~~\le C_6  \sum_{k=1}^{\infty} \alpha(k)^{\frac{\kappa}{4+\kappa}} \big\{\mathbb{E}|\bu_{1}'\tilde{\bK}_{\ba} \tilde{\boldsymbol{\mathcal{E}}}_{\bX,t}|^{\frac{4+\kappa}{2}}\big\}^{\frac{2}{4+\kappa}}\big\{\mathbb{E}|\bu_{2}'\tilde{\bK}_{\ba} \tilde{\boldsymbol{\mathcal{E}}}_{\bX,t-k}|^{\frac{4+\kappa}{2}}\big\}^{\frac{2}{4+\kappa}} \\
    &~~~~~\le C_7  \sum_{k=1}^{\infty}\alpha(k)^{\frac{\kappa}{4+\kappa}} \sup_{\bu} \big\{\mathbb{E}|\bu'\tilde{\bK}_{\ba} \tilde{\boldsymbol{\mathcal{E}}}_{\bX,t}|^{\frac{4+\kappa}{2}}\big\}^{\frac{2}{4+\kappa}} \sup_{\bu} \big\{\mathbb{E}|\bu'\tilde{\bK}_{\ba} \tilde{\boldsymbol{\mathcal{E}}}_{\bX,t-k}|^{\frac{4+\kappa}{2}}\big\}^{\frac{2}{4+\kappa}} < \infty\,.\notag
\end{align}

Using the similar arguments for deriving \eqref{eq:varianceSnp}, we also have
\begin{align*}
    & {\rm Cov}(\sqrt{n}\bu_1'\tilde{\bG}_{{\rm diag},1},\sqrt{n}\bu_2'\tilde{\bG}_{{\rm diag},2})  \\
    &~~~~~~~~~ = \frac{n-1}{n}\bu_1'\tilde{\bK}_{\ba} \bSigma_{\bX,\bE}(0) \tilde{\bK}_{\bb}'\bu_2 + \sum_{k=1}^{n-2}\Big(1-\frac{k+1}{n}\Big)\bu_1'\tilde{\bK}_{\ba} \big\{ \bSigma_{\bX,\bE}(k) + \bSigma_{\bX,\bE}(k)'\big\}\tilde{\bK}_{\bb}'\bu_2\,,\\
    & {\rm Var}(\sqrt{n}\bu_2'\tilde{\bG}_{{\rm band},2}) \\
    &~~~~~~~~~ = \frac{n-1}{n}\bu_2'\tilde{\bK}_{\bb} \bSigma_{\bX,\bE}(0) \tilde{\bK}_{\bb}'\bu_2 + \sum_{k=1}^{n-2}\Big(1-\frac{k+1}{n}\Big)\bu_2'\tilde{\bK}_{\bb} \big\{ \bSigma_{\bX,\bE}(k) + \bSigma_{\bX,\bE}(k)'\big\}\tilde{\bK}_{\bb}'\bu_2\,.
\end{align*}
Analogous to \eqref{eq:longruncov}, it also holds that 
\begin{align*}
    \sum_{k=1}^{\infty}|\bu_1'\tilde{\bK}_{\ba} \big\{ \bSigma_{\bX,\bE}(k) + \bSigma_{\bX,\bE}(k)'\big\}\tilde{\bK}_{\bb}'\bu_2| < \infty 
\end{align*}
and
\begin{align*}
 \sum_{k=1}^{\infty}|\bu_2'\tilde{\bK}_{\bb} \big\{ \bSigma_{\bX,\bE}(k) + \bSigma_{\bX,\bE}(k)'\big\}\tilde{\bK}_{\bb}'\bu_2| < \infty\,.    
\end{align*}
%uniformly in $p$ and $q$. 
It then follows from the dominated convergence theorem that 
\begin{align*}
    {\rm Var}\bigg( \frac{\sqrt{n}\bu'\tilde{\bG}_{\rm diag}}{\sqrt{\bu'\tilde{\bP}_{\rm diag}\bu}} \bigg) \to 1\,,
\end{align*}
where
\begin{align*}
    \tilde{\bP}_{\rm diag} = \begin{pmatrix}
       \tilde{\bK}_{\ba}\tilde{\bSigma}_{\bX,\bE}\tilde{\bK}_{\ba}'\,, & \tilde{\bK}_{\ba}\tilde{\bSigma}_{\bX,\bE}\tilde{\bK}_{\bb}'\\
       \tilde{\bK}_{\bb}\tilde{\bSigma}_{\bX,\bE}\tilde{\bK}_{\ba}'\,, & \tilde{\bK}_{\bb}\tilde{\bSigma}_{\bX,\bE}\tilde{\bK}_{\bb}'
    \end{pmatrix}
\end{align*}
with $\tilde{\bSigma}_{\bX,\bE} = \bSigma_{\bX,\bE}(0) + \sum_{k=1}^{\infty}\big\{ \bSigma_{\bX,\bE}(k) + \bSigma_{\bX,\bE}(k)'\big\}$.

Then, the asymptotic normality of $\sqrt{n}\bu'\tilde{\bG}_{\rm diag}$ using the similar arguments as those in \cite{Gao2019} by employing the small-block and large-block technique. See also \cite{Dou2016}. We omit the details here. It holds that 
\begin{align*}
    \sqrt{n} \tilde{\bP}_{\rm diag}^{-1/2} \bU_{\rm diag}(\hat{\bgamma}-\bgamma) \stackrel{\mathrm{d}}{\to} \mathcal{N}(0,\bI_{2(p+q)})\,.
\end{align*}
%if $d = o(n^{\nu})$ with $\nu = \max\{1/9, 2/(16+13\kappa)\}$.
This completes the proof of Proposition \ref{CLT:diag_fixdim}. $\hfill\Box$

\section{Proof of Theorem \ref{tm:bandwidth}}

We first show $\mathbb{P}(\hat{k}_0^{(\bB)}=k_0^{(\bB)}) \to 1$. It suffices to show that 
\begin{align}\label{eq:kB_1}
    \mathbb{P}(\hat{k}_0^{(\bB)} > k_0^{(\bB)}) \to 0 ~\mbox{and}~ \mathbb{P}(\hat{k}_0^{(\bB)} < k_0^{(\bB)}) \to 0\,,
\end{align}
respectively. For each $i \in [pq]$, let 
\begin{align*}
    (\hat{k}_{i}^{(\bA)}, \hat{k}_{i}^{(\bB)}) = \arg\max_{(k^{(\bA)},k^{(\bB)}) \in [K]} \frac{\Delta\text{RSS}_i(k^{(\bA)}, k^{(\bB)} - 1) + \omega_n}{\Delta\text{RSS}_i(k^{(\bA)}, k^{(\bB)}) + \omega_n}\,,
\end{align*}
where $\Delta\text{RSS}_i(k^{(\bA)}, k^{(\bB)})$ is specified in \eqref{eq:Delta_RSS} in Section \ref{sc:bandwidth}. 
% For each $i \in [pq]$ and $k_0^{(\bA)}$, let 
% \begin{align*}
%     \hat{k}_{i}^{(\bB)}(k^{(\bA)}) = \arg\max_{k \in [K]} \frac{\Delta\text{RSS}_i(k^{(\bA)}, k - 1) + \omega_n}{\Delta\text{RSS}_i(k^{(\bA)}, k) + \omega_n}\,,
% \end{align*}
% where $\Delta\text{RSS}_i(k^{(\bA)}, k^{(\bB)})$ is specified in \eqref{eq:Delta_RSS}. 
Analogous to the proof of Theorem 1 in \cite{Gao2019}, we first investigate the convergence rate of $\Delta\text{RSS}_i(k^{(\bA)}, k^{(\bB)})$. For $i \in [pq]$, recall that the estimates for $\bc_i^{(0)}$ and $\bc_i^{(1)}$, which are the $i$-th row of $\bC_0$ and $\bC_1$, respectively, are obtained by solving the following minimization problem
\begin{align}\label{eq:min_c01}
    \min_{\bc^{(0)}_i,\bc^{(1)}_i} \| \hat \bSigma_{\bX}(1)' \be_i - \hat \bSigma_{\bX}(1)' \bc^{(0)}_i -  \hat \bSigma_{\bX}(0) \bc^{(1)}_i\|_2^2\,,
\end{align}
where 
\[
\hat \bSigma_{\bX}(1) = \frac{1}{n} \sum_{t=2}^n \text{vec}(\bX_t) \text{vec}(\bX_{t-1})', \; \; \hat \bSigma_{\bX}(0) = \frac{1}{n} \sum_{t=2}^n \text{vec}(\bX_{t-1}) \text{vec}(\bX_{t-1})'\,.
\]
We omit the term $\text{vec}(\bX_{n}) \text{vec}(\bX_{n})'$ in the definition of $\hat \bSigma_{\bX}(0)$ for technical convenience. For notation simplicity, write $\bk = (k^{(\bA)}, k^{(\bB)})$ and $\bk_0 = (k_0^{(\bA)}, k_0^{(\bB)})$. Given $\bk$, let $\bbeta_{i,\bk} = (\bc^{(0)'}_{i,\bk},\bc^{(1)'}_{i,\bk})' \in \mathbb{R}^{\tau_{i,\bk}}$, where $\bc_{i,\bk}^{(0)}$ and $\bc_{i,\bk}^{(1)}$ denote the vectors consisting of the non-zero elements of $\bc_{i}^{(0)}$ and $\bc_{i}^{(1)}$ determined by $\bk=(k^{(\bA)},k^{(\bB)})$, respectively, and $\tau_{i,\bk}$ is the total number of non-zero elements in $\bc_{i}^{(0)}$ and $\bc_{i}^{(1)}$. Let $\hat{\bV}_{i,\bk} = (\hat{\bSigma}_{\bX,\bk}(1)', \hat{\bSigma}_{\bX,\bk}(0)) \in \mathbb{R}^{(pq) \times \tau_{i,\bk}}$, with $\hat{\bSigma}_{\bX,\bk}(1)'$ and $\hat{\bSigma}_{\bX,\bk}(0)$ being matrices consisting of the columns of $\hat{\bSigma}_{\bX}(1)'$ and $\hat{\bSigma}_{\bX}(0)$ corresponding to the non-zero elements of $\bc_{i}^{(0)}$ and $\bc_{i}^{(1)}$, respectively.
%let $\tau_{i,\bk}$ be the total number of parameters to be estimated when solving \eqref{eq:min_c01}. 
%Let $\bbeta_{i,\bk}$ denote the $\tau_{i,\bk} \times 1$ vector consisting of the nonzero elements in $\bc^{(0)}_i$ and $\bc^{(1)}_i$ when solving \eqref{eq:min_c01}, and $\hat{\bV}_{i,\bk}$ be the $(pq) \times \tau_{i,\bk}$ matrix consisting of the corresponding columns of $\hat \bSigma_{\bX}(1)'$ and $\hat \bSigma_{\bX}(0)$. 
Write $\hat{\bz}_i = \hat \bSigma_{\bX}(1)' \be_i$. Hence, \eqref{eq:min_c01} is equivalent to 
\begin{align}\label{eq:min_equiv}
    \min_{\bbeta_{i,\bk}} \| \hat{\bz}_i - \hat{\bV}_{i,\bk}\bbeta_{i,\bk}\|_2^2\,,
\end{align}
which leads to the least squares estimator
\begin{align*}
    \hat{\bbeta}_{i,\bk} = (\hat{\bV}_{i,\bk}'\hat{\bV}_{i,\bk})^{-1}\hat{\bV}_{i,\bk}'\hat{\bz}_i\,,
\end{align*}
and the residual sum of squares
\begin{align*}
    \text{RSS}_i(k^{(\bA)}, k^{(\bB)}) =  \| \hat{\bz}_i - \hat{\bV}_{i,\bk}\hat{\bbeta}_{i,\bk}\|_2^2\,.
\end{align*}
Let $\bbeta_{i,\bk_0} = (\bc_{i,\bk_0}^{(0)'}, \bc_{i,\bk_0}^{(1)'})' \in \mathbb{R}^{\tau_{i,\bk_0}}$ and $\hat{\bV}_{i,\bk_0} = (\hat{\bSigma}_{\bX,\bk_0}(1)', \hat{\bSigma}_{\bX,\bk_0}(0))$, where $\bc_{i,\bk_0}^{(0)}$ and $\bc_{i,\bk_0}^{(1)}$ denote the vectors consisting of the non-zero elements in $\bc^{(0)}_i$ and $\bc^{(1)}_i$ determined by the true bandwidth $\bk_0$, $\hat{\bSigma}_{\bX,\bk_0}(1)'$ and $\hat{\bSigma}_{\bX,\bk_0}(0)$ denote matrices consisting of the corresponding columns of $\hat{\bSigma}_{\bX}(1)'$ and $\hat{\bSigma}_{\bX}(0)$, respectively. By \eqref{model_vec_phi}, we have
%When $\bk = \bk_0$, i.e. $k^{(\bA)} = k_0^{(\bA)}$ and $k^{(\bB)} = k_0^{(\bB)}$, by \eqref{model_vec_phi}, we have
\begin{align}\label{eq:true_zi}
    \hat{\bz}_i = \hat{\bV}_{i,\bk_0}\bbeta_{i,\bk_0} + \frac{1}{n}\sum_{t=2}^n \text{vec}(\bX_{t-1})\varepsilon_{i,t}\,,
\end{align}
where $\varepsilon_{i,t}$ is the $i$-th element of $\text{vec}(\bE_{t})$. 
By \eqref{eq:moment_bounded}, we have $\mathbb{E}(|X_{t,i,j}|^{4+\kappa}) < \infty$.
%Under Condition \ref{cond:mixing}(ii), for $\kappa \in (0,2\sqrt{2})$, it holds that 
%\begin{align*}
%    \mathbb{E}(|X_{t,i,j}|^{4+\kappa}) & = (4+\kappa)\int_{0}^{+\infty} x^{3+\kappa}\mathbb{P}(|X_{t,i,j}|>x) \,{\mathrm d}x \lesssim \int_{0}^{+\infty} x^{3+\kappa-\frac{8+4\kappa}{\kappa}} \,{\mathrm d}x < \infty
%\end{align*} 
%for any $i \in [p]$ and $j \in [q]$. 
Then we have
\begin{align*}
    \mathbb{E}|\be_j'\text{vec}(\bX_{t-1})\varepsilon_{i,t}|^{(4+\kappa)/2} \le \{\mathbb{E}|\be_j'\text{vec}(\bX_{t-1})|^{4+\kappa}\}^{1/2}\{\mathbb{E}|\varepsilon_{i,t}|^{4+\kappa}\}^{1/2} < \infty\,,
\end{align*}
where $\be_j$ is the $(pq)$-dimensional unit vector with the $j$-th element being 1.
Hence, by Davydov inequality (Proposition 2.5 of \cite{FanYao2003}), it holds that
\begin{align*}
    & \mathbb{E}\bigg\{\frac{1}{n}\sum_{t=2}^{n}\be_j'\text{vec}(\bX_{t-1})\varepsilon_{i,t} \bigg\}^2  \\
    &~~~~ = \frac{1}{n^2}\sum_{t=2}^{n}{\rm Var}\{\be_j'\text{vec}(\bX_{t-1})\varepsilon_{i,t}\}  + \frac{1}{n^2}\sum_{t\neq s}{\rm Cov}(\be_j'\text{vec}(\bX_{t-1})\varepsilon_{i,t}, \be_j'\text{vec}(\bX_{s-1})\varepsilon_{i,s}) \\
    &~~~~\le \frac{C_1}{n} + \frac{8}{n^2}\sum_{t\neq s}  \alpha(|t-s|)^{\frac{\kappa}{4+\kappa}} \big[\mathbb{E}|\be_j'\text{vec}(\bX_{t-1})\varepsilon_{i,t}|^{\frac{4+\kappa}{2}}\big]^{\frac{2}{4+\kappa}}\big[\mathbb{E}|\be_j'\text{vec}(\bX_{s-1})\varepsilon_{i,s}|^{\frac{4+\kappa}{2}}\big]^{\frac{2}{4+\kappa}} \\
    &~~~~ \le \frac{C_1}{n} + \frac{C_2}{n}\sum_{j=1}^{n}  \alpha(j)^{\frac{\kappa}{4+\kappa}} = O\Big(\frac{1}{n}\Big)\,.
\end{align*} 
% For any $j \in [pq]$, by Conditions \ref{cond:mixing}(ii), \eqref{cond:moments} and Davydov inequality (Proposition 2.5 of \cite{FanYao2003}), it holds that
% \begin{align*}
%     & \mathbb{E}\bigg\{\frac{1}{n}\sum_{t=2}^{n}\be_j'\text{vec}(\bX_{t-1})\varepsilon_{i,t} \bigg\}^2  \\
%     &~~~~ = \frac{1}{n^2}\sum_{t=2}^{n}{\rm Var}\{\be_j'\text{vec}(\bX_{t-1})\varepsilon_{i,t}\}  + \frac{1}{n^2}\sum_{t\neq s}{\rm Cov}(\be_j'\text{vec}(\bX_{t-1})\varepsilon_{i,t}, \be_j'\text{vec}(\bX_{s-1})\varepsilon_{i,s}) \\
%     &~~~~\le \frac{C_1}{n} + \frac{8}{n^2}\sum_{t\neq s}  \alpha(|t-s|)^{\frac{\kappa}{4+\kappa}} \big[\mathbb{E}|\be_j'\text{vec}(\bX_{t-1})\varepsilon_{i,t}|^{\frac{4+\kappa}{2}}\big]^{\frac{2}{4+\kappa}}\big[\mathbb{E}|\be_j'\text{vec}(\bX_{s-1})\varepsilon_{i,s}|^{\frac{4+\kappa}{2}}\big]^{\frac{2}{4+\kappa}} \\
%     &~~~~ \le \frac{C_1}{n} + \frac{C_2}{n^2}\sum_{t\neq s}  \alpha(|t-s|)^{\frac{\kappa}{4+\kappa}} \le \frac{C_1}{n} + \frac{C_3}{n}\sum_{j=1}^{n}  \alpha(j)^{\frac{\kappa}{4+\kappa}} = O\Big(\frac{1}{n}\Big)\,.
% \end{align*}
Then, we have
\begin{align}\label{eq:yt-1et}
    \Big\| \frac{1}{n}\sum_{t=2}^n \text{vec}(\bX_{t-1})\varepsilon_{i,t} \Big\|_2 = O_{\rm p}\Big(\sqrt{\frac{pq}{n}}\Big)\,.
\end{align}
Hence, it holds that 
\begin{align}\label{eq:RSSi_k0}
    \text{RSS}_i(k_0^{(\bA)},k_0^{(\bB)}) & =  \| \hat{\bz}_i - \hat{\bV}_{i,\bk_0}'\hat{\bbeta}_{i,\bk_0}\|_2^2 = \Big\|(\bI_{pq}-\bH_{i,\bk_0}) \cdot \frac{1}{n}\sum_{t=2}^n \text{vec}(\bX_{t-1})\varepsilon_{i,t}\Big\|_2^2 \notag\\
    & \le \|\bI_{pq}-\bH_{i,\bk_0}\|_2^2 \cdot \Big\|\frac{1}{n}\sum_{t=2}^n \text{vec}(\bX_{t-1})\varepsilon_{i,t} \Big\|_2^2 \notag\\
    & \le O_{\rm p}\Big(\frac{pq}{n}\Big)\,,
\end{align}
where $\bH_{i,\bk_0} = \hat{\bV}_{i,\bk_0}(\hat{\bV}_{i,\bk_0}'\hat{\bV}_{i,\bk_0})^{-1}\hat{\bV}_{i,\bk_0}'$ and then $\|\bI_{pq}-\bH_{i,\bk_0}\|_2^2 \le 1$. 

For $\bk = (k^{(\bA)}, k^{(\bB)}) \ne \bk_0$ with $k^{(\bA)} \ge k_0^{(\bA)}$ and $k^{(\bB)} \ge k_0^{(\bB)}$, write 
\begin{align*}
    \hat{\bV}_{i,\bk} = (\bS_{i,\bk}^{(1)}, \hat{\bSigma}_{\bX,\bk_0}(1)',\bS_{i,\bk}^{(2)},\bS_{i,\bk}^{(3)},\hat{\bSigma}_{\bX,\bk_0}(0),\bS_{i,\bk}^{(4)}) ~\mbox{and}~ \bbeta_{i,\bk} = (\bu_{i,\bk}^{(1)'},\bc_{i,\bk_0}^{(0)'},\bu_{i,\bk}^{(2)'},\bv_{i,\bk}^{(1)'},\bc_{i,\bk_0}^{(1)'},\bv_{i,\bk}^{(2)'})'\,.
\end{align*}
%where $\bc_{i,\bk_0}^{(0)}$ and $\bc_{i,\bk_0}^{(1)}$ denote the vectors consisting of the nonzero elements of $\bc_{i}^{(0)}$ and $\bc_{i}^{(1)}$ determined by the bandwidth $\bk_0 = (k_0^{(\bA)}, k_0^{(\bB)})$, respectively. 
If $k^{(\bA)} = k_0^{(\bA)}$ or $k^{(\bB)} = k_0^{(\bB)}$, the corresponding columns of $\bS_{i,\bk}^{(1)}$, $\bS_{i,\bk}^{(1)}$, $\bS_{i,\bk}^{(3)}$, $\bS_{i,\bk}^{(4)}$ and the corresponding elements of $\bu_{i,\bk}^{(1)}$, $\bu_{i,\bk}^{(2)}$, $\bu_{i,\bk}^{(3)}$, $\bu_{i,\bk}^{(4)}$ are omitted accordingly. %Similarly, if $k^{(\bB)} = k_0^{(\bB)}$, the corresponding $\bS_{i,\bk}^{(3)}$, $\bS_{i,\bk}^{(4)}$, $\bv_{i,\bk}^{(1)}$ and $\bv_{i,\bk}^{(2)}$ are omitted accordingly.
%In particular, $\hat{\bV}_{i,\bk} = (\hat{\bSigma}_{\bX,\bk_0}(1)',\bS_{i,\bk}^{(3)},\hat{\bSigma}_{\bX,\bk_0}(0),\bS_{i,\bk}^{(4)})$ and $\bbeta_{i,\bk} = (\bc_{i,\bk_0}^{(0),'},\bv_{i,\bk}^{(1),'},\bc_{i,\bk_0}^{(1),'},\bv_{i,\bk}^{(2),'})'$ when $k^{(\bA)} = k_0^{(\bA)}$, $\hat{\bV}_{i,\bk} = (\bS_{i,\bk}^{(1)}, \hat{\bSigma}_{\bX,\bk_0}(1)',\bS_{i,\bk}^{(2)},\hat{\bSigma}_{\bX,\bk_0}(0))$ and $\bbeta_{i,\bk} = (\bu_{i,\bk}^{(1),'},\bc_{i,\bk_0}^{(0),'},\bu_{i,\bk}^{(1),'},\bc_{i,\bk_0}^{(1),'})'$ when $k^{(\bB)} = k_0^{(\bB)}$.  
Then, we have
\begin{align}\label{eq:RSSi_ks1}
    \text{RSS}_i(k^{(\bA)}, k^{(\bB)})  =  \min_{\bw_1,\bw_2}\| \hat{\bz}_i - \hat{\bV}_{i,\bk_0}\bw_1 - \bS_{i,\bk}\bw_2\|_2^2\,,
\end{align}
where $\hat{\bV}_{i,\bk_0} = (\hat{\bSigma}_{\bX,\bk_0}(1)',\hat{\bSigma}_{\bX,\bk_0}(0))$ and $\bS_{i,\bk} = (\bS_{i,\bk}^{(1)}, \bS_{i,\bk}^{(2)},\bS_{i,\bk}^{(3)}, \bS_{i,\bk}^{(4)})$. Similar to the derivation in the proof of Theorem 1 in \cite{Gao2019}, it holds that 
\begin{align}\label{eq:RSSi_ks1a}
    \text{RSS}_i(k^{(\bA)}, k^{(\bB)}) & =  \|(\bI_{pq}-\bH_{i,\bk_0})\hat{\bz}_i\|_2^2 - \|\tilde{\bS}_{i,\bk}\hat{\bw}_{2,\bk}\|_2^2 \\
    & = \text{RSS}_i(k_0^{(\bA)},k_0^{(\bB)}) - \|\tilde{\bS}_{i,\bk}\hat{\bw}_{2,\bk}\|_2^2\,, \notag
\end{align}
where $\tilde{\bS}_{i,\bk} = (\bI_{pq}-\bH_{i,\bk_0})\bS_{i,\bk}$ and $\hat{\bw}_{2,\bk} = (\tilde{\bS}_{i,\bk}'\tilde{\bS}_{i,\bk})^{-1}\tilde{\bS}_{i,\bk}'\hat{\bz}_i$. Similarly, for $\tilde{\bk}= (k^{(\bA)}+1,k^{(\bB)})$, we have 
\begin{align*}
    \text{RSS}_i(k^{(\bA)}+1, k^{(\bB)}) & =  \text{RSS}_i(k_0^{(\bA)},k_0^{(\bB)}) - \|\tilde{\bS}_{i,\tilde{\bk}}\hat{\bw}_{2,\tilde{\bk}}\|_2^2\,,
\end{align*}
where $\tilde{\bS}_{i,\tilde{\bk}} = (\bI_{pq}-\bH_{i,\bk_0})\bS_{i,\tilde{\bk}}$, $\bS_{i,\tilde{\bk}}$ is defined in the same manner as in \eqref{eq:RSSi_ks1} but with $\bk$ replaced by $\tilde{\bk}$, and $\hat{\bw}_{2,\tilde{\bk}} = (\tilde{\bS}_{i,\tilde{\bk}}'\tilde{\bS}_{i,\tilde{\bk}})^{-1}\tilde{\bS}_{i,\tilde{\bk}}'\hat{\bz}_i$. Hence, it follows that
\begin{align*}
    & \text{RSS}_i(k^{(\bA)}, k^{(\bB)}) - \text{RSS}_i(k^{(\bA)}+1, k^{(\bB)}) \\
    &~~~ = \|\tilde{\bS}_{i,\tilde{\bk}}\hat{\bw}_{2,\tilde{\bk}}\|_2^2 - \|\tilde{\bS}_{i,\bk}\hat{\bw}_{2,\bk}\|_2^2 \\
    &~~~ \le \|\tilde{\bS}_{i,\bk}(\tilde{\bS}_{i,\bk}'\tilde{\bS}_{i,\bk})^{-1}\tilde{\bS}_{i,\bk}'\|_2^2 \cdot \Big\|\frac{1}{n}\sum_{t=2}^n \text{vec}(\bX_{t-1})\varepsilon_{i,t} \Big\|_2^2 + \|\tilde{\bS}_{i,\tilde{\bk}}(\tilde{\bS}_{i,\tilde{\bk}}'\tilde{\bS}_{i,\tilde{\bk}})^{-1}\tilde{\bS}_{i,\tilde{\bk}}'\|_2^2 \cdot \Big\|\frac{1}{n}\sum_{t=2}^n \text{vec}(\bX_{t-1})\varepsilon_{i,t} \Big\|_2^2 \\
    &~~~ = O_{\rm p}\Big(\frac{pq}{n}\Big)\,,
\end{align*}
where the second equality holds since $\|\tilde{\bS}_{i,\bk}(\tilde{\bS}_{i,\bk}'\tilde{\bS}_{i,\bk})^{-1}\tilde{\bS}_{i,\bk}'\|_2^2 \le 1$ and $\|\tilde{\bS}_{i,\tilde{\bk}}(\tilde{\bS}_{i,\tilde{\bk}}'\tilde{\bS}_{i,\tilde{\bk}})^{-1}\tilde{\bS}_{i,\tilde{\bk}}'\|_2^2 \le 1$. In the same way, we can show that $\text{RSS}_i(k^{(\bA)}, k^{(\bB)}) - \text{RSS}_i(k^{(\bA)}, k^{(\bB)}+1) = O_{\rm p}(pq/n)$. Hence, when $\bk = (k^{(\bA)}, k^{(\bB)}) \ne \bk_0$ with $k^{(\bA)} \ge k_0^{(\bA)}$ and $k^{(\bB)} \ge k_0^{(\bB)}$, it holds that 
\begin{align}\label{eq:DeltaRssi_klarge}
    \Delta\text{RSS}_i(k^{(\bA)}, k^{(\bB)}) = O_{\rm p}\Big(\frac{pq}{n}\Big)\,.
\end{align}

% By \eqref{eq:true_zi} and \eqref{eq:RSSi_k0}, we have
% \begin{align*}
%     \text{RSS}_i(\bk) & =  \text{RSS}_i(\bk_0) - |\tilde{\bS}_{i,\bk}\hat{\bw}_2|_2^2 \\
%     & \le \text{RSS}_i(\bk_0) + \|\tilde{\bS}_{i,\bk}(\tilde{\bS}_{i,\bk}'\tilde{\bS}_{i,\bk})^{-1}\tilde{\bS}_{i,\bk}'\|_2^2 \Big|\frac{1}{n}\sum_{t=2}^n \text{vec}(\bX_{t-1})\varepsilon_{i,t} \Big|_2^2 \\
%     & = O_{\rm p}\Big(\frac{pq}{n}\Big)\,,
% \end{align*}
% where the second equality holds since $\|\tilde{\bS}_{i,\bk}(\tilde{\bS}_{i,\bk}'\tilde{\bS}_{i,\bk})^{-1}\tilde{\bS}_{i,\bk}'\|_2^2 \le 1$.

For $\bk = (k^{(\bA)}, k^{(\bB)})$ with $k^{(\bA)} < k_0^{(\bA)}$ and $k^{(\bB)} < k_0^{(\bB)}$, let $\hat{\bV}_{i,\bk} =(\hat{\bSigma}_{\bX,\bk}(1)', \hat{\bSigma}_{\bX,\bk}(0))$ and $\bbeta_{i,\bk} = (\bc_{i,\bk}^{(0)'},\bc_{i,\bk}^{(1)'})'$, where $\bc_{i,\bk}^{(0)}$ and $\bc_{i,\bk}^{(1)}$, respectively, denote the vectors consisting of the nonzero elements of $\bc_{i}^{(0)}$ and $\bc_{i}^{(1)}$ given $\bk=(k^{(\bA)},k^{(\bB)})$, $\hat{\bSigma}_{\bX,\bk}(1)'$ and $\hat{\bSigma}_{\bX,\bk}(0)$ are matrices consisting of the corresponding columns of $\hat{\bSigma}_{\bX}(1)'$ and $\hat{\bSigma}_{\bX}(0)$, respectively. Hence, it holds that
\begin{align}\label{eq:RSSi_ksmall}
    \text{RSS}_i(k^{(\bA)},k^{(\bB)}) & = \| \hat{\bz}_i - \hat{\bV}_{i,\bk}\hat{\bbeta}_{i,\bk}\|_2^2 = \| (\bI_{pq}-\bH_{i,\bk})\hat{\bz}_i\|_2^2\,.
\end{align}
Similar to \eqref{eq:RSSi_ks1}, we have
\begin{align*}
    \text{RSS}_i(k^{(\bA)}+1, k^{(\bB)})  =  \min_{\bw_1,\bw_2}\| \hat{\bz}_i - \hat{\bV}_{i,\bk}\bw_1 - \bR_{i,\tilde{\bk}}\bw_2\|_2^2\,,
\end{align*}
where $\bw_2$ denotes the additional parameters that need to be estimated when the bandwidth is set to $\tilde{\bk}= (k^{(\bA)}+1, k^{(\bB)})$ instead of $\bk= (k^{(\bA)}, k^{(\bB)})$, and $\bR_{i,\tilde{\bk}}$ is the matrix consisting of the corresponding columns of $\hat{\bSigma}_{\bX}(1)'$ and $\hat{\bSigma}_{\bX}(0)$. Using the similar arguments for deriving \eqref{eq:RSSi_ks1a}, we can also show that
\begin{align}\label{eq:RSSi_ksmall2}
    \text{RSS}_i(k^{(\bA)}+1, k^{(\bB)})  =  \|(\bI_{pq}-\bH_{i,\bk})\hat{\bz}_i\|_2^2 - \|\tilde{\bR}_{i,\tilde{\bk}}\hat{\bw}_{2}\|_2^2\,,
\end{align}
where $\bH_{i,\bk} = \hat{\bV}_{i,\bk}(\hat{\bV}_{i,\bk}'\hat{\bV}_{i,\bk})^{-1}\hat{\bV}_{i,\bk}'$, $\tilde{\bR}_{i,\tilde{\bk}} = (\bI_{pq}-\bH_{i,\bk})\bR_{i,\tilde{\bk}}$ and $\hat{\bw}_{2} = (\tilde{\bR}_{i,\tilde{\bk}}'\tilde{\bR}_{i,\tilde{\bk}})^{-1}\tilde{\bR}_{i,\tilde{\bk}}'\hat{\bz}_i$. By \eqref{eq:RSSi_ksmall} and \eqref{eq:RSSi_ksmall2}, we have
\begin{align}\label{eq:DeltaRSSi_ks}
    \text{RSS}_i(k^{(\bA)}, k^{(\bB)}) - \text{RSS}_i(k^{(\bA)}+1, k^{(\bB)}) = \|\tilde{\bR}_{i,\tilde{\bk}}\hat{\bw}_{2}\|_2^2 = \big\|\tilde{\bR}_{i,\tilde{\bk}}(\tilde{\bR}_{i,\tilde{\bk}}'\tilde{\bR}_{i,\tilde{\bk}})^{-1}\tilde{\bR}_{i,\tilde{\bk}}'\hat{\bz}_i\big\|_2^2\,.
\end{align}
% Since $k^{(\bA)} < k_0^{(\bA)}$ and $k^{(\bB)} < k_0^{(\bB)}$, we write
% \begin{align*}
%     \hat{\bV}_{i,\bk_0} = (\bR_{i,\bk}^{(1)}, \hat{\bSigma}_{\bX,\bk}(1)',\bR_{i,\bk}^{(2)},\bR_{i,\bk}^{(3)},\hat{\bSigma}_{\bX,\bk}(0),\bR_{i,\bk}^{(4)}) ~\mbox{and}~ \bbeta_{i,\bk_0} = (\tilde{\bu}_{i,\bk}^{(1),'},\bc_{i,\bk}^{(0),'},\tilde{\bu}_{i,\bk}^{(1),'},\tilde{\bv}_{i,\bk}^{(1),'},\bc_{i,\bk}^{(1),'},\tilde{\bv}_{i,\bk}^{(2),'})'\,,
% \end{align*}
% where $\bc_{i,\bk}^{(0)}$ and $\bc_{i,\bk}^{(1)}$, respectively, denote the vectors consisting of the nonzero elements of $\bc_{i}^{(0)}$ and $\bc_{i}^{(1)}$ given $\bk=(k^{(\bA)},k^{(\bB)})$, $\hat{\bSigma}_{\bX,\bk}(1)'$ and $\hat{\bSigma}_{\bX,\bk}(0)$ are matrices consisting of the corresponding columns of $\hat{\bSigma}_{\bX}(1)'$ and $\hat{\bSigma}_{\bX}(0)$, respectively. 
%Hence, we have $\hat{\bV}_{i,\bk} =(\hat{\bSigma}_{\bX,\bk}(1)', \hat{\bSigma}_{\bX,\bk}(0))$ and $\bbeta_{i,\bk} = (\bc_{i,\bk}^{(0),'},\bc_{i,\bk}^{(1),'})'$. Similarly, 
Since $k^{(\bA)} < k_0^{(\bA)}$ and $k^{(\bB)} < k_0^{(\bB)}$, by \eqref{eq:true_zi}, we have
\begin{align}\label{eq:Zi1}
    \hat{\bz}_i = \hat{\bV}_{i,\bk_0}\bbeta_{i,\bk_0} + \frac{1}{n}\sum_{t=2}^n \text{vec}(\bX_{t-1})\varepsilon_{i,t} = \hat{\bV}_{i,\bk}\bbeta_{i,\bk} + \bJ_{i,\bk}\tilde{\bw}_{i,\bk} + \frac{1}{n}\sum_{t=2}^n \text{vec}(\bX_{t-1})\varepsilon_{i,t}\,,
\end{align}
where $\tilde{\bw}_{i,\bk}$ is the vector containing the non-zero elements in $\bbeta_{i,\bk_0}$ that are not included in $\bbeta_{i,\bk}$, and $\bJ_{i,\bk}$ is the matrix consisting of the corresponding columns of $\hat{\bSigma}_{\bX}(1)'$ and $\hat{\bSigma}_{\bX}(0)$. Hence, it holds that
\begin{align}\label{eq:DeltaRSSi_ks2}
    & \big\|\tilde{\bR}_{i,\tilde{\bk}}(\tilde{\bR}_{i,\tilde{\bk}}'\tilde{\bR}_{i,\tilde{\bk}})^{-1}\tilde{\bR}_{i,\tilde{\bk}}'\hat{\bz}_i\big\|_2^2 \notag\\
    &~~~~~ = \Big\|\tilde{\bR}_{i,\tilde{\bk}}(\tilde{\bR}_{i,\tilde{\bk}}'\tilde{\bR}_{i,\tilde{\bk}})^{-1}\tilde{\bR}_{i,\tilde{\bk}}'\bJ_{i,\bk}\tilde{\bw}_{i,\bk} + \tilde{\bR}_{i,\tilde{\bk}}(\tilde{\bR}_{i,\tilde{\bk}}'\tilde{\bR}_{i,\tilde{\bk}})^{-1}\tilde{\bR}_{i,\tilde{\bk}}' \cdot \frac{1}{n}\sum_{t=2}^n \text{vec}(\bX_{t-1})\varepsilon_{i,t} \Big\|_2^2 \notag\\
    &~~~~~ \le  \big\|\tilde{\bR}_{i,\tilde{\bk}}(\tilde{\bR}_{i,\tilde{\bk}}'\tilde{\bR}_{i,\tilde{\bk}})^{-1}\tilde{\bR}_{i,\tilde{\bk}}'\bJ_{i,\bk}\tilde{\bw}_{i,\bk}\big\|_2^2 + \Big\|\tilde{\bR}_{i,\tilde{\bk}}(\tilde{\bR}_{i,\tilde{\bk}}'\tilde{\bR}_{i,\tilde{\bk}})^{-1}\tilde{\bR}_{i,\tilde{\bk}}' \cdot \frac{1}{n}\sum_{t=2}^n \text{vec}(\bX_{t-1})\varepsilon_{i,t} \Big\|_2^2 \notag\\
    &~~~~~~~~~~~ + 2\tilde{\bw}_{i,\bk}'\bJ_{i,\bk}'\tilde{\bR}_{i,\tilde{\bk}}(\tilde{\bR}_{i,\tilde{\bk}}'\tilde{\bR}_{i,\tilde{\bk}})^{-1}\tilde{\bR}_{i,\tilde{\bk}}' \cdot \frac{1}{n}\sum_{t=2}^n \text{vec}(\bX_{t-1})\varepsilon_{i,t} \notag \\
    &~~~~~ := {\mathrm L}_{i,1} + {\mathrm L}_{i,2} + {\mathrm L}_{i,3} \,.
\end{align}

% \begin{align}\label{eq:RSSi_ksmall}
%     \text{RSS}_i(k^{(\bA)},k^{(\bB)}) & = | \hat{\bz}_i - \hat{\bV}_{i,\bk}\hat{\bbeta}_{i,\bk}|_2^2 = | (\bI_{pq}-\bH_{i,\bk})\hat{\bz}_i|_2^2 \notag \\ 
%     & = \Big|(\bI_{pq}-\bH_{i,\bk})\bR_{i,\bk}\tilde{\bw}_{i,\bk} + (\bI_{pq}-\bH_{i,\bk})\frac{1}{n}\sum_{t=2}^n \text{vec}(\bX_{t-1})\varepsilon_{i,t} \Big|_2^2 \notag \\
%     & = |(\bI_{pq}-\bH_{i,\bk})\bR_{i,\bk}\tilde{\bw}_{i,\bk}|_2^2 + \Big|(\bI_{pq}-\bH_{i,\bk})\frac{1}{n}\sum_{t=2}^n \text{vec}(\bX_{t-1})\varepsilon_{i,t} \Big|_2^2  \notag \\
%     &~~~+ 2\tilde{\bw}_{i,\bk}'\bR_{i,\bk}'(\bI_{pq}-\bH_{i,\bk})\frac{1}{n}\sum_{t=2}^n \text{vec}(\bX_{t-1})\varepsilon_{i,t}\,,
% \end{align}
% where $\bH_{i,\bk} = \hat{\bV}_{i,\bk}(\hat{\bV}_{i,\bk}'\hat{\bV}_{i,\bk})^{-1}\hat{\bV}_{i,\bk}'$.

Analogous to the derivation of \eqref{eq:hatSjk_dev} in the proof of Lemma \ref{lem:Sigmajk} in Section \ref{lem:1}, we have
\begin{align*}
    \| \hat \bSigma_{\bX}(1) - \bSigma_{\bX}(1) \|_2 \le \| \hat \bSigma_{\bX}(1) - \bSigma_{\bX}(1) \|_F = O_{\rm p}\{ (pq)^{(2+3\kappa)/(2+\kappa)}n^{-2/(2+\kappa)}\}\,,
\end{align*}
where $\bSigma_{\bX}(1) = \mathbb{E}\{\text{vec}(\bX_t) \text{vec}(\bX_{t-1})'\}$. Recall $d = p \vee q$.  
Hence, we have $\| \hat \bSigma_{\bX}(1) - \bSigma_{\bX}(1) \|_2 = o_{\rm p}(1)$, provided that $d=o\{n^{1/(2+3\kappa)}\}$. We can also show that the same result holds for $\| \hat \bSigma_{\bX}(0) - \bSigma_{\bX}(0) \|_2$. For ${\mathrm L}_{i,1}$ in \eqref{eq:DeltaRSSi_ks2}, by Condition \ref{cond:proj_matx}, we have 
\begin{align*}
      \lambda_{\min}\{\bJ_{i,\bk}'(\tilde{\bR}_{i,\tilde{\bk}}(\tilde{\bR}_{i,\tilde{\bk}}'\tilde{\bR}_{i,\tilde{\bk}})^{-1}\tilde{\bR}_{i,\tilde{\bk}}')\bJ_{i,\bk}\} \ge \lambda_1 ~\mbox{and}~ \lambda_{\max}\{\bJ_{i,\bk}'\tilde{\bR}_{i,\tilde{\bk}}(\tilde{\bR}_{i,\tilde{\bk}}'\tilde{\bR}_{i,\tilde{\bk}})^{-1}\tilde{\bR}_{i,\tilde{\bk}}'\bJ_{i,\bk}\} \le \lambda_2\,.  
\end{align*}
% \begin{align*}
%     \lambda_{\min}\{\bR_{i,\bk}'(\bI_{pq}-\bH_{i,\bk})\bR_{i,\bk}\} \ge \lambda_1 ~\mbox{and}~ \lambda_{\max}\{\bR_{i,\bk}'(\bI_{pq}-\bH_{i,\bk})\bR_{i,\bk}\} \le \lambda_2\,.
% \end{align*}
Hence, by Condition \ref{cond:min}, we have
\begin{align*}
    \lambda_1 d_{i,\bk_0}^2 \le \big\|\tilde{\bR}_{i,\tilde{\bk}}(\tilde{\bR}_{i,\tilde{\bk}}'\tilde{\bR}_{i,\tilde{\bk}})^{-1}\tilde{\bR}_{i,\tilde{\bk}}'\bJ_{i,\bk}\tilde{\bw}_{i,\bk}\big\|_2^2 \le \lambda_2\|\bbeta_{i,\bk_0}\|_2^2\,,
\end{align*}
where for some $i_1\in [p]$ and $i_2 \in [q]$,
\begin{align*}
    d_{i,\bk_0}^2 & = \{a_{i_1,i_1-k_0^{(\bA)}}^{(0)}b_{i_2,i_2-k_0^{(\bB)}}^{(0)}\}^2 + \{a_{i_1,i_1+k_0^{(\bA)}}^{(0)}b_{i_2,i_2+k_0^{(\bB)}}^{(0)}\}^2 \\
    &~~~~~~~ + \{a_{i_1,i_1-k_0^{(\bA)}}^{(1)}b_{i_2,i_2-k_0^{(\bB)}}^{(1)}\}^2+\{a_{i_1,i_1+k_0^{(\bA)}}^{(1)}b_{i_2,i_2+k_0^{(\bB)}}^{(1)}\}^2 \\
    & \ge \frac{C_3C_n k_0^{(\bA)} k_0^{(\bB)} \log\{(pq) \vee n\}}{n}\,.
\end{align*}
For ${\mathrm L}_{i,2}$ in \eqref{eq:DeltaRSSi_ks2}, by \eqref{eq:yt-1et}, we have 
\begin{align*}
    \Big\|\tilde{\bR}_{i,\tilde{\bk}}(\tilde{\bR}_{i,\tilde{\bk}}'\tilde{\bR}_{i,\tilde{\bk}})^{-1}\tilde{\bR}_{i,\tilde{\bk}}' \cdot \frac{1}{n}\sum_{t=2}^n \text{vec}(\bX_{t-1})\varepsilon_{i,t} \Big\|_2^2 = O_{\rm p}\Big(\frac{pq}{n}\Big)\,.
\end{align*}
Furthermore, by the Cauchy-Schwarz inequality, the ${\mathrm L}_{i,3}$ in \eqref{eq:DeltaRSSi_ks2} is bounded by the sum of ${\mathrm L}_{i,1}$ and ${\mathrm L}_{i,2}$. Thus, we have
\begin{align}\label{eq:bound1}
    \frac{C_4C_n \lambda_1 k_0^{(\bA)} k_0^{(\bB)} \log\{(pq) \vee n\}}{n} + O_{\rm p}\Big(\frac{pq}{n}\Big) & \le \text{RSS}_i(k^{(\bA)},k^{(\bB)}) - \text{RSS}_i(k^{(\bA)}+1,k^{(\bB)}) \notag \\
    &\le C_5\lambda_2 k_0^{(\bA)} k_0^{(\bB)}\,.
\end{align}
We can show that the same bounds hold for $\text{RSS}_i(k^{(\bA)},k^{(\bB)}) - \text{RSS}_i(k^{(\bA)},k^{(\bB)}+1)$ using the similar arguments. Hence, for $k^{(\bA)} < k_0^{(\bA)}$ and $k^{(\bB)} < k_0^{(\bB)}$ it holds that
\begin{align*}%\label{eq:RSSi_ks2}
    \frac{C_6C_n \lambda_1 k_0^{(\bA)} k_0^{(\bB)} \log\{(pq) \vee n\}}{n}+ O_{\rm p}\Big(\frac{pq}{n}\Big) & \le \Delta\text{RSS}_i(k^{(\bA)},k^{(\bB)}) \le C_7\lambda_2 k_0^{(\bA)} k_0^{(\bB)}\,,
\end{align*}
provided that $d=o\{n^{1/(2+3\kappa)}\}$.

For $\bk = (k^{(\bA)},k^{(\bB)})$ with $k^{(\bA)} < k_0^{(\bA)}$ and $k^{(\bB)} \ge k_0^{(\bB)}$, \eqref{eq:DeltaRSSi_ks} still holds. It suffices to bound $\|\tilde{\bR}_{i,\tilde{\bk}}(\tilde{\bR}_{i,\tilde{\bk}}'\tilde{\bR}_{i,\tilde{\bk}})^{-1}\tilde{\bR}_{i,\tilde{\bk}}'\hat{\bz}_i\|_2^2$.
Notice that $\bC_0=\bB_0 \otimes \bA_0$ and $\bC_1 = \bB_1 \otimes \bA_1$. Recall that $\bbeta_{i,\bk_0}$ is the $\tau_{i,\bk_0} \times 1$ vector consisting of the nonzero elements in $\bc_i^{(0)}$ and $\bc_i^{(1)}$ determined by the true bandwidth parameter $k_0^{(\bA)}$ and $k_0^{(\bB)}$. Notice that the elements corresponding to $k^{(\bB)} > k_0^{(\bB)}$ in the true $\bc_i^{(0)}$ and $\bc_i^{(1)}$ are zeros. By \eqref{eq:true_zi}, we have
\begin{align}\label{eq:Zi2}
    \hat{\bz}_i = \hat{\bV}_{i,\bk_0}\bbeta_{i,\bk_0} + \frac{1}{n}\sum_{t=2}^n \text{vec}(\bX_{t-1})\varepsilon_{i,t} = \hat{\bV}_{i,\bk}\bbeta_{i,\bk} + \tilde{\bJ}_{i,\bk}\tilde{\bv}_{i,\bk} + \frac{1}{n}\sum_{t=2}^n \text{vec}(\bX_{t-1})\varepsilon_{i,t}\,,
\end{align}
where $\tilde{\bv}_{i,\bk}$ is the vector containing the nonzero elements in $\bbeta_{i,\bk_0}$ that are not included in $\bbeta_{i,\bk}$, and $\tilde{\bJ}_{i,\bk}$ is the matrix consisting of the corresponding columns of $\hat{\bSigma}_{\bX}(1)'$ and $\hat{\bSigma}_{\bX}(0)$.  Analogous to the derivation of \eqref{eq:DeltaRSSi_ks2}, we then have
\begin{align}\label{eq:DeltaRSSi_ks3}
    & \big\|\tilde{\bR}_{i,\tilde{\bk}}(\tilde{\bR}_{i,\tilde{\bk}}'\tilde{\bR}_{i,\tilde{\bk}})^{-1}\tilde{\bR}_{i,\tilde{\bk}}'\hat{\bz}_i\big\|_2^2 \notag\\
    &~~~~~ = \Big\|\tilde{\bR}_{i,\tilde{\bk}}(\tilde{\bR}_{i,\tilde{\bk}}'\tilde{\bR}_{i,\tilde{\bk}})^{-1}\tilde{\bR}_{i,\tilde{\bk}}'\tilde{\bJ}_{i,\bk}\tilde{\bv}_{i,\bk} + \tilde{\bR}_{i,\tilde{\bk}}(\tilde{\bR}_{i,\tilde{\bk}}'\tilde{\bR}_{i,\tilde{\bk}})^{-1}\tilde{\bR}_{i,\tilde{\bk}}' \, \frac{1}{n}\sum_{t=2}^n \text{vec}(\bX_{t-1})\varepsilon_{i,t} \Big\|_2^2 \notag\\
    &~~~~~ \le  \big\|\tilde{\bR}_{i,\tilde{\bk}}(\tilde{\bR}_{i,\tilde{\bk}}'\tilde{\bR}_{i,\tilde{\bk}})^{-1}\tilde{\bR}_{i,\tilde{\bk}}'\tilde{\bJ}_{i,\bk}\tilde{\bv}_{i,\bk}\big\|_2^2 + \Big\|\tilde{\bR}_{i,\tilde{\bk}}(\tilde{\bR}_{i,\tilde{\bk}}'\tilde{\bR}_{i,\tilde{\bk}})^{-1}\tilde{\bR}_{i,\tilde{\bk}}' \,\frac{1}{n}\sum_{t=2}^n \text{vec}(\bX_{t-1})\varepsilon_{i,t} \Big\|_2^2 \notag\\
    &~~~~~~~~~~~ + 2\tilde{\bv}_{i,\bk}'\tilde{\bJ}_{i,\bk}'\tilde{\bR}_{i,\tilde{\bk}}(\tilde{\bR}_{i,\tilde{\bk}}'\tilde{\bR}_{i,\tilde{\bk}})^{-1}\tilde{\bR}_{i,\tilde{\bk}}' \, \frac{1}{n}\sum_{t=2}^n \text{vec}(\bX_{t-1})\varepsilon_{i,t}\,.
\end{align}
Similar to the derivation of \eqref{eq:bound1}, by Conditions \ref{cond:proj_matx} and \ref{cond:min}, it holds that
\begin{align*}
    \frac{C_8C_n \lambda_1 k_0^{(\bA)} k_0^{(\bB)} \log\{(pq) \vee n\}}{n} + O_{\rm p}\Big(\frac{pq}{n}\Big) & \le \text{RSS}_i(k^{(\bA)},k^{(\bB)}) - \text{RSS}_i(k^{(\bA)}+1,k^{(\bB)}) \notag \\
    &\le C_9\lambda_2 k_0^{(\bA)} k_0^{(\bB)}\,,
\end{align*}
provided that $d=o\{n^{1/(2+3\kappa)}\}$.
We can show that the same bounds hold for $\text{RSS}_i(k^{(\bA)},k^{(\bB)}) - \text{RSS}_i(k^{(\bA)},k^{(\bB)}+1)$ using similar arguments. Note that the above derivation still holds when $k^{(\bA)} < k_0^{(\bA)}$ and $k^{(\bB)} = k_0^{(\bB)}$. Hence, for $k^{(\bA)} < k_0^{(\bA)}$ and $k^{(\bB)} \ge k_0^{(\bB)}$, it holds that
\begin{align}\label{eq:RSSi_ksmalllarge}
    \frac{C_{10}C_n \lambda_1 k_0^{(\bA)} k_0^{(\bB)} \log\{(pq) \vee n\}}{n} + O_{\rm p}\Big(\frac{pq}{n}\Big) & \le \Delta\text{RSS}_i(k^{(\bA)},k^{(\bB)}) \le C_{11}\lambda_2 k_0^{(\bA)} k_0^{(\bB)}\,,
\end{align}
provided that $d=o\{n^{1/(2+3\kappa)}\}$.

When $k^{(\bA)} \ge k_0^{(\bA)}$ and $k^{(\bB)} < k_0^{(\bB)}$, it can be shown that the same convergence rate of $\Delta\text{RSS}_i(k^{(\bA)},k^{(\bB)})$ holds as in \eqref{eq:RSSi_ksmalllarge} in the same manner.

Now we are ready to prove \eqref{eq:kB_1}. Notice that $\mathbb{P}(\hat{k}_0^{(\bB)} > k_0^{(\bB)}) \le \mathbb{P}\{\hat{k}_{i}^{(\bB)} > k_0^{(\bB)}\}$ for some $i \in [pq]$. When $k^{(\bA)} \ge k_0^{(\bA)}$, by \eqref{eq:DeltaRssi_klarge} and \eqref{eq:RSSi_ksmalllarge}, we have
\begin{align}\label{eq:ratio_k0B}
    \frac{\Delta\text{RSS}_i(k^{(\bA)}, k_0^{(\bB)} - 1) + \omega_n}{\Delta\text{RSS}_i(k^{(\bA)}, k_0^{(\bB)}) + \omega_n} \gtrsim \frac{C_n \lambda_1 k_0^{(\bA)} k_0^{(\bB)}\log\{(pq) \vee n\}/n }{O_{\rm p}(pq/n)} \to \infty\,,
\end{align}
and 
\begin{align}\label{eq:Op(1)order}
    \max_{k > k_0^{(\bB)}} \frac{\Delta\text{RSS}_i(k^{(\bA)}, k - 1) + \omega_n}{\Delta\text{RSS}_i(k^{(\bA)}, k) + \omega_n} \le \frac{{O_{\rm p}(pq/n)}+\omega_n}{\omega_n} = O_{\rm p}(1)\,.
\end{align}
When $k^{(\bA)} < k_0^{(\bA)}$, By \eqref{eq:RSSi_ksmalllarge}, it holds that 
\begin{align}\label{eq:order}
    \max_{k > k_0^{(\bB)}} \frac{\Delta\text{RSS}_i(k^{(\bA)}, k - 1) + \omega_n}{\Delta\text{RSS}_i(k^{(\bA)}, k) + \omega_n} \lesssim \frac{\lambda_2 k_0^{(\bA)} k_0^{(\bB)} +\omega_n}{C_n \lambda_1 k_0^{(\bA)} k_0^{(\bB)}\log\{(pq) \vee n\}/n + \omega_n }\,.
\end{align}
Comparing the above upper bound with the lower bound of \eqref{eq:ratio_k0B}, if $C_n^2/(npq) \to \infty$, we have
\begin{align}\label{eq:ratio_to_0}
    & \Big\{\frac{\omega_n + \lambda_2 k_0^{(\bA)} k_0^{(\bB)}}{C_n \lambda_1 k_0^{(\bA)} k_0^{(\bB)}\log\{(pq) \vee n\}/n + O_{\rm p}(pq/n) }\Big\} \Big/ \Big\{\frac{C_n \lambda_1 k_0^{(\bA)} k_0^{(\bB)}\log\{(pq) \vee n\}/n }{O_{\rm p}(pq/n)}\Big\}  \notag\\
    & ~~~~~~~~ \lesssim \frac{ O_{\rm p}(p^2q^2) + O_{\rm p}\{npqk_0^{(\bA)} k_0^{(\bB)}\} }{[C_n \lambda_1 k_0^{(\bA)} k_0^{(\bB)}\log\{(pq) \vee n\}]^2 + C_n pq \lambda_1 k_0^{(\bA)} k_0^{(\bB)}\log\{(pq) \vee n\}} \to 0\,,
\end{align}
Combining \eqref{eq:ratio_k0B}, \eqref{eq:Op(1)order} and \eqref{eq:ratio_to_0}, it implies that
\begin{align*}
    \mathbb{P}\big\{ \hat{k}_{i}^{(\bB)} > k_0^{(\bB)}\big\} & \le \mathbb{P}\Big\{ \max_{k > k_0^{(\bB)}} \frac{\Delta\text{RSS}_i(k^{(\bA)}, k - 1) + \omega_n}{\Delta\text{RSS}_i(k^{(\bA)}, k) + \omega_n} > \frac{\Delta\text{RSS}_i(k^{(\bA)}, k_0^{(\bB)} - 1) + \omega_n}{\Delta\text{RSS}_i(k^{(\bA)}, k_0^{(\bB)}) + \omega_n} \Big\} \\
    & \to 0
\end{align*}
for some $i$ and $k^{(\bA)}$. Hence, it holds that $\mathbb{P}(\hat{k}_0^{(\bB)} > k_0^{(\bB)}) \to 0$. Similarly, to prove $\mathbb{P}(\hat{k}_0^{(\bB)} < k_0^{(\bB)}) \to 0$, it suffices to show
\begin{align}\label{eq:ks}
    \mathbb{P}\Big\{ \max_{k < k_0^{(\bB)}} \frac{\Delta\text{RSS}_i(k^{(\bA)}, k - 1) + \omega_n}{\Delta\text{RSS}_i(k^{(\bA)}, k) + \omega_n} > \frac{\Delta\text{RSS}_i(k^{(\bA)}, k_0^{(\bB)} - 1) + \omega_n}{\Delta\text{RSS}_i(k^{(\bA)}, k_0^{(\bB)}) + \omega_n} \Big\} \to 0
\end{align}
for some $i \in [pq]$ and some $k^{(\bA)}$. Similar to \eqref{eq:order}, for any $k^{(\bA)}$, it holds that
\begin{align*}
    \max_{k < k_0^{(\bB)}} \frac{\Delta\text{RSS}_i(k^{(\bA)}, k - 1) + \omega_n}{\Delta\text{RSS}_i(k^{(\bA)}, k) + \omega_n} \lesssim \frac{\lambda_2 k_0^{(\bA)} k_0^{(\bB)}+\omega_n}{C_n \lambda_1 k_0^{(\bA)} k_0^{(\bB)}\log\{(pq) \vee n\}/n + \omega_n }\,.
\end{align*}
By \eqref{eq:ratio_to_0}, we have \eqref{eq:ks} holds. 
% Comparing the above upper bound with the lower bound of \eqref{eq:ratio_k0B}, if $C_n^2/(npq) \to \infty$, we have
% \begin{align*}
%     & \Big\{\frac{\omega_n + \lambda_2 k_0^{(\bA)} k_0^{(\bB)}}{C_n \lambda_1 k_0^{(\bA)} k_0^{(\bB)}\log\{(pq) \vee n\}/n + O_{\rm p}(pq/n) }\Big\} \Big/ \Big\{\frac{C_n \lambda_1 k_0^{(\bA)} k_0^{(\bB)}\log\{(pq) \vee n\}/n }{O_{\rm p}(pq/n)}\Big\} \\
%     & ~~~~~~~~ \lesssim \frac{ O_{\rm p}(p^2q^2) + O_{\rm p}\{npqk_0^{(\bA)} k_0^{(\bB)}\} }{[C_n \lambda_1 k_0^{(\bA)} k_0^{(\bB)}\log\{(pq) \vee n\}]^2 + C_n pq \lambda_1 k_0^{(\bA)} k_0^{(\bB)}\log\{(pq) \vee n\}} \to 0\,,
% \end{align*}
% which implies that \eqref{eq:ks} holds. 
Therefore, we have $\mathbb{P}(\hat{k}_0^{(\bB)} = k_0^{(\bB)}) \to 1$ as $n \to \infty$.

Then, we show that $\mathbb{P}(\hat{k}_0^{(\bA)} = k_0^{(\bA)}, \hat{k}_0^{(\bB)} = k_0^{(\bB)}) \to 1$ as $n \to \infty$. By the above results, it then suffices to show $\mathbb{P}(\hat{k}_0^{(\bA)} = k_0^{(\bA)}\,|\, \hat{k}_0^{(\bB)} = k_0^{(\bB)}) \to 1$. Analogous to \eqref{eq:kB_1}, it suffices to show that 
\begin{align*}
    \mathbb{P}(\hat{k}_0^{(\bA)} < k_0^{(\bA)}\,|\, \hat{k}_0^{(\bB)} = k_0^{(\bB)}) \to 0 ~\mbox{and}~ \mathbb{P}(\hat{k}_0^{(\bA)} > k_0^{(\bA)}\,|\, \hat{k}_0^{(\bB)} = k_0^{(\bB)}) \to 0\,, 
\end{align*}
respectively. 
For each $i \in [pq]$, let 
\begin{align*}
    \hat{k}_{i}^{(\bA)} = \arg\max_{1 \le  k \le K} \frac{\Delta\text{RSS}_i(k-1, \hat{k}_0^{(\bB)}) + \omega_n}{\Delta\text{RSS}_i(k,\hat{k}_0^{(\bB)}) + \omega_n}\,.
\end{align*}
To prove $\mathbb{P}(\hat{k}_0^{(\bA)} < k_0^{(\bA)}\,|\, \hat{k}_0^{(\bB)} = k_0^{(\bB)}) \to 0$, we only need to show that $\mathbb{P}(\hat{k}_i^{(\bA)} < k_0^{(\bA)}\,|\, \hat{k}_0^{(\bB)} = k_0^{(\bB)}) \to 0$ for some $i \in [pq]$. 
% Notice that 
% \begin{align}\label{eq:kA}
%     \mathbb{P}(\hat{k}_i^{(\bA)} < k_0^{(\bA)}) &  = \mathbb{P}(\hat{k}_i^{(\bA)} < k_0^{(\bA)}, \hat{k}_i^{(\bB)} = k_0^{(\bB)}) + \mathbb{P}(\hat{k}_i^{(\bA)} < k_0^{(\bA)}, \hat{k}_i^{(\bB)} \neq k_0^{(\bB)}) \\
%     & \le \mathbb{P}(\hat{k}_i^{(\bA)} < k_0^{(\bA)}\,|\,\hat{k}_i^{(\bB)} = k_0^{(\bB)})\mathbb{P}(\hat{k}_i^{(\bB)} = k_0^{(\bB)}) + \mathbb{P}(\hat{k}_i^{(\bB)} \neq k_0^{(\bB)})\,.
% \end{align}
Define 
\begin{align*}
    \tilde{k}_{i}^{(\bA)} = \arg\max_{1 \le  k \le K} \frac{\Delta\text{RSS}_i(k-1, k_0^{(\bB)}) + \omega_n}{\Delta\text{RSS}_i(k,k_0^{(\bB)}) + \omega_n}\,.
\end{align*}
Hence, we have $\mathbb{P}(\hat{k}_i^{(\bA)} < k_0^{(\bA)}\,|\,\hat{k}_i^{(\bB)} = k_0^{(\bB)}) = \mathbb{P}(\tilde{k}_i^{(\bA)} < k_0^{(\bA)})$. Using the similar arguments for deriving $\mathbb{P}(\hat{k}_i^{(\bB)} < k_0^{(\bB)}) \to 0$, we can show that $\mathbb{P}(\tilde{k}_i^{(\bA)} < k_0^{(\bA)}) \to 0$. 
%By \eqref{eq:kA}, it holds that $\mathbb{P}(\hat{k}_i^{(\bA)} < k_0^{(\bA)}) \to 0$, and then $\mathbb{P}(\hat{k}_0^{(\bA)} < k_0^{(\bA)}) \to 0$. 
Similarly, we can show that $\mathbb{P}(\hat{k}_0^{(\bA)} > k_0^{(\bA)}\,|\,\hat{k}_i^{(\bB)} = k_0^{(\bB)}) \to 0$. 
%It implies that $\mathbb{P}(\hat{k}_0^{(\bA)} = k_0^{(\bA)}) \to 1$ as $n \to \infty$. 
This completes the proof of Theorem \ref{tm:bandwidth}. $\hfill\Box$

\section{Proof of Proposition \ref{pn:consistency_banded}}

By Theorem \ref{tm:bandwidth}, it suffices to show that Proposition \ref{pn:consistency_banded} holds for true $k_0^{(\bA)}$ and $k_0^{(\bB)}$, provided that $C_n\log(d\vee n)k_{0,max}^{2} = o(n)$ and $d=o\{n^{1/(2+3\kappa)}\}$. 
Recall that the true sets of indices of nonzero elements in $\bA_k$ and $\bB_k$ are $\mathcal{A}_{1}^{(k)}, \ldots, \mathcal{A}_{p}^{(k)}$ and $\mathcal{B}_{1}^{(k)}, \ldots, \mathcal{B}_{q}^{(k)}$, for $k=0,1$, respectively. For simplicity, we begin by introducing some notations that will be used throughout this section. Write $\bar{\bA}_k = (\bar{a}_{i,j}^{(k)})_{p \times p}$ and $\bar{\bB}_k = (\bar{b}_{i,j}^{(k)})_{q \times q}$ for $k=0,1$.
% $\bar{\bA}_0 = (\bar{a}_{i,j}^{(0)})_{p \times p} \in \tilde{\mathcal{B}}(k_0^{(\bA)})$, $\bar{\bA}_1 = (\bar{a}_{i,j}^{(1)})_{p \times p} \in \mathcal{B}(k_0^{(\bA)})$ and $\bar{\bB}_k = (\bar{b}_{i,j}^{(k)})_{q \times q} \in \mathcal{B}(k_0^{(\bB)})$ for $k=0,1$, 
%where $\tilde{\mathcal{B}}(k)$ and $\mathcal{B}(k)$ are specified in \eqref{eq:Banded_opt1}.
%$\bar{a}_{i,j}^{(k)}=0$ for all $|i-j|>k_0^{(\bA)}$ and $\bar{b}_{i,j}^{(k)}=0$ for all $|i-j|>k_0^{(\bB)}$.
Recall $r_{m} = r_{m,0}+r_{m,1}$ and $\ell_{j} = \ell_{j,0}+\ell_{j,1}$, where $r_{m,k}$ is the cardinality of the set $\mathcal{A}_{m}^{(k)}$ and $\ell_{j,k}$ is the cardinality of the set $\mathcal{B}_{j}^{(k)}$, $k=0,1$, and $s= \sum_{m=1}^p r_{m} + \sum_{j=1}^q \ell_{j}$.
% $s_{\ba_m} = |S_{\ba_m^{(0)}}|+|S_{\ba_m^{(1)}}|$ and $s_{\bb_j} = |S_{\bb_j^{(0)}}|+|S_{\bb_j^{(1)}}|$ and $s= \sum_{m=1}^p s_{\ba_m} + \sum_{j=1}^q s_{\bb_j}$. 
Write $\bar{\btheta} = (\bar{\tilde{\ba}}_1',\ldots,\bar{\tilde{\ba}}_p',\bar{\tilde{\bb}}_1',\ldots,\bar{\tilde{\bb}}_q')'$, which is a $s$-dimensional parameter with $\bar{\tilde{\ba}}_m = (\bar{\tilde{\ba}}_m^{(0)'},\bar{\tilde{\ba}}_m^{(1)'})'$, $\bar{\tilde{\bb}}_j = (\bar{\tilde{\bb}}_j^{(0)'},\bar{\tilde{\bb}}_j^{(1)'})'$, where $\bar{\tilde{\ba}}_m^{(k)}$ and $\bar{\tilde{\bb}}_j^{(k)}$ representing the vectors consisting of the nonzero entries of the $m$-th row of $\bar{\bA}_k$ and the $j$-th row of $\bar{\bB}_k$, $k=0,1$, respectively.
% $\bar{\tilde{\ba}}_m^{(k)}=(\bar{a}_{m,l}^{(k)})_{l \in S_{\ba_m^{(k)}}}$, $\bar{\tilde{\bb}}_j^{(k)}=(\bar{b}_{j,i}^{(k)})_{i \in S_{\bb_j^{(k)}}}$ for $k=0,1$, 
Let 
\begin{align*}
    \hat{L}_{n}(\bar{\btheta}) = \frac{1}{p^2q^2}\sum_{j,k=1}^q\|\hat{\bT}_{j,k}(\bar{\btheta})\|_F^2\,,
\end{align*}
where
\begin{align}\label{eq:hatTjk}
   \hat{\bT}_{j,k}(\bar{\btheta}) & = \hat\bSigma_{jk}(1) - \sum_{i=1}^q \bar{\bA}_0 \hat \bSigma_{ik}(1) \bar{b}_{j, i}^{(0)} - \sum_{i=1}^q  \bar{\bA}_1 \hat \bSigma_{ik}(0) \bar{b}_{j, i}^{(1)} \\
   & = \frac{1}{n}\sum_{t=2}^n \Big\{ \bX_{t,\cdot j}\bX_{t-1,\cdot k}' - \sum_{i=1}^q \bar{\bA}_0 \bX_{t,\cdot i}\bX_{t-1,\cdot k}' \bar{b}_{j, i}^{(0)} - \sum_{i=1}^q \bar{\bA}_1 \bX_{t-1,\cdot i}\bX_{t-1,\cdot k}' \bar{b}_{j, i}^{(1)} \Big\}\,.\notag
\end{align}
By \eqref{eq:Banded_opt1}, the generalized Yule-Walker estimator 
\begin{align}\label{eq:Banded_Opt2}
    \hat{\btheta} = \arg\min_{\bar{\btheta}\in \boldsymbol{\Omega}} \hat{L}_{n}(\bar{\btheta})\,.
\end{align}
Since $\{{\rm vec}(\bX_t)\}$ is stationary, for any $\bar{\btheta}$, we define $\tilde{L}(\bar{\btheta}) = p^{-2}q^{-2}\sum_{j,k=1}^q\|\bT_{j,k}(\bar{\btheta})\|_F^2$ with
\begin{align}\label{eq:Tjk}
    \bT_{j,k}(\bar{\btheta}) & = \mathbb{E}\{\hat{\bT}_{j,k}(\bar{\btheta})\}  = \bSigma_{jk}(1) - \sum_{i=1}^q \bar{\bA}_0 \bSigma_{ik}(1) \bar{b}_{j, i}^{(0)} - \sum_{i=1}^q \bar{\bA}_1 \bSigma_{ik}(0) \bar{b}_{j, i}^{(1)} \,.
\end{align}
Recall that $\btheta = (\tilde{\ba}_1',\ldots,\tilde{\ba}_p',\tilde{\bb}_1',$ $\ldots,\tilde{\bb}_q')'$ is the true parameter. By \eqref{YW}, it holds that $\bT_{j,k}(\btheta) = 0$ and then $\tilde{L}(\btheta) = 0$.
%, and $\bgamma$ is the unique solution of the optimization problem $\min_{\bar{\bgamma}}\tilde{Q}(\bar{\bgamma})$.

The proof of Proposition \ref{pn:consistency_banded} is similar to that of Proposition \ref{pn:consistency}. It suffices to show the uniform convergence of $\hat{L}_{n}(\bar{\btheta})$ over the compact and convex set $\boldsymbol{\Omega}$, i.e.
\begin{align}\label{eq:uniformConverge_banded}
    \sup_{\bar{\btheta} \in \boldsymbol{\Omega}} |\hat{L}_{n}(\bar{\btheta}) - \tilde{L}(\bar{\btheta})| = o_{\rm p}(1)\,,
\end{align}
provided that $p^{-1}(k_0^{(\bA)})^3(k_0^{(\bB)})^4 + q^{-1}(k_0^{(\bA)})^4(k_0^{(\bB)})^3 = O(1)$ and $d^{(2+11\kappa)/4}k_{0,max}^{7(2+\kappa)/4} = o(n)$.
%$d^{(22+21\kappa)/4}\{k_0^{(\bA)}\}^{(6+3\kappa)/4}\{k_0^{(\bB)}\}^{(2+\kappa)/2} = o(n)$.
% The proof of Proposition \ref{pn:consistency_banded} is similar to that of Proposition \ref{pn:consistency}. It suffices to show that, for any given $\bar{\btheta} \in \boldsymbol{\Omega}$, it holds that
% \begin{align}\label{eq:uniformConverge_banded}
%     \hat{L}_{n}(\bar{\btheta}) - \tilde{L}(\bar{\btheta}) = o_{\rm p}(1)\,,
% \end{align}
% provided that $d^{(22+21\kappa)/4}\{k_0^{(\bA)}\}^{(6+3\kappa)/4}\{k_0^{(\bB)}\}^{(2+\kappa)/2} = o(n)$.
%$d^{(18+19\kappa)/4}\{k_0^{(\bA)}\}^{(2+\kappa)/4}\{k_0^{(\bB)}\}^{(2+\kappa)/2} = o(n)$.

We first show that 
\begin{align*}
    |\hat{L}_{n}(\bar{\btheta}) - \tilde{L}(\bar{\btheta})| = o_{\rm p}(1)
\end{align*}
for any given $\bar{\btheta} \in \boldsymbol{\Omega}$, provided that $d^{(2+11\kappa)/4}(k_{0,max})^{7(2+\kappa)/4} = o(n)$.
Analogous to the proof of Proposition \ref{pn:consistency}, it suffices to derive the upper bound of $\mathbb{P}(\|\hat{\bT}_{j,k}(\bar{\btheta}) - \bT_{j,k}(\bar{\btheta})\|_F > z)$ for any $z>0$. 
By \eqref{eq:hatTjk} and \eqref{eq:Tjk}, it holds that 
\begin{align*}
    \hat{\bT}_{j,k}(\bar{\btheta}) - \bT_{j,k}(\bar{\btheta}) &= \{\hat\bSigma_{jk}(1) - \bSigma_{jk}(1)\} - \sum_{i=1}^q \bar{\bA}_0 \{\hat{\bSigma}_{ik}(1) - \bSigma_{ik}(1)\} \bar{b}_{j, i}^{(0)}  \\
    &~~~~~~~~- \sum_{i=1}^q \bar{\bA}_1 \{\hat{\bSigma}_{ik}(0) - \bSigma_{ik}(0)\} \bar{b}_{j, i}^{(1)} \,.
\end{align*}
Then we have
\begin{align*}
    \|\hat{\bT}_{j,k}(\bar{\btheta}) - \bT_{j,k}(\bar{\btheta})\|_F^2 & \le 3 \|\hat\bSigma_{jk}(1) - \bSigma_{jk}(1)\|_F^2 + 3\Big\|\sum_{i=1}^q \bar{\bA}_0 \{\hat{\bSigma}_{ik}(1) - \bSigma_{ik}(1)\} \bar{b}_{j, i}^{(0)} \Big\|_F^2 \\
    &~~~~~~ + 3\Big\| \sum_{i=1}^q \bar{\bA}_1 \{\hat{\bSigma}_{ik}(0) - \bSigma_{ik}(0)\} \bar{b}_{j, i}^{(1)} \Big\|_F^2\,.
    % & \le 3 \|\hat\bSigma_{jk}(1) - \bSigma_{jk}(1)\|_F^2 + 3q \sum_{i=1}^q \{v_{j, i}^{(0)}\}^2 \{\bar{\beta}_{jj}^{(0)}\}^2\|D(\bar{\balpha}_0) \bW_0\|_F^2 \|\hat{\bSigma}_{ik}(1) - \bSigma_{ik}(1)\|_F^2 \\
    % &~~~~~~~~ + 3q \sum_{i=1}^q \{v_{j, i}^{(1)}\}^2 \{\bar{\beta}_{jj}^{(1)}\}^2 \|D(\bar{\balpha}_1) \bW_1\|_F^2 \|\hat{\bSigma}_{ik}(0) - \bSigma_{ik}(0)\|_F^2\,. 
\end{align*}
For any $z>0$, it then holds that 
\begin{align}\label{eq:hatTjk_dev1}
    \mathbb{P}(\|\hat{\bT}_{j,k}(\bar{\bgamma}) - \bT_{j,k}(\bar{\bgamma})\|_F^2 > z) & \le \mathbb{P}(3 \|\hat\bSigma_{jk}(1) - \bSigma_{jk}(1)\|_F^2 > z/3) \notag\\
    & ~~~~~ + \mathbb{P}\Big( 3\Big\|\sum_{i=1}^q \bar{\bA}_0 \{\hat{\bSigma}_{ik}(1) - \bSigma_{ik}(1)\} \bar{b}_{j, i}^{(0)} \Big\|_F^2 > z/3 \Big) \notag\\
    &~~~~~ + \mathbb{P}\Big( 3\Big\| \sum_{i=1}^q \bar{\bA}_1 \{\hat{\bSigma}_{ik}(0) - \bSigma_{ik}(0)\} \bar{b}_{j, i}^{(1)} \Big\|_F^2 > z/3 \Big) \notag\\
    & := \mathrm{M}_1 + \mathrm{M}_2 + \mathrm{M}_3\,.
\end{align}
It suffices to derive the upper bound of $\mathrm{M}_1$, $\mathrm{M}_2$, and $\mathrm{M}_3$, respectively. By the derivation of the term $\mathrm{I}_1$ in the proof of Proposition \ref{pn:consistency}, we have
\begin{align*}
    \mathrm{M}_1 \lesssim p^{(2+3\kappa)/\kappa}n^{-2/\kappa}z^{-(2+\kappa)/(2\kappa)}\,.
\end{align*}
Notice that $\bar{\bA}_{k}$ and $\bar{\bB}_k$ are banded with the bandwidths $k_0^{(\bA)}$ and  $k_0^{(\bB)}$, respectively. We then have
\begin{align*}
    &\Big\|\sum_{i=1}^q \bar{\bA}_0 \{\hat{\bSigma}_{ik}(1) - \bSigma_{ik}(1)\} \bar{b}_{j, i}^{(0)} \Big\|_F^2 \\ 
    &~~~~~~~~~~~~~ \le k_0^{(\bB)} \sum_{i \in \mathcal{B}_{j}^{(0)}} \{\bar{b}_{j, i}^{(0)}\}^2 \|\bar{\bA}_0\|_F^2 \|\hat{\bSigma}_{ik}(1) - \bSigma_{ik}(1)\|_F^2 \\
    &~~~~~~~~~~~~~  \le C_1\{k_0^{(\bB)}\}^2 \|\bar{\bA}_0\|_F^2 \max_{i \in [q]} \|\hat{\bSigma}_{ik}(1) - \bSigma_{ik}(1)\|_F^2  \,.
\end{align*}
Since $\|\bar{\bA}_0\|_F^2 \le C_2 k_0^{(\bA)} p$, by using the similar arguments as the derivation of the term $\mathrm{I}_2$ in the proof of Proposition \ref{pn:consistency}, it holds that
\begin{align*}
    \mathrm{M}_2 & \le \mathbb{P}\Big\{ \max_{i \in [q]}\|\hat{\bSigma}_{ik}(1) - \bSigma_{ik}(1)\|_F^2 > \frac{C_3 z}{\{k_0^{(\bB)}\}^2k_0^{(\bA)}p} \Big\} \\
    & \lesssim p^{(6+7\kappa)/(2\kappa)}q\{k_0^{(\bB)}\}^{(2+\kappa)/\kappa}\{k_0^{(\bA)}\}^{(2+\kappa)/(2\kappa)}n^{-2/\kappa}z^{-(2+\kappa)/(2\kappa)}\,.
\end{align*}
Using the similar arguments, we can show that the same upper bound holds for $\mathrm{M}_3$. By \eqref{eq:hatTjk_dev1}, for any $z >0$, we have
\begin{align}\label{eq:hatTjk_dev2}
    & \mathbb{P}(\|\hat{\bT}_{j,k}(\bar{\btheta}) - \bT_{j,k}(\bar{\btheta})\|_F^2 > z) \\
    & ~~~~~~~\lesssim p^{(2+3\kappa)/\kappa}n^{-2/\kappa}z^{-(2+\kappa)/(2\kappa)} + p^{(6+7\kappa)/(2\kappa)}q\{k_0^{(\bB)}\}^{(2+\kappa)/\kappa}\{k_0^{(\bA)}\}^{(2+\kappa)/(2\kappa)}n^{-2/\kappa}z^{-(2+\kappa)/(2\kappa)} \,. \notag
    %p^{4(1+\kappa)/\kappa}n^{-(4+\kappa)/\kappa}z^{-(2+\kappa)/\kappa} + p^{(6+5\kappa)/\kappa}\{k_0^{(\bB)}\}^{(4+3\kappa)/\kappa}\{k_0^{(\bA)}\}^{(2+\kappa)/\kappa}n^{-(4+\kappa)/\kappa}z^{-(2+\kappa)/\kappa}\,. \notag
\end{align}
Notice that $\|\bT_{j,k}(\bar{\btheta})\|_F^2 \lesssim p^2\{k_0^{(\bA)}\}^2\{k_0^{(\bB)}\}^2$. Using the similar arguments as in \eqref{eq:Q_dev1} in the proof of Proposition \ref{pn:consistency}, by choosing some sufficiently large constant $\tilde{C}$, for any $x>0$, it holds that 
\begin{align*}
    & \mathbb{P}\Big\{\big| \|\hat{\bT}_{j,k}(\bar{\btheta})\|_F^2 - \|\bT_{j,k}(\bar{\btheta})\|_F^2 \big| > p^2 x \Big\} \notag \\
    & ~~~~~~ \le \mathbb{P}\Big(\|\hat{\bT}_{j,k}(\bar{\btheta}) - \bT_{j,k}(\bar{\btheta})\|_F > \frac{p^2 x}{\tilde{C}pk_0^{(\bA)}k_0^{(\bB)}} \Big) + \mathbb{P}( \|\hat{\bT}_{j,k}(\bar{\btheta})+\bT_{j,k}(\bar{\btheta})\|_F > \tilde{C}pk_0^{(\bA)}k_0^{(\bB)} )\,.
\end{align*}
By \eqref{eq:hatTjk_dev2}, we have
\begin{align*}
    &\mathbb{P}( \|\hat{\bT}_{j,k}(\bar{\bgamma})+\bT_{j,k}(\bar{\bgamma})\|_F^2 > \tilde{C}^2p^2\{k_0^{(\bA)}\}^2\{k_0^{(\bB)}\}^2 ) \\
    &~~~~~~ \lesssim p^2\{k_0^{(\bA)}\}^{-(2+\kappa)/\kappa}\{k_0^{(\bB)}\}^{-(2+\kappa)/\kappa}n^{-2/\kappa} + p^{(2+5\kappa)/(2\kappa)}q\{k_0^{(\bA)}\}^{-(2+\kappa)/(2\kappa)}n^{-2/\kappa} \\
    &~~~~~~ \lesssim p^2n^{-2/\kappa} + p^{(2+5\kappa)/(2\kappa)}qn^{-2/\kappa}
    %&~~~~~~ \lesssim p^2q^{-(2+\kappa)/\kappa}\{k_0^{(\bA)}\}^{-(2+\kappa)/\kappa}n^{-2/\kappa} + p^{(2+5\kappa)/(2\kappa)}q^{-2/\kappa}\{k_0^{(\bB)}\}^{(2+\kappa)/\kappa}\{k_0^{(\bA)}\}^{-(2+\kappa)/(2\kappa)}n^{-2/\kappa}
    %&~~~~~~ \lesssim p^2n^{-2/\kappa} + p^{(2+5\kappa)/(2\kappa)}q\{k_0^{(\bB)}\}^{(2+\kappa)/\kappa}\{k_0^{(\bA)}\}^{(2+\kappa)/(2\kappa)}n^{-2/\kappa}
\end{align*}
and 
\begin{align*}
    & \mathbb{P}\Big( \|\hat{\bT}_{j,k}(\bar{\btheta}) - \bT_{j,k}(\bar{\btheta})\|_F^2 >  \frac{C_4 p^2 x^2}{\{k_0^{(\bA)}\}^2\{k_0^{(\bB)}\}^2} \Big) \\
    & ~~~~~\lesssim p^{2}\{k_0^{(\bA)}\}^{(2+\kappa)/\kappa}\{k_0^{(\bB)}\}^{(2+\kappa)/\kappa}n^{-2/\kappa}x^{-(2+\kappa)/\kappa} \\
    & ~~~~~~~~~~ + p^{(2+5\kappa)/(2\kappa)}q\{k_0^{(\bB)}\}^{(4+2\kappa)/\kappa}\{k_0^{(\bA)}\}^{(6+3\kappa)/(2\kappa)}n^{-2/\kappa}x^{-(2+\kappa)/\kappa}\,.
    %& ~~~~~\lesssim p^{(4+4\kappa)/\kappa}q^{(6+3\kappa)/\kappa}\{k_0^{(\bA)}\}^{(2+\kappa)/\kappa}n^{-2/\kappa}x^{-(2+\kappa)/\kappa} \\
    %& ~~~~~~~~~~ + p^{(10+9\kappa)/(2\kappa)}q^{(6+4\kappa)/\kappa}\{k_0^{(\bB)}\}^{(2+\kappa)/\kappa}\{k_0^{(\bA)}\}^{(6+3\kappa)/(2\kappa)}n^{-2/\kappa}x^{-(2+\kappa)/\kappa}\,.
\end{align*}
Hence, it holds that
\begin{align*}
    & \mathbb{P}\{|\hat{L}_{n}(\bar{\btheta}) - \tilde{L}(\bar{\btheta})| > x\} \notag \\
    & ~~~~~~\le \sum_{j,k=1}^q \mathbb{P} \Big\{\big| \|\hat{\bT}_{j,k}(\bar{\btheta})\|_F^2 - \|\bT_{j,k}(\bar{\btheta})\|_F^2 \big| > p^2 x \Big\} \notag \\
    & ~~~~~~\lesssim p^2q^2n^{-2/\kappa} + p^{(2+5\kappa)/(2\kappa)}q^3n^{-2/\kappa} \\
    & ~~~~~~~~~~~+ p^{2}q^2\{k_0^{(\bA)}\}^{(2+\kappa)/\kappa}\{k_0^{(\bB)}\}^{(2+\kappa)/\kappa}n^{-2/\kappa}x^{-(2+\kappa)/\kappa} \\
    & ~~~~~~~~~~~ + p^{(2+5\kappa)/(2\kappa)}q^3\{k_0^{(\bB)}\}^{(4+2\kappa)/\kappa}\{k_0^{(\bA)}\}^{(6+3\kappa)/(2\kappa)}n^{-2/\kappa}x^{-(2+\kappa)/\kappa}\,.
    %& ~~~~~~\lesssim  p^{(2+5\kappa)/(2\kappa)}q^{(2\kappa-2)/\kappa}\{k_0^{(\bB)}\}^{(2+\kappa)/\kappa}\{k_0^{(\bA)}\}^{-(2+\kappa)/(2\kappa)}n^{-2/\kappa}  + \notag \\
    %& ~~~~~~~~~~~+ p^{(10+9\kappa)/(2\kappa)}q^{(6+6\kappa)/\kappa}\{k_0^{(\bB)}\}^{(2+\kappa)/\kappa}\{k_0^{(\bA)}\}^{(6+3\kappa)/(2\kappa)}n^{-2/\kappa}x^{-(2+\kappa)/\kappa} \\
    %& ~~~~~~~~~~~ + p^{2}q^{(\kappa-2)/\kappa}n^{-2/\kappa} + p^{(4+4\kappa)/\kappa}q^{(6+5\kappa)/\kappa}\{k_0^{(\bA)}\}^{(2+\kappa)/\kappa}n^{-2/\kappa}x^{-(2+\kappa)/\kappa}\,. 
\end{align*}
Recall $d=p \vee q$ and $k_{0,max} = k_0^{(\bA)} \vee k_0^{(\bB)}$. Hence, for any given $\bar{\btheta} \in \boldsymbol{\Omega}$,  we have $|\hat{L}_{n}(\bar{\btheta}) - \tilde{L}(\bar{\btheta})| = o_{\rm p}(1)$, provided that $d^{(2+11\kappa)/4}(k_{0,max})^{7(2+\kappa)/4} = o(n)$.
%$d^{(22+21\kappa)/4}\{k_0^{(\bA)}\}^{(6+3\kappa)/4}\{k_0^{(\bB)}\}^{(2+\kappa)/2} = o(n)$. 
%$d^{(18+19\kappa)/4}\{k_0^{(\bA)}\}^{(2+\kappa)/4}\{k_0^{(\bB)}\}^{(2+\kappa)/2} = o(n)$.

We then show the uniform convergence of $\hat{L}_{n}(\bar{\btheta})$ over the compact and convext set $\bOmega$. Recall $\tilde{L}(\bar{\btheta}) = p^{-2}q^{-2}\sum_{j,k=1}^q\|\bT_{j,k}(\bar{\btheta})\|_F^2$, where $\bT_{j,k}(\bar{\btheta})$ is specified in \eqref{eq:Tjk}. Write $\bar{\bA}_k = (\bar{a}_{i,j}^{(k)})_{p \times p}$ and $\bar{\bB}_k = (\bar{b}_{i,j}^{(k)})_{q \times q}$ for $k =0$, $1$. Recall $\bSigma_{jk}(1) = (\sigma_{l,m}^{(j,k,1)})_{p \times p}$ and $\bSigma_{jk}(0) = (\sigma_{l,m}^{(j,k,0)})_{p \times p}$ for any given $j,k \in [q]$. Then, for any $m \in [p]$, $h \in \mathcal{A}_{m}^{(0)}$, $\tilde{h} \in \mathcal{A}_{m}^{(1)}$, $v \in [q]$, $u \in \mathcal{B}_{j}^{(0)}$ and $\tilde{u} \in \mathcal{B}_{j}^{(1)}$,  we have
\begin{align*}
    \frac{\partial \tilde{L}(\bar{\btheta})}{\partial \bar{a}_{m,h}^{(0)}} & = \frac{-2}{p^2 q^2}\sum_{j,k=1}^q \sum_{m_2=1}^p t_{m,m_2}^{j,k}(\bar{\btheta}) \bigg\{ \sum_{i=1}^q \sigma_{h,m_2}^{(i,k,1)}\bar{b}_{j,i}^{(0)} \bigg\}\,, \\
    \frac{\partial \tilde{L}(\bar{\btheta})}{\partial \bar{a}_{m,\tilde{h}}^{(1)}} & = \frac{-2}{p^2 q^2}\sum_{j,k=1}^q \sum_{m_2=1}^p t_{m,m_2}^{j,k}(\bar{\btheta}) \bigg\{ \sum_{i=1}^q \sigma_{\tilde{h},m_2}^{(i,k,0)}\bar{b}_{j,i}^{(1)} \bigg\}\,, \\
    \frac{\partial \tilde{L}(\bar{\btheta})}{\partial \bar{b}_{v,u}^{(0)}} & = \frac{-2}{p^2 q^2}\sum_{k=1}^q \sum_{m_1,m_2=1}^p t_{m_1,m_2}^{v,k}(\bar{\btheta}) \bigg\{ \sum_{l=1}^p \bar{a}_{m_1,l}^{(0)} \sigma_{l,m_2}^{(u,k,1)} \bigg\}\,, \\
    \frac{\partial \tilde{L}(\bar{\btheta})}{\partial \bar{b}_{v,\tilde{u}}^{(1)}} & = \frac{-2}{p^2 q^2}\sum_{k=1}^q \sum_{m_1,m_2=1}^p t_{m_1,m_2}^{v,k}(\bar{\btheta}) \bigg\{ \sum_{l=1}^p \bar{a}_{m_1,l}^{(1)}\sigma_{l,m_2}^{(\tilde{u},k,0)} \bigg\}\,,
\end{align*}
where 
\begin{align*}
    t_{m_1,m_2}^{j,k}(\bar{\btheta}) & = \sigma_{m_1,m_2}^{(j,k,1)} - \sum_{i=1}^q\sum_{l=1}^p\bar{a}_{m_1,l}^{(0)} \sigma_{l,m_2}^{(i,k,1)} \bar{b}_{j,i}^{(0)} - \sum_{i=1}^q\sum_{l=1}^p \bar{a}_{m_1,l}^{(1)} \sigma_{l,m_2}^{(i,k,0)} \bar{b}_{j,i}^{(1)} \,.
\end{align*}
It can be derived that $\max_{m_1,m_2 \in [p], j,k \in [q]}|t_{m_1,m_2}^{j,k}(\bar{\btheta})| \le k_0^{(\bA)}k_0^{(\bB)}$. Hence, it holds that 
\begin{align}\label{eq:gradient}
    \bigg\| \frac{\nabla \tilde{L}(\bar{\btheta})}{\nabla \bar{\btheta}}  \bigg\|_2 \lesssim \Big[p^{-1}\big\{k_0^{(\bA)}\big\}^3\big\{k_0^{(\bB)}\big\}^4 + q^{-1}\big\{k_0^{(\bA)}\big\}^4\big\{k_0^{(\bB)}\big\}^3 \Big]^{1/2}\,,
\end{align}
which implies $\|\nabla \tilde{L}(\bar{\btheta})/\nabla \bar{\btheta}\|_2 = O(1)$ if $p^{-1}(k_0^{(\bA)})^3(k_0^{(\bB)})^4 + q^{-1}(k_0^{(\bA)})^4(k_0^{(\bB)})^3 = O(1)$. Hence, $\tilde{L}(\bar{\btheta})$ is equicontinuous on $\bOmega$.

Recall $\hat{L}_{n}(\bar{\btheta}) = p^{-2}q^{-2}\sum_{j,k=1}^q\|\hat{\bT}_{j,k}(\bar{\btheta})\|_F^2$, where $\hat{\bT}_{j,k}(\bar{\bgamma})$ is specified in \eqref{eq:hatTjk}. Hence, we have
\begin{align*}
    \frac{\partial \hat{L}_{n}(\bar{\btheta})}{\partial \bar{a}_{m,h}^{(0)}} & = \frac{-2}{p^2 q^2}\sum_{j,k=1}^q \sum_{m_2=1}^p \hat{t}_{m,m_2}^{j,k}(\bar{\btheta}) \bigg\{ \sum_{i=1}^q \hat{\sigma}_{h,m_2}^{(i,k,1)}\bar{b}_{j,i}^{(0)} \bigg\}\,,
\end{align*}
where 
\begin{align*}
    \hat{t}_{m_1,m_2}^{j,k}(\bar{\bgamma}) & = \frac{1}{n}\sum_{t=2}^n \Big\{ X_{t,m_1,j}X_{t-1,m_2, k} - \sum_{i=1}^q\sum_{l=1}^p \bar{a}_{m_1,l}^{(0)} X_{t,l, i}X_{t-1,m_2, k}  \bar{b}_{j,i}^{(0)} \notag\\
    &~~~~~~~~~~~~~~~~~~~~~~~- \sum_{i=1}^q\sum_{l=1}^p \bar{a}_{m_1,l}^{(1)} X_{t-1,l, i}X_{t-1,m_2, k}  \bar{b}_{j,i}^{(1)} \Big\}
    %&~~~~~~~~~~~~~~~~~~~\hat{\sigma}_{l,m_2}^{(i,k,1)} = \frac{1}{n}\sum_{t=2}^n X_{t,l, i}X_{t-1,m_2, k}\notag
\end{align*}
and $\hat{\sigma}_{l,m_2}^{(i,k,1)} = n^{-1}\sum_{t=2}^n X_{t,l, i}X_{t-1,m_2, k}$ for any $i,j,k \in [q]$ and $m_1,m_2,l \in [p]$. Similar to the derivation of \eqref{eq:T_decompose}, it holds that
\begin{align}\label{eq:L_decompose}
    & \left| \frac{\partial \hat{L}_{n}(\bar{\btheta})}{\partial \bar{a}_{m,h}^{(0)}} - \frac{\partial \tilde{L}(\bar{\btheta})}{\partial \bar{a}_{m,h}^{(0)}} \right| \notag\\
    &~~~ \le \frac{2}{p^2q^2} \sum_{j,k=1}^q \sum_{m_2=1}^p \big| t_{m,m_2}^{j,k}(\bar{\btheta}) \big| \bigg\{ \sum_{i=1}^q \big|\hat{\sigma}_{h,m_2}^{(i,k,1)}-\sigma_{h,m_2}^{(i,k,1)}\big|\big|\bar{b}_{j,i}^{(0)}\big| \bigg\}  \notag\\
    &~~~~~~~~~ + \frac{2}{p^2q^2} \sum_{j,k=1}^q \sum_{m_2=1}^p \big|\hat{t}_{m,m_2}^{j,k}(\bar{\btheta}) -  t_{m,m_2}^{j,k}(\bar{\btheta}) \big| \bigg\{ \sum_{i=1}^q  \big|\sigma_{h,m_2}^{(i,k,1)}\bar{b}_{j,i}^{(0)}\big| \bigg\} \notag\\
    &~~~~~~~~~ + \frac{2}{p^2q^2} \sum_{j,k=1}^q \sum_{m_2=1}^p \big|\hat{t}_{m,m_2}^{j,k}(\bar{\bgamma}) -  t_{m,m_2}^{j,k}(\bar{\bgamma}) \big| \bigg\{ \sum_{i=1}^q  \big|\hat{\sigma}_{h,m_2}^{(i,k,1)}-\sigma_{h,m_2}^{(i,k,1)}\big|\big|\bar{b}_{j,i}^{(0)}\big| \bigg\} \notag\\
    &~~~ = \tilde{T}_{1,m,h,j,k} + \tilde{T}_{2,m,h,j,k} + \tilde{T}_{3,m,h,j,k}\,.
\end{align}
For $\tilde{T}_{1,m,h,j,k}$, we have 
\begin{align*}
    \max_{m,h \in [p],j,k\in [q]}|\tilde{T}_{1,m,h,j,k}| \lesssim &~ \frac{k_0^{(\bA)}\{k_0^{(\bB)}\}^2}{p}\max_{j,k\in [q]}\|\hat{\bSigma}_{jk}(1)-\bSigma_{jk}(1)\|_{\infty}\,.
\end{align*}
By Lemma \ref{lem:Sigmajk}, it holds that $\max_{m,h \in [p],j,k\in [q]}|\tilde{T}_{1,m,h,j,k}| = O_{\rm p}\{k_0^{(\bA)}(k_0^{(\bB)})^2d^{(4\kappa)/(2+\kappa)}n^{-2/(2+\kappa)}\}$. Hence, we have $\tilde{T}_1 = o_{\mathrm{p}}(s^{-1/2})$ if $k_{0,max}^{7(2+\kappa)/4}d^{(2+9\kappa)/4}=o(n)$.
%$\{k_0^{(\bA)}\}^{(2+\kappa)/2}\{k_0^{(\bB)}\}^{2+\kappa}d^{(2+5\kappa)/(4\kappa)}=o(n)$. 
For $\tilde{T}_{2,m,h,j,k}$, notice that
\begin{align*}
    \max_{m_1,m_2 \in [p],j,k\in [q]}\big|\hat{t}_{m_1,m_2}^{j,k}(\bar{\bgamma}) -  t_{m_1,m_2}^{j,k}(\bar{\bgamma}) \big| & \lesssim   k_0^{(\bA)}k_0^{(\bB)}\Big\{ \max_{j,k\in [q]}\|\hat{\bSigma}_{jk}(1)-\bSigma_{jk}(1)\|_{\infty}  \\
    &~~~~~~~~~~~~~~~~~~~~~~~~~~+ \max_{j,k\in [q]}\|\hat{\bSigma}_{jk}(0)-\bSigma_{jk}(0)\|_{\infty}\Big\}\,.
\end{align*}
Hence, we also have $\max_{m,h \in [p],j,k\in [q]}|\tilde{T}_{2,m,h,j,k}| = o_{\mathrm{p}}(s^{-1/2})$ if $k_{0,max}^{7(2+\kappa)/4}d^{(2+9\kappa)/4}=o(n)$.
%$\{k_0^{(\bA)}\}^{(2+\kappa)/2}\{k_0^{(\bB)}\}^{2+\kappa}d^{(2+5\kappa)/(4\kappa)}=o(n)$. 
It can be shown that the same conclusion holds for $\max_{m,h \in [p],j,k\in [q]}|\tilde{T}_{3,m,h,j,k}|$. By \eqref{eq:L_decompose}, we have 
\begin{align*}
    \left| \frac{\partial \hat{L}_{n}(\bar{\btheta})}{\partial \bar{a}_{m,h}^{(0)}} \right| \le \left| \frac{\partial \tilde{L}(\bar{\btheta})}{\partial \bar{a}_{m,h}^{(0)}} \right| + o_{\mathrm{p}}(s^{-1/2})\,.
\end{align*}
The partial derivatives of $\hat{L}_{n}(\bar{\theta})$ with respect to $\bar{a}_{m,\tilde{h}}^{(1)}$, $\bar{b}_{v,u}^{(0)}$ and $\bar{b}_{v,\tilde{u}}^{(1)}$ can be obtained in a similar manner, and the corresponding results hold.
By \eqref{eq:gradient}, we have
\begin{align*}
    \bigg\| \frac{\nabla \hat{L}_n(\bar{\btheta})}{\nabla \bar{\btheta}}  \bigg\|_2 \le O_{\mathrm{p}}(1)
\end{align*}
if $p^{-1}(k_0^{(\bA)})^3(k_0^{(\bB)})^4 + q^{-1}(k_0^{(\bA)})^4(k_0^{(\bB)})^3 = O(1)$ and $k_{0,max}^{7(2+\kappa)/4}d^{(2+9\kappa)/4}=o(n)$.
%$\{k_0^{(\bA)}\}^{(2+\kappa)/2}\{k_0^{(\bB)}\}^{2+\kappa}d^{(2+5\kappa)/(4\kappa)}=o(n)$. 
By Corollary 2.2 of \cite{Newey1991}, the uniform convergence of $\hat{L}_{n}(\bar{\btheta})$ over $\bOmega$ holds. By using the similar arguments as in the proof of Proposition \ref{pn:consistency}, we have Proposition \ref{pn:consistency_banded} holds.
%This completes the proof of Proposition \ref{pn:consistency_banded}. 
$\hfill\Box$

\section{Proof of Theorem \ref{tm:mixing_banded}}

By Theorem \ref{tm:bandwidth}, it suffices to prove \eqref{eq:CLT_band} over the set $\mathcal{A}_n = \{\hat{k}_0^{(\bA)} = k_0^{(\bA)}, \hat{k}_0^{(\bB)} = k_0^{(\bB)}\}$ since $\mathbb{P}(\mathcal{A}_n) \to 1$. Over the set $\mathcal{A}_n$, it suffices to show that \eqref{eq:CLT_band} holds for  $k_0^{(\bA)}$ and $k_0^{(\bB)}$, provided that $C_n\log(d\vee n)k_{0,max}^{2} = o(n)$ and $d=o\{n^{1/(2+3\kappa)}\}$. Recall that the generalized Yule-Walker estimator aims to minimize 
\begin{align}\label{eq:Ln_full}
\hat{L}_n(\bar{\bA}_0,\bar{\bA}_1, \bar{\bB}_0, \bar{\bB}_1) = \frac{1}{p^2q^2}\Big\|\hat{\bSigma}_{jk}(1) - \sum_{i=1}^q \bar{\bA}_0\hat{\bSigma}_{ik}(1)\bar{b}_{j,i}^{(0)} - \sum_{i=1}^q \bar{\bA}_1 \hat{\bSigma}_{ik}(0) \bar{b}_{j,i}^{(1)} \Big\|_F^2
\end{align}
with respect to $\bar{\bA}_k = (\bar{a}_{i,j}^{(k)})_{p \times p}$ and $\bar{\bB}_k = (\bar{b}_{i,j}^{(k)})_{q \times q}$ for $k=0,1$, where $\bar{a}_{i,j}^{(k)}=0$ for all $|i-j|>k_0^{(\bA)}$, $\bar{a}_{i,i}^{(0)}=0$, and $\bar{b}_{i,j}^{(k)}=0$ for all $|i-j|>k_0^{(\bB)}$. Write $\bar{\bA}_k = (\bar{\ba}_{1}^{(k)}, \ldots, \bar{\ba}_{p}^{(k)})'$ and $\bar{\bB}_k = (\bar{\bb}_{1}^{(k)}, \ldots, \bar{\bb}_{q}^{(k)})'$ for $k=0,1$, and let $\bar{\tilde{\ba}}_{m}^{(k)}$ and $\bar{\tilde{\bb}}_{j}^{(k)}$ denote the vectors of nonzero elements in $\bar{\ba}_{m}^{(k)}$ and $\bar{\bb}_{j}^{(k)}$, respectively. Hence, minimizing \eqref{eq:Ln_full} is equivalent to minimizing
\begin{align}\label{eq:Ltheta_est_min}
    \hat{L}_n(\bar{\btheta}) = \frac{1}{p^2q^2}\sum_{j, k=1}^q\sum_{m=1}^p\Big\|\hat{\bSigma}_{jk}(1)'\be_m - \sum_{i \in \mathcal{B}_{j}^{(0)}} \bar{b}_{j, i}^{(0)} \hat{\bSigma}_{ik}(1)'\bE_m^{(0)} \bar{\tilde{\ba}}_m^{(0)} - \sum_{i \in \mathcal{B}_{j}^{(1)}} \bar{b}_{j, i}^{(1)} \hat{\bSigma}_{ik}(0)' \bE_m^{(1)} \bar{\tilde{\ba}}_m^{(1)} \Big\|_2^2
\end{align}
with respect to $\bar{\btheta} = (\bar{\tilde{\ba}}_1',\ldots,\bar{\tilde{\ba}}_p',\bar{\tilde{\bb}}_1',\ldots,\bar{\tilde{\bb}}_q')'$, where $\bar{\tilde{\ba}}_m = (\bar{\ba}_{m}^{(0)'},\bar{\ba}_{m}^{(1)'})' \in \mathbb{R}^{r_{m}}$  with $r_{m} = r_{m,0}+r_{m,1}$ and $\bar{\tilde{\bb}}_j = (\bar{\bb}_{j}^{(0)'},\bar{\bb}_{j}^{(1)'})' \in \mathbb{R}^{\ell_{j}}$  with $\ell_{j} = \ell_{j,0}+\ell_{j,1}$.
Let $\hat{\btheta} = (\hat{\tilde{\ba}}_1',\ldots, \hat{\tilde{\ba}}_p', \hat{\tilde{\bb}}_1',\ldots, \hat{\tilde{\bb}}_q')'$ denote the generalized Yule-Walker estimator obtained by minimizing \eqref{eq:Ltheta_est_min}. Recall $\hat{\bA}_k = (\hat{a}_{i,j}^{(k)})_{p \times p}$ and $\hat{\bB}_k = (\hat{b}_{i,j}^{(k)})_{q \times q}$ for $k=0,1$, where $\hat{a}_{i,j}^{(k)}=0$ for all $|i-j|>k_0^{(\bA)}$, $\hat{a}_{i,i}^{(0)}=0$, and $\hat{b}_{i,j}^{(k)}=0$ for all $|i-j|>k_0^{(\bB)}$, and the nonzero elements of $\hat{\bA}_k$ and $\hat{\bB}_k$ take their corresponding values from $\hat{\tilde{\ba}}_1,\ldots, \hat{\tilde{\ba}}_p$ and $\hat{\tilde{\bb}}_1,\ldots, \hat{\tilde{\bb}}_q$, respectively. 

To establish the asymptotic normality of $\hat{\btheta}$, we first derive the first-order gradient condition by taking partial derivatives of \eqref{eq:Ltheta_est_min} with respect to the elements of $\bar{\btheta}$. 
%Define $\mathcal{F}_{\bf C}(k_0) = \{{\bf C}=(c_{i,j})_{d \times d}: c_{i,j}=0 ~\mbox{for all}~|i-j|>k_0\}$ for any $d \times d$ matrix ${\bf C}$ and positive integer $k_0$.
For each $m \in [p]$, taking the partial derivative of $\hat{L}_n(\bar{\btheta})$ with respect to $\bar{\tilde{\ba}}_m$ is equivalent to taking the partial derivative of 
\begin{align}\label{eq:Ltheta_est_min_a}
     \sum_{j, k=1}^q\Big\|\hat{\bSigma}_{jk}(1)'\be_m - \sum_{i=1}^q \bar{b}_{j, i}^{(0)} \hat{\bSigma}_{ik}(1)'\bE_m^{(0)} \bar{\tilde{\ba}}_m^{(0)} - \sum_{i=1}^q \bar{b}_{j, i}^{(1)} \hat{\bSigma}_{ik}(0)' \bE_m^{(1)} \bar{\tilde{\ba}}_m^{(1)} \Big\|_2^2\,,
\end{align}
where $\bar{b}_{j,i}^{(k)}=0$ for all $i \notin \mathcal{B}_{j}^{(k)}$, $k=0,1$. Hence, the generalized Yule-Walker estimator satisfies
\begin{align*}
    {\hat{\bX}_{\tilde{\ba}_m}}'\hat{\bX}_{\tilde{\ba}_m}\hat{\tilde{\ba}}_m - {\hat{\bX}_{\tilde{\ba}_m}}'\bY_{\tilde{\ba}_m} = {\bf 0}_{r_{m}\times 1}\,,
\end{align*}
where 
\begin{align*}
    \hat\bX_{\tilde \ba_m} = 
\begin{pmatrix}
\sum_{i=1}^q \hat{b}_{1,i}^{(0)}\hat \bSigma_{i1}(1)' \bE_m^{(0)}\,, & \sum_{i=1}^q  \hat{b}_{1,i}^{(1)}\hat \bSigma_{i1}(0)' \bE_m^{(1)} \\
\sum_{i=1}^q \hat{b}_{1,i}^{(0)}\hat \bSigma_{i2}(1)' \bE_m^{(0)}\,, & \sum_{i=1}^q  \hat{b}_{1,i}^{(1)}\hat \bSigma_{i2}(0)' \bE_m^{(1)} \\
\vdots & \vdots \\
\sum_{i=1}^q \hat{b}_{q,i}^{(0)}\hat \bSigma_{iq}(1)' \bE_m^{(0)}\,, & \sum_{i=1}^q  \hat{b}_{q,i}^{(1)}\hat \bSigma_{iq}(0)' \bE_m^{(1)} 
\end{pmatrix}
\end{align*}
is a $pq^2 \times  r_{m}$ matrix.

Analogously, for each $j \in [q]$, taking the partial derivative of $\hat{L}_n(\bar{\btheta})$ with respect to $\bar{\tilde{\bb}}_j$ is equivalent to the partial derivative of 
\begin{align*}
    \sum_{k=1}^q \Big\|\text{vec}\{\hat\bSigma_{jk}(1)\} - \sum_{i \in  \mathcal{B}_{j}^{(0)}} \text{vec}\{\bar{\bA}_0\hat\bSigma_{ik}(1)\}\bar{b}_{j,i}^{(0)} - \sum_{i \in \mathcal{B}_{j}^{(1)}} \text{vec}\{\bar{\bA}_1 \hat\bSigma_{ik}(0)\} \bar{b}_{j,i}^{(1)} \Big\|_2^2\,,
\end{align*}
where $\bar{a}_{m,l}^{(k)}=0$ for all $l \notin \mathcal{A}_{m,k}$, $m\in [p]$ and $k=0,1$. Hence, the generalized Yule-Walker estimator satisfies that
\begin{align*}
    {\hat{\bX}_{\tilde{\bb}_j}}'\hat{\bX}_{\tilde{\bb}_j}\hat{\tilde{\bb}}_j - {\hat{\bX}_{\tilde{\bb}_j}}'\bY_{\tilde{\bb}_j} = {\bf 0}_{\ell_{j}\times 1}\,,
\end{align*}
where
\begin{align*}
    \hat{\bX}_{\tilde{\bb}_j} = 
\begin{pmatrix}
\text{vec}\{\hat\bA_0\hat\bSigma_{j_{1}^{(0)}1}(1)\}\,, \cdots, \text{vec}\{\hat\bA_0\hat\bSigma_{j_{\ell_{j,0}}^{(0)}1}(1)\}\,, \text{vec}\{\hat\bA_1\hat\bSigma_{j_{1}^{(1)}1}(0)\}\,, \cdots, \text{vec}\{\hat\bA_1\hat\bSigma_{j_{\ell_{j,1}}^{(1)}1}(0)\} \\
\text{vec}\{\hat\bA_0\hat\bSigma_{j_{1}^{(0)}2}(1)\}\,, \cdots, \text{vec}\{\hat\bA_0\hat\bSigma_{j_{\ell_{j,0}}^{(0)}2}(1)\}\,, \text{vec}\{\hat\bA_1\hat\bSigma_{j_{1}^{(1)}2}(0)\}\,, \cdots, \text{vec}\{\hat\bA_1\hat\bSigma_{j_{\ell_{j,1}}^{(1)}2}(0)\}\\
\vdots ~~~~~~~~~~~~~~~~~~~~~~~~~~~~~~\vdots~~~~~~~~~~~~~~~~~~~~~~~~~~\vdots~~~~~~~~~~~~~~~~~~~~~~~~~~~~~\vdots \\
\text{vec}\{\hat\bA_0\hat\bSigma_{j_{1}^{(0)}q}(1)\}\,, \cdots, \text{vec}\{\hat\bA_0\hat\bSigma_{j_{\ell_{j,0}}^{(0)}q}(1)\}\,, \text{vec}\{\hat\bA_1\hat\bSigma_{j_{1}^{(1)}q}(0)\}\,, \cdots, \text{vec}\{\hat\bA_1\hat\bSigma_{j_{\ell_{j,1}}^{(1)}q}(0)\}
\end{pmatrix}
\end{align*}
is a $p^2 q \times \ell_{j}$ matrix.

Hence, we obtain the first-order gradient condition:
\begin{equation}\label{eq:gradient_condition_banded}
\begin{pmatrix}
{\hat{\bX}_{\tilde{\ba}_1}}'\hat{\bX}_{\tilde{\ba}_1}\hat{\tilde{\ba}}_1 - {\hat{\bX}_{\tilde{\ba}_1}}'\bY_{\tilde{\ba}_1} \\
\vdots \\
{\hat{\bX}_{\tilde{\ba}_p}}'\hat{\bX}_{\tilde{\ba}_p}\hat{\tilde{\ba}}_p - {\hat{\bX}_{\tilde{\ba}_p}}'\bY_{\tilde{\ba}_p} \\
{\hat{\bX}_{\tilde{\bb}_1}}'\hat{\bX}_{\tilde{\bb}_1}\hat{\tilde{\bb}}_1 - {\hat{\bX}_{\tilde{\bb}_1}}'\bY_{\tilde{\bb}_1} \\
\vdots \\
{\hat{\bX}_{\tilde{\bb}_q}}'\hat{\bX}_{\tilde{\bb}_q}\hat{\tilde{\bb}}_q - {\hat{\bX}_{\tilde{\bb}_q}}'\bY_{\tilde{\bb}_q}
\end{pmatrix}   = {\bf 0}_{s\times 1}\,, 
\end{equation}
where $s=\sum_{m=1}^p r_{m} + \sum_{j=1}^q \ell_{j}$. 

Then, we perform a Taylor expansion on the first-order gradient condition \eqref{eq:gradient_condition_banded}. Recall $\btheta = (\tilde{\ba}_1',\ldots,\tilde{\ba}_p',\tilde{\bb}_1',\ldots,\tilde{\bb}_q')'$ is the true parameter. By Proposition \ref{pn:consistency_banded}, it holds that $\|\hat{\btheta} - \btheta\|_2 = o_{\rm p}(1)$, provided that $C_n\log(d\vee n)k_{0,max}^{2} = o(n)$, $p^{-1}(k_0^{(\bA)})^3(k_0^{(\bB)})^4 + q^{-1}(k_0^{(\bA)})^4(k_0^{(\bB)})^3 = O(1)$, $d^{2+3\kappa}=o(n)$ and $d^{(2+11\kappa)/4}(k_{0,max})^{7(2+\kappa)/4} = o(n)$.
%$k_{0,max}^{(10+5\kappa)/4}d^{(22+21\kappa)/4} = o(n)$.
%$k_{0,max}^{(6+3\kappa)/4}d^{(18+19\kappa)/4} = o(n)$.
%$d^{18+13\kappa}\{k_0^{(\bA)}\}^{4+3\kappa}\{k_0^{(\bB)}\}^{2+\kappa} = o(n^{4+\kappa})$. 
For any $m \in [p]$ and $j \in [q]$, define ${\bf f}_m(\bar{\btheta}) = \bar{\bX}_{\tilde{\ba}_m}'\bar{\bX}_{\tilde{\ba}_m}\bar{\tilde{\ba}}_m - \bar{\bX}_{\tilde{\ba}_m}'\bY_{\tilde{\ba}_m}$ and ${\bf f}_{p+j}(\bar{\btheta}) = \bar{\bX}_{\tilde{\bb}_j}'\bar{\bX}_{\tilde{\bb}_j}\bar{\tilde{\bb}}_j - \bar{\bX}_{\tilde{\bb}_j}'\bY_{\tilde{\bb}_j}$, where $\bar{\bX}_{\tilde{\ba}_m}$ and $\bar{\bX}_{\tilde{\bb}_j}$ are defined in the same way as $\bX_{\tilde{\ba}_m}$ and $\bX_{\tilde{\bb}_j}$ in \eqref{LS_A} and \eqref{LS_B}, respectively, except that the parameter $\btheta$ is replaced with $\bar{\btheta}$. By Taylor's Theorem, if $C_n\log(d\vee n)k_{0,max}^{2} = o(n)$, $p^{-1}(k_0^{(\bA)})^3(k_0^{(\bB)})^4 + q^{-1}(k_0^{(\bA)})^4(k_0^{(\bB)})^3 = O(1)$, $d^{2+3\kappa}=o(n)$ and $d^{(2+11\kappa)/4}(k_{0,max})^{7(2+\kappa)/4} = o(n)$.
%$k_{0,max}^{(10+5\kappa)/4}d^{(22+21\kappa)/4} = o(n)$,
%if $k_{0,max}^{(6+3\kappa)/4}d^{(18+19\kappa)/4} = o(n)$,
%$d^{18+13\kappa}\{k_0^{(\bA)}\}^{4+3\kappa}\{k_0^{(\bB)}\}^{2+\kappa} = o(n^{4+\kappa})$,  
we have 
\begin{align}\label{eq:Taylor_band}
    \bG_{\rm band} =  \bH_{\rm band} (\hat{\btheta} - \btheta) + o_{\rm p}(\|\hat{\btheta} - \btheta\|_2)\,,
\end{align}
where 
\begin{align*}
    \bG_{\rm band} = \begin{pmatrix}
        \bX_{\tilde{\ba}_1}'\bY_{\tilde{\ba}_1} - \bX_{\tilde{\ba}_1}'\bX_{\tilde{\ba}_1}\bar{\tilde{\ba}}_1  \\
        \vdots \\
        \bX_{\tilde{\ba}_p}'\bY_{\tilde{\ba}_p} - \bX_{\tilde{\ba}_p}'\bX_{\tilde{\ba}_p}\bar{\tilde{\ba}}_p  \\
        \bX_{\tilde{\bb}_1}'\bY_{\tilde{\bb}_1} - \bX_{\tilde{\bb}_1}'\bX_{\tilde{\bb}_1}\bar{\tilde{\bb}}_1  \\
        \vdots \\
        \bX_{\tilde{\bb}_q}'\bY_{\tilde{\bb}_q} - \bX_{\tilde{\bb}_q}'\bX_{\tilde{\bb}_q}\bar{\tilde{\bb}}_q 
        \end{pmatrix}\,,~
    \bH_{\rm band} = \begin{pmatrix}
        \frac{\partial {\bf f}_1(\btheta)}{\partial \bar{\btheta}} \\
        \vdots \\
        \frac{\partial {\bf f}_p(\btheta)}{\partial \bar{\btheta}} \\
        \frac{\partial {\bf f}_{p+1}(\btheta)}{\partial \bar{\btheta}} \\
        \vdots \\
        \frac{\partial {\bf f}_{p+q}(\btheta)}{\partial \bar{\btheta}}
    \end{pmatrix}\,,
\end{align*}
with $\bX_{\tilde{\ba}_m}$ and $\bX_{\tilde{\bb}_j}$ specified in \eqref{LS_A} and \eqref{LS_B}, respectively. 

Now we derive the asymptotic normality of $\bG_{\rm band}$. Similar to the derivation of \eqref{eq:Sigmajk_1}, we have 
\begin{equation}\label{eq:Sigmajk_1_band}
\hat\bSigma_{jk}(1) =  \sum_{i=1}^q \bA_0 \hat{\bSigma}_{ik}(1) b_{j, i}^{(0)} + \sum_{i=1}^q \bA_1 \hat{\bSigma}_{ik}(0) b_{j, i}^{(1)} + \frac{1}{n}\sum_{t=2}^n \bE_{t, \cdot j}\bX_{t-1, \cdot k}' \,.
\end{equation}
For $m \in [p]$ and $j \in [q]$, recall 
\begin{align*}
\bY_{\tilde{\ba}_m} = 
\begin{pmatrix}
\hat \bSigma_{11}(1)' \be_m \\
\hat \bSigma_{12}(1)' \be_m \\
\vdots \\
\hat \bSigma_{qq}(1)' \be_m
\end{pmatrix} ~ \mbox{and} ~
\bY_{\tilde{\bb}_j} = 
\begin{pmatrix}
\text{vec}\{\hat \bSigma_{j1}(1)\} \\
\text{vec}\{\hat \bSigma_{j2}(1)\} \\
\vdots\\
\text{vec}\{\hat \bSigma_{jq}(1)\}
\end{pmatrix}\,.
\end{align*}
Hence, it holds that 
\begin{align*}
\bY_{\tilde{\ba}_m} = \bX_{\tilde{\ba}_m}\tilde{\ba}_m + \boldsymbol{\mathcal{E}}_{\ba_m}
\end{align*}
and 
\begin{align*}
\bY_{\tilde{\bb}_j} = \bX_{\tilde{\bb}_j}\tilde{\bb}_j +\boldsymbol{\mathcal{E}}_{\bb_j}\,,
\end{align*}
where $\boldsymbol{\mathcal{E}}_{\ba_m}$ and $\boldsymbol{\mathcal{E}}_{\bb_j}$ are specified in \eqref{eq:YamYbj}. We then have 
\begin{align}\label{eq:CLT_term_band}
    \bG_{\rm band} = \begin{pmatrix}
        \bX_{\tilde{\ba}_1}'\boldsymbol{\mathcal{E}}_{\ba_1} \\
        \vdots \\
        \bX_{\tilde{\ba}_p}'\boldsymbol{\mathcal{E}}_{\ba_p} \\
        \bX_{\tilde{\bb}_1}'\boldsymbol{\mathcal{E}}_{\bb_1} \\
        \vdots \\
        \bX_{\tilde{\bb}_q}'\boldsymbol{\mathcal{E}}_{\bb_q}
    \end{pmatrix} = \begin{pmatrix}
        \tilde{\bX}_{\tilde{\ba}_1}'\boldsymbol{\mathcal{E}}_{\ba_1} \\
        \vdots \\
        \tilde{\bX}_{\tilde{\ba}_p}'\boldsymbol{\mathcal{E}}_{\ba_p} \\
        \tilde{\bX}_{\tilde{\bb}_1}'\boldsymbol{\mathcal{E}}_{\bb_1} \\
        \vdots \\
        \tilde{\bX}_{\tilde{\bb}_q}'\boldsymbol{\mathcal{E}}_{\bb_q}
    \end{pmatrix} + \begin{pmatrix}
        (\bX_{\tilde{\ba}_1}-\tilde{\bX}_{\tilde{\ba}_1})'\boldsymbol{\mathcal{E}}_{\ba_1} \\
        \vdots \\
        (\bX_{\tilde{\ba}_p}-\tilde{\bX}_{\tilde{\ba}_p})'\boldsymbol{\mathcal{E}}_{\ba_p} \\
        (\bX_{\tilde{\bb}_1}-\tilde{\bX}_{\tilde{\bb}_1})'\boldsymbol{\mathcal{E}}_{\bb_1} \\
        \vdots \\
        (\bX_{\tilde{\bb}_q}-\tilde{\bX}_{\tilde{\bb}_q})'\boldsymbol{\mathcal{E}}_{\bb_q}
    \end{pmatrix} := \tilde{\bG}_{\rm band} + \tilde{\boldsymbol{\mathcal{U}}}_{\rm band}\,,
\end{align}
where $\tilde{\bX}_{\tilde{\ba}_m}$ and $\tilde{\bX}_{\tilde{\bb}_j}$ be $pq^2 \times r_{m}$ and $p^2q \times \ell_{j}$ matrices obtained from $\bX_{\tilde{\ba}_m}$ and $\bX_{\tilde{\bb}_j}$, respectively, by replacing each $\hat{\bSigma}_{ik}(1)$ and $\hat{\bSigma}_{ik}(0)$ with their population counterparts $\bSigma_{ik}(1)$ and $\bSigma_{ik}(0)$.
%are specified in \eqref{eq:Xam_tilde} and \eqref{eq:Xbj_tilde}, respectively. 

Analogous to the derivation of \eqref{eq:Xam_1}, by Lemma \ref{lem:Sigmajk}, for any $x > 0$, we have
\begin{align*}
    \mathbb{P}(\|\bX_{\tilde{\ba}_m}-\tilde{\bX}_{\tilde{\ba}_m}\|_{F}^2 > x) & \lesssim pq^2 k_0^{(\bA)} \mathbb{P}\Big( \max_{i,k} \|\hat\bSigma_{ik}(1) - \bSigma_{ik}(1)\|_{\infty} > \frac{C_1x^{1/2}}{p^{1/2}q \{k_0^{(\bA)} \}^{1/2}k_0^{(\bB)}} \Big) \\
    &~~~~~~ + pq^2 k_0^{(\bA)} \mathbb{P}\Big( \max_{i,k} \|\hat\bSigma_{ik}(0) - \bSigma_{ik}(0)\|_{\infty} > \frac{C_1x^{1/2}}{p^{1/2}q \{k_0^{(\bA)} \}^{1/2}k_0^{(\bB)}} \Big) \\
    & \lesssim x^{-(2+\kappa)/(2\kappa)}\{k_0^{(\bA)}\}^{(2+3\kappa)/(2\kappa)}\{k_0^{(\bB)}\}^{(2+\kappa)/\kappa}p^{(2+7\kappa)/(2\kappa)}q^{(2+5\kappa)/\kappa}n^{-2/\kappa}\,.
\end{align*}
It then holds that 
\begin{align*}
    & \mathbb{P}(\max_{m}\|\bX_{\tilde{\ba}_m}-\tilde{\bX}_{\tilde{\ba}_m}\|_{F}^2 > x) \\
    &~~~~~~~~ \lesssim x^{-(2+\kappa)/(2\kappa)}\{k_0^{(\bA)}\}^{(2+3\kappa)/(2\kappa)}\{k_0^{(\bB)}\}^{(2+\kappa)/\kappa}p^{(2+3\kappa)/(2\kappa)}q^{(2+5\kappa)/\kappa}n^{-2/\kappa}\,,
    %x^{-(2+\kappa)/\kappa}\{k_0^{(\bA)}\}^{(2+\kappa)/\kappa}\{k_0^{(\bB)}\}^{(4+2\kappa)/\kappa}p^{(2+5\kappa)/\kappa}q^{(4+6\kappa)/\kappa}n^{-(4+\kappa)/\kappa}\,,
\end{align*}
which implies that 
\begin{align*}
    \max_{m \in [p]}\|\bX_{\tilde{\ba}_m}-\tilde{\bX}_{\tilde{\ba}_m}\|_{F}^2 = O_{\rm p}\big[\{k_0^{(\bA)}\}^{(2+3\kappa)/(2+\kappa)}\{k_0^{(\bB)}\}^2p^{(2+9\kappa)/(2+\kappa)}q^{(4+10\kappa)/(2+\kappa)}n^{-4/(2+\kappa)} \big]\,.
    %O_{\rm p}\big[k_0^{(\bA)}\{k_0^{(\bB)}\}^2p^{(2+5\kappa)/(2+\kappa)}q^{(4+6\kappa)/(2+\kappa)}n^{-(4+\kappa)/(4+2\kappa)} \big]\,.
\end{align*}
Analogously, we have 
\begin{align*}
    \max_{j \in [q]}\|\bX_{\tilde{\bb}_j}-\tilde{\bX}_{\tilde{\bb}_j}\|_{F}^2 = O_{\rm p} \big[ \{\{k_0^{(\bA)}\}^2\{k_0^{(\bB)}\}^{(2+3\kappa)/(2+\kappa)}p^{(4+10\kappa)/(2+\kappa)}q^{(2+9\kappa)/(2+\kappa)}n^{-4/(2+\kappa)} \big]\,.
    %O_{\rm p} \big[ \{\{k_0^{(\bA)}\}^2k_0^{(\bB)}p^{(4+6\kappa)/(2+\kappa)}q^{(2+5\kappa)/(2+\kappa)}n^{-(4+\kappa)/(4+2\kappa)} \big]\,.
\end{align*}
Recall $d = p \vee q$ and $k_{0,max} = k_0^{(\bA)} \vee k_0^{(\bB)}$. 
By \eqref{eq:Epsilon_am}, we have $\max_{m \in [p]}\|(\bX_{\tilde{\ba}_m}-\tilde{\bX}_{\tilde{\ba}_m})'\boldsymbol{\mathcal{E}}_{\ba_m}\|_2^2 = O_{\rm p}\{k_{0,max}^{(6+5\kappa)/(2+\kappa)}d^{(14+23\kappa)/(2+\kappa)}n^{-(6+\kappa)/(4+2\kappa)}\} = \max_{j \in [q]}\|(\bX_{\tilde{\bb}_j}-\tilde{\bX}_{\tilde{\bb}_j})'\boldsymbol{\mathcal{E}}_{\bb_j}\|_2^2$. Hence, we have $\|\tilde{\boldsymbol{\mathcal{U}}}_{\rm band}\|_2^2 = O_{\rm p}\{k_{0,max}^{(6+5\kappa)/(2+\kappa)}d^{(16+24\kappa)/(2+\kappa)}n^{-(6+\kappa)/(2+\kappa)}\}$,
which implies $\bG_{\rm band} = \tilde{\bG}_{\rm band} + o_{\rm p}(n^{-1/2})$ provided that $k_{0,max}^{6+5\kappa}d^{16+24\kappa} = o(n^{8+2\kappa})$.

As we will show in Sections \ref{subsec:BE_bound_band} and \ref{subsec:HtoU_band}, as $n \to \infty$, the following two assertions holds:
%Hence, by \eqref{eq:Taylor_band}, it suffices to prove the assertions \eqref{eq:CLT_mixing_band} and \eqref{eq:HtoU_band} below:
\begin{align}
    & \sqrt{n} \bT_{n}^{-1/2} \tilde{\bG}_{\rm band} \stackrel{\mathrm{d}}{\to} \mathcal{N}(0,\bI_{s})\,, \label{eq:BE_bound_band} \\
    &~~~~~~~ \bU_{\rm band}\bH_{\rm band}^{-1} \stackrel{\mathrm{p}}{\to} \bI_{s}\,, \label{eq:HtoU_band}
\end{align}
provided that $k_{0,max}^{\delta_1}d^{\delta_2} = o(n)$ with $\delta_1 = \max\{ 5, (6+3\kappa)/2 \}$, $\delta_2 = \max\{ 11, (10+9\kappa)/2 \}$ and $\eta_n = o\{n^{(\beta_2-1)(\rho_2-2)/(2\beta_2+2)}\}$.
%$k_{0,max}^{(6+3\kappa)/4}d^{(18+19\kappa)/4} = o(n)$, %$k_{0,max}^{6+4\kappa}d^{18+13\kappa} = o(n^{4+\kappa})$, 
%where $\bU_{\rm band}$ is specified in \eqref{eq:Umatrix_band}. 
By \eqref{eq:Taylor_band}, the conclusion of Theorem \ref{tm:mixing_banded} holds. We complete the proof of Theorem \ref{tm:mixing_banded}. $\hfill\Box$

\subsection{Proof of \eqref{eq:BE_bound_band}}\label{subsec:BE_bound_band}
% To prove \eqref{eq:CLT_mixing_band}, we need to show that $\sqrt{n}\bu'\tilde{\bG}_{\rm band}$ is asymptotically normal for any nonzero vector $\bu = (\bu_1',\bu_2')'\in \mathbb{R}^s$, where $\bu_1\in \mathbb{R}^{s_1}$, $\bu_2\in \mathbb{R}^{s_2}$, $s_1=\sum_{m=1}^p(|S_{\ba_m^{(0)}}|+|S_{\ba_m^{(1)}}|)$ and $s_2=\sum_{j=1}^q(|S_{\bb_j^{(0)}}|+|S_{\bb_j^{(1)}}|)$. 
Recall $\tilde{\mathcal{E}}_{\bX,t} = ({\rm vec}'(\bE_{t,\cdot 1}\bX_{t-1,\cdot 1}'),{\rm vec}'(\bE_{t,\cdot 1}\bX_{t-1,\cdot 2}'),\ldots,{\rm vec}'(\bE_{t,\cdot q}\bX_{t-1,\cdot q}') )'$. Then we have
\begin{align*}
    \tilde{\bX}_{\tilde{\ba}_m}'\boldsymbol{\mathcal{E}}_{\ba_m} & =     \begin{pmatrix} 
        \sum_{j=1}^q\sum_{k=1}^q \big\{ \sum_{i=1}^q b_{j,i}^{(0)}\bE_m^{(0)'}\bSigma_{ik}(1) \cdot \frac{1}{n}\sum_{t=2}^n \bX_{t-1, \cdot k}\be_m'\bE_{t,\cdot j} \big\} \\
        \sum_{j=1}^q\sum_{k=1}^q \big\{ \sum_{i=1}^q b_{j,i}^{(1)} \bE_m^{(1)'}\bSigma_{ik}(0) \cdot \frac{1}{n}\sum_{t=2}^n \bX_{t-1,\cdot k}\be_m' \bE_{t,\cdot j} \big\}
    \end{pmatrix} \\
    & = \frac{1}{n}\sum_{t=2}^n \tilde{\bX}_{\tilde{\ba}_m}'(\bI_{pq^2}\otimes\be_m') \cdot \tilde{\boldsymbol{\mathcal{E}}}_{\bX,t}\,,
\end{align*}
where $\be_m$ is the $p$-dimensional unit vector with the $m$-th element being 1. Write 
\begin{align*}
    \tilde{\bJ}_{\ba} = \begin{pmatrix}
        \tilde{\bX}_{\tilde{\ba}_1}'(\bI_{pq^2}\otimes\be_1')  \\
        \vdots \\
        \tilde{\bX}_{\tilde{\ba}_p}'(\bI_{pq^2}\otimes\be_p')
    \end{pmatrix} \in \mathbb{R}^{s_1\times (pq)^2}\,,
\end{align*}
where $s_1 = \sum_{m=1}^p r_m$.
Hence, it holds that
\begin{align}\label{eq:XEam_reformulate2}
    \tilde{\bG}_{{\rm band},1} := \begin{pmatrix}
        \tilde{\bX}_{\tilde{\ba}_1}'\boldsymbol{\mathcal{E}}_{\ba_1} \\
        \vdots \\
        \tilde{\bX}_{\tilde{\ba}_p}'\boldsymbol{\mathcal{E}}_{\ba_p}
    \end{pmatrix} & = \frac{1}{n}\sum_{t=2}^n \tilde{\bJ}_{\ba}  \tilde{\boldsymbol{\mathcal{E}}}_{\bX,t}\,.
\end{align}
Analogously, we have
\begin{align*}
    \tilde{\bX}_{\tilde{\bb}_j}'\boldsymbol{\mathcal{E}}_{\bb_j} & = \frac{1}{n}\sum_{t=2}^n \tilde{\bX}_{\tilde{\bb}_j}' \begin{pmatrix}
        {\rm vec}(\bE_{t,\cdot j}\bX_{t-1,\cdot 1}') \\
        \vdots \\
        {\rm vec}(\bE_{t,\cdot j}\bX_{t-1,\cdot q}')
    \end{pmatrix}\,.
\end{align*}
Write 
\begin{align*}
    \tilde{\bJ}_{\bb} = \begin{pmatrix}
        \tilde{\bX}_{\tilde{\bb}_1}' &   &  \\
        &   \ddots  & \\
        &   &  \tilde{\bX}_{\tilde{\bb}_q}'
    \end{pmatrix} \in \mathbb{R}^{s_2\times (pq)^2}\,,
\end{align*}
where $s_2 = \sum_{j=1}^q \ell_j$.
Hence, we have
\begin{align}\label{eq:XEbj_reformulate2}
    \tilde{\bG}_{{\rm band},2} := \begin{pmatrix}
        \tilde{\bX}_{\tilde{\bb}_1}'\boldsymbol{\mathcal{E}}_{\bb_1} \\
        \vdots \\
        \tilde{\bX}_{\tilde{\bb}_q}'\boldsymbol{\mathcal{E}}_{\bb_q}
    \end{pmatrix} & = \frac{1}{n}\sum_{t=2}^n \tilde{\bJ}_{\bb} \tilde{\boldsymbol{\mathcal{E}}}_{\bX,t}\,.
\end{align}
By the definition of $\tilde{\bG}_{\rm band}$, it holds that
\begin{align*}
    \tilde{\bG}_{\rm band} = \begin{pmatrix} \tilde{\bG}_{{\rm band},1} \\ \tilde{\bG}_{{\rm band},2} \end{pmatrix} =  \frac{1}{n}\sum_{t=2}^n \begin{pmatrix} \tilde{\bJ}_{\ba} \\ \tilde{\bJ}_{\bb}\end{pmatrix} \tilde{\boldsymbol{\mathcal{E}}}_{\bX,t} = \frac{1}{n}\sum_{t=2}^n \tilde{\bJ}\tilde{\boldsymbol{\mathcal{E}}}_{\bX,t} \,.
\end{align*} 
Given a vector $\bu \in \mathbb{R}^s$, it suffices to show that $\sqrt{n}\bu'\tilde{\bG}_{\rm band}$ follow a standard normal distribution asymptotically. By Assumption \ref{cond:mmt_bandedcase}, we have  $\mathbb{E}|\bu'\tilde{\bJ}\tilde{\boldsymbol{\mathcal{E}}}_{\bX,t}|^{\rho_2} < \eta_n$ and $\bu'\bT_n\bu \ge c_2$. By the Berry-Esseen bound in Theorem 2 of \cite{Sunklodas1984}, we have
\begin{align*}
    \sup_{z \in \mathbb{R}}\left| \mathbb{P}\bigg( \frac{\sqrt{n}\bu'\tilde{\bG}_{\rm band}}{\sqrt{\bu'\bT_n\bu}} \le z\bigg) - \Phi(z)  \right| \le \eta_n n^{-(\beta_2-1)(\rho_2-2)/(2\beta_2+2)}\,,
\end{align*}
where $\Phi(\cdot)$ is the cumulative distribution function of $\mathcal{N}(0,1)$.
It implies that \eqref{eq:BE_bound_band} holds, provided that %$d= o\{n^{(6+\kappa)/(18+25\kappa)}\}$ and 
$\eta_n = o\{n^{(\beta_2-1)(\rho_2-2)/(2\beta_2+2)}\}$. $\hfill\Box$

\subsection{Proof of \eqref{eq:HtoU_band}}\label{subsec:HtoU_band}
To prove \eqref{eq:HtoU_band}, it suffices to show 
% \begin{align*}
%     \|\bU_{\rm band} - \bH_{\rm band}\|_{2} = o_{\rm p}(1)\,,
% \end{align*}
% where $\|\cdot\|_2$ denotes the matrix operator norm, which is implied by 
\begin{align}\label{eq:infty_norm_band}
    \|\bU_{\rm band} - \bH_{\rm band}\|_{\infty} = o_{\rm p}(s^{-1})\,.
\end{align}
% Hence, it suffices to show \eqref{eq:infty_norm_band}. 

by the definition of $\bH_{\rm band}$, we first calculate $\partial (\bar{\bX}_{\tilde{\ba}_m}'\bar{\bX}_{\tilde{\ba}_m}\bar{\tilde{\ba}}_m - \bar{\bX}_{\tilde{\ba}_m}'\bY_{\tilde{\ba}_m})/\partial \bar{\btheta}$ and $\partial (\bar{\bX}_{\tilde{\bb}_j}'\bar{\bX}_{\tilde{\bb}_j}\bar{\tilde{\bb}}_j - \bar{\bX}_{\tilde{\bb}_j}'\bY_{\tilde{\bb}_j})/\partial \bar{\btheta}$ for any $m \in [p]$ and $j \in [q]$. It follows from the direct computation that 
\begin{align*}
    \frac{\partial \bar{\bX}_{\tilde{\ba}_m}}{\partial \bar{b}_{j,l}^{(0)}} = \begin{pmatrix}
        {\bf 0}_{(j-1)pq \times r_{m,0}}  & {\bf 0}_{(j-1)pq \times r_{m,1}} \\
        \hat{\bSigma}_{l1}(1)' \bE_{m}^{(0)}\,, & {\bf 0}_{p \times r_{m,1}} \\
        \vdots & \vdots \\
        \hat{\bSigma}_{lq}(1)' \bE_{m}^{(0)}\,, & {\bf 0}_{p \times r_{m,1}} \\
        {\bf 0}_{(q-j)pq \times r_{m,0}}  & {\bf 0}_{(q-j)pq \times r_{m,1}}
    \end{pmatrix}
\end{align*}
for any $m \in [p]$, $j \in [q]$ and $l \in \mathcal{B}_{j}^{(0)}$. Then, it holds that
\begin{align}\label{eq:partial_band_am_1}
    \frac{\partial (\bar{\bX}_{\tilde{\ba}_m}'\bar{\bX}_{\tilde{\ba}_m}\bar{\tilde{\ba}}_m)}{\partial \bar{b}_{j,l}^{(0)} } & = \bigg( \frac{\partial\bar{\bX}_{\tilde{\ba}_m}}{\partial\bar{b}_{j,l}^{(0)}} \bigg)'\bar{\bX}_{\tilde{\ba}_m}\bar{\tilde{\ba}}_m + \bar{\bX}_{\tilde{\ba}_m}'\bigg( \frac{\partial\bar{\bX}_{\tilde{\ba}_m}}{\partial\bar{b}_{j,l}^{(0)}} \bigg)\bar{\tilde{\ba}}_m \notag\\
    & = \begin{pmatrix}
        \{\bzeta_{\tilde{\ba}_m,l}^{(0)}+\bzeta_{\tilde{\ba}_m,l}^{(0)'}\}\bar{\tilde{\ba}}_{m}^{(0)}+\bzeta_{\tilde{\ba}_m,l}^{(1)}\bar{\tilde{\ba}}_{m}^{(1)} \\
        \bzeta_{\tilde{\ba}_m,l}^{(1)'}\bar{\tilde{\ba}}_{m}^{(0)}
    \end{pmatrix}\,,
\end{align}
where 
\begin{align*}
    \bzeta_{\tilde{\ba}_m,l}^{(0)} & = \sum_{k=1}^q \Big\{ \bE_{m}^{(0)'}\hat{\bSigma}_{lk}(1) \Big\}\Big\{ \sum_{i=1}^q \bar{b}_{j, i}^{(0)}\hat{\bSigma}_{ik}(1)'\bE_{m}^{(0)} \Big\}\,, \\
   \bzeta_{\tilde{\ba}_m,l}^{(1)} & = \sum_{k=1}^q \Big\{ \bE_{m}^{(0)'}\hat{\bSigma}_{lk}(1) \Big\}\Big\{ \sum_{i=1}^q \bar{b}_{j, i}^{(1)}\hat{\bSigma}_{ik}(0)'\bE_{m}^{(1)} \Big\}\,. 
\end{align*}
In the same manner, we can show that 
\begin{align}\label{eq:partial_band_am_2}
    \frac{\partial (\bar{\bX}_{\tilde{\ba}_m}'\bY_{\tilde{\ba}_m})}{\partial \bar{b}_{j,l}^{(0)} } & = \bigg( \frac{\partial\bar{\bX}_{\tilde{\ba}_m}}{\partial\bar{b}_{j,l}^{(0)}} \bigg)'\bY_{\tilde{\ba}_m}  = \begin{pmatrix}
        \sum_{k=1}^q \bE_{m}^{(0)'}\hat{\bSigma}_{lk}(1) \hat\bSigma_{jk}(1)'\be_m \\
        {\bf 0}_{r_{m,1} \times 1}
    \end{pmatrix}\,.
\end{align}
Combining \eqref{eq:partial_band_am_1} and \eqref{eq:partial_band_am_2}, by \eqref{eq:Sigmajk_1_band}, we have
\begin{align*}
        \frac{\partial {\bf f}_m(\btheta)}{\partial \bar{b}_{j,l}^{(0)}} = 
        & \frac{\partial (\bar{\bX}_{\tilde{\ba}_m}'\bar{\bX}_{\tilde{\ba}_m}\bar{\tilde{\ba}}_m - \bar{\bX}_{\tilde{\ba}_m}'\bY_{\tilde{\ba}_m})}{\partial \bar{b}_{j,l}^{(0)}}\bigg|_{\bar{\btheta} = \btheta} \notag\\
    & = 
    \begin{pmatrix}
        \sum_{k=1}^q \big\{ \sum_{i=1}^q \bE_{m}^{(0)'}\hat{\bSigma}_{ik}(1) b_{j, i}^{(0)} \big\} \hat{\bSigma}_{lk}(1)'\bE_{m}^{(0)} \big\} \\
        \sum_{k=1}^q \big\{ \sum_{i=1}^q \bE_{m}^{(1)'}\hat{\bSigma}_{ik}(0) b_{j, i}^{(1)} \big\} \hat{\bSigma}_{lk}(1)'\bE_{m}^{(0)} \big\}
    \end{pmatrix}\tilde{\ba}_m^{(0)} \notag\\
    &~~~~~~~ - \begin{pmatrix}
        \sum_{k=1}^q \bE_{m}^{(0)'}\hat{\bSigma}_{lk}(1) \big\{\frac{1}{n}\sum_{t=2}^n E_{t, m, j}\bX_{t-1, \cdot k} \big\} \\
        {\bf 0}_{r_{m,1} \times 1}
    \end{pmatrix}
    %&~~~~~~~~~~ = \bX_{\ba_m}'\bpsi_{\ba_m,j}^{(0)} + (\mathcal{E}_{\ba_m}\,, ~ {\bf 0}_{pq^2 \times 1})'\bpsi_{\ba_m,j}^{(0)}\,,
\end{align*}
Analogous to the derivation of \eqref{eq:dev1}, we have
\begin{align}\label{eq:dev_band1}
    \sum_{k=1}^q \bE_{m}^{(0)'}\hat{\bSigma}_{lk}(1) \Big\{\frac{1}{n}\sum_{t=2}^n \bX_{t-1, \cdot k}E_{t, m, j} \Big\} & = \sum_{k=1}^q \bE_{m}^{(0)'}\bSigma_{lk}(1) \Big\{\frac{1}{n}\sum_{t=2}^n \bX_{t-1, \cdot k}E_{t, m, j}\Big\} \notag\\
    &~~~~~~ +  \sum_{k=1}^q  \bE_{m}^{(0)'}\big\{\hat{\bSigma}_{lk}(1) - \bSigma_{lk}(1) \big\} \Big\{\frac{1}{n}\sum_{t=2}^n \bX_{t-1, \cdot k}E_{t, m, j}\Big\} \notag\\
    & := \tilde{\mathrm{S}}_{1,m,j,l}+\tilde{\mathrm{S}}_{2,m,j,l}\,.
\end{align}
We then derive the stochastic orders of $\max_{m\in[p],j\in[q],l\in \mathcal{B}_{j}^{(0)}}\|\tilde{\mathrm{S}}_{1,m,j,l}\|_2$ and $\max_{m\in[p],j\in[q],l\in \mathcal{B}_{j}^{(0)}}\|\tilde{\mathrm{S}}_{2,m,j,l}\|_2$, respectively. We notice that
\begin{align*}
    & \tilde{\mathrm{S}}_{1,m,j,l} = \bE_{m}^{(0)'} \check{\bD}_j \boldsymbol{\mathcal{E}}_{\ba_m}\,,
\end{align*}
where $\boldsymbol{\mathcal{E}}_{\ba_m}$ is specified in \eqref{eq:YamYbj} and 
\begin{align*}
    \check{\bD}_j = \Big(\mathbf{0}_{ p \times (j-1)pq}\,, ~ \bSigma_{l1}(1)\,, \ldots, \bSigma_{lq}(1)\,, ~ \mathbf{0}_{ p \times (q-j)pq} \Big)\,.
\end{align*}
Under Conditions \ref{cond:mixing}, by \eqref{eq:Epsilon_am}, we have shown that $\|\boldsymbol{\mathcal{E}}_{\ba_m}\|_2^2 = O_{\rm p}(pq^2/n)$. Since $\|\bE_{m}^{(0)'} \check{\bD}_j\|_F^2 = O_{\rm p}\{k_0^{(\bA)}pq\}$, then it holds that 
\begin{align*}
    \|\tilde{\mathrm{S}}_{1,m,j,l}\|_2 \le \|\bE_{m}^{(0)'} \check{\bD}_j\|_F \cdot \|\boldsymbol{\mathcal{E}}_{\ba_m}\|_2 \le O_{\rm p}\Big(\frac{ \{k_0^{(\bA)}\}^{1/2}pq^{3/2}}{n^{1/2}} \Big)\,.
\end{align*}
Hence, we have  
\begin{align*}
    \max_{m\in[p],j\in[q],l\in \mathcal{B}_{j}^{(0)}}\|\tilde{\mathrm{S}}_{1,m,j,l}\|_2 \le \sum_{m\in[p],j\in[q],l\in \mathcal{B}_{j}^{(0)}} \|\tilde{\mathrm{S}}_{1,m,j,l}\|_2 \le O_{\rm p}\Big(\frac{ \{k_0^{(\bA)}\}^{1/2}k_0^{(\bB)}p^2q^{5/2}}{n^{1/2}} \Big)\,,
\end{align*}
which implies $\max_{m\in[p],j\in[q],l\in \mathcal{B}_{j}^{(0)}}\|\tilde{\mathrm{S}}_{1,m,j,l}\|_2 = o_{\rm p}(s^{-1})$, where $s \lesssim dk_{0,max}$, provided that $k_{0,max}^5d^{11} = o(n)$. 
Using the similar arguments, by Lemma \ref{lem:Sigmajk}, it holds that $\max_{m\in[p],j\in[q],l\in \mathcal{B}_{j}^{(0)}}\|\tilde{\mathrm{S}}_{2,m,j,l}\|_2 = o_{\rm p}(s^{-1})$ provided that $k_{0,max}^{6(2+\kappa)}d^{16+16\kappa} = o(n^{6+\kappa})$.
% Under Conditions \ref{cond:mixing} and \ref{cond:mmt_bandedcase}, 
% by \eqref{eq:longruncov_1}, we show that $\bu'\boldsymbol{\mathcal{E}}_{\ba_m} = O_{p}(n^{-1/2})$ for any $\bu \in \mathbb{R}^{pq^2}$. Since $\max_{l,k \in [q]}\|\bSigma_{lk}(1)\|_{\infty} = O(1)$, we have $\tilde{\mathrm{S}}_1 = o_{\rm p}(1)$. Using the similar arguments, by Lemma \ref{lem:Sigmajk}, it holds that $\tilde{\mathrm{S}}_2 = o_{\rm p}(1)$ provided that $d = o\{n^{(3+\kappa)/(2\kappa)}\}$. 
 By \eqref{eq:dev_band1}, it then holds that 
 % Similar to the derivation of \eqref{eq:varianceSnp}, under Conditions \ref{cond:mixing} and \ref{cond:mmt_diagonalcase}, we can show the variance of the term $\sum_{k=1}^q \{ \sum_{i=1}^q \be_m'\bW_0\bSigma_{ik}(1) v_{j, i}^{(0)} \}\{n^{-1}\sum_{t=2}^n \bX_{t-1, \cdot k}E_{t, m, j}\}$ is $O(n^{-1})$. By Lemma \ref{lem:Sigmajk}, it then implies that
 \begin{align}\label{eq:derv_am_bjl0}
         \frac{\partial {\bf f}_m(\btheta)}{\partial \bar{b}_{j,l}^{(0)}} = \begin{pmatrix}
         \sum_{k=1}^q \big\{ \sum_{i=1}^q \bE_{m}^{(0)'}\hat{\bSigma}_{ik}(1) b_{j, i}^{(0)} \big\} \hat{\bSigma}_{lk}(1)'\bE_{m}^{(0)} \big\} \\
         \sum_{k=1}^q \big\{ \sum_{i=1}^q \bE_{m}^{(1)'}\hat{\bSigma}_{ik}(0) b_{j, i}^{(1)} \big\} \hat{\bSigma}_{lk}(1)'\bE_{m}^{(0)} \big\}
     \end{pmatrix}\tilde{\ba}_m^{(0)} + o_{\rm p}(s^{-1})\,.
 \end{align}
provided that $k_{0,max}^5d^{11} = o(n)$.
%provided that $d=o\{n^{(3+\kappa)/(2\kappa)}\}$. 

% % Similar to the derivation of \eqref{eq:derivative_am_j0_pre}, under Conditions \ref{cond:mixing}, \ref{cond:moments} and \ref{cond:mmt_bandedcase}, we can show the variance of the term $\sum_{k=1}^q \bE_{m}^{(0)'}\bSigma_{lk}(1) \{\frac{1}{n}\sum_{t=2}^n E_{t, m, j}\bX_{t-1, \cdot k} \}$ is $O(n^{-1})$. By Lemma \ref{lem:Sigmajk}, it then implies that 
% % \begin{align}\label{eq:derv_am_bjl0}
% %         \frac{\partial {\bf f}_m(\btheta)}{\partial \bar{b}_{j,l}^{(0)}} = \begin{pmatrix}
% %         \sum_{k=1}^q \big\{ \sum_{i=1}^q \bE_{m}^{(0)'}\hat{\bSigma}_{ik}(1) b_{j, i}^{(0)} \big\} \hat{\bSigma}_{lk}(1)'\bE_{m}^{(0)} \big\} \\
% %         \sum_{k=1}^q \big\{ \sum_{i=1}^q \bE_{m}^{(1)'}\hat{\bSigma}_{ik}(0) b_{j, i}^{(1)} \big\} \hat{\bSigma}_{lk}(1)'\bE_{m}^{(0)} \big\}
% %     \end{pmatrix}\tilde{\ba}_m^{(0)} + o_{\rm p}(1)\,.
% % \end{align}

Analogously, for any $m \in [p]$, $j \in [q]$ and $l \in \mathcal{B}_{j}^{(1)}$, we can show that the derivative of $\bar{\bX}_{\tilde{\ba}_m}$ with respect to $\bar{b}_{j,l}^{(1)}$ is
\begin{align*}
    \frac{\partial \bar{\bX}_{\tilde{\ba}_m}}{\partial \bar{b}_{j,l}^{(1)}} = \begin{pmatrix}
        {\bf 0}_{(j-1)pq \times r_{m,0}}  & {\bf 0}_{(j-1)pq \times r_{m,1}} \\
        {\bf 0}_{p \times r_{m,0}}\,, & \hat{\bSigma}_{l1}(0)' \bE_{m}^{(1)} \\
        \vdots & \vdots \\
        {\bf 0}_{p \times r_{m,0}}\,, & \hat{\bSigma}_{lq}(0)' \bE_{m}^{(1)} \\
        {\bf 0}_{(q-j)pq \times r_{m,0}}  & {\bf 0}_{(q-j)pq \times r_{m,1}}
    \end{pmatrix}\,.
\end{align*}
Using the similar arguments for the derivation of \eqref{eq:derv_am_bjl0}, it holds that
\begin{align}\label{eq:derv_am_bjl1}
       \frac{\partial {\bf f}_m(\btheta)}{\partial \bar{b}_{j,l}^{(1)}} & =  \frac{\partial (\bar{\bX}_{\tilde{\ba}_m}'\bar{\bX}_{\tilde{\ba}_m}\bar{\tilde{\ba}}_m-\bar{\bX}_{\tilde{\ba}_m}'\bY_{\tilde{\ba}_m})}{\partial \bar{b}_{j,l}^{(1)}}\bigg|_{\bar{\btheta} = \btheta} \notag\\
    & =  \begin{pmatrix}
        \sum_{k=1}^q \big\{ \sum_{i=1}^q \bE_{m}^{(0)'}\hat{\bSigma}_{ik}(1) b_{j, i}^{(0)} \big\} \hat{\bSigma}_{lk}(0)'\bE_{m}^{(1)} \big\} \\
        \sum_{k=1}^q \big\{ \sum_{i=1}^q \bE_{m}^{(1)'}\hat{\bSigma}_{ik}(0) b_{j, i}^{(1)} \big\} \hat{\bSigma}_{lk}(0)'\bE_{m}^{(1)} \big\}
    \end{pmatrix}\tilde{\ba}_m^{(1)} + o_{\rm p}(s^{-1})\,,
\end{align}
provided that $k_{0,max}^5d^{11} = o(n)$.
%provided that $d=o\{n^{(3+\kappa)/(2\kappa)}\}$. 
Combining \eqref{eq:derv_am_bjl0} and \eqref{eq:derv_am_bjl1}, we have
\begin{align*}%\label{eq:derv_am_j_band}
    \frac{\partial {\bf f}_m(\btheta)}{\partial \bar{\tilde{\bb}}_{j}} = \bX_{\tilde{\ba}_m}'\bZ_{\tilde{\ba}_m,j} + o_{\rm p}(s^{-1})\,,
\end{align*}
provided that $k_{0,max}^5d^{11} = o(n)$, where
%provided that $d=o\{n^{(3+\kappa)/(2\kappa)}\}$, where
\begin{align*}
    \bZ_{\tilde{\ba}_m,j} = \begin{pmatrix}
        \boldsymbol{\psi}_{\tilde{\ba}_m,j,j_1^{(0)}}^{(0)}\,, & \cdots\,, & \boldsymbol{\psi}_{\tilde{\ba}_m,j,j_{\ell_{j,0}}^{(0)}}^{(0)}\,, &  \boldsymbol{\psi}_{\tilde{\ba}_m,j,j_1^{(1)}}^{(1)}\,, & \cdots\,, & \boldsymbol{\psi}_{\tilde{\ba}_m,j,j_{\ell_{j,1}}^{(1)}}^{(1)}
    \end{pmatrix}
\end{align*}
is a $pq^2\times \ell_{j}$ matrix with 
\begin{align*}
    \boldsymbol{\psi}_{\tilde{\ba}_m,j,l}^{(0)} = \big({\bf 0}_{1 \times (j-1)pq}\,, ~ \tilde{\ba}_m^{(0)'}\bE_{m}^{(0)'}\hat{\bSigma}_{l1}(1)\,,\cdots\,,\tilde{\ba}_m^{(0)'}\bE_{m}^{(0)'}\hat{\bSigma}_{lq}(1)\,,~ {\bf 0}_{1 \times (q-j)pq} \big)'\,,\\
    \boldsymbol{\psi}_{\tilde{\ba}_m,j,l}^{(1)} = \big({\bf 0}_{1 \times (j-1)pq}\,, ~ \tilde{\ba}_m^{(1)'}\bE_{m}^{(1)'}\hat{\bSigma}_{l1}(0)\,,\cdots\,,\tilde{\ba}_m^{(1)'}\bE_{m}^{(1)'}\hat{\bSigma}_{lq}(0)\,,~ {\bf 0}_{1 \times (q-j)pq} \big)'\,.
\end{align*}
%{\color{red} By Lemma \ref{lem:Sigmajk}, it holds that $\mathcal{E}_{\ba_m}'\bZ_{\ba_m,j}= \mathcal{E}_{\ba_m}'\tilde{\bZ}_{\ba_m,j} + o_{\rm p}(1)$, where $\tilde{\bZ}_{\ba_m,j}$ is defined in the same manner as $\bZ_{\ba_m,j}$ except that $\hat{\bSigma}_{jk}(1)$ and $\hat{\bSigma}_{jk}(0)$ are replaced by $\bSigma_{jk}(1)$ and $\bSigma_{jk}(0)$, respectively. Since $\mathbb{E}(\mathcal{E}_{\ba_m}'\tilde{\bZ}_{\ba_m,j}) = 0$, by calculating its variance under Condition \ref{cond:mixing}, it can be shown that $\mathcal{E}_{\ba_m}'\tilde{\bZ}_{\ba_m,j}=o_{\rm p}(1)$. }
Furthermore, we have
\begin{align*}
    \frac{\partial {\bf f}_m(\btheta)}{\partial \bar{\tilde{\ba}}_m} = \bX_{\tilde{\ba}_m}'\bX_{\tilde{\ba}_m}\,, ~\mbox{and}~ \frac{\partial {\bf f}_m(\btheta)}{\partial \bar{\tilde{\ba}}_k} = {\bf 0}_{r_{m} \times r_{m} } ~\mbox{for}~ k \neq m\,.
\end{align*}
Hence, if $k_{0,max}^5d^{11} = o(n)$, it holds that
%if $d=o\{n^{(3+\kappa)/(2\kappa)}\}$, it holds that 
\begin{align}\label{eq:der_am_band}
    &\frac{\partial {\bf f}_m(\btheta)}{\partial \bar{\btheta}} = \big({\bf 0}_{r_{1} \times r_{1}}\,, \cdots\,, \bX_{\tilde{\ba}_m}'\bX_{\tilde{\ba}_m}\,, \cdots\,, {\bf 0}_{r_{p} \times r_{p}}\,,  \bX_{\tilde{\ba}_m}'\bZ_{\tilde{\ba}_m,1}\,, \cdots\,, \bX_{\tilde{\ba}_m}'\bZ_{\tilde{\ba}_m,q}\big) + o_{\rm p}(s^{-1})\,.
\end{align} 

On the other hand, for any $j \in [q]$, $m \in [p]$ and $l \in \mathcal{A}_{m}^{(0)}$, it follows from the direct computation that 
\begin{align*}
    \frac{\partial \bar{\bX}_{\tilde{\bb}_j}}{\partial \bar{a}_{m,l}^{(0)}} = \begin{pmatrix}
        \{\hat{\bSigma}_{j_1^{(0)}1}(1)' \be_l\}\otimes \be_m\,, & \cdots\,, & \{\hat{\bSigma}_{j_{\ell_{j,0}}^{(0)}1}(1)' \be_l\}\otimes \be_m\,, & {\bf 0}_{p^2 \times \ell_{j,1}} \\
        \vdots &  &  \vdots  & \vdots \\
        \{\hat{\bSigma}_{j_1^{(0)}q}(1)' \be_l\}\otimes \be_m\,, & \cdots\,, & \{\hat{\bSigma}_{j_{\ell_{j,0}}^{(0)}q}(1)' \be_l\}\otimes \be_m\,, & {\bf 0}_{p^2 \times \ell_{j,1}} \\
    \end{pmatrix}\,.
\end{align*}
Then, we have
\begin{align}\label{eq:band_bj_1}
    \bigg( \frac{\partial\bar{\bX}_{\tilde{\bb}_j}}{\partial\bar{a}_{m,l}^{(0)}} \bigg)'\bar{\bX}_{\tilde{\bb}_j}\bar{\tilde{\bb}}_j  = \begin{pmatrix}
        \phi_{j_1^{(0)},m,l}^{(j)} \\
        \vdots \\
        \phi_{j_{\ell_{j,0}}^{(0)},m,l}^{(j)} \\
        {\bf 0}_{\ell_{j,1} \times 1}
    \end{pmatrix}\,,
\end{align}
where 
\begin{align*}
    & \phi_{v,m,l}^{(j)}  = \sum_{i \in \mathcal{B}_{j}^{(0)}} \sum_{k=1}^q \big[\{\be_l'\hat{\bSigma}_{vk}(1) \}\otimes \be_m' \big]\text{vec}\{ \bar{\bA}_0 \hat{\bSigma}_{ik}(1)\} \bar{b}_{j, i}^{(0)} \\
    &~~~~~~~~~~~~~~~~~~~~~~~ + \sum_{i \in \mathcal{B}_{j}^{(1)}}  \sum_{k=1}^q \big[\{\be_l'\hat{\bSigma}_{vk}(1) \}\otimes \be_m' \big]\text{vec}\{ \bar{\bA}_1 \hat{\bSigma}_{ik}(0)\} \bar{b}_{j, i}^{(1)}\,.
\end{align*}
Similarly, it holds that 
\begin{align}\label{eq:band_bj_2}
    \bar{\bX}_{\tilde{\bb}_j}'\bigg( \frac{\partial\bar{\bX}_{\tilde{\bb}_j}}{\partial\bar{a}_{m,l}^{(0)}} \bigg)\bar{\tilde{\bb}}_j  = \begin{pmatrix}
        \sum_{i \in \mathcal{B}_{j}^{(0)}} \big[ \sum_{k=1}^q \text{vec}'\{ \bar{\bA}_0 \hat{\bSigma}_{j_1^{(0)}k}(1)\} \big\{(\hat{\bSigma}_{ik}(1)'\be_l )\otimes \be_m\big\} \big] \bar{b}_{j, i}^{(0)} \\
        \vdots \\
        \sum_{i \in \mathcal{B}_{j}^{(0)}} \big[ \sum_{k=1}^q \text{vec}'\{ \bar{\bA}_0 \hat{\bSigma}_{j_{\ell_{j,0}}^{(0)}k}(1)\} \big\{(\hat{\bSigma}_{ik}(1)'\be_l )\otimes \be_m\big\} \big] \bar{b}_{j, i}^{(0)} \\
        \sum_{i \in \mathcal{B}_{j}^{(0)}} \big[ \sum_{k=1}^q \text{vec}'\{ \bar{\bA}_1 \hat{\bSigma}_{j_1^{(1)}k}(0)\} \big\{(\hat{\bSigma}_{ik}(1)'\be_l )\otimes \be_m\big\} \big] \bar{b}_{j, i}^{(0)} \\
        \vdots \\
        \sum_{i \in \mathcal{B}_{j}^{(0)}} \big[ \sum_{k=1}^q \text{vec}'\{ \bar{\bA}_1 \hat{\bSigma}_{j_{\ell_{j,1}}^{(1)}k}(0)\} \big\{(\hat{\bSigma}_{ik}(1)'\be_l )\otimes \be_m\big\} \big] \bar{b}_{j, i}^{(0)}
    \end{pmatrix}\,.
\end{align}
In the same manner, we can show that 
\begin{align}\label{eq:band_bj_3}
    \frac{\partial (\bar{\bX}_{\tilde{\bb}_j}'\bY_{\tilde{\bb}_j})}{\partial \bar{a}_{m,l}^{(0)} } & = \bigg( \frac{\partial\bar{\bX}_{\tilde{\bb}_j}}{\partial\bar{a}_{m,l}^{(0)}} \bigg)'\bY_{\tilde{\bb}_j}  = \begin{pmatrix}
        \sum_{k=1}^q \big\{ (\be_l'\hat{\bSigma}_{j_1^{(0)}k}(1) )\otimes \be_m' \big\}\text{vec}\{\hat \bSigma_{jk}(1)\} \\
        \vdots \\
        \sum_{k=1}^q \big\{ (\be_l'\hat{\bSigma}_{j_{\ell_{j,0}}^{(0)}k}(1) )\otimes \be_m' \big\}\text{vec}\{\hat \bSigma_{jk}(1)\} \\
        {\bf 0}_{\ell_{j,1} \times 1}
    \end{pmatrix}\,.
\end{align}
Hence, combining \eqref{eq:band_bj_1} - \eqref{eq:band_bj_3}, by \eqref{eq:Sigmajk_1_band} and Lemma \ref{lem:Sigmajk}, using similar arguments for deriving \eqref{eq:derv_am_bjl0},we have
\begin{align}\label{eq:derv_bj_aml0}
    \frac{\partial {\bf f}_{p+j}(\btheta)}{\partial \bar{a}_{m,l}^{(0)}} & = \frac{\partial (\bar{\bX}_{\tilde{\bb}_j}'\bar{\bX}_{\tilde{\bb}_j}\bar{\tilde{\bb}}_j-\bar{\bX}_{\tilde{\bb}_j}'\bY_{\tilde{\bb}_j})}{\partial \bar{a}_{m,l}^{(0)}}\bigg|_{\bar{\btheta} = \btheta} \notag\\
    & = 
    \begin{pmatrix}
        \sum_{i \in \mathcal{B}_{j}^{(0)}} \big[ \sum_{k=1}^q \text{vec}'\{ \bA_0 \hat{\bSigma}_{j_1^{(0)}k}(1)\} \big\{(\hat{\bSigma}_{ik}(1)'\be_l )\otimes \be_m\big\} \big] b_{j, i}^{(0)} \\
        \vdots \\
        \sum_{i \in \mathcal{B}_{j}^{(0)}} \big[ \sum_{k=1}^q \text{vec}'\{ \bA_0 \hat{\bSigma}_{j_{\ell_{j,0}}^{(0)}k}(1)\} \big\{(\hat{\bSigma}_{ik}(1)'\be_l )\otimes \be_m\big\} \big] b_{j, i}^{(0)} \\
        \sum_{i \in \mathcal{B}_{j}^{(0)}} \big[ \sum_{k=1}^q \text{vec}'\{ \bA_1 \hat{\bSigma}_{j_1^{(1)}k}(0)\} \big\{(\hat{\bSigma}_{ik}(1)'\be_l )\otimes \be_m\big\} \big] b_{j, i}^{(0)} \\
        \vdots \\
        \sum_{i \in \mathcal{B}_{j}^{(0)}} \big[ \sum_{k=1}^q \text{vec}'\{ \bA_1 \hat{\bSigma}_{j_{\ell_{j,1}}^{(1)}k}(0)\} \big\{(\hat{\bSigma}_{ik}(1)'\be_l )\otimes \be_m\big\} \big] b_{j, i}^{(0)}
    \end{pmatrix} +o_{\rm p}(s^{-1})\,,
    % &~~~~~~ - \begin{pmatrix}
    %     \sum_{k=1}^q \big\{ (\be_l'\hat{\bSigma}_{j_1^{(0)}k}(1) )\otimes \be_m' \big\}\big\{\frac{1}{n}\sum_{t=2}^n {\rm vec}(\bE_{t, \cdot j}\bX_{t-1, \cdot k}')\big\} \\
    %     \vdots \\
    %     \sum_{k=1}^q \big\{ (\be_l'\hat{\bSigma}_{j_{S_{\bb_j}^{(0)}}^{(0)}k}(1) )\otimes \be_m' \big\}\big\{\frac{1}{n}\sum_{t=2}^n {\rm vec}(\bE_{t, \cdot j}\bX_{t-1, \cdot k}')\big\}\\
    %     {\bf 0}_{|S_{\bb_j}^{(1)}| \times 1}
    % \end{pmatrix}\,.
\end{align}
provided that $k_{0,max}^5d^{11} = o(n)$.
Analogously, we can show that the derivative of $\bar{\bX}_{\tilde{\bb}_j}$ with respect to $\bar{\tilde{a}}_{m,l}^{(1)}$ is
\begin{align*}
    \frac{\partial \bar{\bX}_{\tilde{\bb}_j}}{\partial \bar{a}_{m,l}^{(1)}} = \begin{pmatrix}
        {\bf 0}_{p^2 \times \ell_{j,0}}\,, & \{\hat{\bSigma}_{j_1^{(1)}1}(0)' \be_l\}\otimes \be_m\,, & \cdots\,, & \{\hat{\bSigma}_{j_{\ell_{j,1}}^{(1)}1}(0)' \be_l\}\otimes \be_m  \\
        \vdots &  &  \vdots  & \vdots \\
        {\bf 0}_{p^2 \times \ell_{j,0}}\,, & \{\hat{\bSigma}_{j_1^{(1)}q}(0)' \be_l\}\otimes \be_m\,, & \cdots\,, & \{\hat{\bSigma}_{j_{\ell_{j,1}}^{(1)}q}(0)' \be_l\}\otimes \be_m  \\
    \end{pmatrix}\,,
\end{align*}
provided that $k_{0,max}^5d^{11} = o(n)$.
%provided that $d=o\{n^{(3+\kappa)/(2\kappa)}\}$. 
Using the similar arguments for the derivation of \eqref{eq:derv_bj_aml0}, it holds that
\begin{align}\label{eq:derv_bj_aml1}
    \frac{\partial {\bf f}_{p+j}(\btheta)}{\partial \bar{a}_{m,l}^{(1)}} & =\frac{\partial (\bar{\bX}_{\tilde{\bb}_j}'\bar{\bX}_{\tilde{\bb}_j}\bar{\tilde{\bb}}_j-\bar{\bX}_{\tilde{\bb}_j}'\bY_{\tilde{\bb}_j})}{\partial \bar{a}_{m,l}^{(1)}}\bigg|_{\bar{\btheta} = \btheta} \notag\\
    & = 
    \begin{pmatrix}
        \sum_{i \in \mathcal{B}_{j}^{(1)}} \big[ \sum_{k=1}^q \text{vec}'\{ \bA_0 \hat{\bSigma}_{j_1^{(0)}k}(1)\} \big\{(\hat{\bSigma}_{ik}(0)'\be_l )\otimes \be_m\big\} \big] b_{j, i}^{(1)} \\
        \vdots \\
        \sum_{i \in \mathcal{B}_{j}^{(1)}} \big[ \sum_{k=1}^q \text{vec}'\{ \bA_0 \hat{\bSigma}_{j_{\ell_{j,0}}^{(0)}k}(1)\} \big\{(\hat{\bSigma}_{ik}(0)'\be_l )\otimes \be_m\big\} \big] b_{j, i}^{(1)} \\
        \sum_{i \in \mathcal{B}_{j}^{(1)}} \big[ \sum_{k=1}^q \text{vec}'\{ \bA_1 \hat{\bSigma}_{j_1^{(1)}k}(0)\} \big\{(\hat{\bSigma}_{ik}(0)'\be_l )\otimes \be_m\big\} \big] b_{j, i}^{(1)} \\
        \vdots \\
        \sum_{i \in \mathcal{B}_{j}^{(1)}} \big[ \sum_{k=1}^q \text{vec}'\{ \bA_1 \hat{\bSigma}_{j_{\ell_{j,1}}^{(1)}k}(0)\} \big\{(\hat{\bSigma}_{ik}(0)'\be_l )\otimes \be_m\big\} \big] b_{j, i}^{(1)}
    \end{pmatrix} +o_{\rm p}(s^{-1}) \,,
    % &~~~~~~ - \begin{pmatrix}
    %     {\bf 0}_{|S_{\bb_j}^{(0)}| \times 1} \\
    %     \sum_{k=1}^q \big\{ (\be_l'\hat{\bSigma}_{j_1^{(1)}k}(0) )\otimes \be_m' \big\}\big\{\frac{1}{n}\sum_{t=2}^n {\rm vec}(\bE_{t, \cdot j}\bX_{t-1, \cdot k}')\big\} \\
    %     \vdots \\
    %     \sum_{k=1}^q \big\{ (\be_l'\hat{\bSigma}_{j_{|S_{\bb_j}^{(1)}|}^{(1)}k}(0) )\otimes \be_m' \big\}\big\{\frac{1}{n}\sum_{t=2}^n {\rm vec}(\bE_{t, \cdot j}\bX_{t-1, \cdot k}')\big\}
    % \end{pmatrix}\,.
\end{align}
provided that $k_{0,max}^5d^{11} = o(n)$.
%provided that $d=o\{n^{(3+\kappa)/(2\kappa)}\}$.
Combining \eqref{eq:derv_bj_aml0} and \eqref{eq:derv_bj_aml1}, we have
\begin{align*}
    \frac{\partial {\bf f}_{p+j}(\btheta)}{\partial \bar{\tilde{\ba}}_{m}} = \bX_{\tilde{\bb}_j}'\bZ_{\tilde{\bb}_j,m} + o_{\rm p}(s^{-1})\,,
\end{align*}
provided that $k_{0,max}^5d^{11} = o(n)$, where 
%provided that $d=o\{n^{(3+\kappa)/(2\kappa)}\}$, where 
\begin{align*}
    \bZ_{\tilde{\bb}_j,m} = \begin{pmatrix}
        \boldsymbol{\psi}_{\tilde{\bb}_j,m,m_1^{(0)}}^{(0)}\,, & \cdots\,, & \boldsymbol{\psi}_{\tilde{\bb}_j,m,m_{r_{m,0}}^{(0)}}^{(0)}\,, &  \boldsymbol{\psi}_{\tilde{\bb}_j,m,m_1^{(1)}}^{(1)}\,, & \cdots\,, & \boldsymbol{\psi}_{\tilde{\bb}_j,m,m_{r_{m,1}}^{(1)}}^{(1)}
    \end{pmatrix}
\end{align*}
is a $p^2 q \times r_{m}$ matrix with 
\begin{align*}
    \boldsymbol{\psi}_{\tilde{\bb}_j,m,l}^{(0)} = \begin{pmatrix}
        (\hat{\bSigma}_{j_1^{(0)}1}(1)'\be_l)\otimes \be_m\,, & \cdots\,, & (\hat{\bSigma}_{j_{\ell_{j,0}}^{(0)}1}(1)'\be_l)\otimes \be_m \\
        \vdots  &  &  \vdots \\
        (\hat{\bSigma}_{j_1^{(0)}q}(1)'\be_l)\otimes \be_m\,, & \cdots\,, & (\hat{\bSigma}_{j_{\ell_{j,0}}^{(0)}q}(1)'\be_l)\otimes \be_m
    \end{pmatrix}\tilde{\bb}_j^{(0)}\,,\\
    \boldsymbol{\psi}_{\tilde{\bb}_j,m,l}^{(1)} = \begin{pmatrix}
        (\hat{\bSigma}_{j_1^{(1)}1}(0)'\be_l)\otimes \be_m\,, & \cdots\,, & (\hat{\bSigma}_{j_{\ell_{j,1}}^{(1)}1}(0)'\be_l)\otimes \be_m \\
        \vdots  &  &  \vdots \\
        (\hat{\bSigma}_{j_1^{(1)}q}(0)'\be_l)\otimes \be_m\,, & \cdots\,, & (\hat{\bSigma}_{j_{\ell_{j,1}}^{(1)}q}(0)'\be_l)\otimes \be_m
    \end{pmatrix}\tilde{\bb}_j^{(1)}\,.
\end{align*}
%{\color{red} Analogously, by Condition \ref{cond:mixing} and Lemma \ref{lem:Sigmajk}, it holds that $\mathcal{E}_{\bb_j}'\bZ_{\bb_j,m}=o_{\rm p}(1)$.}
Furthermore, we have
\begin{align*}
    \frac{\partial {\bf f}_{p+j}(\btheta)}{\partial \bar{\tilde{\bb}}_{j}} = \bX_{\tilde{\bb}_j}'\bX_{\tilde{\bb}_j}\,, ~\mbox{and}~ \frac{\partial {\bf f}_{p+j}(\btheta)}{\partial \bar{\tilde{\bb}}_k} = {\bf 0}_{s_{\bb_j} \times s_{\bb_j}} ~\mbox{for}~ k \neq j\,.
\end{align*}
Hence, if $k_{0,max}^5d^{11} = o(n)$, it holds that
%if $d=o\{n^{(3+\kappa)/(2\kappa)}\}$, it holds that 
\begin{align}\label{eq:der_bj_band}
    \frac{\partial {\bf f}_{p+j}(\btheta)}{\partial \bar{\btheta}} = \big(\bX_{\tilde{\bb}_j}'\bZ_{\tilde{\bb}_j,1}\,,  \cdots\,,\bX_{\tilde{\bb}_j}'\bZ_{\tilde{\bb}_j,p}\,, {\bf 0}_{\ell_{1} \times \ell_{1}}\,, \cdots\,, \bX_{\tilde{\bb}_j}'\bX_{\tilde{\bb}_j}\,, \cdots\,, {\bf 0}_{\ell_{q} \times \ell_{q}}\big) + o_{\rm p}(s^{-1})\,.
\end{align} 
Recall that $\bH_{\rm band}$ is specified in \eqref{eq:Taylor_band}. 

For any $m \in [p]$, we have
\begin{align*}
    & \|\bX_{\tilde{\ba}_m}'\bX_{\tilde{\ba}_m} - \tilde{\bX}_{\tilde{\ba}_m}'\tilde{\bX}_{\tilde{\ba}_m}\|_{\infty} \\
    &~~~~~ \le \Big\| \sum_{i_1,i_2,j,k=1}^q  b_{j, i_1}^{(0)}b_{j, i_2}^{(0)} \bE_m^{(0)'} \{\hat\bSigma_{i_1k}(1)\hat\bSigma_{i_2k}(1)'- \bSigma_{i_1k}(1)\bSigma_{i_2k}(1)'\} \bE_m^{(0)}  \Big\|_{\infty} \\
    & ~~~~~~~~~ + \Big\| \sum_{i_1,i_2,j,k=1}^q  b_{j, i_1}^{(1)} b_{j, i_2}^{(1)} \bE_m^{(1)'} \{\hat\bSigma_{i_1k}(0)\hat\bSigma_{i_2k}(0)'- \bSigma_{i_1k}(0)\bSigma_{i_2k}(0)'\} \bE_m^{(1)}  \Big\|_{\infty} \\
    & ~~~~~~~~~ + 2 \Big\| \sum_{i_1,i_2,j,k=1}^q  b_{j, i_1}^{(0)}b_{j, i_2}^{(1)}  \bE_m^{(0)'} \{\hat\bSigma_{i_1k}(1)\hat\bSigma_{i_2k}(0)'- \bSigma_{i_1k}(1)\bSigma_{i_2k}(0)'\} \bE_m^{(1)}  \Big\|_{\infty} \\
    &~~~~~ := \tilde{\mathrm{R}}_{1,m} + \tilde{\mathrm{R}}_{2,m} + \tilde{\mathrm{R}}_{3,m}\,.
\end{align*}
For $\tilde{\mathrm{R}}_{1,m}$, for any $j \in q$, since $b_{j,i}^{(0)}= 0$ for $i \notin \mathcal{B}_{j}^{(0)}$ and $b_{j,i}^{(1)}= 0$ for $i \notin \mathcal{B}_{j}^{(1)}$, it holds that
\begin{align}\label{eq:R1_band}
    \max_{m \in [p]}|\tilde{\mathrm{R}}_{1,m}| & \lesssim \{k_0^{(\bB)}\}^2pq^2 \max_{i_1,i_2,k \in [q]}\|\{\hat\bSigma_{i_1k}(1)\hat\bSigma_{i_2k}(1)'- \bSigma_{i_1k}(1)\bSigma_{i_2k}(1)'\} \|_{\infty} \,,
\end{align}
since $|\mathcal{B}_{j}^{(k)}| \lesssim k_0^{(\bB)}$ for $k=0,1$. Recall $d = p \vee q$, $k_{0,max} = k_0^{(\bA)} \vee k_0^{(\bB)}$, and $s \lesssim dk_{0,max}$. By \eqref{eq:XX_consistency} and \eqref{eq:R1_band}, we have $\max_{m \in [p]}|\mathrm{R}_{1,m}| = o_{\rm p}(s^{-1})$ provided that $k_{0,max}^{(6+3\kappa)/2}d^{(10+9\kappa)/2} = o(n)$. Using the similar arguments, we show that the same results hold for $\max_{m \in [p]}|\mathrm{R}_{2,m}|$ and $\max_{m \in [p]}|\mathrm{R}_{3,m}|$, which yields $\max_{m \in [p]}\|\bX_{\tilde{\ba}_m}'\bX_{\tilde{\ba}_m} - \tilde{\bX}_{\tilde{\ba}_m}'\tilde{\bX}_{\tilde{\ba}_m}\|_{\infty} = o_{\rm p}(s^{-1})$ provided that $k_{0,max}^{(6+3\kappa)/2}d^{(10+9\kappa)/2} = o(n)$. Using the similar arguments, we can show that $\max_{m \in [p], j\in[q]}\|\bX_{\ba_m}'\bZ_{\ba_m,j} - \tilde{\bX}_{\ba_m}'\tilde{\bZ}_{\ba_m,j}\|_{\infty} = o_{\rm p}(s^{-1})$, provided that $k_{0,max}^{(6+3\kappa)/2}d^{(4+4\kappa)} = o(n)$. Similarly, it holds that $\max_{j \in [q]}\|\bX_{\bb_j}'\bX_{\bb_j} - \tilde{\bX}_{\bb_j}'\tilde{\bX}_{\bb_j}\|_{\infty} = o_{\rm p}(s^{-1}) = \max_{j \in [q], m \in [p]}\|\bX_{\bb_j}'\bZ_{\bb_j,m} - \tilde{\bX}_{\bb_j}'\tilde{\bZ}_{\bb_j,m}\|_{\infty}$, provided that $k_{0,max}^{(6+3\kappa)/2}d^{(10+9\kappa)/2} = o(n)$. We complete the proof of \eqref{eq:HtoU_band}.
$\hfill\Box$

\section{Proposition \ref{CLT:band_fixdim}}\label{sec:band_fixdim}

\begin{proposition}\label{CLT:band_fixdim}
    Assume that Conditions \ref{cond:mixing} and \ref{cond:ident_2}-\ref{cond:min} hold. When $p,q \ge 1$ are fixed, if
    \begin{align*}
         & \sup_{\bu \in \mathbb{R}^{s_1}}\mathbb{E}|\bu'\tilde{\bJ}_{\ba}\tilde{\boldsymbol{\mathcal{E}}}_{\bX,t}|^{\frac{4+\kappa}{2}} < \infty\,, ~ \sup_{\bu \in \mathbb{R}^{s_2}}\mathbb{E}|\bu'\tilde{\bJ}_{\bb}\tilde{\boldsymbol{\mathcal{E}}}_{\bX,t}|^{\frac{4+\kappa}{2}} < \infty\,,
    \end{align*}
    then we have 
    \begin{align*}%\label{eq:CLT_band}
    \sqrt{n} \tilde{\bP}_{\rm band}^{-1/2} \bU_{\rm band}(\hat{\btheta}-\btheta) \stackrel{\mathrm{d}}{\to} \mathcal{N}(0,\bI_{s})\,,
    \end{align*}
    where 
\begin{align*}%\label{eq:Pband}
    \tilde{\bP}_{\rm band} = \begin{pmatrix}
       \tilde{\bJ}_{\ba}\tilde{\bSigma}_{\bX,\bE}\tilde{\bJ}_{\ba}'\,, & \tilde{\bJ}_{\ba}\tilde{\bSigma}_{\bX,\bE}\tilde{\bJ}_{\bb}'\\
       \tilde{\bJ}_{\bb}\tilde{\bSigma}_{\bX,\bE}\tilde{\bJ}_{\ba}'\,, & \tilde{\bJ}_{\bb}\tilde{\bSigma}_{\bX,\bE}\tilde{\bJ}_{\bb}'
    \end{pmatrix}\,.
\end{align*}
\end{proposition}

Proposition \ref{CLT:band_fixdim} can be proved using arguments similar to those for Proposition \ref{CLT:diag_fixdim}, and the details are therefore omitted.

\section{Proof of Lemma \ref{lem:Sigmajk}}\label{lem:1}

Recall $\bSigma_{jk}(1) = \cov(\bX_{t, \cdot j}, \bX_{t-1, \cdot k}) = \mathbb{E}(\bX_{t, \cdot j}\bX_{t-1, \cdot k}')$ since $\mathbb{E}(\bX_t) = {\bf 0}$, and
\begin{equation*}
\hat\bSigma_{jk}(1) = \frac{1}{n}\sum_{t=2}^n \bX_{t, \cdot j} \bX_{t-1, \cdot k}'\,.
\end{equation*}
Write $\bSigma_{jk}(1) = (\sigma_{i_1,i_2}^{(j,k,1)})_{p \times p}$ and $\hat\bSigma_{jk}(1) = (\hat{\sigma}_{i_1,i_2}^{(j,k,1)})_{p \times p}$. Then we have
\begin{align*}
    \hat{\sigma}_{i_1,i_2}^{(j,k,1)} - \sigma_{i_1,i_2}^{(j,k,1)} = \frac{1}{n}\sum_{t=2}^n \{X_{t,i_1,j}X_{t-1,i_2,k} - \mathbb{E}(X_{t,i_1,j}X_{t-1,i_2,k})\}\,.
\end{align*}
Under Condition \ref{cond:mixing}{\rm (ii)}, using the similar arguments of proving Lemma 2 of \cite{Chang2013}, it holds that 
\begin{align*}
    \mathbb{P}\{|X_{t,i_1,j}X_{t-1,i_2,k} - \mathbb{E}(X_{t,i_1,j}X_{t-1,i_2,k})| > x \} \le C_1 x^{-(4+\kappa)/\kappa}
\end{align*}
for any $x>0$. Notice that $\{X_{t,i_1,j}X_{t-1,i_2,k} - \mathbb{E}(X_{t,i_1,j}X_{t-1,i_2,k})\}_{t=2}^n$ is also an $\alpha$-mixing sequence with $\alpha$-mixing coefficients $\{\alpha(|k-1|_{+})\}_{k \geq 1}$, where $|\cdot|_{+} = \max(\cdot, 0)$ and $\alpha(\cdot)$ is given in \eqref{eq:mixingcoeff} at the end of Section \ref{sec_intro}. Applying the Fuk-Nagaev inequality for $\alpha$-mixing sequences (Theorem 6.2 of \cite{Rio2000}), since $\alpha(k) \ll k^{-(4+\kappa)/\kappa}$ under Condition \ref{cond:mixing}{\rm (i)}, by setting $r=(2+\kappa)/\kappa > 1$ in Theorem 6.2 of \cite{Rio2000}, we then have
\begin{align}\label{eq:elementwise}
    \mathbb{P}(|\hat{\sigma}_{i_1,i_2}^{(j,k,1)} - \sigma_{i_1,i_2}^{(j,k,1)} | > x ) \lesssim x^{-(2+\kappa)/\kappa}n^{-2/\kappa}\,.
\end{align}
Hence, by the Bonferroni inequality, for any $x >0$, we have
\begin{align}\label{eq:InfNorm}
    \mathbb{P}(\max_{j,k \in [q]}\max_{i_1,i_2 \in [p]}|\hat{\sigma}_{i_1,i_2}^{(j,k,1)} - \sigma_{i_1,i_2}^{(j,k,1)} | > x ) & \le \sum_{j,k= 1}^q \sum_{i_1,i_2= 1}^p 
    %\max_{j,k \in [q]}\max_{i_1,i_2 \in [p]} 
    \mathbb{P}(|\hat{\sigma}_{i_1,i_2}^{(j,k,1)} - \sigma_{i_1,i_2}^{(j,k,1)} | > x ) \notag\\
     & \lesssim x^{-(2+\kappa)/\kappa}p^2q^2n^{-2/\kappa}\,.
\end{align}
It implies that $\max_{j,k \in [q]}\|\hat{\bSigma}_{jk}(1)-\bSigma_{jk}(1)\|_{\infty} = O_{\rm p}\{(pq)^{2\kappa/(2+\kappa)}n^{-2/(2+\kappa)}\}$. Using the similar arguments, we can also show that $\max_{j,k \in [q]}\|\hat{\bSigma}_{jk}(0)-\bSigma_{jk}(0)\|_{\infty} = O_{\rm p}\{(pq)^{2\kappa/(2+\kappa)}n^{-2/(2+\kappa)}\}$.

On the other hand, for any $x >0$, we have
\begin{align}\label{eq:hatSjk_dev}
    \mathbb{P} \Big\{ \max_{j,k \in [q]}\|\hat{\bSigma}_{jk}(1)-\bSigma_{jk}(1)\|_F^2 > x \Big\} & \le \sum_{j,k \in [q]} \mathbb{P}(\|\hat{\bSigma}_{jk}(1)-\bSigma_{jk}(1)\|_F^2 > x ) \notag \\
    & \le q^2 \max_{j,k \in [q]}  \mathbb{P}\Big(\sum_{i_1,i_2=1}^p \{\hat{\sigma}_{i_1,i_2}^{(j,k,1)} - \sigma_{i_1,i_2}^{(j,k,1)}\}^2 > x \Big) \notag \\
    & \le q^2 \sum_{i_1,i_2=1}^p \mathbb{P}(|\hat{\sigma}_{i_1,i_2}^{(j,k,1)} - \sigma_{i_1,i_2}^{(j,k,1)} | > x^{1/2}p^{-1} )\notag \\
    & \lesssim p^2q^2 (x^{1/2}p^{-1})^{-(2+\kappa)/\kappa}n^{-2/\kappa} \notag\\
    & = p^{(2+3\kappa)/\kappa}q^2n^{-2/\kappa}x^{-(2+\kappa)/(2\kappa)}\,,
\end{align}
which implies that $\max_{j,k \in [q]} \|\hat{\bSigma}_{jk}(1)-\bSigma_{jk}(1)\|_F = O_{\rm p}\{ p^{(2+3\kappa)/(2+\kappa)}q^{2\kappa/(2+\kappa)}n^{-2/(2+\kappa)}\}$. Applying the similar arguments, we can also show that the same result holds for $\max_{j,k \in [q]}  \|\hat{\bSigma}_{jk}(0)-\bSigma_{jk}(0)\|_F$.
%$= O_{\rm p}\{p^{4(1+\kappa)/(2+\kappa)}q^{(2\kappa)/(2+\kappa)}n^{-(4+\kappa)/(2+\kappa)} \}$. 
%$\|\hat{\bSigma}_{jk}(1)-\bSigma_{jk}(1)\|_F = O_{\rm p}\{p^{2(1+\kappa)/(\kappa+2)}n^{-(4+\kappa)/(2\kappa+4)}\}$ for any $j,k \in [q]$. Applying the similar arguments, we can also show that $\|\hat{\bSigma}_{jk}(0)-\bSigma_{jk}(0)\|_F = O_{\rm p}\{p^{2(1+\kappa)/(\kappa+2)}n^{-(4+\kappa)/(2\kappa+4)}\}$. 
This completes the proof of Lemma \ref{lem:Sigmajk}.  $\hfill\Box$

\section{Uncovering Latent Banded Structures via Conditional Dependence and Optimal Ordering of Spatial and Panel Variables} \label{sec:ordering_var}

\subsection{Justification of Banded Structure for Panel Variables}
\label{justify_panel}

Before presenting empirical evidence for a banded structure in our panel variables, we first clarify the interpretation of the panel-side weight matrix and outline the logic of our approach. Specifically, we explain the meaning of the entries in the weight matrix $\mathbf{B}_0$, then examine the conditional dependence structure among the five Japanese stocks using daily trading volume data. Based on this empirical analysis, we apply the Reverse Cuthill-McKee algorithm to reorder the stocks and reveal a latent banded pattern. Finally, we quantify the parameter reduction achieved by imposing a banded structure and assess how effectively such a band captures the conditional dependencies present in the data.

\subsubsection{Interpreting the Panel-Side Weight Matrix}

It is important to note that the entry $(\mathbf{B}_0)_{ij}$ in the panel-side weight matrix (regardless of its banded structure) encodes the direct linear dependence of variable/stock $i$ and variable/stock $j$, conditional on all other variables/stocks. Therefore:

\begin{itemize}
    \item A non-zero $(\mathbf{B}_0)_{ij}$ implies a direct connection (i.e., conditional correlation).
    \item A zero $(\mathbf{B}_0)_{ij}$ does not imply unconditional independence; the variables may still be correlated through pathways involving intermediate variables. However, there is no direct connection between variables $i$ and $j$.  
\end{itemize}

\subsubsection{Empirical Analysis of Conditional Dependence of Panel Variables} \label{G::panel}

Let $\mathbf{A} \in \{0,1\}^{q \times q}$ denote the conditional adjacency matrix for $q$ randomly ordered panel variables. For $i \neq j$,
\[
\mathbf{A}_{ij} = \mathbb{I}\bigl\{\text{Corr}(X_i, X_j \mid X_{-ij}) \neq 0\bigr\},
\]
where $X_{-ij}$ denotes all panel variables excluding $X_i$ and $X_j$. Diagonal entries are defined as one.

We examine five Japanese stocks — ordered randomly as 6758, 7974, 8035, 6723, 9984 — over the period from June to December 2023. To characterize the direct linear relationships among these panel variables, which inform the structure of the weight matrix $\mathbf{B}_0$, we apply conditional independence tests to the daily trading volume series. Specifically, for each stock, we consider the total daily trading volume time series. A bucket-by-bucket calculation using a simple average across all buckets can also be applied and yields similar conclusions, though it tends to be noisier. This yields a conditional adjacency matrix in which pairwise dependencies are assessed conditional on all other stocks. The resulting matrix at the $5\%$ significance level is:

\[
\mathbf{A} = 
\begin{array}{c|ccccc}
& \textbf{6758} & \textbf{7974} & \textbf{8035} & \textbf{6723} & \textbf{9984}\\
\hline
\textbf{6758} & 1 & 0 & 1 & 0 & 1 \\
\textbf{9984} & 0 & 1 & 0 & 1 & 0 \\
\textbf{7974} & 1 & 0 & 1 & 1 & 1 \\
\textbf{8035} & 0 & 1 & 1 & 1 & 0 \\
\textbf{6723} & 1 & 0 & 1 & 0 & 1
\end{array}
\]

The adjacency matrix exhibits considerable sparsity, indicating numerous conditionally independent stock pairs, and displays no apparent banded pattern in its original order.

\subsubsection{Optimal Ordering via Reverse Cuthill-McKee}\label{G::order}

To uncover a latent banded structure, we seek a permutation $\pi$ of the stocks that minimizes the bandwidth:

\[
\text{BW}(\mathbf{A}) = \max\{|i-j| : \mathbf{A}_{ij}=1\}
\]

Equivalently, we aim for an ordering that places most non-zero entries near the diagonal. To obtain an optimal stock ordering that yields a banded adjacency matrix, we apply the Reverse Cuthill-McKee (RCM) algorithm; see \cite{Cuthill1969}. Given $\mathbf{A} \in \{0,1\}^{q \times q}$, RCM finds a permutation $\pi$ that minimizes:

\[
\text{BW}(\mathbf{A},\pi) = \max_{\{i,j: \mathbf{A}_{ij}=1\}} |\pi(i) - \pi(j)|
\]
where $\pi(i)$ denotes the new position of original stock $i$ after reordering, so $|\pi(i) - \pi(j)|$ measures the distance between two connected stocks in the new ordering, and the bandwidth is the maximum such distance over all edges. After applying RCM, we obtain the permuted adjacency matrix with assets ordered as 9984, 6758, 8035, 6723, 7974:

\[
\mathbf{A}_{\pi} = 
\begin{array}{c|ccccc}
& \textbf{9984} & \textbf{6758} & \textbf{8035} & \textbf{6723} & \textbf{7974} \\
\hline
\textbf{9984} & 1 & 1 & 1 & 0 & 0 \\
\textbf{6758} & 1 & 1 & 1 & 0 & 0 \\
\textbf{8035} & 1 & 1 & 1 & 1 & 1 \\
\textbf{6723} & 0 & 0 & 1 & 1 & 1 \\
\textbf{7974} & 0 & 0 & 1 & 1 & 1
\end{array}
\]

This matrix exhibits a clear banded structure with optimally ordered assets. Crucially, the banded pattern of $\mathbf{A}_{\pi}$ directly implies an identical banded structure for the panel-side weight matrix $\mathbf{B}_0$, demonstrating that an appropriate variable ordering is essential for revealing this structure.

\subsubsection{Parameter Reduction under Banded Structure}

When imposing a band constraint with bandwidth $k$, we estimate all parameters within the $k$-diagonal band, regardless of whether the corresponding adjacency entry is $0$ or $1$. Specifically:

\begin{itemize}
\item \textbf{Parameter counts under banded constraints:}
\begin{itemize}
    \item Full matrix (no band): $25$ parameters (all entries)
    \item Banded $k=1$: $13$ parameters (all entries within $|i-j|\leq 1$)
    \item Banded $k=2$: $19$ parameters (all entries within $|i-j| \leq 2$)
\end{itemize}
This represents parameter reductions of $48\%$ for $k=1$ and $24\%$ for $k=2$ relative to the full specification.

\item \textbf{Conditional dependency capture:}
\begin{itemize}
    \item With $k=2$: All $6$ conditional dependencies lie within the band (100\% captured)
    \item With $k=1$: $4$ out of $6$ dependencies lie within the band (67\% captured)
\end{itemize}
\end{itemize}

\subsection{A Practical Procedure for Ordering Panel Variables}
\label{ordering_method}

The preceding analysis demonstrates that a banded structure for the panel-side weight matrix achieves two key objectives: (1) it preserves the essential conditional dependencies present in the data, and (2) it substantially reduces the parameter count, thereby enhancing model interpretability — a critical advantage as the dimensionality of the panel variables increases. However, realizing these benefits requires an appropriate variable ordering. We therefore propose a practical two-step procedure for ordering panel variables. This procedure is general in nature:

\begin{enumerate}
    \item Compute the conditional adjacency matrix from the panel variables, irrespective of initial ordering.
    \item Apply the RCM algorithm to this conditional adjacency matrix to obtain the optimal variable ordering for banded structure.
\end{enumerate}

This ordered sequence is then supplied to the proposed model, ensuring that the banded structure aligns with the inherent dependency pattern. Although the exact bandwidth parameter remains to be determined, it can be selected automatically using the model framework and estimation method outlined in our paper. Furthermore, the above procedure also helps inform the choice of bandwidth, and its output further supports our real data analysis.

\subsection{Justification of Banded Structure for Spatial Variables}\label{G::spatial}

We now examine the banded structure of the spatial-side weight matrix $\mathbf{A}_0$, following an approach analogous to that used for the panel variables. In this setting, the spatial units correspond to intraday time buckets. To mitigate estimation noise, we aggregate the original 60 five-minute buckets into 30 ten-minute buckets by averaging consecutive observations. Table~\ref{tab:spatial_density} reports, for each asset, the proportion of nonzero entries in the first ten sub-diagonals of its conditional adjacency matrix based on the 30 ten-minute bucket representation.

Applying a $20\%$ nonzero proportion cutoff (i.e., a sub-diagonal is retained if more than $20\%$ of its entries are nonzero), we find that only the first two sub-diagonals satisfy this criterion in the 30-bucket representation. These two dense sub-diagonals indicate that each ten-minute bucket is conditionally dependent on its immediate predecessor (sub-diagonal 1) and its second predecessor (sub-diagonal 2) in the aggregated time scale.

To translate this finding back to the original 60 five-minute bucket specification, we account for the aggregation factor: each ten-minute bucket consists of two consecutive five-minute buckets. Therefore, a dependency between ten-minute buckets separated by $k$ intervals implies a dependency between five-minute buckets separated by up to $2k$ intervals. Specifically, the first dense sub-diagonal in the 30-bucket representation (adjacent ten-minute buckets) corresponds to dependencies spanning at most two five-minute buckets. Similarly, the second dense sub-diagonal in the aggregated representation corresponds to dependencies spanning at most four five-minute buckets. Thus, the effective bandwidth for the original 60-bucket specification is $2 \times 2 = 4$, meaning that nonzero entries are allowed up to the fourth sub-diagonal. 

We also note that a direct examination of the full $60 \times 60$ conditional correlation matrix (without aggregation) yields comparable results, though these are noisier and are therefore not reported here.

\begin{table}[h]
\centering
\caption{Percentage of Nonzero Entries by Sub-diagonal for Each Asset}
\label{tab:spatial_density}
\vspace{0.5em}
\begin{tabular}{lcccccccccc}
\toprule
\multirow{2}{*}{\textbf{Asset}} & \multicolumn{10}{c}{\textbf{Sub-diagonal index (distance from diagonal)}} \\
\cmidrule(lr){2-11}
 & \textbf{1} & \textbf{2} & \textbf{3} & \textbf{4} & \textbf{5} & \textbf{6} & \textbf{7} & \textbf{8} & \textbf{9} & \textbf{10} \\
\midrule
\textbf{9984} & 89\% & 25\% & 18\% & 11\% & 0\% & 4\% & 0\% & 4\% & 0\% & 5\% \\
\textbf{6758} & 62\% & 28\% & 11\% & 0\% & 12\% & 4\% & 4\% & 9\% & 4\% & 0\% \\
\textbf{8035} & 79\% & 14\% & 7\% & 3\% & 8\% & 12\% & 4\% & 4\% & 9\% & 15\% \\
\textbf{6723} & 89\% & 14\% & 11\% & 7\% & 4\% & 0\% & 13\% & 13\% & 4\% & 0\% \\
\textbf{7974} & 72\% & 10\% & 18\% & 3\% & 4\% & 0\% & 30\% & 9\% & 4\% & 5\% \\
\bottomrule
\end{tabular}
\vspace{0.3em}
\end{table}

\subsection{A Correlation-Based Approach to Specifying Spatial Weight Matrices}\label{G::known_weight_mat_spec}

Similar to the earlier analyses in (\ref{G::panel}) and (\ref{G::spatial}), which yielded conditional adjacency matrices for both the spatial and panel dimensions, we compute the corresponding conditional correlation matrices. For the panel side, this calculation is straightforward, yielding the $5 \times 5$ conditional correlation matrix $\mathbf{V}^*$ shown below:

\[
\mathbf{V}^* = 
\begin{array}{c|ccccc}
& \textbf{9984} & \textbf{6758} & \textbf{8035} & \textbf{6723} & \textbf{7974}\\
\hline
\textbf{9984} & 1.00 & 0.21 & 0.29 & 0.16 & 0.02 \\
\textbf{6758} & 0.21 & 1.00 & 0.22 & -0.01 & 0.10 \\
\textbf{8035} & 0.29 & 0.22 & 1.00 & 0.39 & 0.18 \\
\textbf{6723} & 0.16 & -0.01 & 0.39 & 1.00 & 0.21 \\
\textbf{7974} & 0.02 & 0.10 & 0.18 & 0.21 & 1.00 \\
\end{array}
\]

For the spatial component, we begin by computing a separate $60 \times 60$ conditional correlation matrix for each of the five assets. These five asset-specific matrices are then averaged element-wise to form the final spatial weight matrix $\mathbf{W}^*$. The resulting conditional correlation matrix is shown below, where we display only the first eight rows and columns to improve readability.

\[
\mathbf{W}^* = 
\begin{array}{c|cccccccc}
& \textbf{B1} & \textbf{B2} & \textbf{B3} & \textbf{B4} & \textbf{B5} & \textbf{B6} & \textbf{B7} & \textbf{B8} \\
\hline
\textbf{B1} & 1.00 & 0.30 & 0.06 & 0.07 & 0.14 & 0.10 & 0.12 & -0.06 \\
\textbf{B2} & 0.30 & 1.00 & 0.31 & 0.11 & 0.02 & -0.03 & -0.07 & 0.07 \\
\textbf{B3} & 0.06 & 0.31 & 1.00 & 0.26 & 0.03 & 0.10 & 0.13 & -0.11 \\
\textbf{B4} & 0.07 & 0.11 & 0.26 & 1.00 & 0.29 & 0.09 & -0.00 & 0.09 \\
\textbf{B5} & 0.14 & 0.02 & 0.03 & 0.29 & 1.00 & 0.22 & 0.05 & 0.08 \\
\textbf{B6} & 0.10 & -0.03 & 0.10 & 0.09 & 0.22 & 1.00 & 0.13 & 0.12 \\
\textbf{B7} & 0.12 & -0.07 & 0.13 & -0.00 & 0.05 & 0.13 & 1.00 & 0.28 \\
\textbf{B8} & -0.06 & 0.07 & -0.11 & 0.09 & 0.08 & 0.12 & 0.28 & 1.00 \\
\end{array}
\]

For completeness, in our real data analysis, the spatio-temporal model takes $\mathbf{W}^*$ and $\mathbf{V}^*$ as inputs, treating them as known weight matrices. Additionally, we set $\mathbf{W}_0$ so that its main diagonal and upper triangular entries are zeros, thereby avoiding lookahead bias in prediction.


\begin{thebibliography}{}
\bibitem[Andersen(1996)]{Andersen1996} Andersen, T. G. (1996).
\newblock {Return volatility and trading volume: An information flow interpretation of stochastic volatility.}
\newblock {\sl Journal of Finance} {\bf 51(1)}, 169-204.

%\bibitem[Andersen and Bollerslev(1997)]{Andersen1997} Andersen, T. G., and Bollerslev, T. (1997). \newblock {Intraday periodicity and volatility persistence in financial markets.}
%\newblock {\sl Journal of Empirical Finance} {\bf 4}, 115-158.

\bibitem[Andersen et al.(2019)]{Andersen2019} Andersen, T. G., Thyrsgaard, M., and Todorov, V. (2019). \newblock {Time-varying periodicity in intraday volatility.}
\newblock {\sl Journal of American Statistical Association} {\bf 114(528)}, 1695-1707.

\bibitem[Andersen et al.(2024)]{Andersen2024} Andersen, T. G., Su, T., Todorov, V., and Zhang, Z. (2024). \newblock {Intraday periodic volatility curves.} 
\newblock {\sl Journal of American Statistical Association} {\bf 119(546)}, 1181-1191.

\bibitem[Andersen et al.(2025a)]{Andersen2025a} Andersen, T. G., Tan, Y., Todorov, V., and Zhang, Z. (2025a). \newblock {Testing mean stationarity of intraday volatility curves.}
\newblock {\sl Quantitative Economics} {\bf 16(3)}, 1059-1091.

\bibitem[Andersen et al.(2025b)]{Andersen2025b} Andersen, T. G., Tan, Y., Todorov, V., and Zhang, Z. (2025b). \newblock {On-line detection of changes in the shape of intraday volatility curves.}
\newblock {\sl Journal of Econometrics} {\bf 252}, 106089.

\bibitem[Białkowski et al.(2008)]{Bialkowski2008} Białkowski, J., Darolles, S., and Gaëlle, L. (2008).
\newblock {Improving VWAP strategies: A dynamic volume approach.}
\newblock {\sl Journal of Banking and Finance} {\bf 32(9)}, 1709-1722.

\bibitem[Chang et al.(2015)]{Chang2015} Chang, J., Guo, B., and Yao, Q. (2015).
\newblock {High dimensional stochastic regression with latent factors, endogeneity and nonlinearity.} \newblock {\sl Journal of Econometrics} {\bf 189(2)}, 297–312.

\bibitem[Chang et al.(2023)]{Chang2023} Chang, J., He, J., Yang, L., and Yao, Q. (2023). 
\newblock {Modelling matrix time series via a tensor CP-decomposition.}
\newblock {\sl Journal of the Royal Statistical Society, Series B} {\bf 85(1)}, 127-148.

\bibitem[Chang et al.(2025)]{Chang2025} Chang, J., Du, Y., Huang, G., and Yao, Q. (2025). 
\newblock {Identification and estimation for matrix time series CP-factor models.}
\newblock {\sl The Annals of Statistics} forthcoming.
%\newblock {\sl arXiv preprint arXiv:2410.05634.}

\bibitem[Chen et al.(2020a)]{Chen2020a} Chen, E. Y., Tsay, R. S., and Chen, R. (2020a). 
\newblock {Constrained factor models for high-dimensional matrix-variate time series.}
\newblock {\sl Journal of the American Statistical Association} {\bf 115(530)}, 775-793.

\bibitem[Chen et al.(2020b)]{Chen2020b} Chen, E. Y., Yun, X., Chen, R., and Yao, Q. (2020b). 
\newblock {Modeling multivariate spatial-temporal data with latent low-dimensional dynamics.}
\newblock {\sl arXiv preprint arXiv: 2002.01305.} 

\bibitem[Chen et al.(2021)]{Chen2021} Chen, R., Xiao, H., and Yang, D. (2021). 
\newblock {Autoregressive models for matrix-valued time series.}
\newblock {\sl Journal of Econometrics} {\bf 222(1)}, 539-560.

\bibitem[CLSA(2018)]{CLSA2018} CLSA. (2018). 
\textit{CLSA Execution Services Factsheet}. Retrieved from \url{https://www.clsa.com/wp-content/uploads/2016/07/CLSA-Liquidity-Services-ADAPTIVE-inserts-2018.pdf}

\bibitem[Conley(1999)]{Conley1999} Conley, T. G. (1999). 
\newblock {GMM estimation with cross sectional dependence.}
\newblock {\sl Journal of Econometrics} {\bf 92(1)}, 1-45.

\bibitem[Dou et al.(2016)]{Dou2016} Dou, B., Parrella, M. L., and Yao, Q. (2016). 
\newblock {Generalized Yule-Walker estimation for spatio-temporal models with unknown diagonal coefficients.}
\newblock {\sl Journal of Econometrics} {\bf 194(2)}, 369-382.

\bibitem[Fan and Yao(2003)]{Fan2003} Fan, J., and Yao, Q. (2003).
\newblock  {\em Nonlinear Time Series Analysis: Nonparametric and Parametric Methods}. \newblock Springer, New York.\\[-0.8cm]

\bibitem[Gao et al.(2019)]{Gao2019} Gao, Z., Ma, Y., Wang, H., and Yao, Q. (2019). 
\newblock {Banded spatio-temporal autoregressions.}
\newblock {\sl Journal of Econometrics} {\bf 208}, 211-230.

\bibitem[Gao and Tsay(2023)]{Gao2023} Gao, Z. and Tsay, R. S. (2023).
\newblock {A two-way transformed factor model for matrix-variate time series.}
\newblock {\sl Econometrics and Statistics} {\bf 27}, 83-101.

\bibitem[Ge et al.(2023)]{Ge2023} Ge, S., Li, S., and Linton, O. (2023). 
\newblock {News-implied linkages and local dependency in the equity market.}
\newblock {\sl Journal of Econometrics} {\bf 235(2)}, 779-815.

\bibitem[Guo et al.(2016)]{Guo2016} Guo, S., Wang, Y., and Yao, Q. (2016). 
\newblock {High-dimensional and banded vector autoregressions.}
\newblock {\sl Biometrika} {\bf 103(4)}, 889-903.



\bibitem[Han et al.(2024a)]{Han2024a} Han, Y., Chen, R., Yang, D., and Zhang, C.-H. (2024a).
\newblock {Tensor factor model estimation by iterative projection.}
\newblock {\sl The Annals of Statistics} {\bf 52(6)}, 2641–2667.

\bibitem[Han et al.(2024b)]{Han2024b} Han, Y., Yang, D., Zhang, C.-H., and Chen, R. (2024b). 
\newblock {CP factor model for dynamic tensors.}
\newblock {\sl Journal of the Royal Statistical Society Series B} {\bf 86(5)}, 1384–1413.

\item[] Hannan, E. J. (1970).
\newblock {Multiple time series.}
\newblock {\sl John Wiley and Sons, Inc., New York-London-Sydney.} 

\bibitem[Jiang et al.(2025)]{Jiang2025} Jiang, H., Shen, B., Li, Yu., and Gao, Z. (2025).
\newblock {Regularized estimation of high-dimensional matrix-variate autoregressive models.}
\newblock {\sl Statistica Sinica} {forthcoming}.

\bibitem[Kelejian and Prucha(2001)]{Kelejian2001} Kelejian, H. H., and Prucha, I. R. (2001). 
\newblock {On the asymptotic distribution of the Moran I test statistic with applications.} \newblock {\sl Journal of Econometrics} {\bf 104(2)}, 219-257.

\bibitem[Lam and Yao(2012)]{Lam2012} Lam, C., and Yao, Q. (2012). 
\newblock {Factor modeling for high-dimensional time seires: Inference for the number of factors.} \newblock {\sl The Annals of Statistics} {\bf 40(2)}, 694-726.

\bibitem[Lam et al.(2011)]{Lam2011} Lam, C., Yao, Q., and Bethia, N. (2011). 
\newblock {Estimating of latent factors for high-dimensional time series.}
\newblock {\sl Biometrika} {\bf 98(4)}, 901-918.

\bibitem[Lee and Yu(2010a)]{Lee2010a} Lee, L.-F., and Yu, J. (2010a).
\newblock {Some recent developments in spatial panel data models.}
\newblock {\sl Regional Science and Urban Economics} {\bf 40(5)}, 255-271.

\bibitem[Lee and Yu(2010b)]{Lee2010b} Lee, L.-F., and Yu, J. (2010b).
\newblock {Estimation of spatial autoregressive panel data models with fixed effects.}
\newblock {\sl Journal of Econometrics} {\bf 154(2)}, 165–185.

\bibitem[Qiao et al.(2025)]{Qiao2025} Qiao, X., Wang, Z., Yao, Q., and Zhang, B. (2025). 
\newblock {Weight-calibrated estimation for factor models of high-dimensional time series.}
\newblock {\sl arXiv preprint arXiv: 2505.01357.} 

\bibitem[Todorov and Tauchen(2012)]{Todorov2012} Todorov, V., and Tauchen, G. (2012)
\newblock {The realized Laplace transform of volatility.}
\newblock {\sl Econometrica} {\bf 80}, 1105-1127.

%\item[] Tsay, R. S. (2014).
%\newblock {Multivariate time series analysis.}
%\newblock {\sl Wiley Series in Probability and Statistics. John Wiley \& Sons, Inc., Hoboken, NJ.} 

\bibitem[UBS(2015)]{UBS2015} UBS. (2015). 
\textit{UBS Algo Factsheet}. Retrieved from \url{https://a.c-dn.net/c/content/dam/publicsites/1429182809655/igcom/uk/files/other/UBS%20Algo%20factsheet_VolumeInline_EMEA_IG_180315.pdf}

\bibitem[Van Loan(2000)]{VanLoan2000} Van Loan, C. F. (2000).
\newblock {The ubiquitous Kronecker product.}
\newblock {\sl Journal of Computational and Applied Mathematics} {\bf 123(1)}, 85-100. 
%Numerical Analysis 2000.Vol. III: Linear Algebra.

\bibitem[Wang et al.(2019)]{Wang2019} Wang, D., Liu, X., and Chen, R. (2019).
\newblock {Factor models for matrix-valued high-dimensional time series.}
\newblock {\sl Journal of Econometrics} {\bf 208(1)}, 231–248.

\bibitem[Xiao et al.(2022)]{Xiao2022} Xiao, H., Han, Y., and Chen, R. (2022).
\newblock {Reduced rank autoregressive models for matrix time series. }
\newblock {\sl Journal of Business and Economic Statistics} {forthcoming.}

\bibitem[Yu et al.(2008)]{Yu2008} Yu, J., De Jong, R., and Lee, L.-F. (2008). 
\newblock {Quasi-maximum likelihood estimators for spatial dynamic panel data with fixed effects when both $n$ and $T$ are large.}
\newblock {\sl Journal of Econometrics} {\bf 146(1)}, 118-134.

\bibitem[Yu et al.(2012)]{Yu2012} Yu, J., De Jong, R., and Lee, L.-F. (2012). 
\newblock {Estimation for spatial dynamic panel data with fixed effects: The case of spatial cointegration.}
\newblock {\sl Journal of Econometrics} {\bf 167(1)}, 16-37.

\bibitem[Yu et al.(2022)]{Yu2022} Yu, L., He, Y., Kong, X., and Zhang, X. (2022). 
\newblock {Projected estimation for large-dimensional matrix factor models.}
\newblock {\sl Journal of Econometrics} {\bf 229(1)}, 201-217.

\bibitem[Zhang and Yu(2018)]{Zhang2018} Zhang, X., and Yu, J. (2018).
\newblock {Spatial weights matrix selection and model averaging for spatial autoregressive models.}
\newblock {\sl Journal of Econometrics} {\bf 203(1)}, 1-18.

\end{thebibliography}

\begin{thebibliography}{}
	
	%\bibitem[Bierens(2004)]{Bierens2004}	
	%Azadkia, M. and Chatterjee, S. (2021). \newblock A simple measure of conditional dependence. \newblock {\em  The Annals of Statistics}, 49, 3070-3102.\\[-0.8cm]
    %Bierens, H. J. (2004). \newblock {\em Introduction to the Mathematical and Statistical Foundations of Econometrics}. \newblock Cambridge University Press, Cambridge.\\[-0.8cm]

    \bibitem[Chang et al.(2013)]{Chang2013}	
    Chang, J., Tang, C. Y., and Wu, Y. (2013). \newblock  Marginal empirical likelihood and sure independence feature screening. {\em The Annals of statistics}, 41, 2123-2148.\\[-0.8cm]

    \bibitem[Dou et al.(2016)]{Dou2016}	
	Dou, B., Parrella, M. L., and Yao, Q. (2016).\newblock  Generalized Yule-Walker estimation for spatio-temporal models with unknown diagonal coefficients. \newblock  {\em Journal of Econometrics}, 194, 369-382.\\[-0.8cm]

    \bibitem[Cuthill et al.(1969)]{Cuthill1969}	
    Cuthill, E and McKee, J. (1969).\newblock Reducing the Bandwidth of Sparse Symmetric Matrices. \newblock {\em Proceedings of the 1969 24th National Conference, ACM}, 157–172.\\[-0.8cm]

    \bibitem[Fan and Yao(2003)]{FanYao2003}	
	Fan, J. and Yao, Q. (2003).\newblock  {\em Nonlinear Time Series Analysis: Nonparametric and Parametric Methods}. \newblock Springer, New York.\\[-0.8cm]

    \bibitem[Gao et al.(2019)]{Gao2019}	
    Gao, Z., Ma, Y., Wang, H., and Yao, Q. (2019). \newblock Banded spatio-temporal autoregressions. \newblock {\em Journal of Econometrics}, 208, 211-230.

    %\bibitem[Hall and Heyde(1980)]{Hall1980}	
	%Hall, P. and Heyde, C. C. (1980). \newblock  {\em Martingale Limit Theory and its Application}. \newblock Academic Press, New York.\\[-0.8cm]

    \bibitem[Newey(1991)]{Newey1991}	
    Newey, W. K. (1991). \newblock  Uniform convergence in probability and stochastic equicontinuity. \newblock {\em Econometrica}, 59, 1161-1167.\\[-0.8cm]

   \bibitem[Rio(2000)]{Rio2000}	
    Rio, E. (2000). \newblock  {\em Asymptotic Theory of Weakly Dependent Random Processes}. \newblock Springer, New York.\\[-0.8cm]

    \bibitem[Sunklodas(1984)]{Sunklodas1984}	
    Sunklodas, J. (1984). \newblock  Rate of convergence in the central limit theorem for random variables with strong mixing. \newblock {\em Lithuanian Mathematical Journal}, 24, 182-190.\\[-0.8cm]

\end{thebibliography}
\end{document}